\documentclass[twocolumn,times]{aastex631}

\usepackage[para,online,flushleft]{threeparttable}

\usepackage{subfigure}
\usepackage{url}
\colorlet{juan}{Fuchsia!50!WildStrawberry}
\graphicspath{{./}{figs/}}

\begin{document}

\title{SOFIA observations of far-IR fine-structure lines in galaxies to measure metallicity}

\correspondingauthor{Luigi Spinoglio}
\email{luigi.spinoglio@inaf.it}

\author[0000-0001-8840-1551]{Luigi Spinoglio}
\affiliation{Istituto di Astrofisica e Planetologia Spaziale - INAF, Via Fosso del cavaliere 100, I-00133, Roma, Italy}
\author[0000-0001-9490-899X]{Juan Antonio Fern\'andez-Ontiveros}
\affiliation{Istituto di Astrofisica e Planetologia Spaziale - INAF, Via Fosso del cavaliere 100, I-00133, Roma, Italy}
\affiliation{Centro de Estudios de F\'isica del Cosmos de Arag\'on, Unidad Asociada al CSIC, Plaza San Juan 1, E--44001 Teruel, Spain}
\author{Matthew A. Malkan}
\author{Suyash Kumar}
\affiliation{Department of Physics and Astronomy, UCLA, Los Angeles, CA 90095-1547, USA}
\author{Miguel Pereira-Santaella}
\affiliation{Centro de Astrobiología (CSIC-INTA), Ctra. de Ajalvir, Km 4, 28850, Torrej\'on de Ardoz, Madrid, Spain}
\author{Borja P\'erez-D\'iaz}
\author{Enrique P\'erez-Montero}
\affiliation{Instituto de Astrof\'isica de Andaluc\'ia (IAA-CSIC), Glorieta de la Astronom\'ia s/n, E--18008 Granada, Spain}
\author{Alfred Krabbe}
\affiliation{Deutsches SOFIA Institut, University of Stuttgart, Pfaffenwaldring 29, D-70569 Stuttgart, Germany}
\author{William Vacca}
\affiliation{SOFIA/USRA, NASA Ames Research Center, MS N232-12, Moffett Field, CA 94035-1000}
\author{Sebastian Colditz}
\author{Christian Fischer}
\affiliation{Deutsches SOFIA Institut, University of Stuttgart, Pfaffenwaldring 29, D-70569 Stuttgart, Germany}


\begin{abstract}

We present new and archival SOFIA FIFI-LS far-IR spectroscopic observations of 25 local galaxies of either the [OIII]52$\mu$m and/or the [NIII]57$\mu$m lines.
Including other 31 galaxies from {\it Herschel}-PACS, we discuss a local sample of 47 galaxies, including HII region, luminous IR, low-metallicity dwarf and Seyfert galaxies. Analyzing the mid- to far-IR fine-structure lines of this sample, we assess the metallicity  and compare with the optical spectroscopy estimates. Using the IR, we find a similar O/H--N/O relation to that known in the optical. As opposite, we find systematically lower N/O IR abundances when compared to the optical determinations, especially at high values of N/O ($\log(\rm{N/O}) > -0.8$). 
We explore various hypotheses to account for this difference: (i) difference in ionization structure traced by optical (O$^+$, N$^+$ regions) versus IR lines (O$^{2+}$, N$^{2+}$ regions); (ii) contamination of diffuse ionized gas affecting the optical lines used to compute the N/O abundance; (iii) dust obscuration affecting the optical-based determinations. However, we have not found any correlation of the $\rm  \Delta(N/O)= (N/O)_{OPT}-(N/O)_{IR}$ 
with either ionization, or electron density, or optical extinction. 
We speculatively suggest that accretion of metal-poor gas from the circumgalactic medium could provide an explanation for this difference, because the rapid decrease of total abundances during infall is followed by a N/O ratio decrease due to primary production of young - possibly embedded - massive stars, are preferentially traced by the IR diagnostics, while optical diagnostics would better trace the secondary production, when both N/O and O/H abundance ratios will increase.

\end{abstract}

\keywords{Galaxy abundances, Active galaxies, Starburst galaxies, Dwarf galaxies, Far infrared astronomy, Spectroscopy} 

\section{Introduction} \label{intro}
The 
content of heavy elements in galaxies is a key physical diagnostic of galaxy evolution, because metals are a by-product 
of the star formation activity 
Additionally, the metal abundances are modulated by inflow and outflow events, which involves both starburst and active galactic nuclei feeding and feedback processes. 
It has long been known that massive galaxies have higher metallicities, compared to galaxies with lower stellar mass \citep{lequeux1979,tremonti2004}. 
More recently the cosmic evolution of this correlation has been measured out to high redshifts (\citealt{henry2013, ly2016}; see \citealt{maiolino2019} and references therein for a review). 
Besides the global metallicity, as measured by the O/H ratio, the relative abundances between heavy elements are also a key tool to understand the chemical evolution of galaxies \citep[e.g.][]{perez-montero2013}. One of the main problems that still need to be explored is for instance the behavior of the N/O ratio, which is dominated by primary oxygen and nitrogen production in young stellar populations. When these evolve during the lifetime of the galaxy, the nitrogen abundance is increased by the secondary production due to the yield of intermediate mass stars \citep{dopita1986,pilyugin2003,vila-costas1993,vincenzo2016}. Furthermore, the N/O ratio depends on the star formation efficiency \citep[e.g.][]{molla2009}, the accretion of pristine external gas \citep{amorin2010,perez2011,torrey2012}, the efficiency of galactic winds in removing metals from the ISM \citep[e.g.][]{hogarth2020}, and on the shape of the IMF function \citep[e.g.][]{tsujimoto2011}. Therefore, while O/H provides information on the total amount of heavy elements produced, the N/O ratio tells us how these elements have been formed.

The earliest galaxies are likely to have very strong high-ionization emission lines. At their high redshifts ($z > 4$), their key diagnostic fine structure forbidden lines can be detected at sub-millimeter wavelengths by ALMA \citep{debreuck2019}. To understand the abundances in the ionized ISM in these high-redshift galaxies, we need more and better observations of the same far-IR fine-structure lines in local galaxies and especially in nearby dwarf galaxies, characterized by low metal abundances.

Nearly all studies of the gas metallicity in galaxies have used rest-frame optical and UV  diagnostics \citep[e.g.,][and references therein]{nagao2006,kewley2008}. However, optical/UV spectroscopic methods to derive the gas metallicity have several limitations: (i) optical/UV diagnostics cannot probe the metallicity of the regions affected by significant dust extinction, or of dust-obscured galaxies; (ii) the emissivity of optical and UV permitted or forbidden lines depends strongly on the gas temperature, since the atomic levels involved in the transitions are highly excited above the ground level; (iii) nebular diagnostics using the optical [NII] lines can introduce biases, e.g. in the O/H abundances and the ionization parameter, for sources with high or low N/O abundances \citep{perez-montero2009a}; (iv) when applied to nebular lines in active galactic nuclei (AGN), the temperature method --\,widely used in star-forming regions\,-- underestimates the metallicities by typically $\sim 0.8\, \rm{dex}$ when compared to estimates from strong-line methods \citep{dors2015}. Additionally, a large fraction of oxygen in AGN is expected to be highly ionized (O$^{>2+}$) and therefore not traced by the optical transitions \citep{perez-montero2019}. 
All these limitations are overcome by using the IR fine structure lines, as was shown using observations of planetary nebulae \citep{pottasch1999, bernard-salas2001, liu2001} by the Short Wavelength Spectrometer (SWS; \citealt{degraauw1996}) and the Long Wavelength Spectrometer (LWS; \citealt{clegg1996}) onboard the Infrared Space Observatory (ISO; \citealt{kessler1996}).

The use of IR fine structure line ratios for gas metallicity diagnostics and in particular for measuring abundance ratios different from the solar values has been pioneered by \citet{spinoglio1992}, who considered in their study HII region abundances and highly dust depleted abundances using a standard photoionization code. More recently, metallicity diagnostics based on the far-IR fine structure lines of [OIII]$51.8\mu$m, [OIII]$88.3\mu$m, and [NIII]$57.2\mu$m have been proposed as metallicity tracers by \citet{nagao2011} and applied to {\it Herschel}-PACS observations of Ultra-Luminous IR Galaxies (ULIRG) by \citet{pereira2017} and \citep{herrera2018}. The nebular ionization structure of oxygen and nitrogen species is expected to be nearly identical, due to the very similar ionization potential values of these elements ($13.6$ and $14.5\, \rm{eV}$ for O$^+$ and N$^+$, respectively, $35.1$ and $29.6\, \rm{eV}$ for O$^{2+}$ and N$^{2+}$). This implies that the intensity ratios of O$^{2+}$ and N$^{2+}$ transitions can be used as a proxy for the global N/O abundance ratio. Nevertheless, this approach requires a previous knowledge of the O/H--N/O relation to derive metallicities, which may present large deviations from the relation observed in the local Universe, especially under extreme star formation conditions \citep[e.g.][]{amorin2010}. 

Alternative IR-based tracers on the mid-IR lines of neon and sulfur were introduced by \citet{fernandez2016,fernandez2017} as potential tools to study the chemical evolution of galaxies at the {\it Cosmic Noon} through space-born IR spectroscopy. However, a large improvement in the determination of IR abundances was achieved by exploiting the full suite of mid- to far-IR lines. This was recently implemented by \citet{fernandez2021} to determine chemical abundances for star-forming galaxies by applying Bayesian techniques to a grid of photoionization models covering a wide range in O/H, N/O, and U. Specifically, the N/O abundance can be fixed from the far-IR transitions of O$^{2+}$ and N$^{2+}$, while the ionization-sensitive ratios of neon and sulfur lines can be used to determine the metallicity using known empirical correlations between the ionization parameter (U) and O/H \citep{perez-montero2014} as priors. 

In this work we present the largest available sample of galaxies for which new far-IR spectroscopic observations of [OIII]$52\mu$m and [NIII]$57\mu$m lines have been collected with SOFIA, complemented by the work of {\it Herschel}-PACS. This unique spectroscopic catalog allows us to develop and test new IR diagnostic for the global and relative abundances of heavy elements.

The paper is organized as follows. Section \ref{obs} presents the observations carried out with the Stratospheric Observatory for Infrared Astronomy (SOFIA; \citealt{temi2014}), their data analysis procedure and the ancillary data that we have assembled to complement and analyze the results. Section \ref{class} used IR line ratio diagrams to classify the galaxies of our sample according to the ionization and density of their line emission regions, using standard photoionization models. Section \ref{abund} outlines the two methods used to derive the abundances using optical lines and IR fine structure lines. Section \ref{results} give the results of this work using various IR line ratio diagrams to measure the metallicity and a discussion is presented in section \ref{discussion}. Finally, section \ref{sum} gives the summary of this work.

\section{Observations} \label{obs}
We present new observations with the SOFIA Far Infrared Field-Imaging Line Spectrometer \citep[FIFI-LS][]{fischer2018} for 25 galaxies. FIFI-LS is a mid-IR medium resolution (R $\sim$ 500-2000) integral field spectrograph. It consists of two independently and simultaneously operated spectrographs and uses a dichroic to split incoming light into two channels: a blue side (50-125 $\mu$m) and a red side (100-205 $\mu$m). Each spectrograph contains an image slicer and a disperser that enables the simultaneous acquisition of spectra from 25 spatial positions. The detectors in both spectrographs consist of 400 (5 spatial x 5 spatial x 16 spectral) Gallium-doped Germanium pixels. The instantaneous spectral coverage of the 16 pixels ranges from 1000-3000 km s$^{-1}$. The blue side spectrograph has an instantaneous field of view of 30 x 30 arcsec with 6 arcsec spatial pixels (spaxels) while the red side has a field of view of 60 x 60 arcsec with 12 arcsec spaxels. The field of view of the blue side is completely contained within that of the red side. A description of the properties of FIFI-LS can be found in \citet{fischer2018} and \citet{colditz2020}, and the SOFIA website\footnote{\url{https://www.sofia.usra.edu/science/proposing-and-observing/observers-handbook-cycle-7/3-fifi-ls}}.

Among the SOFIA  FIFI-LS observations presented here, the only published data are those relative to Arp~299, Haro~3, IIZw40, M83, MCG+12-02-001, NGC~4194 and NGC~4214 and have been reported in \citet{peng2021}, however we performed a new analysis also of these data, in order to have an homogeneous data set. In most cases we have found results consistent (within the errors) with the values published by these authors, as can be seen in Tables \ref{tab:samplefir} and \ref{tab:sample1bis}. 
The only exceptions are for MCG+12-02-001, for which we adopted the [OIII]52$\mu$m and [NIII]57$\mu$m line fluxes published in \citet{peng2021}, because their results show a higher signal to noise ratio and for IIZw40 for which we find a significantly lower (by a factor $\sim$2) [OIII]52$\mu$m line flux, because of the more compact emission region that we have chosen ($\sim$15 $\arcsec$ in diameter) in this very compact galaxy  compared to the more extended area sampled by \citet{peng2021} ($\gtrsim$20 $\arcsec$ in diameter). In this latter case we report our flux measurement for the [OIII]52$\mu$m line and we give an upper limit for the [NIII]57$\mu$m line.

For NGC~2146 we adopted the line flux from \citet{brauher2008}, who used the ISO-LWS spectrometer.

 The journal of the SOFIA FIFI-LS observations that we have reduced is presented in Table \ref{tab:journal}, where the detailed information on the SOFIA programs, including the AOR identification, the Mission-ID, the PI of the program, the observed spectral line and the total integration time. In total we have analyzed the spectra of 25 galaxies. For 8 of these galaxies we have SOFIA detections of both the [OIII]52$\mu$m and [NIII]57$\mu$m lines, for another 8 galaxies we have combined the new [OIII]52$\mu$m line detection from SOFIA with the [NIII]57$\mu$m line observation from {\it Herschel}-PACS  (see Table \ref{tab:samplefir}), while for the remaining 9 galaxies we present detections of the [OIII]52$\mu$m line only, which are not complemented by any observation of the [NIII]57$\mu$m line (Table \ref{tab:sample1bis}). 
The data have been reduced with the FIFI-LS pipeline \citep{vacca2020} and the reduced datacubes have been retrieved from the SOFIA archive at IRSA\footnote{\url{https://irsa.ipac.caltech.edu/Missions/sofia.html}}. 

\begin{figure*}[ht!]
    \centering
    \subfigure[]{\includegraphics[width=0.498\textwidth]{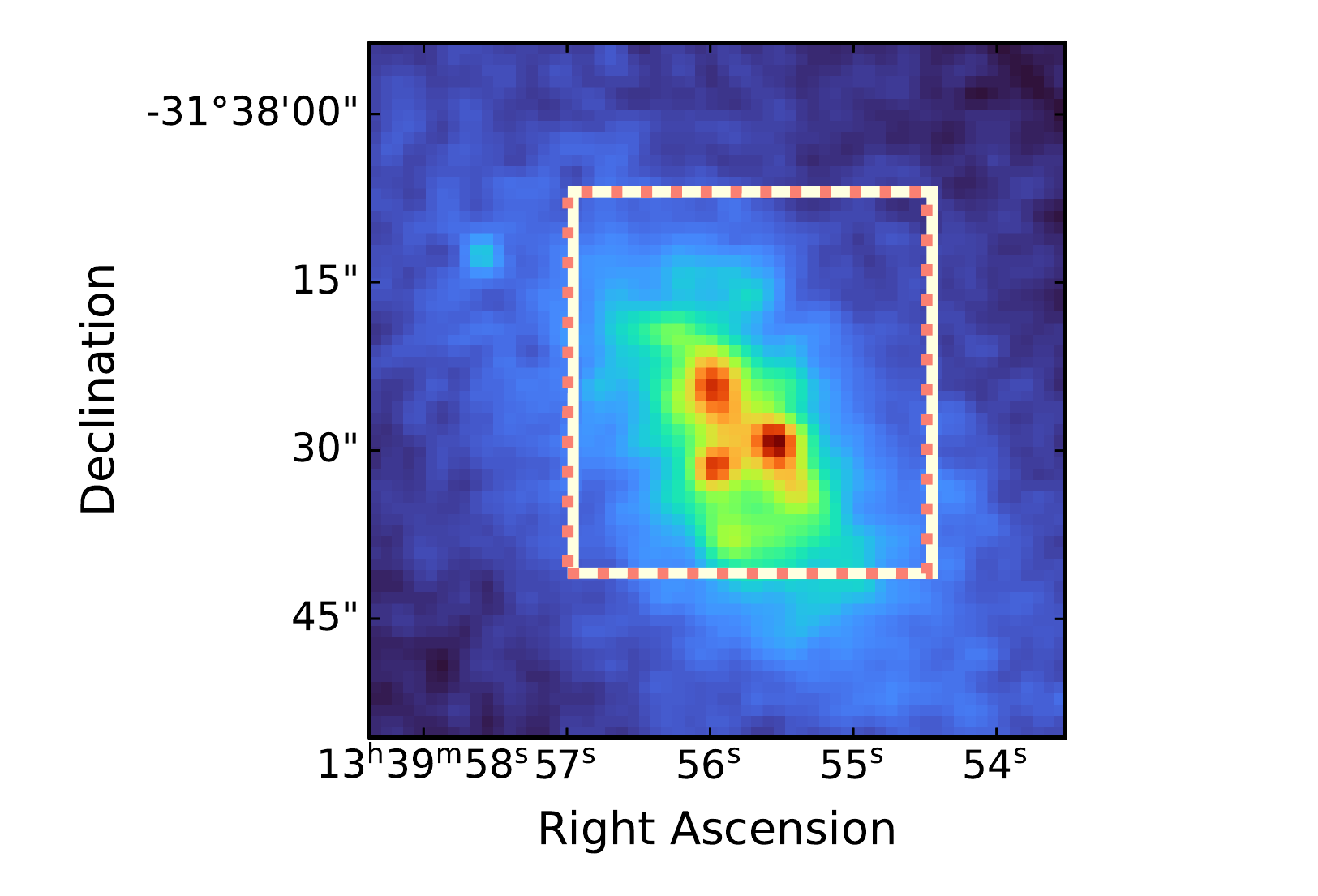}\label{fig:NGC5253-A}}~\\
    \subfigure[]{\includegraphics[width=0.498\textwidth]{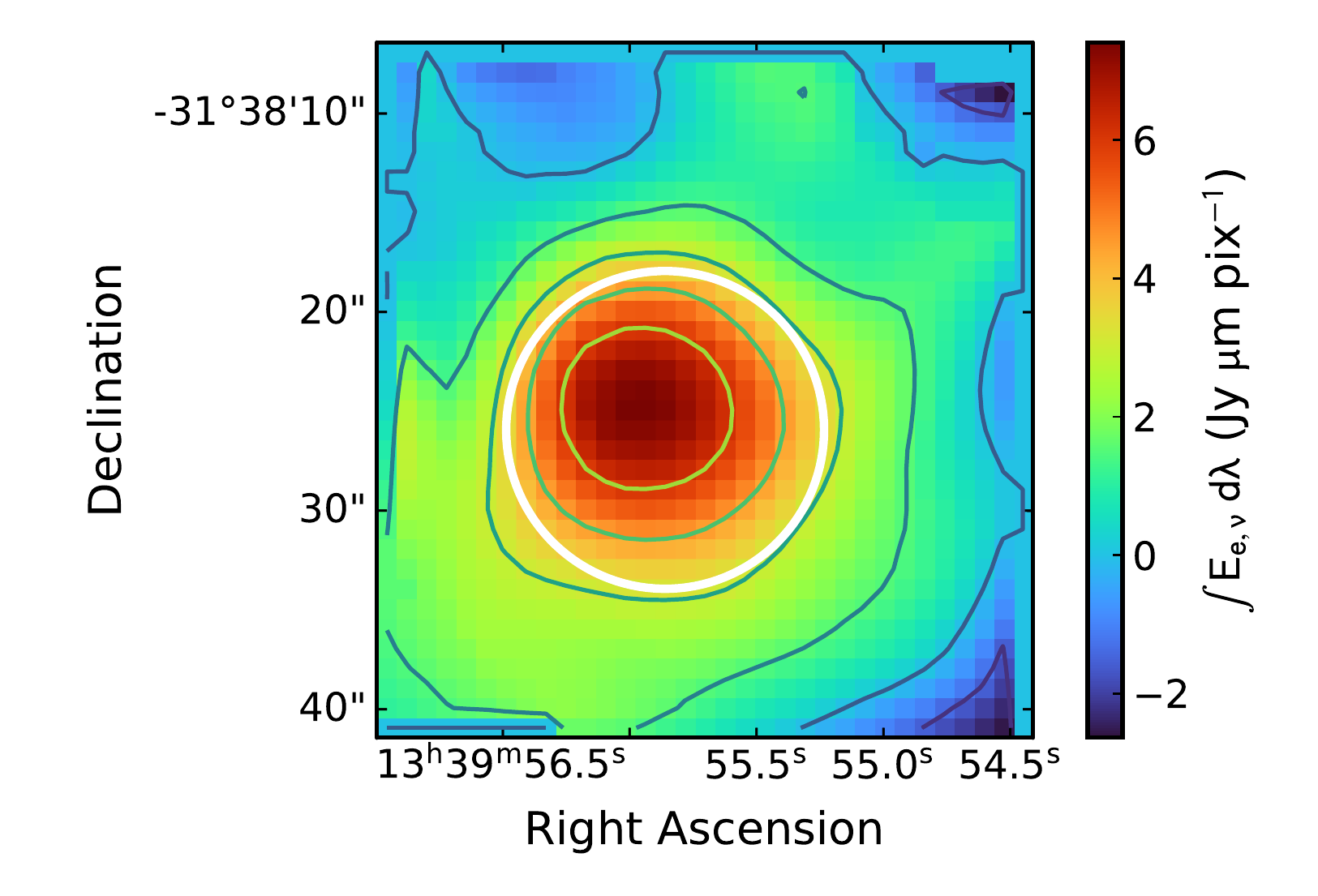}\label{fig:NGC5253-B}}~
    \subfigure[]{\includegraphics[width=0.49\textwidth]{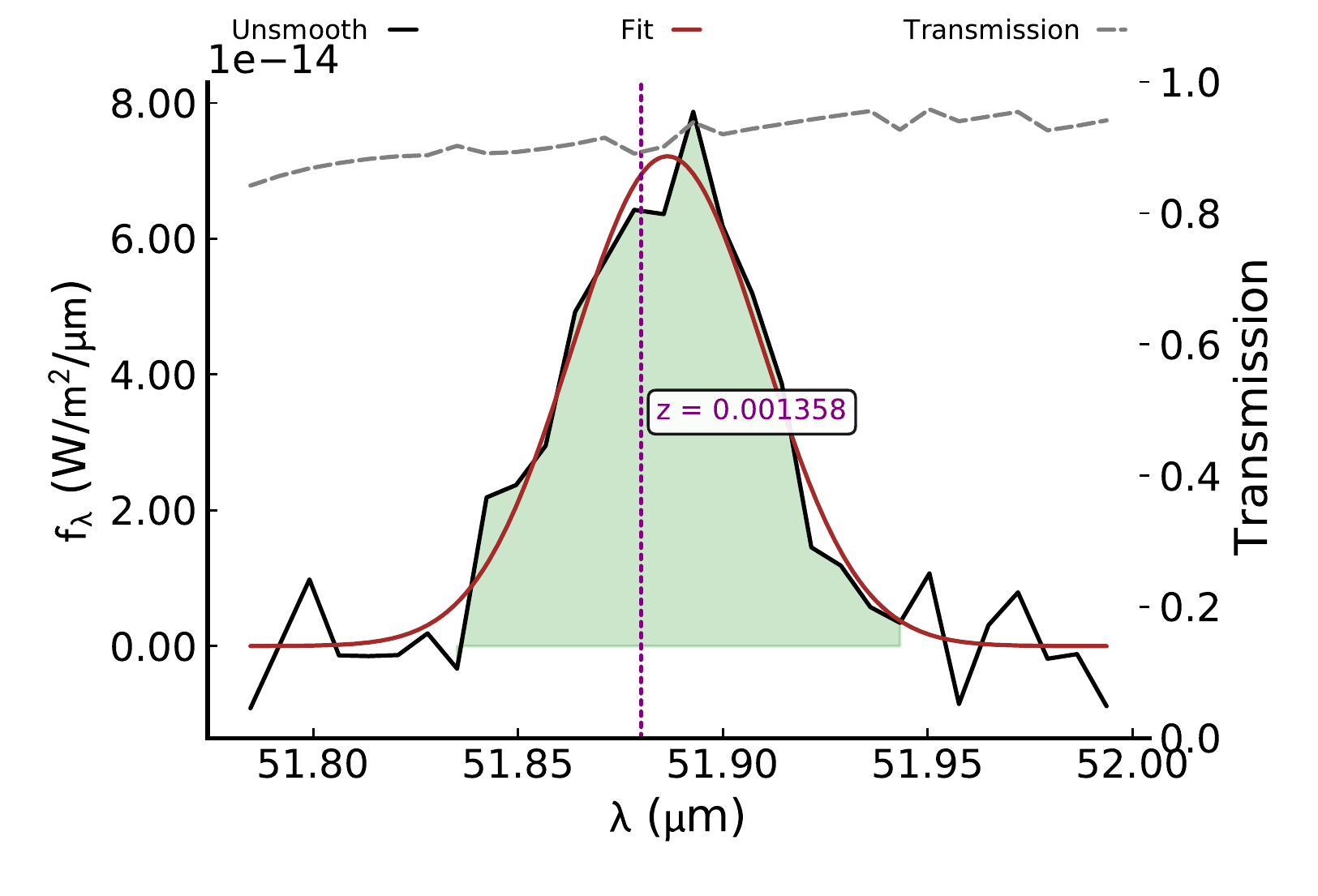}\label{fig:NGC5253-C}}\\
    \subfigure[]{\includegraphics[width=0.498\textwidth]{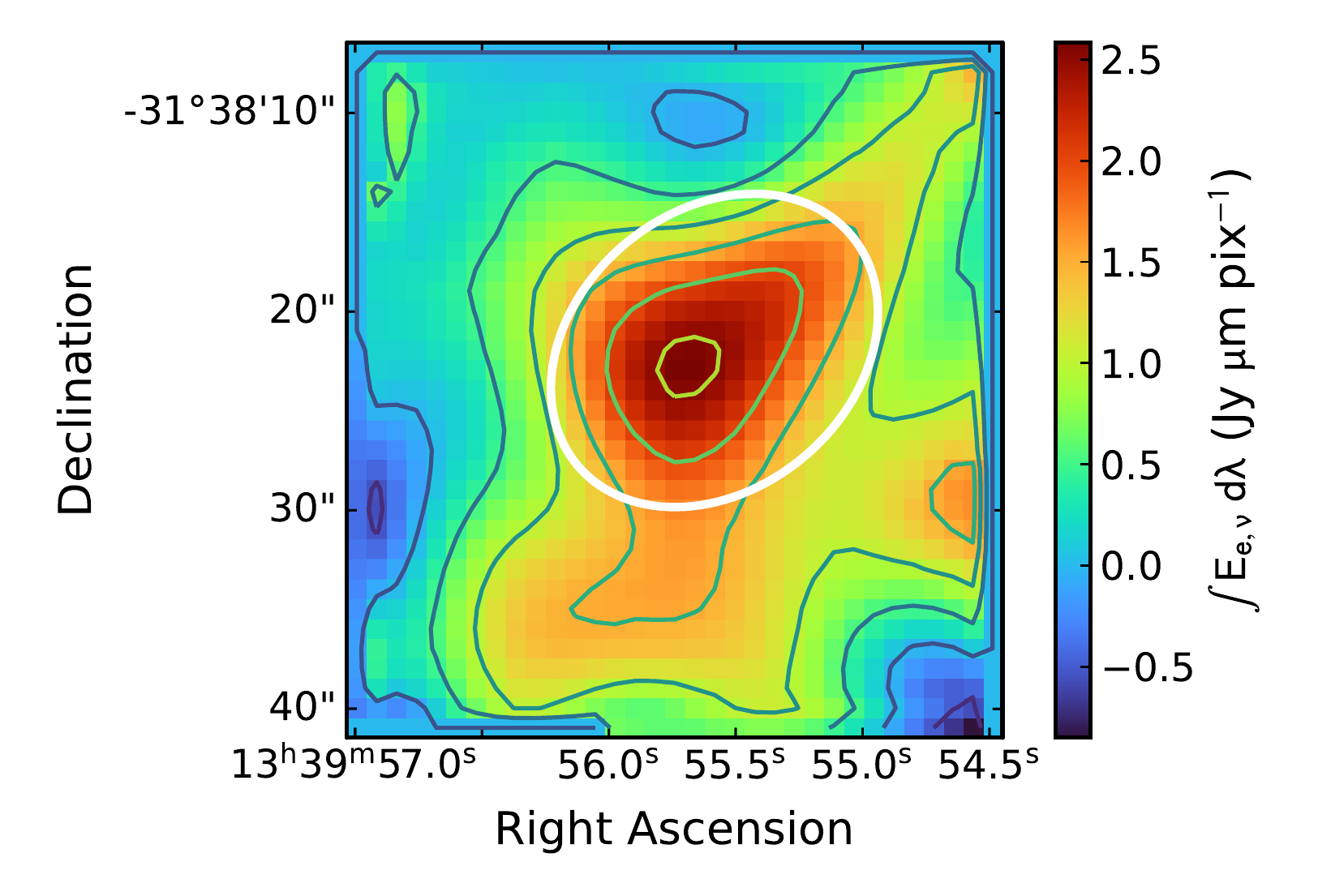}\label{fig:NGC5253-D}}~
    \subfigure[]{\includegraphics[width=0.49\textwidth]{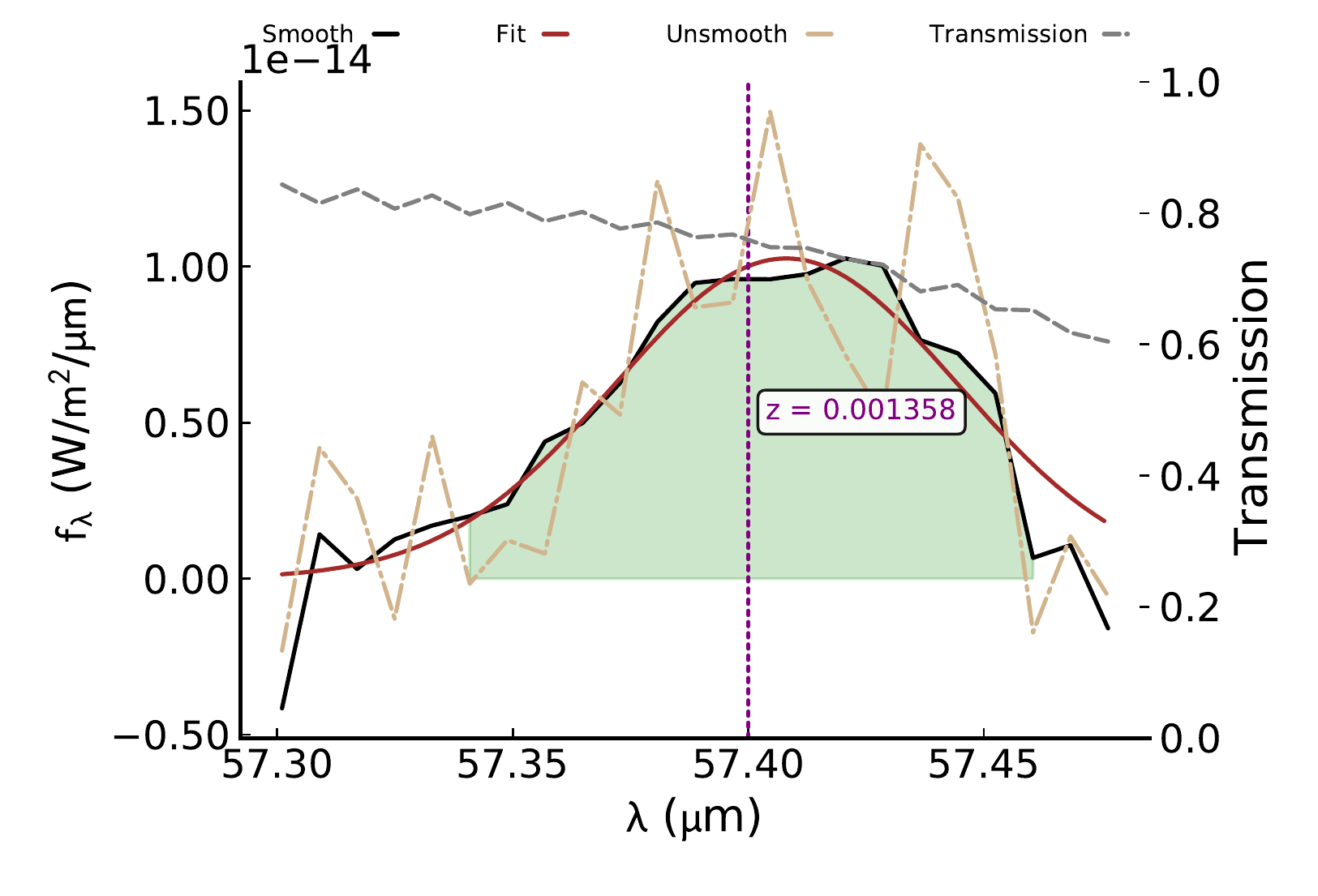}\label{fig:NGC5253-E}}
\caption{The 2MASS H-band image (Figure \ref{fig:NGC5253-A}), 2-D Linemaps and 1-D Spectra for [OIII]52$\mu$m  (Figures \ref{fig:NGC5253-B} and \ref{fig:NGC5253-C}, respectively) and [NIII]57$\mu$m  (Figures \ref{fig:NGC5253-D} and \ref{fig:NGC5253-E}, respectively) in NGC\,5253. The [NIII]57$\mu$m profiles here have not been corrected for atmospheric transmission.}\label{fig:NGC5253}
\end{figure*}

\subsection{Data Analysis}

We present below our procedure for extracting emission line fluxes and their uncertainties for the SOFIA FIFI-LS data.
We could not use the standard paxel by paxel fitting procedure, used normally for Integral Field Unit (IFU) data, because the high background contribution from the atmosphere drastically reduces the signal-to-noise ratio of our data sets.

We use here the example of NGC5253, a galaxy observed within our program (PI M.A. Malkan; AOR ID 07\_0239\_4). All other galaxies spectra have been reduced using the same method. The datacubes used for these reductions have all been obtained from the SOFIA data archive, unless otherwise indicated.

To illustrate the position of the FIFI-LS observations, we obtained a 2MASS H-band All-Sky Release Survey Atlas Image of each galaxy over approximately the same spatial scale as that covered by FIFI-LS on SOFIA. In the special cases of NGC2366 and UGC5189, we present WISE W1 band images of these galaxies instead of the 2MASS images due to the lack of significant emission in the latter; refer Figure \ref{fig:NGC2366-A} and \ref{fig:UGCS5189-A}. This is shown in Figure \ref{fig:NGC5253-A}. The exact areas covered by FIFI are shown by the solid yellow and dotted salmon boxes (representing the regions covered for the [OIII]52$\mu$m and [NIII]57$\mu$m detections respectively).

The first step to determine the line fluxes of each source if to define a useful aperture. We do have a spatial resolution of about FWHM $\sim$ 6 arcsec and compact but still not point like extension of the sources. We use the following procedure to find an ellipsoid shaped aperture for each data set to capture the whole flux while minimizing the noise contribution of pixels without flux from the source.

We locate the emitting region of the galaxy in the far-IR using SOFIA linemaps. The reduced datacube is a stack of 2-D flux channel maps at various wavelengths. We integrate these maps across $\sim$10 wavelength channels surrounding the central redshifted wavelength of the emission line of interest. This gives us a preliminary linemap for the galaxy that helps us locate its emitting region. After choosing a center (which often aligns closely with the optical center of the galaxy), we identify an elliptical aperture around the center that captures the majority of the emitting region. We add up the flux densities from all spaxels within this elliptical aperture, and repeat this procedure separately for all wavelengths to obtain the wavelength profile of the spectral density of the emission line.

Before making the emission line flux calculation, we trim our spectrum by masking out noisy wavelength channels on the red/blue side of the central wavelength. If the spectrum is still noisy, we apply a Wiener filter to it (this is the case of the [NIII]57$\mu$m line, but not the [OIII]52$\mu$m line in NGC5253, see the solid black ``Smooth" line in Figure \ref{fig:NGC5253}).


Thereafter, we identify typically 5 to 10 continuum points on either side of the central peak in the trimmed spectrum. We use the median of these values as continuum signal and subtract it from all channels. We then add all the channels that were considered to be part of detection in the trimmed spectrum to create the continuum subtracted linemap (Figures \ref{fig:NGC5253-B} and \ref{fig:NGC5253-D}).
After obtaining this linemap, we may find that it sometimes differs from the preliminary continuum map we made for the galaxy. In such cases, we choose a new center and elliptical aperture that better matches the continuum-subtracted linemap, and repeat the above procedure. It is possible that in doing so, the choice of continuum points changes. But the difference is usually minor, and after a few iterations, we converge to a choice of center, elliptical aperture, and continuum points that together give the best possible continuum-subtracted linemap, with the chosen elliptical aperture best identifying the emitting region of the galaxy. Figures (\ref{fig:NGC5253-B} and \ref{fig:NGC5253-D}) were obtained through this iterative procedure.

We point out two caveats at this stage. First, while we generally use the flux channel from a datacube that has been corrected for atmospheric transmission (this is the case of the [OIII]52$\mu$m line in NGC5253), occasionally the atmospheric transmission in the wavelength range of interest can be low (notice the grey-dashed line in Figure \ref{fig:NGC5253-E} with an atmospheric transmission as low as $\sim$0.60). In these cases, we use the flux channel that has \textit{not} been corrected for atmospheric transmission (this is the case of the  [NIII]57$\mu$m line in NGC5253). We do so because the SOFIA archive's automated correction procedure does not handle low transmissions. Second, while circular apertures may suit compact objects which are nearly point-like (e.g., the [OIII]52$\mu$m line in NGC5253), in other cases there is extended emission in certain directions, requiring an elliptical aperture to enclose most of the line emitting region and maximize the signal-to-noise ratio. Standard software, like QFitsView\footnote{\url{https://www.mpe.mpg.de/~ott/QFitsView}}, support only using circular apertures, so we wrote our own Python routine that can use elliptical apertures. The robustness of our routine has been verified by calculating fluxes independently for identical circular apertures created by QFitsView and our routine (using the procedure that will be described hereafter), and checking that they are equal.

We now return to the trimmed spectrum that was obtained with an optimal choice of emission center, elliptical aperture, and continuum points to create the best continuum-subtracted linemap for the galaxy. We then subtract off the continuum from the trimmed spectrum and fit a Gaussian to the resulting emission line spectrum (solid brown lines in Figures \ref{fig:NGC5253-C} and \ref{fig:NGC5253-E}). This fitted Gaussian is overplotted with the solid black line in Figures \ref{fig:NGC5253-C} and \ref{fig:NGC5253-E}. For Figure \ref{fig:NGC5253-E}, where we applied the Wiener filter, we also include the unfiltered trimmed spectrum in the tan dot-dashed line). Finally, we overplot the transmission line (grey, dashed).

We then measure the line flux, $F$, by integrating the area under the continuum-subtracted trimmed spectrum (green shaded area in Figures \ref{fig:NGC5253-C} and \ref{fig:NGC5253-E}). We exclude from this integration the spectral range used to estimate the continuum level. We then compute the formal statistical error $\delta F$ in the flux as:

\begin{equation}
    \delta F = \Delta \lambda \sqrt{\sum_{P} (f(P) - y_P)^2} 
\end{equation}

Here, $P$ is a continuum point, $y_P$ is its spectral density, and $f(P)$ is its spectral density as predicted by the fitted continuum line. $\Delta \lambda$ is the wavelength pixel width (which is 0.007 $\mu$m for SOFIA). Calculating $\delta F$ amounts to adding the errors between the actual spectrum and the continuum line in quadrature. Note at this stage that if the flux channel used for the galaxy was \textit{not} corrected for atmospheric transmission, we replace our calculated flux $F$ with $F/<T>$, where $<T>$ is the average transmission in the wavelength range we integrate across to calculate $F$.

The formal statistical error does not take into account the uncertainty in setting the continuum level. To compute a more realistic error bar for our flux measurement, we make a slightly different choice of the continuum points in our trimmed spectrum (or alternatively make a somewhat different selection for the noise to discard from the untrimmed spectrum). This subjective estimate gives us a different, but still ``nearly correct" continuum level. We repeat the aforementioned process to compute the flux with the formal statistical error again.

This leaves us with two flux readings, $F_1 \pm \delta F_1$ and $F_2 \pm \delta F_2$, from the two iterations. We are now ready to estimate the actual flux measurement as $F \pm{\Delta F}$, where:

\begin{equation}
     F = \frac{F_1 + F_2}{2}
\end{equation}

and:

\begin{equation}
    \Delta F = max(F_1, F_2) - min(F_1, F_2) + \delta F_1 + \delta F_2
\end{equation}

Essentially, we are deriving our flux as the average of both estimates, and the error as the difference of the upper bound and lower bound as dictated by the formal statistical error. This completes our data reduction process.

\subsection{Ancillary data}\label{anci}

We have used the literature data to complement the observations obtained with SOFIA FIFI-LS. In particular we have taken the catalog of {\it Herschel} \citep{pilbratt2010} PACS \citep{poglitsch2010} and SPIRE \citep{griffin2010} observations of AGN and Starburst galaxies of \citet{fernandez2016} and the observations of the dwarf galaxies of \citet{cormier2015}. We also included the ISO-LWS observations reported from \citet{brauher2008} 
of NGC\,2146 and NGC\,4194. 
Our sample of local galaxies contains 47 galaxies divided in 21 HII region/(U)LIRG galaxies, 19 Seyfert galaxies, including one LINER and 7 dwarf galaxies. The sample is presented in Table \ref{tab:sample1}, with all the relevant information, including coordinates, redshift, optical metallicities and N/O ratio, as computed from optical emission line observations \citep{perez-montero2014}, as well as the abundances and the N/O ratio computed from the IR lines \citep{fernandez2021}. The full dataset of far-IR spectroscopy of the sample of 48 galaxies is given in Table \ref{tab:samplefir}.  


We have also compiled the observations of the mid-IR fine-structure lines, mainly collected from the {\it Spitzer} \citep{werner2004} Infrared Spectrograph (IRS; \citealt{houck2004}), as shown in Table \ref{tab:sample2}. We note that this table contains 51 galaxies, because we have also included 3 objects for which we do not have full far-IR lines coverage: IC342, NGC2976 and NGC4536. 

Table \ref{tab:sample1bis} presents the SOFIA FIFI-LS observations of the [OIII]52$\mu$m line of an additional list of 11 galaxies, for which we do not have detections of the [NIII]57$\mu$m line, and therefore we are not able to discuss them further. 

\section{Sample characterization}\label{class}

\begin{figure}[ht!]
\centering
\includegraphics[width=\columnwidth]{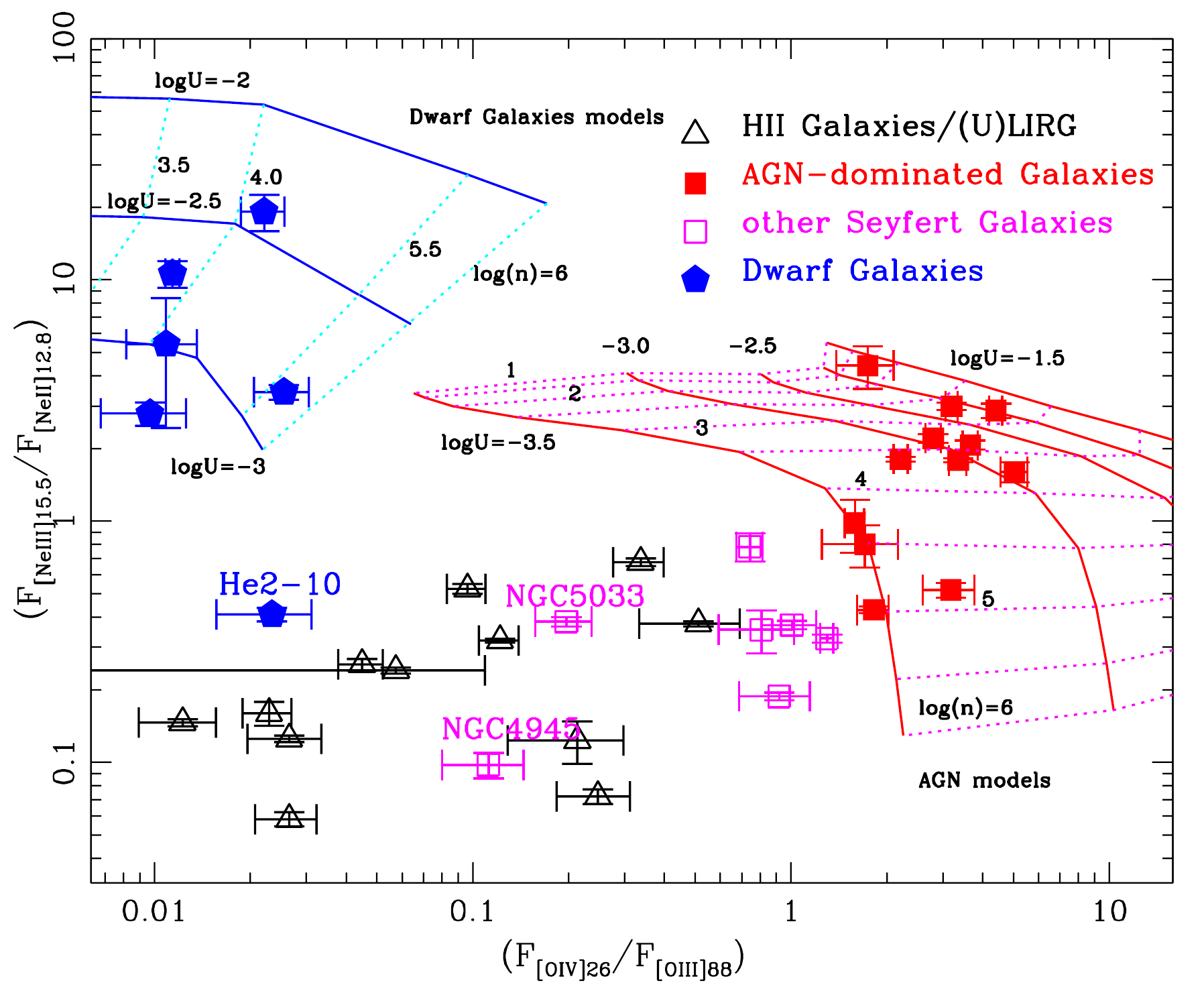}
\caption{[NeIII]15.5$\mu$m/[NeII]12.8$\mu$m line ratio versus the [OIV]25.9$\mu$m/[OIII]88$\mu$m line ratio, the so-called {\it IR BPT diagram} \citep{fernandez2016}, showing the separation between the galaxy types: on the right AGN, on the bottom left Starburts /HII region galaxies and ULIRGs and on the upper left part the dwarf galaxies, characterized by low metallicity. The two Seyfert galaxies NGC4945 and NGC5033 have line ratios more typical of HII region galaxies, and also the dwarf galaxy He2-10 is similar to them. A grid of AGN photoionization models  (models A in Section \ref{abund}) spans the parameters value of: $ 1.0<{\rm log(n)}<6.0$ and $-3.5<{\rm log(U)}<-1.5$. Similarly a grid of low-metallicity starburst galaxies models  (models A in Section \ref{abund} with a metallicity of Z = 0.004 (1/5 Z$_{\odot}$)), the ranges of $ 3.5<{\rm log(n)}<6.0$ and $-3.0<{\rm log(U)}<-2.0$. }\label{bpt1}
\end{figure}

\begin{figure}[ht!]
\centering
\includegraphics[width=\columnwidth]{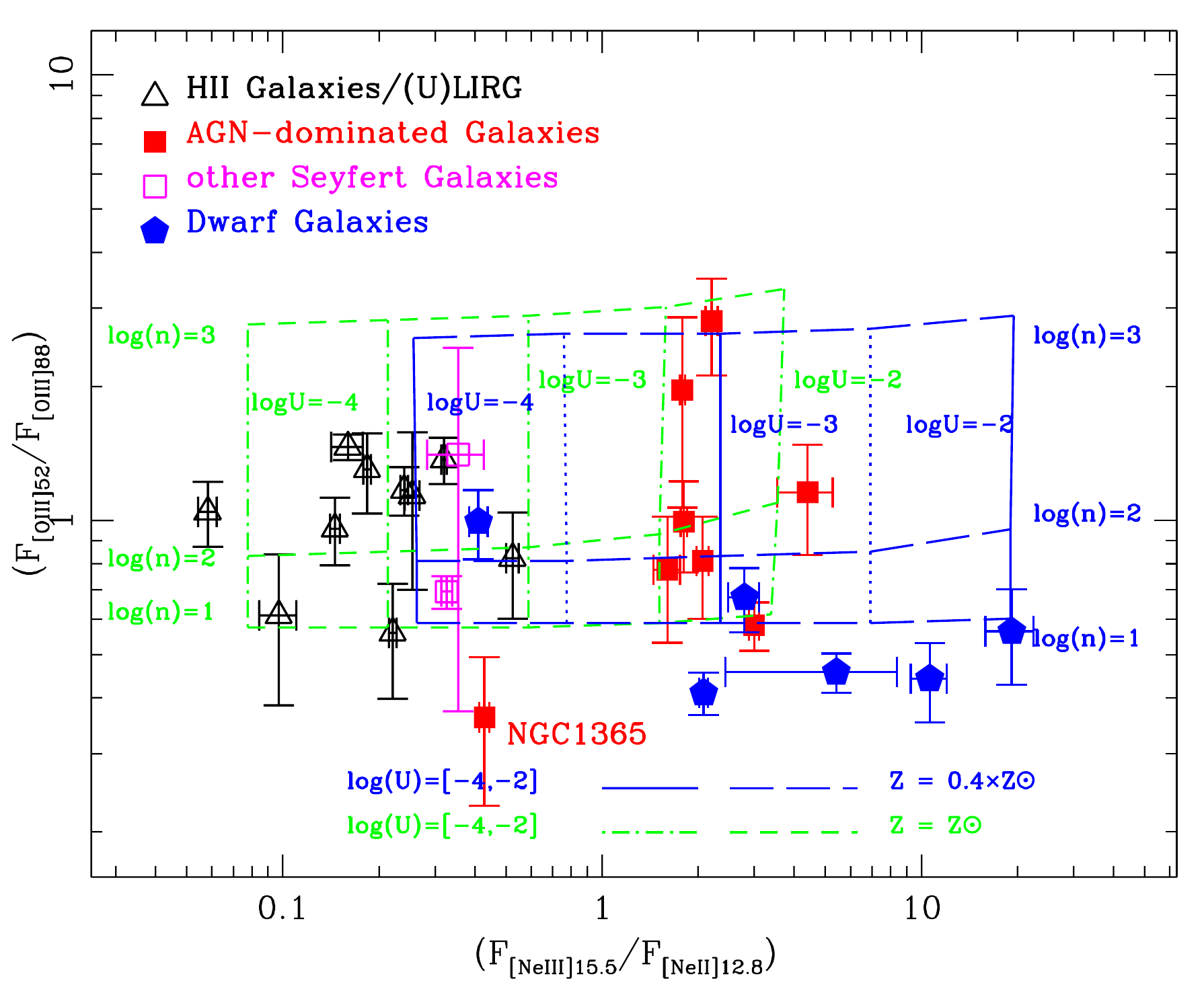}
\caption{Ionization density diagram made with the [OIII]52$\mu$m/88$\mu$m ratio versus the [NeIII]15.5$\mu$m/[NeII]12.8$\mu$m ratio. The grid of models  (models B in Section \ref{abund}) represent Starburst models with density log(n)=[1,3] and ionization potential of log(U)=[-2,-4]. Two types of abundances have been assumed: solar (grid with density as short-dashed line and ionization parameter as dot-dashed line) and subsolar with Z=0.4$\times$Z$\odot$ (grid with density as long-dashed line and ionization parameter as solid line). The effect of sub-solar abundances shifts the models to the right, and intercepts a low metallicity dwarf galaxy.}\label{iondens1}
\end{figure}

\begin{figure}[h!]
\centering
\includegraphics[width=\columnwidth]{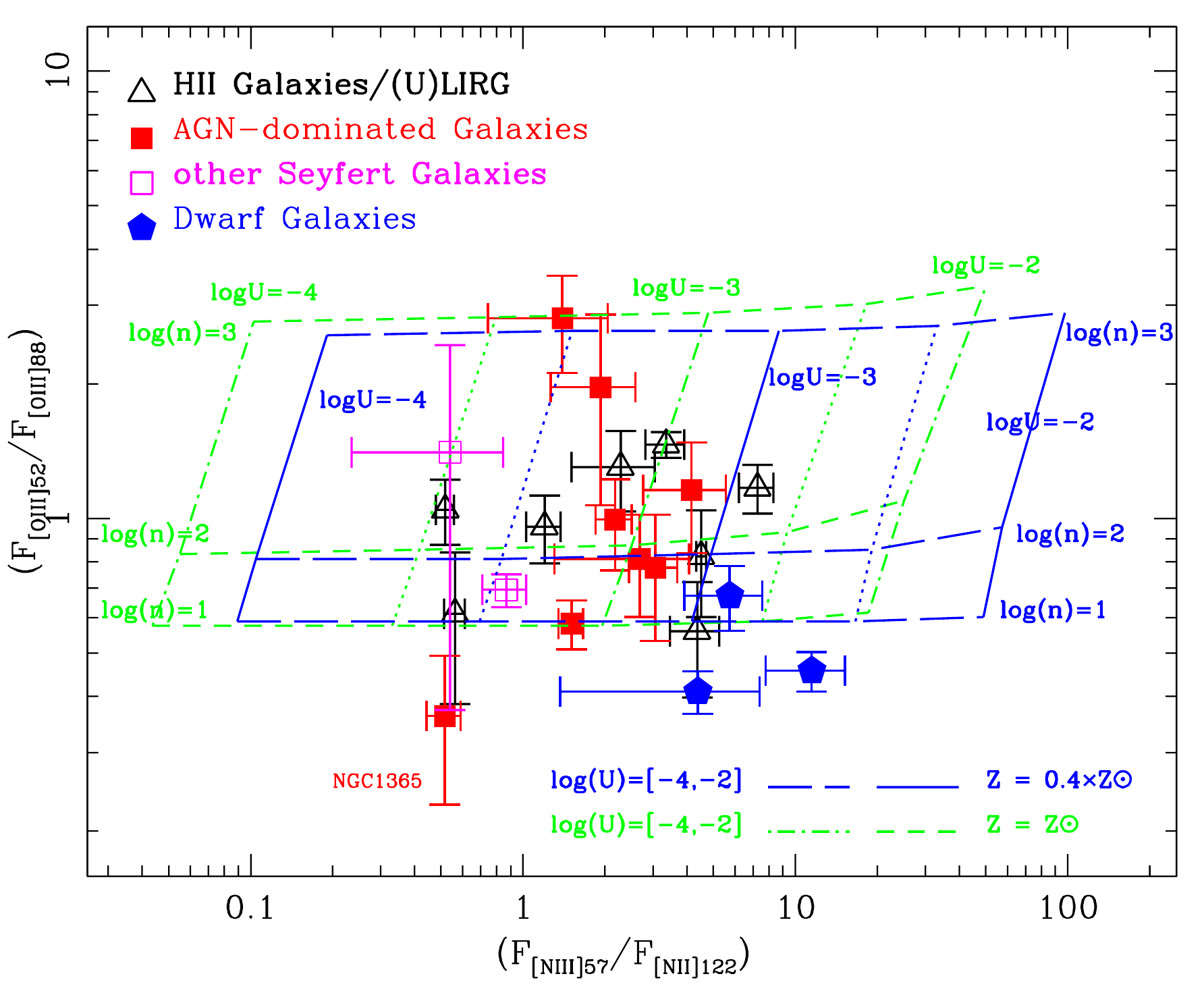}
\caption{Ionization density diagram made with the [OIII]52$\mu$m/88$\mu$m ratio versus the [NIII]57$\mu$m/[NII]122$\mu$m ratio. The models are the same as in Fig. \ref{iondens1}, again  with the two types of abundances: solar and sub-solar (Z=0.4$\times$Z$\odot$).}\label{iondens2}
\end{figure}

We present in Fig.\,\ref{bpt1} the so-called ``IR BPT'' diagram proposed for the first time by \citet{fernandez2016}. In contrast with the optical BPT, the diagram in Fig.\,\ref{bpt1} can be regarded as a ``softness diagram'', because the axes ([OIV]$25.9 \mu$m/[OIII]$88\mu$m and [NeIII]$15.6\mu$m/[NeII]$12.8\mu$m) are independent of the chemical abundances, and thus represent a more selective measure of the strength and shape of the radiation field in the $\sim 20$--$80\, \rm{eV}$ range \citep[see also][]{vilchez1988,perez-montero2009b,melendez2011}. Here the separation of the different types of galaxies is largely independent of photoionization models to define the boundaries, mostly due to the high ionization potential of O$^{3+}$ ($54.9\, \rm{eV}$). This is beyond the double ionization edge of helium ($54.4\, \rm{eV}$), where the continuum emission of any stellar population drops sharply while the AGN power-law continuum remains unaffected. Consequently, AGN are in the right part of the diagram, characterized by a higher value of the ionization of the gas responsible for the emission, the HII region galaxies and (U)LIRG are on the lower left part of the diagram, while the dwarf galaxies at the upper left region, having a metallicity significantly lower than solar. An additional advantage of IR diagnostics is that they are insensitive to dust extinction and temperature effects, in contrast with optical tracers.

The IR BPT represents a very powerful tool to separate galaxies not only by the shape of their primary ionizing spectrum, i.e. star formation or AGN dominated, using the [OIV]$25.9 \mu$m/[OIII]$88\mu$m line ratio, but also for the different metallicities, through the [NeIII]$15.6\mu$m/[NeII]$12.8\mu$m line ratio. We use this diagram to further classify the Seyfert galaxies into two separate classes: those within the grid of AGN photoionization models (models A in Section \ref{abund}) are classified hereafter as ``AGN-dominated'' galaxies, while those outside these models are called ``other Seyfert galaxies''. These latter galaxies include: NGC\,4945, NGC\,5033, CenA (NGC\,5128), NGC\,6240, NGC\,7130 (IC\,5135), NGC\,7469 and NGC\,7582. We note that for these galaxies the IR lines excitation can originate from, or at least be contaminated by, HII regions in the host galaxy \citep{xia2018}. 

In Fig.\,\ref{iondens1} we show an ionization-density diagram made using the line ratios of [OIII]52$\mu$m/88$\mu$m and [NeIII]15.5$\mu$m/[NeII]12.8$\mu$m. Two grids of starburst photoionization models  (models B in Section \ref{abund}) are shown, one with solar metal abundances and the other with sub-solar abundances. The diagram shows that the typical density is $\sim 1.0<{\rm log(n_{e}/cm^{-3})}<3.0$ , while the ionization parameter is $\sim -4<{\rm log(U)}<-2$ for HII region galaxies/ULIRG. 
The effect of decreasing the metal abundances is to shift the model grid to the right side of the diagram, because the line ratio tracing the ionization parameter is sensitive to the metal abundances. The [NeIII]15.5$\mu$m/[NeII]12.8$\mu$m ratio increases by an average factor of $\gtrsim$ 5, from solar to 0.4 $\times$ Z$_\odot$ abundances. We notice in Fig. \ref{iondens1} that the three galaxies in this diagram classified as ``other Seyfert'' galaxies are well within the area populated by the HII region/ULIRG galaxies, confirming that most of their ionized emission does not originate in the AGN Narrow Line Region (NLR), as indicated by the low value of the ionization potential (logU$\sim$-3.5).

Figure \ref{iondens2} shows another ionization-density diagram, made purely with far-IR lines, the [OIII]52$\mu$m/88$\mu$m ratio versus the [NIII]57$\mu$m/[NII]122$\mu$m ratio. Similarly to the previous diagram, this diagram also indicates that the gas density is $\sim 1.0 < {\rm log(n_{e}/ \rm{cm^{-3}})} < 3.0$, and the ionization parameter is $\sim -4<{\rm log(U)}<-2$. The effect of decreasing the metal abundances is to increase the ratio of [NIII]57$\mu$m/[NII]122$\mu$m by an average factor of $\gtrsim$ 2.5, from solar to 0.4 $\times$ Z$\odot$ abundances.

\section{Photoionization models and abundance determinations}\label{abund}

In this work we have used various photoionization models using the CLOUDY code \citep{ferland2017}, also with the aim to demonstrate their overall consistency with each other, which does not depend on the particular choices of the parameters details. One set of models (hereafter models A) has been taken from \citet{fernandez2016} which includes both AGN models and starburst galaxies models. For the AGN, a grid of constant density models (with log(n$_{H}$) (cm$^{-3}$) = 1 to 6) in a plane-parallel geometry has been built using an AGN ionizing continuum of a power law from the optical to the X-rays with a slope of $\alpha = -1.4~~(S_{\nu} \propto \nu^{\alpha}$), and ionization parameters (U) with values of log U = -1.5, -2.0, -2.5, -3.0, -3.5. For these models, the maximum column density of N$_{H}$ = 10$^{23}$ cm$^{-2}$ was used as the stopping criterion of the spatial integration, representative of the column density found in NLR clouds \citep{moore1994}. 

For the star forming galaxies models, the ionizing spectrum was simulated using the STARBURST99 code \citep{leitherer1999} for two cases: (1) a young burst of star formation with an age of 1 Myr and a metallicity of $Z = 0.004 (1/5 Z_{\odot}$), in order to produce the hard UV ionizing spectrum in a low-metallicity environment, such the one typical of dwarf galaxies, and (2) a continuous burst of star formation with an age of 20 Myr and solar metallicity, to model normal star forming galaxies.
We use models with plane-parallel geometry, constant pressure, initial densities in the log(n$_{H}$) (cm$^{-3}$) = 1 to 6 range, and ionization parameters in the log U = -2.0 to -4.5 range. We assumed two intervals for the Kroupa initial mass function (IMF; with exponents 1.3, 2.3 and mass boundaries of 0.1, 0.5, and 100 M$_{\odot}$), the 1994 Geneva tracks with standard mass-loss rates, and the Pauldrach/Hillier atmospheres, which take into account the effects of non-LTE and radiation driven winds.

Another set of models (hereafter models B) has been taken from \citet{pereira2017} and includes both starburst galaxies and AGN models. 
In starburst models, they assumed a constant pressure slab model illuminated by the spectrum of a continuous burst of SF. This illuminating spectrum was calculated using STARBURST99 \citep{leitherer1999} assuming continuous SF with a \citet{kroupa2001} initial mass function with an upper stellar mass boundary of 100 M$_{\odot}$. This is an average spectrum representing the integrated emission of a galaxy with stellar populations of different ages. They produced the spectra for five different stellar metallicities (Z$_{\star}$ = 0.05 Z$_{\odot}$, 0.2 Z$_{\odot}$, 0.4 Z$_{\odot}$, Z$_{\odot}$ and 2 Z$_{\odot}$) available for the Geneva evolutionary tracks \citep{meynet1994}.
AGN photoionization models follows the prescription of starburst models, but with ionizing spectrum with a broken power-law with an index $\alpha = -1.4 ~(S_{\nu} \propto \nu^{\alpha}$) between 10$\mu$m and 50 keV, $\alpha$ = 2.5 for $\lambda$ $>$ 10$\mu$m and $\alpha$ = -2.0 for E $>$ 50 keV. The range of the ionization parameters, log U = -3.0 to -1.6, is that of typical AGN. The remaining input parameters of the model (gas-phase abundances, stopping criteria, gas density range, dust grains, etc.) are the same that we used for the starburst models.

Two independent abundance determinations were obtained for both the O/H and N/O ratios using the optical and the infrared nebular lines. The estimates were derived with the \textsc{Hii-Chi-mistry} \citep[hereafter \textsc{HCm};][]{perez-montero2014} and the \textsc{Hii-Chi-mistry-IR} codes\footnote{Available at: \url{https://www.iaa.csic.es/~epm/HII-CHI-mistry.html}} \citep[\textsc{HCm-IR};][]{fernandez2021}. Both are based on the same grid of photoionization models (hereafter models C), computed using \textsc{Cloudy} \citep{ferland2017} adopting simple stellar population models from \textsc{Popstar} as incident radiation field \citep{molla2009}. The models sample a wide range in oxygen abundance ($6.9 < 12 + \log (\rm{O/H}) < 9.1$), nitrogen abundance ($-2.0 < \log (\rm{N/O}) < 0.0$), and ionization parameter ($-4.0 < \log \rm{U} < -1.5$), assuming a filling factor of 0.1 and a constant electron density of $n_{\rm e} = 100\, \rm{cm^{-3}}$. To compute the optical-based abundances for AGN we used photoionization models assuming a power-law ionizing continuum ($F_\nu \propto \nu^{\alpha_{\rm OX}}$; $\alpha_{\rm OX} = -1.2$) and $n_{\rm e} = 500\, \rm{cm^{-3}}$ \citep[see][]{perez-montero2019}.
We refer to \citet{fernandez2021} for a comparison of the predicted emission line ratios with the observed ones for a sample of star forming galaxies, showing that the range of variations of the main emission-line ratios of this sample are within the ranges covered by the models.  

\textsc{HCm} and \textsc{HCm-IR} perform a Bayesian-like calculation of the $12 + \log (\rm{O/H})$ and the $\rm{N/O}$ abundances, and the ionization parameter, by comparing the optical (reddening corrected) and infrared nebular lines, respectively, with the predicted values in the grids of photoionization models. 
The abundance determination does not rely on a single best-fit model, but it is instead based on the  Bayesian calculation where all the models contribute to the computation of the abundances. The estimated abundance and its associated uncertainty is then computed from the weighted grid of models, as discussed in \citet{fernandez2021}.
The photoionization models provide abundance estimates consistent with the direct method --\,based on the detection of auroral nebular lines in the optical range\,-- within a scatter of $0.1\, \rm{dex}$ and $0.2\, \rm{dex}$, respectively \citep{perez-montero2014,fernandez2021}.

For the case of AGN, the abundance determinations could not be obtained using \textsc{HCm-IR} due to the lack of hydrogen recombination line measurements in the mid-IR. Humphreys-$\alpha$ is the brightest recombination line in the range covered by \textit{Spitzer}/IRS in the high-spectral resolution mode ($9.9$--$37.1\,\rm{\micron}$), however it is a weak line ($\sim 1/100$th of H$\beta$) and thus it is not reported in most of the works discussed in Section\,\ref{anci}, where these spectra were analyzed. Therefore, IR-based $12 + \log (\rm{O/H})$ determinations are not available for the AGN in our sample. In this regard, IR-based abundance determinations for AGN will be addressed in a forthcoming work (P\'erez-D\'iaz et\,al. in prep.). Nevertheless, a robust estimate of the N/O abundance ratio can be obtained from the [OIII]$52,88\mu$m and [NIII]$57\mu$m lines, due to the similar ionization structure of these two elements. For this purpose we define the N3O3 parameter, based on the relative intensities of the nitrogen and oxygen lines in the far-IR:
\begin{equation}\label{eq_n3o3}
{\rm N3O3} = \log \left( \frac{\rm I([NIII]_{57 \mu{\rm m}})}{\rm I([OIII]_{52 \mu{\rm m}}) + I ([OIII]_{88 \mu{\rm m}})} \right)
\end{equation}

This parameter was used by \citet{peng2021} and \citet{fernandez2021} for the derivation of the N/O relative abundances in star-forming galaxies, and can be also be applied to the case of Narrow Line Region (NLR) in AGN, due to the very low dependence on the excitation conditions of the ionized gas. This is due to the similar ionization structure of the oxygen and nitrogen elements in the nebula. To this aim, we analyzed the predicted values of N3O3 for the grid of AGN photoionization models described in \citet{perez-montero2019}. As input ionizing source we adopted a double-peak power law with index $\alpha_{\rm UV} = -1.5$, also considering different values for $\alpha_{\rm OX} = -0.8$ and $-1.2$, with an electron density of $n_{\rm e} = 500\, \rm{cm^{-3}}$, sampling the same range in metallicity, ionization parameter and N/O mentioned earlier. The N3O3 parameter presents a very tight relation with the N/O abundances in star forming regions \citep{peng2021,fernandez2021}. This is also true under harder radiation fields in AGN, as shown by the linear sequence of AGN models at a fixed $12 + \log (\rm{O/H}) = 8.8$ and $\alpha_{\rm OX} = -1.2$ in Appendix\,\ref{app_NO}. Thus, a robust linear relation between these parameters can be extracted. The linear fit to the whole set of models for this value of $\alpha_{\rm OX}$ has a correlation coefficient of $0.91$ and results in the following expression:
\begin{equation}\label{eq_NO_n3o3}
    \log(\rm{N/O}) = (0.97 \pm 0.01) \times \rm{N3O3} - (0.01 \pm 0.01)
\end{equation}
Additionally, the analysis for the model grid with $\alpha_{\rm OX} = -0.8$ lead to similar results within the typical observational errors.

\section{Results} \label{results}
We discuss the results of the abundance determination through both the optical and IR spectral lines. We use in the following a few diagrams showing various IR line ratios as a function of the metallicity. We also compare both the (O/H) and (N/O) abundance determinations using the two methods.

\subsection{Far-IR line ratios as a function of gas-phase metallicity}

\begin{figure}
\centering
\includegraphics[width=\columnwidth]{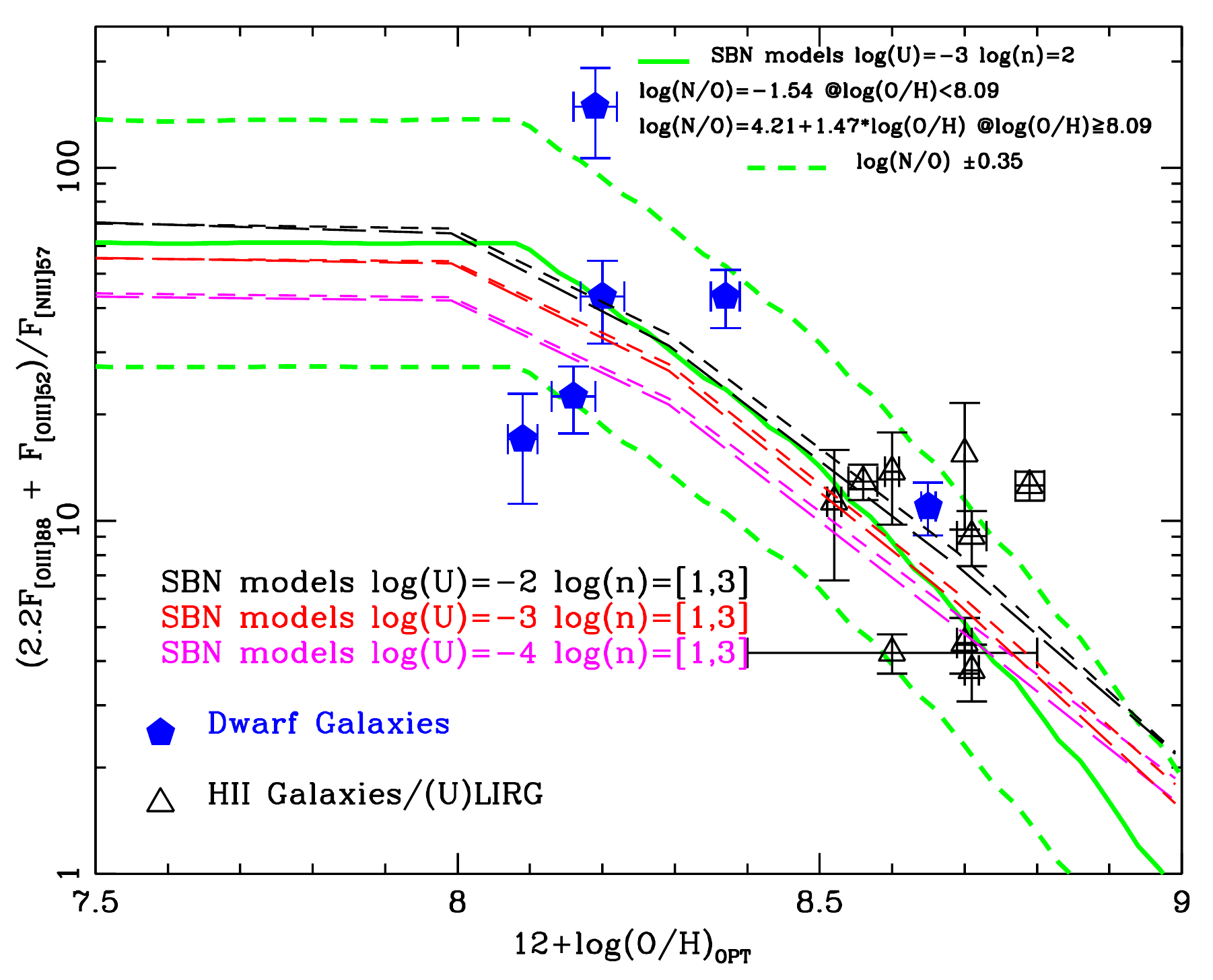}
\caption{Ratio of (2.2 $\times$ [OIII]88$\mu$m + [OIII]52$\mu$m)/[NIII]57$\mu$m  as a function of the gas-phase metallicity for the HII region/ULIRG galaxies and the dwarf galaxies of our sample. The photoionization models  (models B in Section \ref{abund}) are taken from \citet{pereira2017} and include values of ionization parameter log(U)=-2, -3 and -4 from the top to the bottom and densities with log(n$_e$/cm$^{-3}$) = [1,3]. The solid line (in green) represents a photoionization model (models C in Section \ref{abund}) with log(U)=-3, log(n$_e$/cm$^{-3}$)=2 and a log(N/O) = -1.54 for metallicities of [12+log(O/H)] $<$ 8.09 and log(N/O) = 4.21 +1.47$\times$log(O/H) for metallicities of [12+log(O/H)] $>$ 8.09. The two broken lines (in green) show the effect of increasing and decreasing, respectively, the log(N/O) by 0.35 dex. 
}
\label{o3n3_new_sbn}
\end{figure}

\begin{figure}
\centering
\includegraphics[width=\columnwidth]{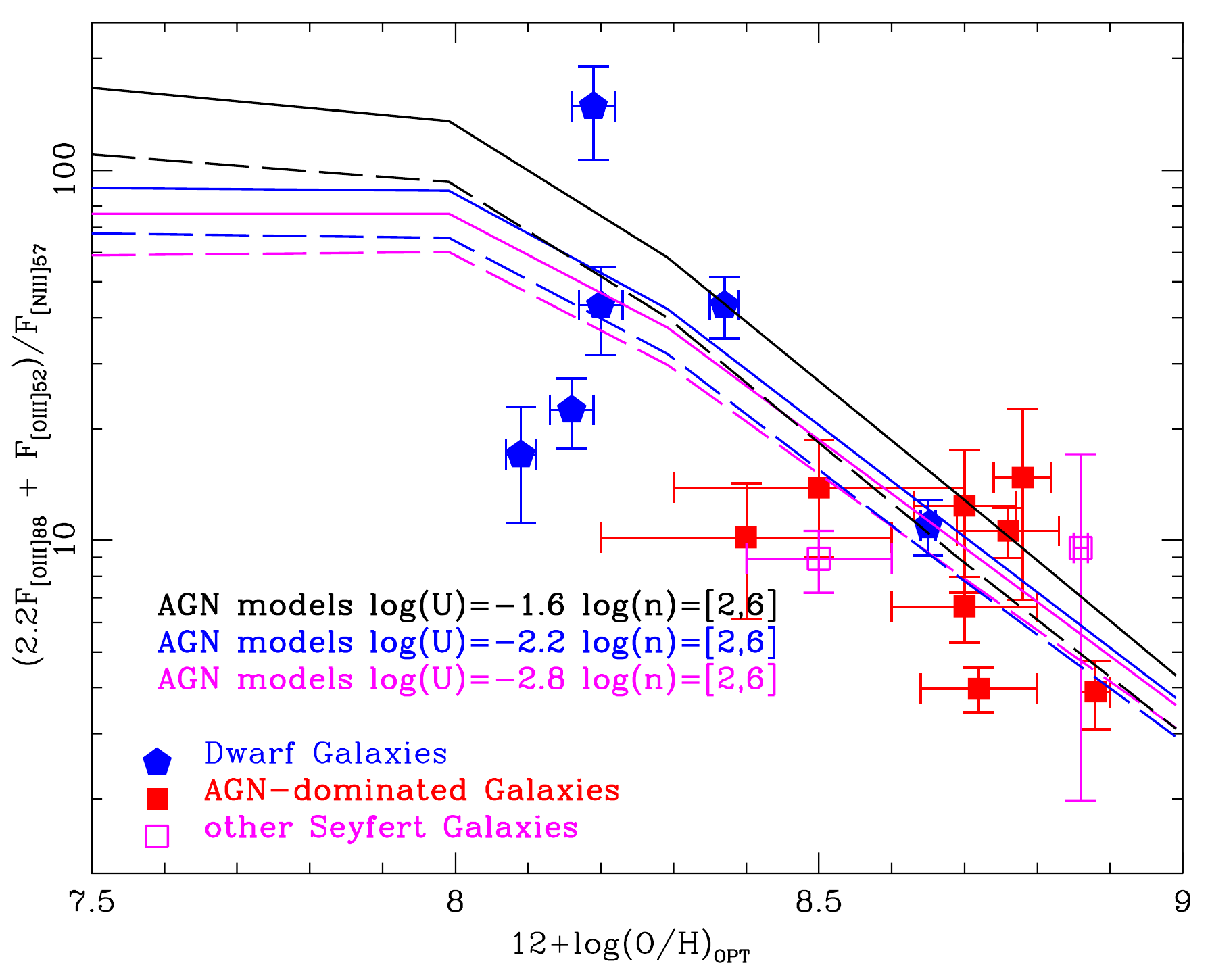}
\caption{Same as Fig. \ref{o3n3_new_sbn}, but with the AGN observations and the relative photoionization models  (models B in Section \ref{abund}) from \citet{pereira2017}. The dwarf galaxies observations have been included for comparison.
}\label{fig_o3n3_1_agn}
\end{figure}

Following the work of \citet{nagao2011} and \citet{pereira2017} we present for the HII region/ULIRG galaxies and the dwarf galaxies of our sample, in Fig. \ref{o3n3_new_sbn}, the ratio of (2.2 $\times$ [OIII]88$\mu$m + [OIII]52$\mu$m)/[NIII]57$\mu$m as a function of the gas-phase metallicity as measured in the optical and expressed in terms of ${\rm 12 + log(O/H)}$. The starburst galaxies photoionization models are taken from \citet{pereira2017} (models B in Section \ref{abund}) and include values of ionization parameter log(U)=-2, -3 and -4 and electron densities with log(n$_e$/cm$^{-3}$)=[1,3]. These models follow the same relation between N/O and O/H derived by \citet{pilyugin2014}, adopted in \citet{pereira2017} and reported in Eq. (2) of this latter study.

\begin{figure}
\centering
\includegraphics[width=\columnwidth]{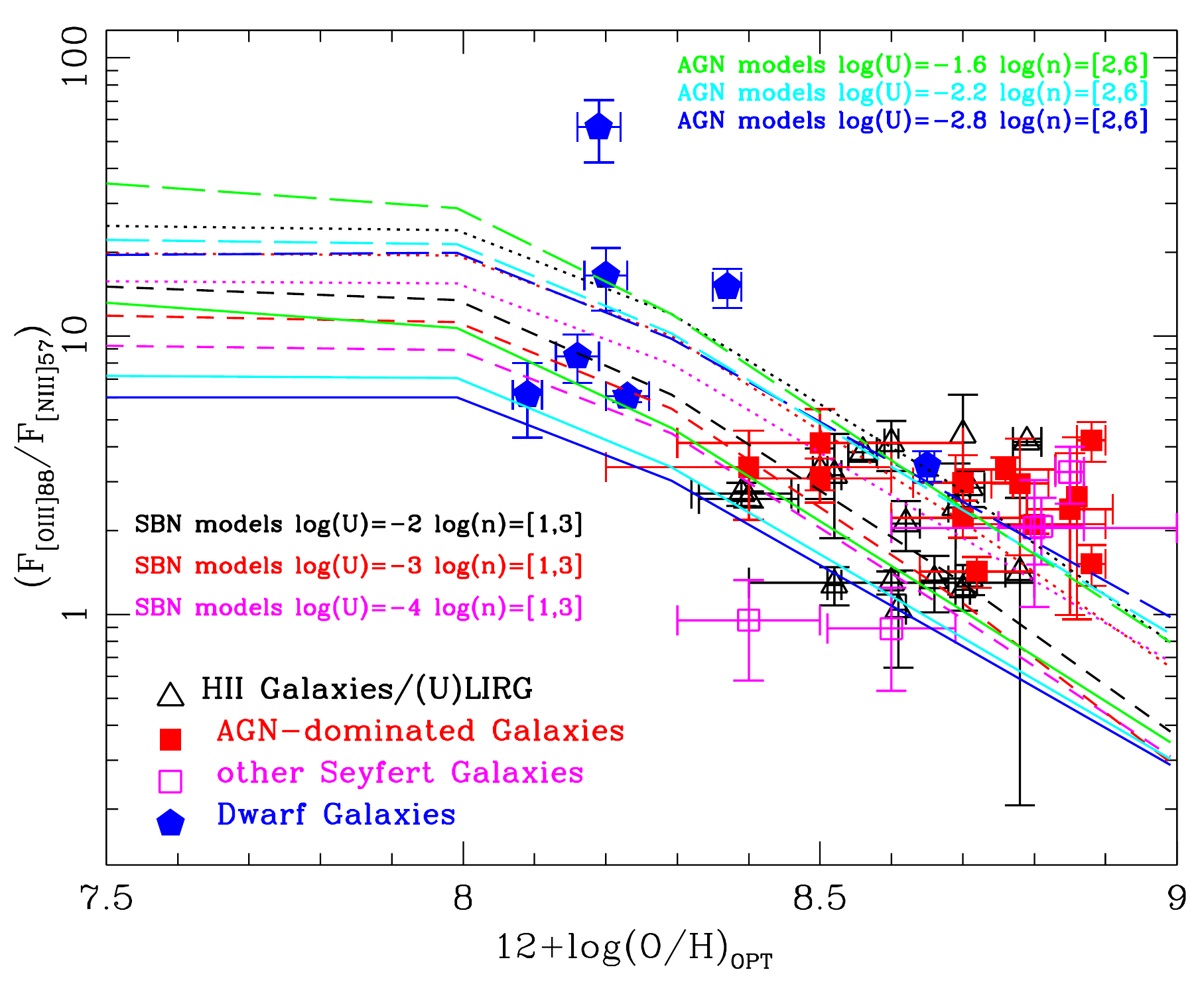}
\caption{Ratio of [OIII]88$\mu$m/[NIII]57$\mu$m as a function of the gas-phase metallicity. The photoionization models (models B in Section \ref{abund}) are taken from \citet{pereira2017} and include values of ionization potential log(U)=-2, -3 and -4 from the top to the bottom and densities with log(n$_e$/cm$^{-3}$) =[1,3] for the starburst models and ionization potential log(U)=-1.6, -2.2 and -2.8 from the top to the bottom and densities with log(n$_e$/cm$^{-3}$) =[2,6] for the AGN models.}\label{fig_o3n3_2_sbn_agn}
\end{figure}

We also show in the same figure (solid line (in green)) a photoionization model (from models C in Section \ref{abund}) with log(U)=-3, log(n$_e$/cm$^{-3}$)=2 and a log(N/O) = -1.54 for metallicities of [12+log(O/H)] $<$ 8.09 and log(N/O) = 4.21 +1.47$\times$log(O/H) for metallicities of [12+log(O/H)] $>$ 8.09, which corresponds to the \citet{pilyugin2014} relation between N and O, as well as (green broken lines) the effect of increasing and decreasing, in the lower and upper curve respectively, the log(N/O) by 0.35 dex. It appears clear from this plot that, to fit most galaxies, an N/O ratio of a factor $\sim$2 higher or lower than the solar value is needed.  As anticipated in Section \ref{abund}, the different photoionization models shown in Fig.\ref{o3n3_new_sbn} do agree well, and a slight discrepancy only appears for high metallicities [12+log(O/H)] $\gtrsim$ 8.7. This shows that only minor differences are present in the N/O abundances using the constant pressure models in \citet{pereira2017} and the constant density models used in \textsc{HCm} and \textsc{HCm-IR}.

The same line ratio is presented for the AGN, including also the dwarf galaxies for comparison, in Fig. \ref{fig_o3n3_1_agn}. Here the photoionization model grid (models B in Section \ref{abund}) includes models with ionization parameter $\sim -2.8<{\rm log(U)}<-1.6$ and densities of $\sim 2.0<{\rm log(n_{e})}<6.0$ cm$^{-3}$. Also in this case the large scatter of the points around the given models could be due to a different value of the N/O ratio.


In Fig. \ref{fig_o3n3_2_sbn_agn}, we show the ratio ${\rm [OIII]88.3{\mu}m/[NIII]57.2{\mu}m}$ versus the metallicity for all the galaxies of the sample, including HII region/ULIRG galaxies, dwarf galaxies and AGN. Here both starburst and AGN models (models B in Section \ref{abund}) are included and the intrinsic spread of the models is larger than in the case of the previous composite line ratio. Also these photoionization models follow the same relation between N/O and O/H derived by \citep{pilyugin2014} adopted in \citet{pereira2017}.



\begin{figure}
\centering
\includegraphics[width=\columnwidth]{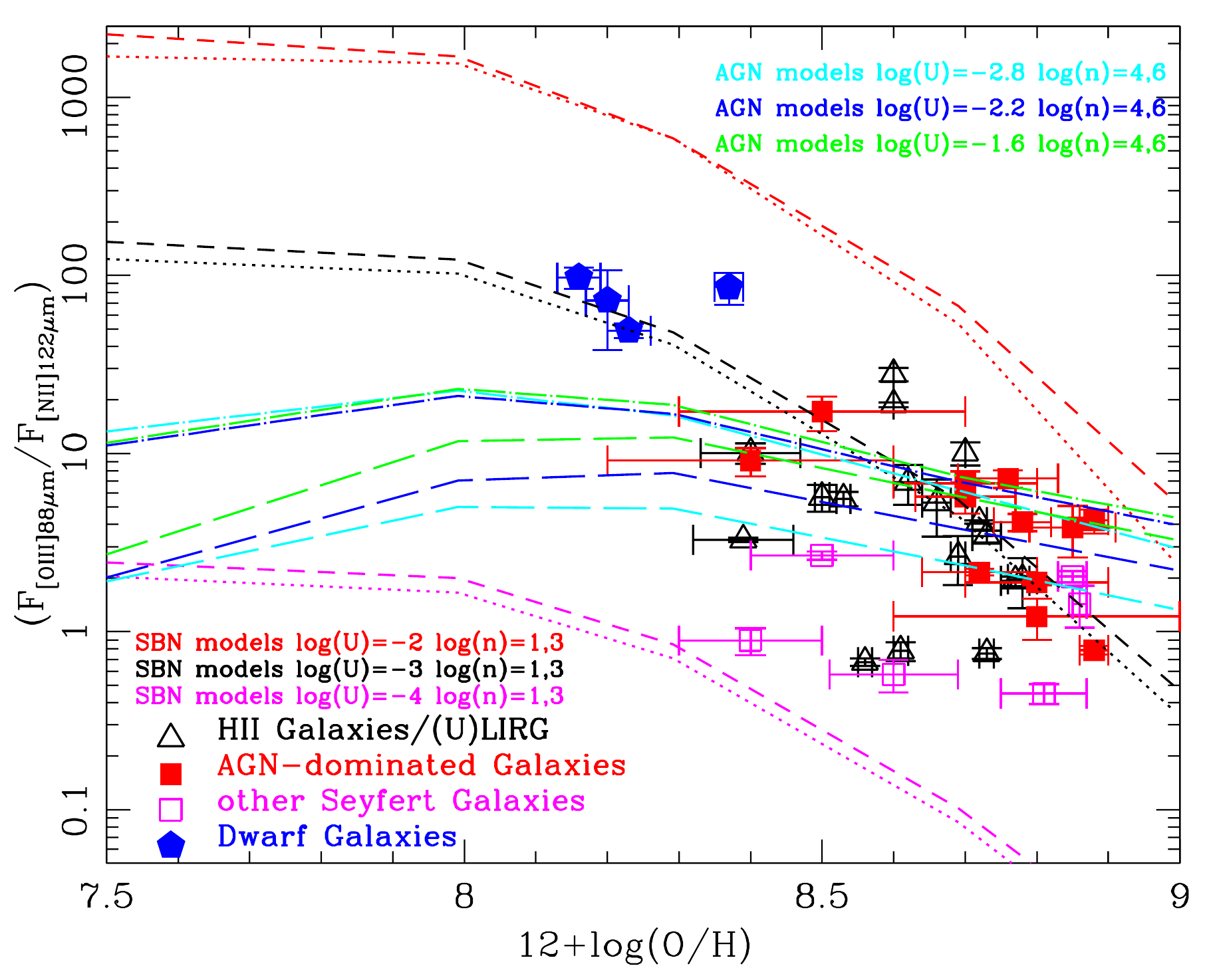}
\caption{Ratio of [OIII]88$\mu$m/[NII]122$\mu$m as a function of the gas-phase metallicity. The photoionization models (models B in Section \ref{abund}) are taken from \citet{pereira2017} and include values of ionization potential log(U)=-2, -3 and -4 from the top to the bottom and densities with log(n$_e$/cm$^{-3}$) =[1,3] for the starburst models and ionization potential log(U)=-1.6, -2.2 and -2.8 from the top to the bottom and densities with log(n$_e$/cm$^{-3}$) =[4,6] for the AGN models.}\label{o3n2_sbn_agn}
\end{figure}

Fig.\,\ref{o3n2_sbn_agn} shows the [OIII]88$\mu$m/[NII]122$\mu$m ratio as a function of the gas-phase metallicity for all galaxies of our sample, compared to the Starburst photoionization models (models B in Section \ref{abund}). Also these photoionization models follow the same relation between N/O and O/H derived by \citet{pilyugin2014} adopted in \citet{pereira2017}.
One can see from this figure a large spread in the models produced by different ionization parameter. In other words, the determination of the metallicity from a given line ratio is only reliable if we can measure the ionization parameter through other line ratios.

\begin{figure}
\centering
\includegraphics[width=\columnwidth]{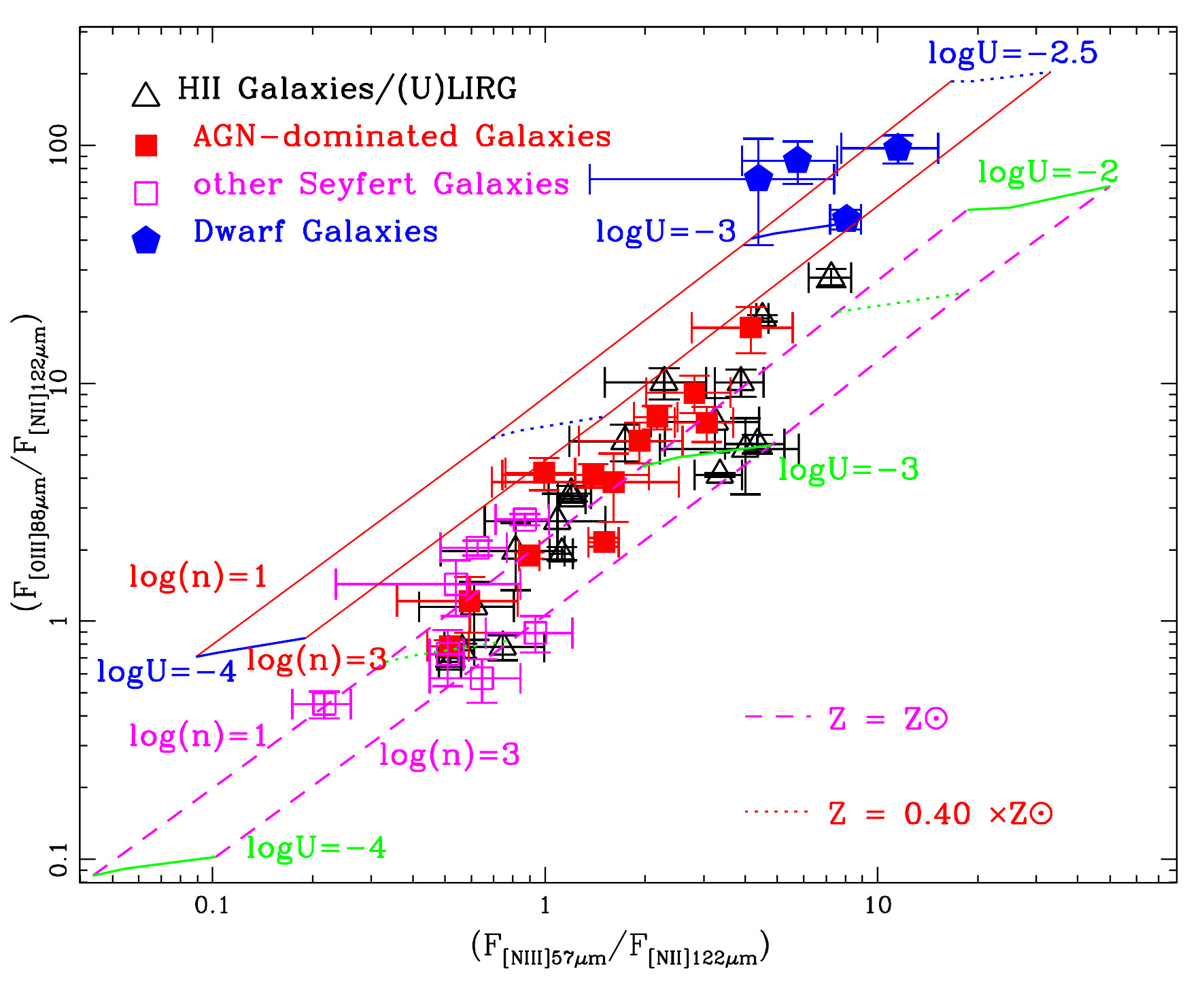}
\caption{Ratio of [OIII]88$\mu$m/[NII]122$\mu$m as a function of the [NIII]57$\mu$m/[NII]122$\mu$m line ratio. The photoionization models (models B in Section \ref{abund}) are taken from \citet{pereira2017} and include values of ionization potential log(U)=-2, -3, -3.5 and -4 from the top to the bottom and densities with log(n) =[1,3] for the starburst models. For comparison, also the AGN data points have been included. the lower grid assumes solar abundances, while the upper grid indicates models with a sub-solar metallicity of Z = 0.4 $\times$ Z${\odot}$.}\label{fig_o3n2_n32}
\end{figure}

In Fig. \ref{fig_o3n2_n32} we show the [OIII]88$\mu$m/[NII]122$\mu$m line ratio as a function of the [NIII]57$\mu$m/[NII]122$\mu$m line ratio for both solar and sub-solar abundances (models B in Section \ref{abund}). As before, the sub-solar value has been set to Z = 0.4 $\times$ Z${\odot}$. We can see from this diagram that only a few objects (4/5 dwarf galaxies) need sub-solar abundances. We also notice that we can read out from the vertical axis a precise estimate of the ionization parameter: assuming, e.g., solar abundances, we can associate a value of log(U) = -4 at a [OIII]88$\mu$m/[NII]122$\mu$m=0.1, a value of  log(U) = -3 for a ratio $\sim$ 5 and a value of Log(U) = -2 at a ratio $\gtrsim$50.
It follows that observations of the three far-IR lines of [NIII]57$\mu$m, [OIII]88$\mu$m and [NII]122$\mu$m can break the degeneracy due to the ionization parameter. In other words, placing an observed galaxy in this diagram will give an estimate of the ionization parameter and therefore make possible an estimate of the metallicity, through the use of the diagram in Fig. \ref{o3n2_sbn_agn}.

\subsection{Mid-IR line ratios as a function of gas-phase metallicity}

\begin{figure}
\centering
\includegraphics[width=\columnwidth]{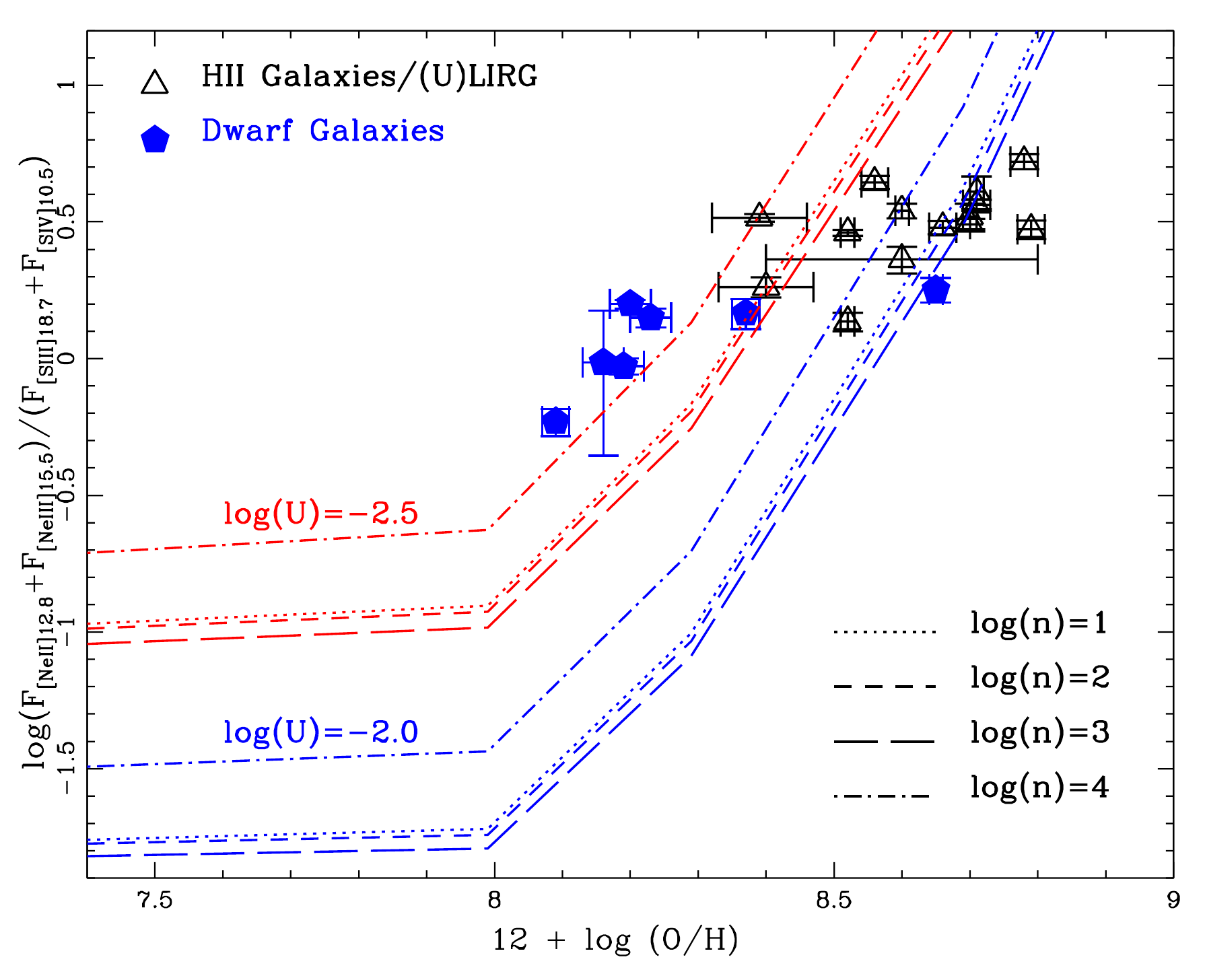}
\caption{Logarithm of the observed ratio of the sum of ${\rm ([NeII]12.8{\mu}m +[NeIII]15.6{\mu}m)}$ to the sum of ${\rm ([SIII]18.7{\mu}m +[SIV]10.5{\mu}m)}$ versus the optical metallicity \citep{fernandez2016} for the starburst and dwarf galaxies of our sample. Two sets of CLOUDY photoionization models (models A in Section \ref{abund}) with log(U)=-2.5 (upper models) and log(U)=-2.0 (lower models) are compared to the data, assuming sulphur depletion at [12 + log(O/H)]$>$ 8, for electron densities in the range of log(n)=[1,4].}\label{fig_ne_sulfur}
\end{figure}

Following the work presented in \citet{fernandez2016, fernandez2017}, we have also considered  the ratio of ${\rm ([NeII]12.8{\mu}m +[NeIII]15.6{\mu}m)}$ to ${\rm ([SIII]18.7{\mu}m +[SIV]10.5{\mu}m)}$ which, as a first approximation, is equal to the total neon to total sulfur ratio if one excludes active galaxies, which can ionize also the [NeV] IR fine structure lines. 
In Fig. \ref{fig_ne_sulfur} this ratio is shown as a function of the metallicity as computed from the optical emission lines.  Starburst galaxies photoionization models (models A in Section \ref{abund}) with a ionization parameter of -2.5$<$log(U)$<$-2.0 and densities 1$<$log(n$_e$/cm$^{-3}$)$<$4 have been included. These models assume sulphur depletion for [12 + log(O/H)]$>$ 8. It appears evident from this figure that there is a correlation between the neon to sulfur ratio, measured by the ${\rm Ne_{23}S_{34}}$ index, and the gas-phase metallicity. In conclusion, we confirm the results of \citet{fernandez2016, fernandez2017} that these line fluxes can be used to measure the metallicity. This is important also because the upcoming mission of the {\it James Webb Space Telescope} (JWST, \citet{gardner2006}) will be able to detect these IR fine-structure lines in the local universe and at low redshift ($z$ $\leq$ 0.8).

\subsection{Metallicities from optical lines versus IR lines }

\begin{figure}
\centering
\includegraphics[width=\columnwidth]{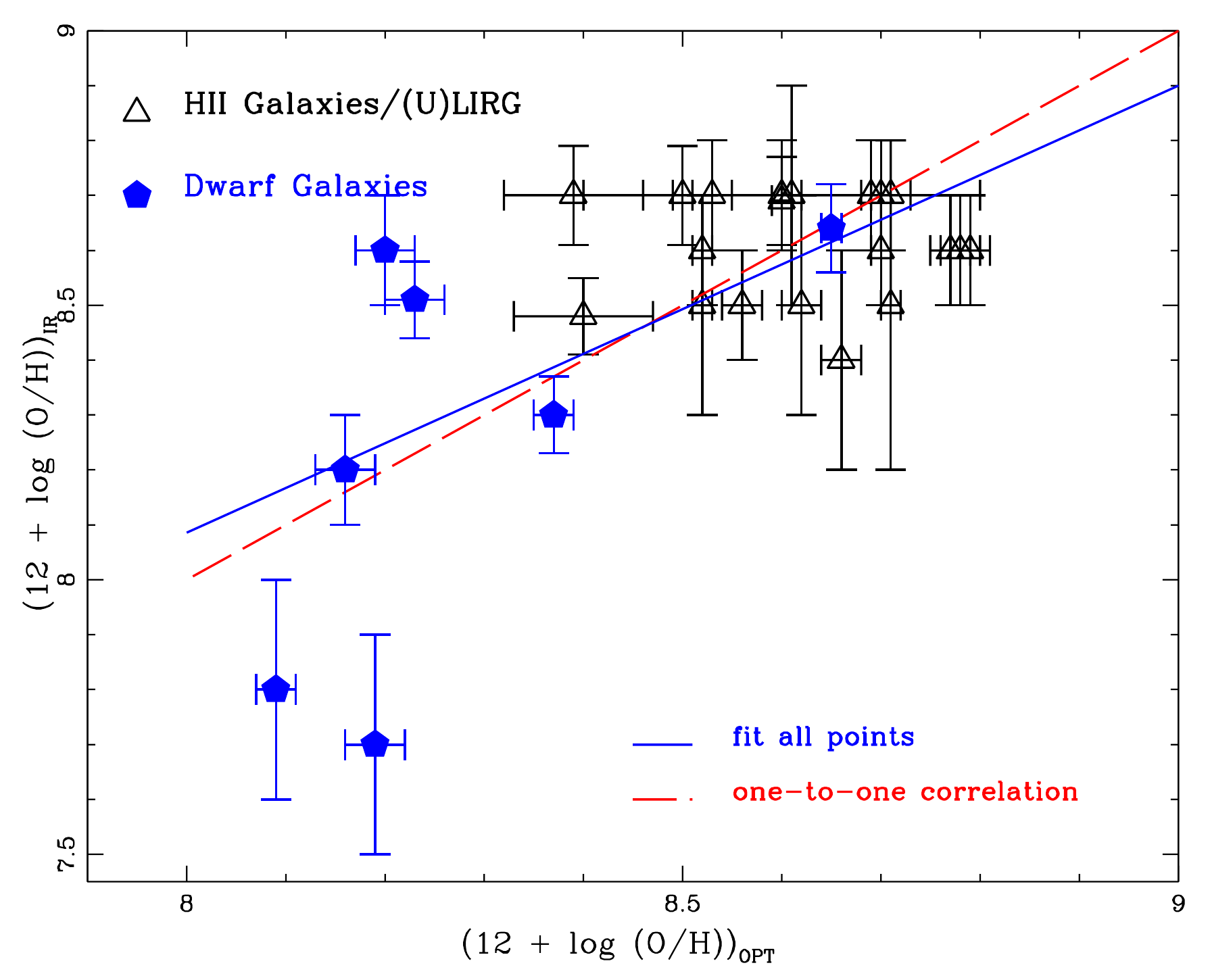}
\caption{Metallicity derived from IR lines, using the method from \citet{fernandez2021}, is compared to the metallicity computed from the optical lines using \citet{perez-montero2014}. The dashed line shows the one-to-one correlation, while the solid line gives the fit: $y = (0.89 \pm 0.15)\cdot x - (0.83 \pm 1.32) (\chi^2=0.53, R=0.77)$.}\label{met_IR_opt}
\end{figure}

In Fig. \ref{met_IR_opt} the metallicity derived from IR lines, using the method from \citet{fernandez2021} is compared to the metallicity computed from the optical lines using \citet{perez-montero2014}. It is shown in the figure a fit to the data, which agrees within the errors with the one-to-one correlation. It follows that, from the data of our sample of galaxies, there is no significant difference between the determination of the metallicity through optical lines and that one through IR lines.

\subsection{N/O ratio from optical lines versus IR lines }

\begin{figure}
\centering
\includegraphics[width=\columnwidth]{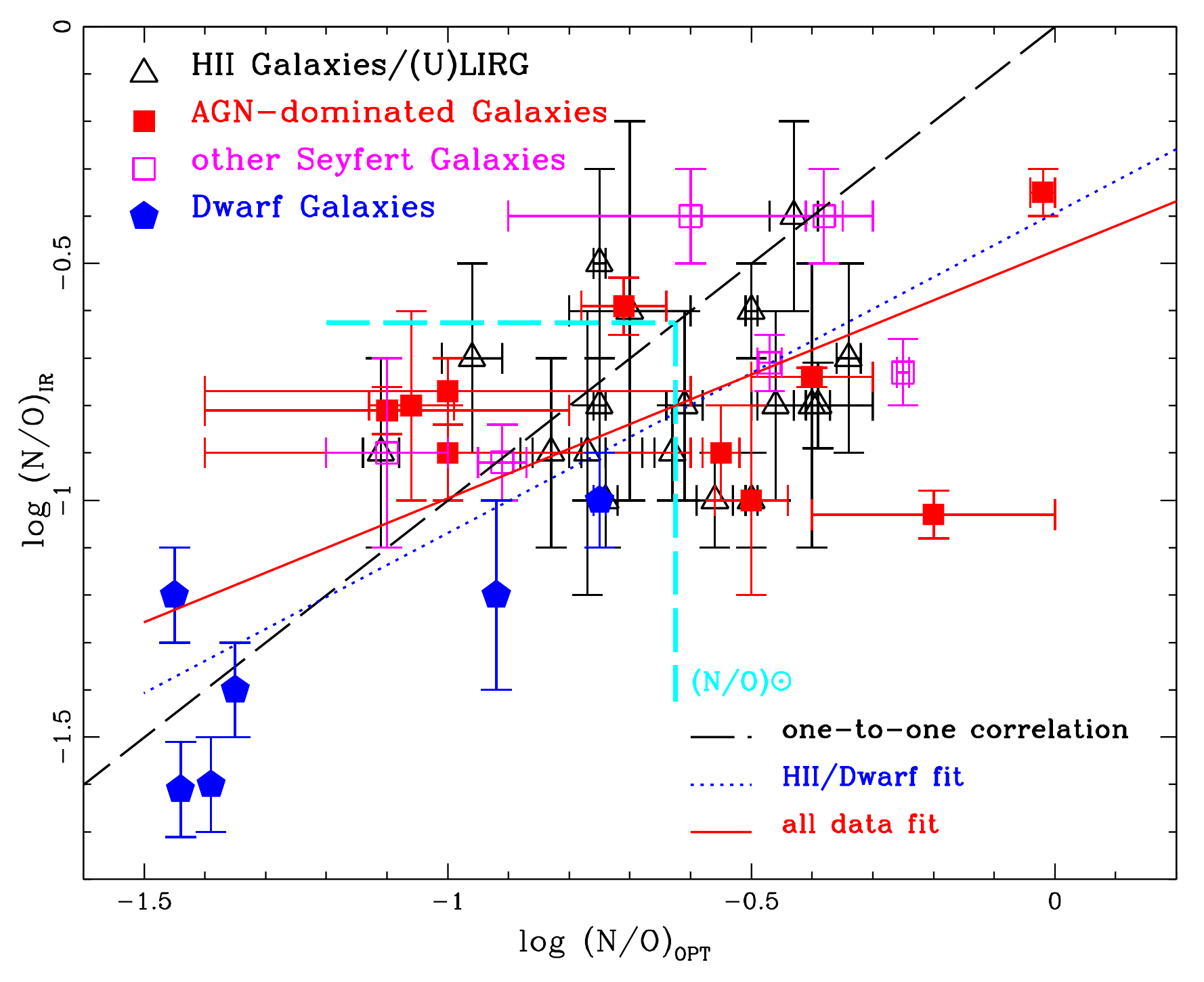}
\caption{N/O ratio derived from IR, using the method from IR lines of \citet{fernandez2021} is compared to the N/O ratio computed from the optical lines using \citet{perez-montero2014}. The dashed line shows the one-to-one correlation, while the dotted line gives the fit for HII/ULIRG and dwarf galaxies: $y = (0.68 \pm 0.12)\cdot x - (0.39 \pm 0.10) (\chi^2=0.93, R=0.76)$. A fit to all the data gives (shown as a solid line): $y = (0.52 \pm 0.10)\cdot x - (0.47 \pm 0.08) (\chi^2=1.84, R=0.64)$. The solar value of ${\rm (N/O)}$=0.24 is also indicated. 
}\label{NO_IR_opt}
\end{figure}

In Fig. \ref{NO_IR_opt} the (N/O)$_{IR}$ ratio derived from IR lines, using the method of \citet{fernandez2021} is compared to the (N/O)$_{OPT}$ ratio computed from the optical lines using the work of \citet{perez-montero2014}. It appears from this comparison that the N/O ratio computed from the IR lines is on average lower than that one derived from optical lines, especially at high values of N/O
(N/O$>$-1). 
The solar value measured with optical lines (N/O)$_{\rm OPT}$ ($\odot$) $\sim$  0.24 corresponds, according to the fit, to an IR determined value of (N/O)$_{\rm IR}$($\odot$) $\sim$  0.12, i.e. a factor two lower. 
A least squares fit to the data results in a slope much flatter than the value of $\alpha$ = 1. Considering all star forming galaxies, i.e. the HII galaxies and ULIRGs together with the dwarf galaxies, the fit gives a slope of  $\alpha$ = 0.68$\pm$0.12, while using all the galaxies of our sample the slope is $\alpha$ = 0.52$\pm$0.10.

\subsection{N/O abundance ratio}

\begin{figure}
\centering
\includegraphics[width=\columnwidth]{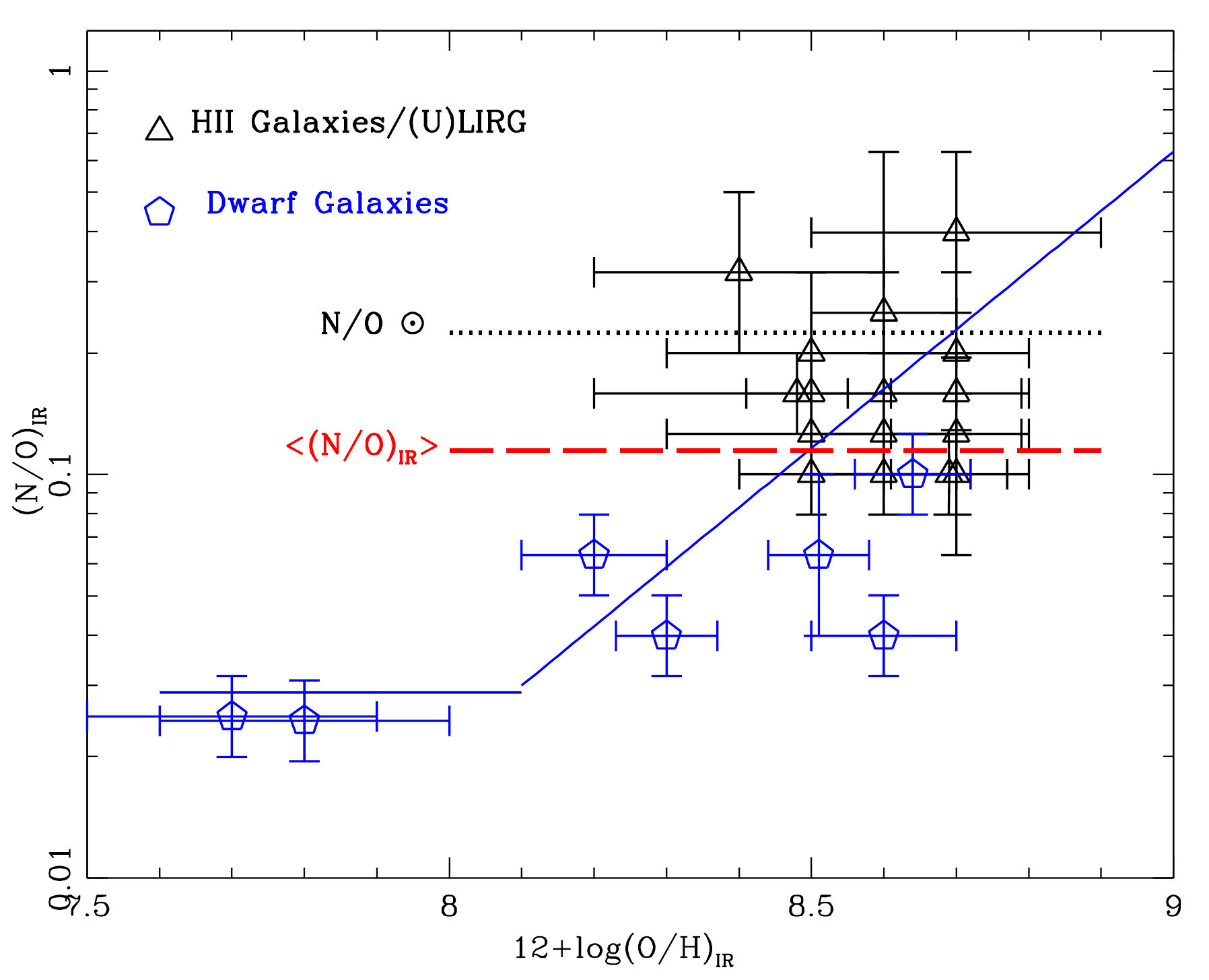}
\caption{N/O ratio computed from IR emission lines versus the metallicity similarly computed. The solar value of N/O = 0.24 has been indicated as well as the average value of the log(N/O)$_{IR}$=-0.94$\pm$0.30. The \citet{pilyugin2014} relation between N and O is shown as a solid line, for comparison, which assumes log(N/O) = -1.54 for: [12+log(O/H)]$<$ 8.1. and log(N/O) = 4.21 + 1.47$\times$[log(O/H)] for: [12+log(O/H)]$>$ 8.1.}\label{met_NO_IR}
\end{figure}

In Fig. \ref{met_NO_IR} the N/O ratio, computed from the IR emission lines using the method developed by \citet{fernandez2021} and reported in Table \ref{tab:sample1}, is plotted as a function of the metallicity, computed also from the IR emission lines. 

\begin{figure}
\centering
\includegraphics[width=\columnwidth]{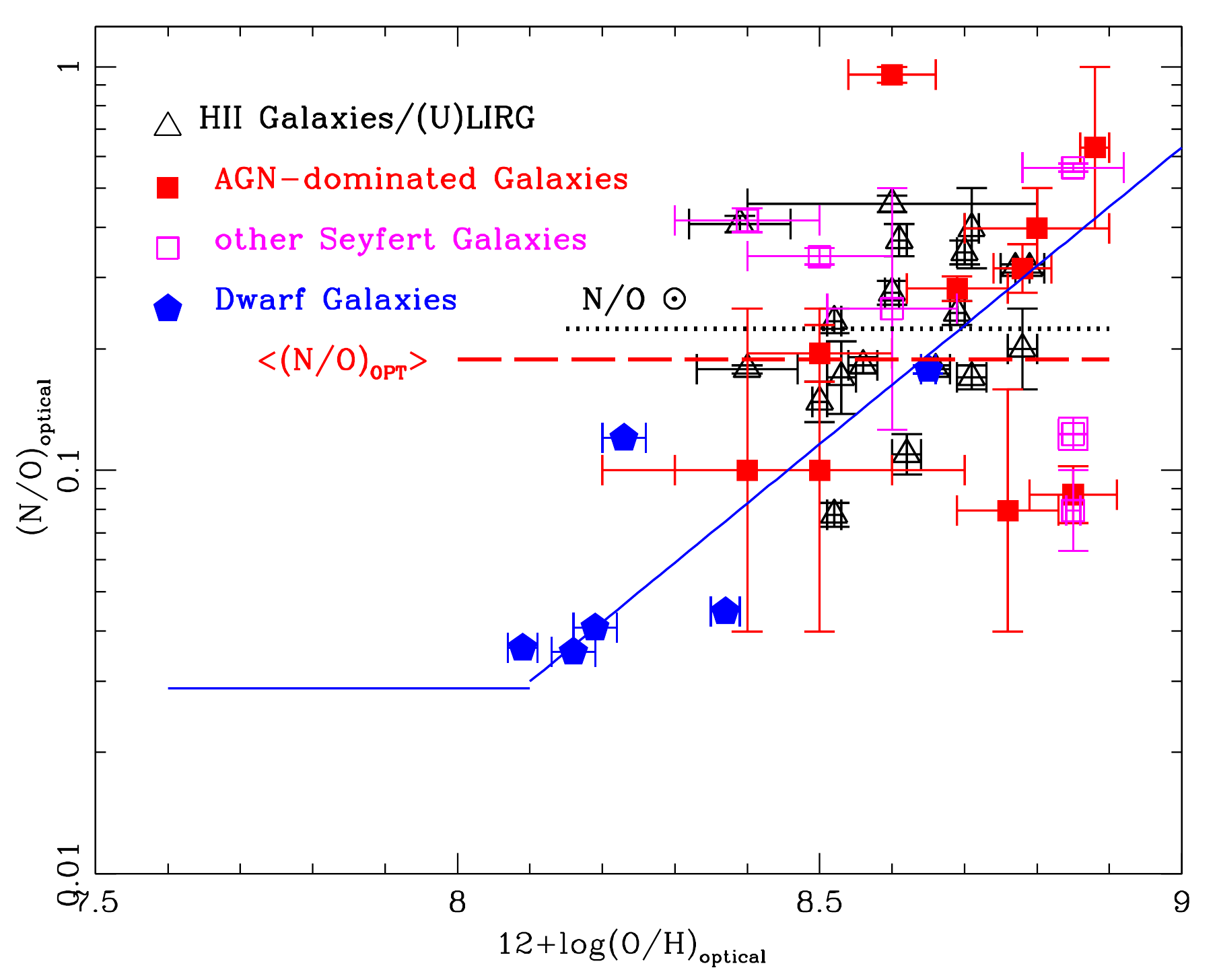}
\caption{N/O ratio computed from optical emission lines versus the metallicity similarly computed. The solar value of N/O = 0.24 has been indicated as the average value of the log(N/O)$_{OPT}$=-0.73$\pm$0.34. The \citet{pilyugin2014} relation between N and O is shown as a solid line, for comparison, as given in the caption of Fig.\,\ref{met_NO_IR}. 
}\label{met_no_opt}
\end{figure}

In Fig. \ref{met_no_opt} the N/O ratio, computed from the optical emission lines using the method developed by \citet{perez-montero2014} and reported in Table \ref{tab:sample1}, is plotted as a function of the metallicity, computed also from the optical emission lines.

If we compare the average value of the log(N/O)$_{IR}$ = -0.73$\pm$0.34 (Fig.\ref{met_NO_IR}) with the average value of the log(N/O)$_{OPT}$ = -0.94$\pm$0.30 (Fig.\ref{met_no_opt}), we see a large difference of $\sim$0.2 dex. 



\begin{figure}
\centering
\includegraphics[width=\columnwidth]{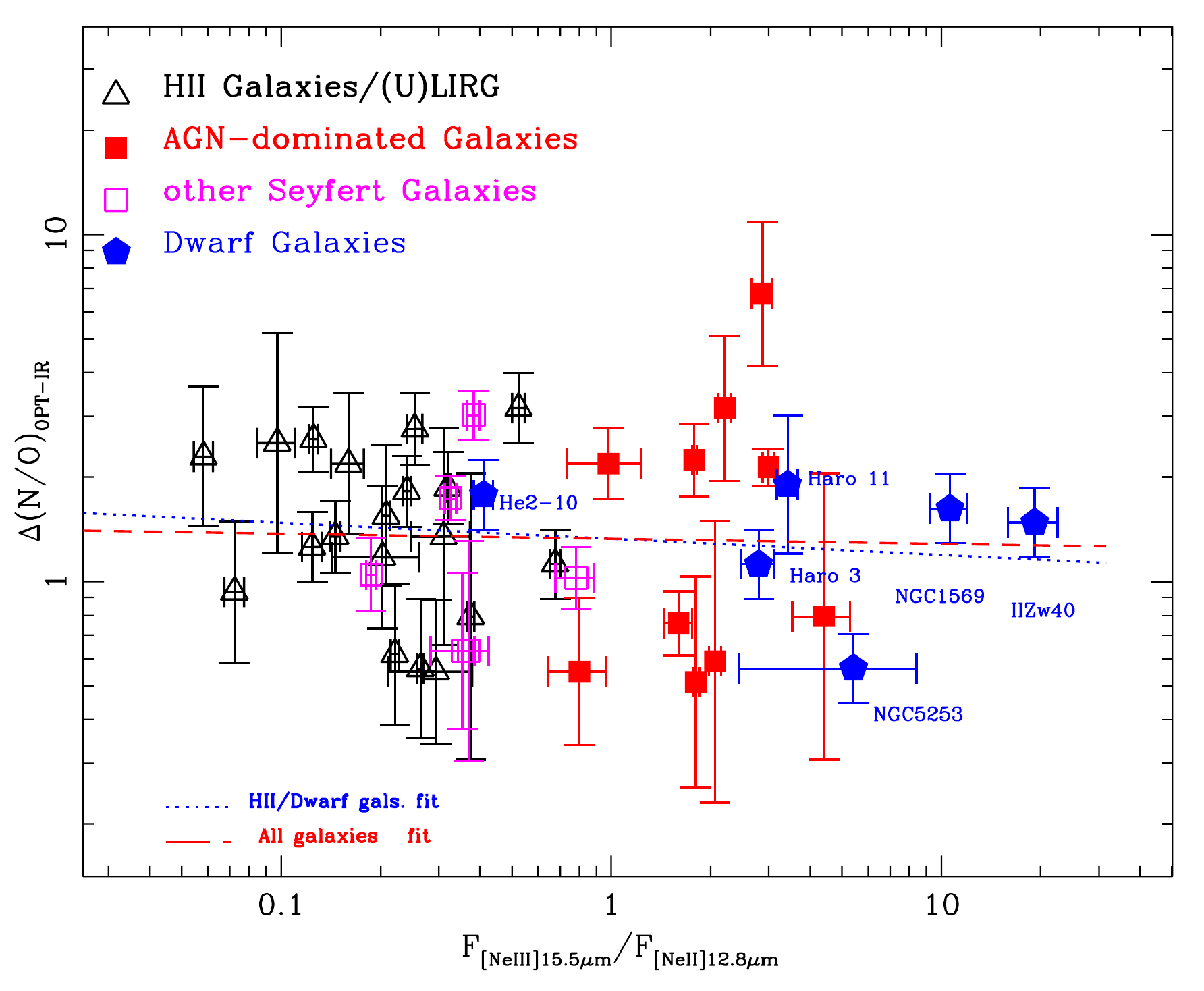}
\caption{Logarithmic difference between the N/O ratio computed from IR emission lines and the N/O from optical lines versus the [NeIII]15.5$\mu$m/[NeII]12.8$\mu$m line ratio, which  measures the gas ionization. The dotted line gives the fit for HII/ULIRG and dwarf galaxies: $y = (-0.05 \pm 0.07)\cdot x +(0.12 \pm 0.05) (\chi^2=1.19, R=-0.14)$. A fit to all the data gives (shown as a dashed line): $y = (-0.01 \pm 0.07)\cdot x +(0.12 \pm 0.05) (\chi^2=2.92, R=-0.03)$.
}\label{deltaNO_vs_ne}
\end{figure}

In the following sections we study the possible dependence of the $\Delta$(N/O) with ionization, density and extinction.

\subsubsection{$\Delta$(N/O) versus ionization}\label{extra_ion}

\begin{figure}
\centering
\includegraphics[width=\columnwidth]{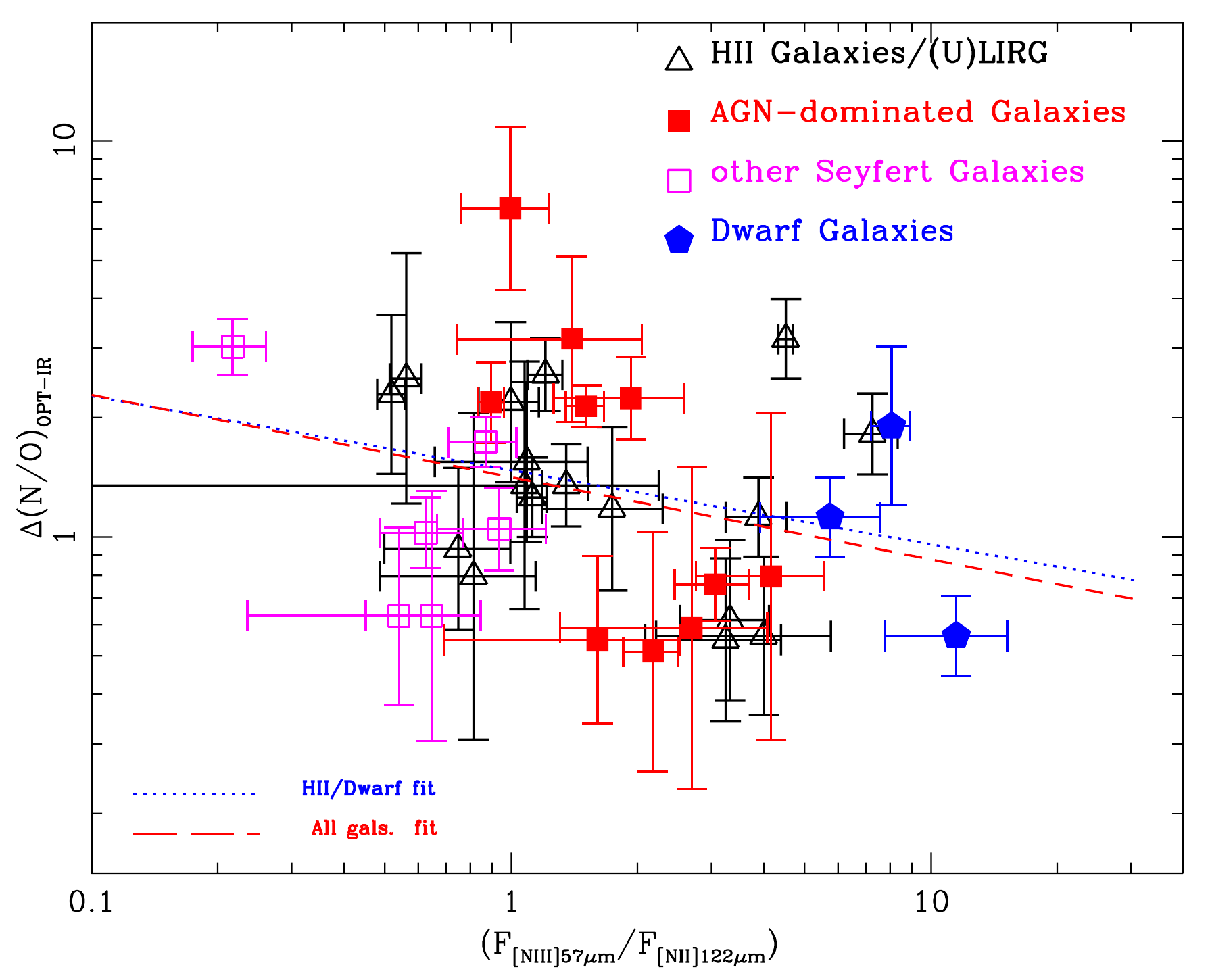}
\caption{Logarithmic difference between the N/O ratio computed from IR emission lines and the N/O from optical lines versus the [NIII]57$\mu$m/[NII]122$\mu$m line ratio, which measures the gas ionization.
The dotted line gives the fit for HII/ULIRG and dwarf galaxies: $y = (-0.18 \pm 0.13)\cdot x +(0.16 \pm 0.06) (\chi^2=0.97, R=-0.31)$. A fit to all the data gives (shown as a dashed line): $y = (-0.21 \pm 0.12)\cdot x +(0.15 \pm 0.05) (\chi^2=2.54, R=-0.29)$.
}\label{deltaNO_vs_N}
\end{figure}

Fig.\,\ref{deltaNO_vs_ne} shows the logarithmic difference between the N/O ratio computed from IR emission lines (N/O)$_{\rm IR}$ and the N/O from optical lines (N/O)$_{\rm OPT}$ (hereafter $\rm \Delta (N/O)= (N/O)_{OPT}-(N/O)_{IR}$) versus the [NeIII]15.5$\mu$m/[NeII]12.8$\mu$m line ratio. In this diagram a slight decreasing trend could be present due to the higher $\Delta$(N/O) values in HII galaxies and (U)LIRGs, albeit the correlation is not statistically significant. Dwarf galaxies are in average consistent with an equal value for the (N/O) and Seyfert galaxies have a high dispersion. A similar diagram is shown in Fig.\,\ref{deltaNO_vs_N}, using the [NIII]57$\mu$m/[NII]122$\mu$m line ratio, however the scatter is larger here due to the lower ionization potential of these lines, which results in a larger confusion along the horizontal axis.

\begin{figure}[ht!]
\centering
\includegraphics[width=\columnwidth]{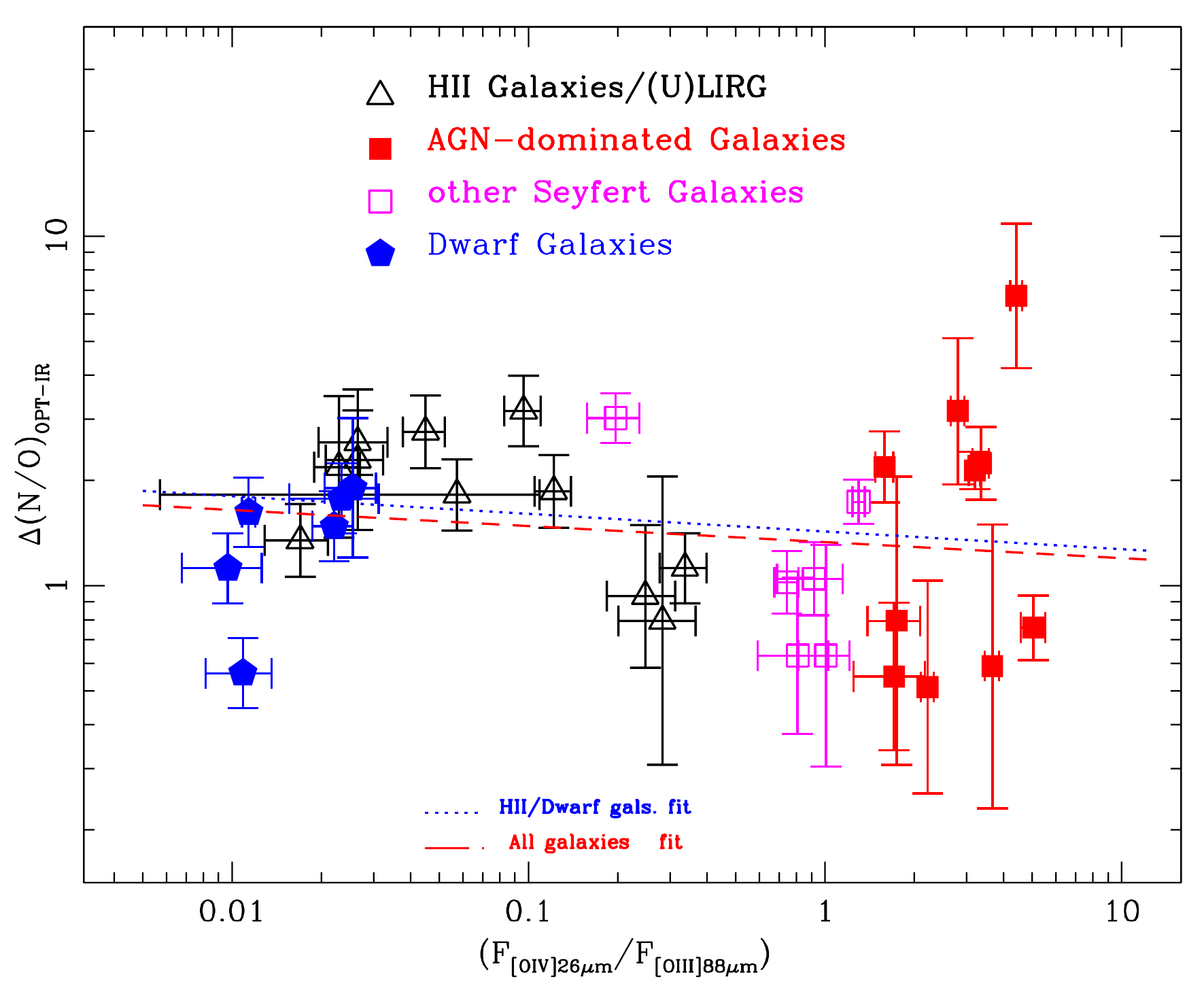}
\caption{Logarithmic difference between the N/O ratio computed from IR emission lines and the N/O from optical lines versus the [OIV]26$\mu$m/[OIII]8$\mu$m line ratio, which measures the gas ionization.
The dotted line gives the fit for HII/ULIRG and dwarf galaxies: $y = (-0.04 \pm 0.10)\cdot x +(0.14 \pm 0.15) (\chi^2=0.64, R=-0.10)$. A fit to all the data gives (shown as a dashed line): $y = (-0.05 \pm 0.05)\cdot x +(0.12 \pm 0.06) (\chi^2=2.36, R=-0.15)$.
}\label{deltaNO_vs_O}
\end{figure}

We also present in Fig.\ref{deltaNO_vs_O} the $\rm  \Delta(N/O)$ plotted as a function of the [OIV]26$\mu$m/[OIII]88$\mu$m line ratio, which covers the largest range in ionization and almost three orders of magnitude in line ratio value from low metallicity dwarf galaxies to AGN. Also in this case the search for a correlation fails because of the large spread of the data, however a slight trend of increasing $\rm \Delta(N/O)$ at low gas ionization remains for HII galaxies and (U)LIRGs.

\subsubsection{$\Delta$(N/O) versus density}\label{extra_dens}

\begin{figure}
\centering
\includegraphics[width=\columnwidth]{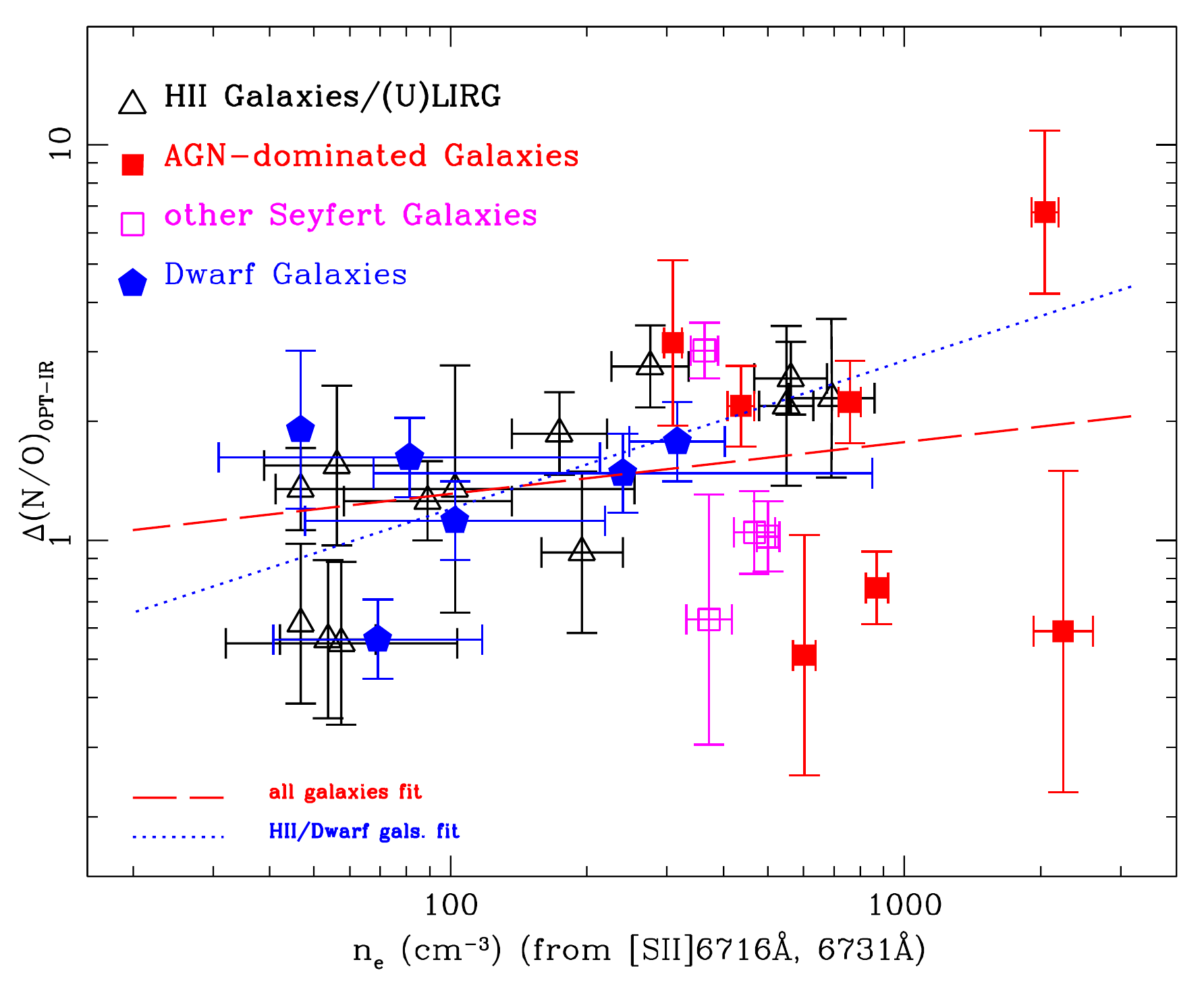}
\caption{Logarithmic difference between the N/O ratio computed from IR emission lines and the N/O from optical lines versus the electron density of the gas, as measured from the optical [SII]${\lambda}{\lambda}$6716,6731 doublet.
The dotted line gives the fit for HII/ULIRG and dwarf galaxies: $y = (0.37 \pm 0.10)\cdot x -(0.67 \pm 0.22) (\chi^2=0.53, R=0.66)$. A fit to all the data gives (shown as a dashed line): $y = (0.13 \pm 0.09)\cdot x -(0.14 \pm 0.23) (\chi^2=2.63, R=0.25)$.}\label{deltaNO_vs_SIIdens}
\end{figure}

In Fig. \ref{deltaNO_vs_SIIdens} we explore the dependence of the $\rm  \Delta(N/O)$ with the electron density of the gas as measured from the optical [SII]{$\lambda$}{$\lambda$}6716,6731 doublet. One can see from this figure that the star forming galaxies (HII region/ULIRGs and dwarf galaxies) show an increasing correlation between the $\rm  \Delta(N/O)$ and the electron density, showing the highest value of the $\rm  \Delta(N/O)$ for the highest densities. The inclusion of the Seyfert galaxies breaks the correlation, probably due to the presence of gas in the Narrow Line Region (NLR) of these galaxies.

We refer to Appendix \ref{extra_dens} for two similar diagrams of the $\rm  \Delta(N/O)$ as a function of the electron density as derived from the [SIII]18.7$\mu$m and 33.5$\mu$m lines and from the [OIII]52$\mu$m and 88$\mu$m lines.
Also using as density tracers the IR fine-structure lines of [SIII] and [OIII], we do not find any correlation between the $\Delta$(N/O) and the electron density. This also demonstrates that there is no influence of the lower critical density of IR fine-structure lines as compared to the optical lines.

\subsubsection{$\Delta$(N/O) versus extinction}\label{extra_ext}

\begin{figure}
\centering
\includegraphics[width=\columnwidth]{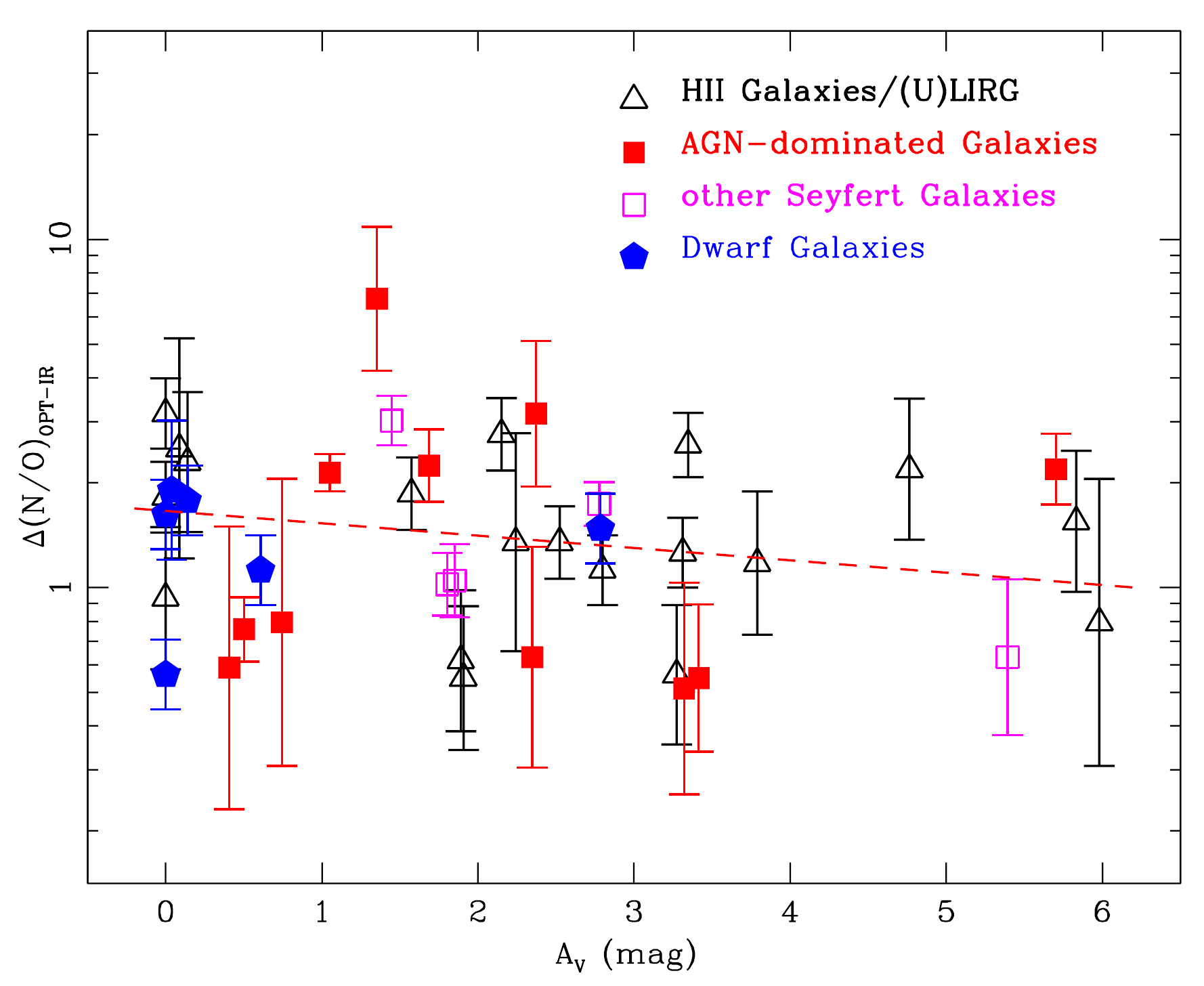}
\caption{Logarithmic difference between the N/O ratio computed from IR emission lines and the N/O from optical lines versus the galaxy extinction A$_V$ in mag. 
The dashed line gives the fit all galaxies: $y = (-0.03 \pm 0.02)\cdot x +(0.22 \pm 0.07) (\chi^2=3.51, -R=0.21)$.
}\label{deltaNO_vs_ext}
\end{figure}

We explored if the difference between the optical and IR determination of the (N/O) ratio could be due to an inaccurate extinction correction of the optical lines used to derive the optical (N/O) ratio. If this were the case, we would expect that a higher value of the $\rm  \Delta(N/O)$ value would originate in galaxies affected by an higher dust extinction.
In Fig.\ref{deltaNO_vs_ext} we have plotted the $\rm  \Delta(N/O)$ value as a function of the optical extinction A$_V$ in magnitudes. It can be seen from the figure that no correlation is apparent and the trend shows a decreasing $\rm  \Delta(N/O)$ value as a function of extinction. Therefore we can rule out the hypothesis that this is due to extinction.

\subsubsection{Density corrected N/O abundance ratio}

\begin{figure}
\centering
\includegraphics[width=\columnwidth]{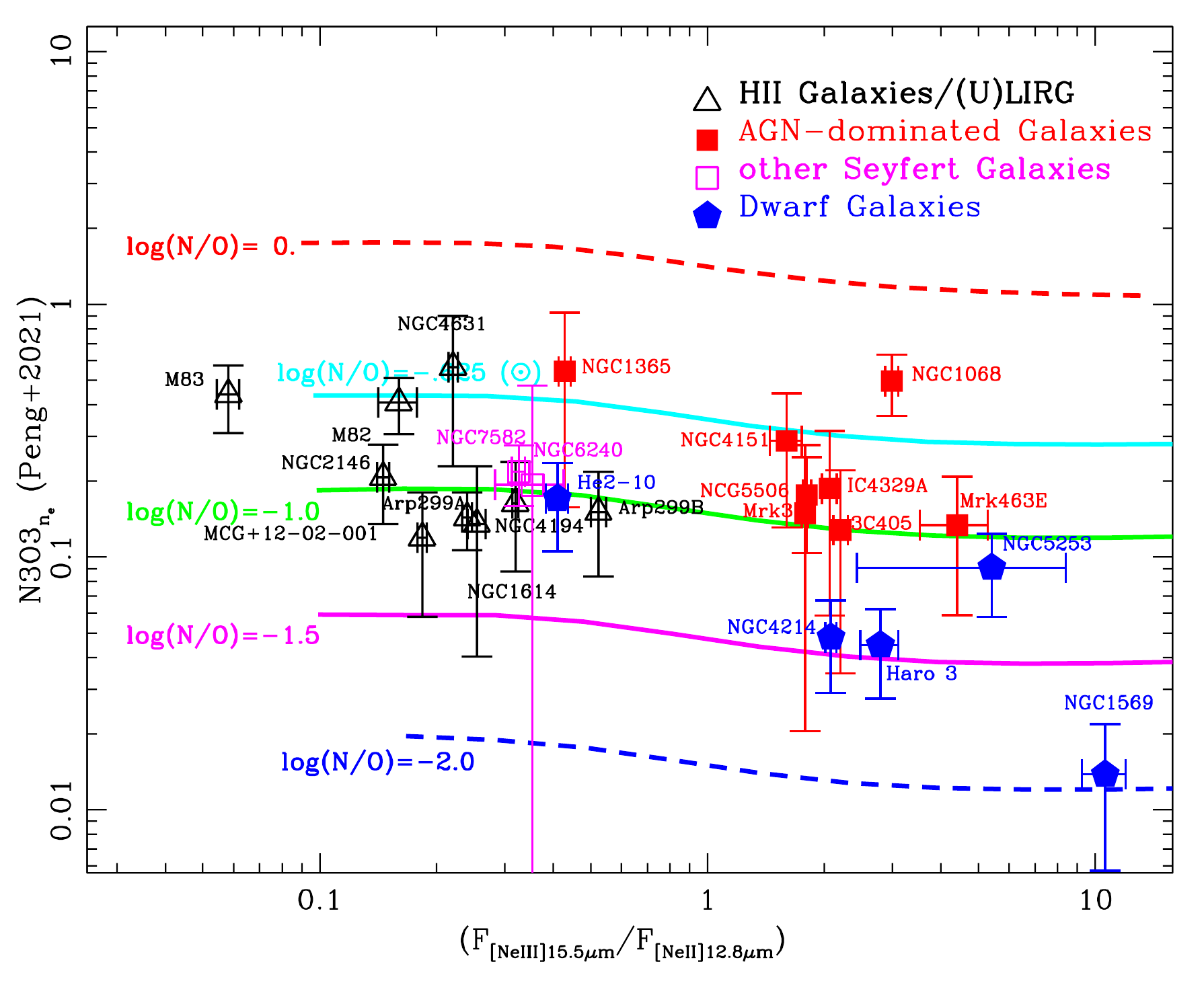}
\caption{Density-corrected N3O3$n_{e}$ \citep{peng2021} of the sample galaxies as a function of the [NeIII]15.5$\mu$m/[NeII]12.8$\mu$m line ratio. The galaxies are divided in the three classes, HII region galaxies/(U)LIRGs, Seyfert galaxies and Dwarf Galaxies and are labeled with their names. The horizontal lines give the positions of constant N/O ratio as computed from photoionization models.}\label{N3O3}
\end{figure}

\begin{figure}
\centering
\includegraphics[width=\columnwidth]{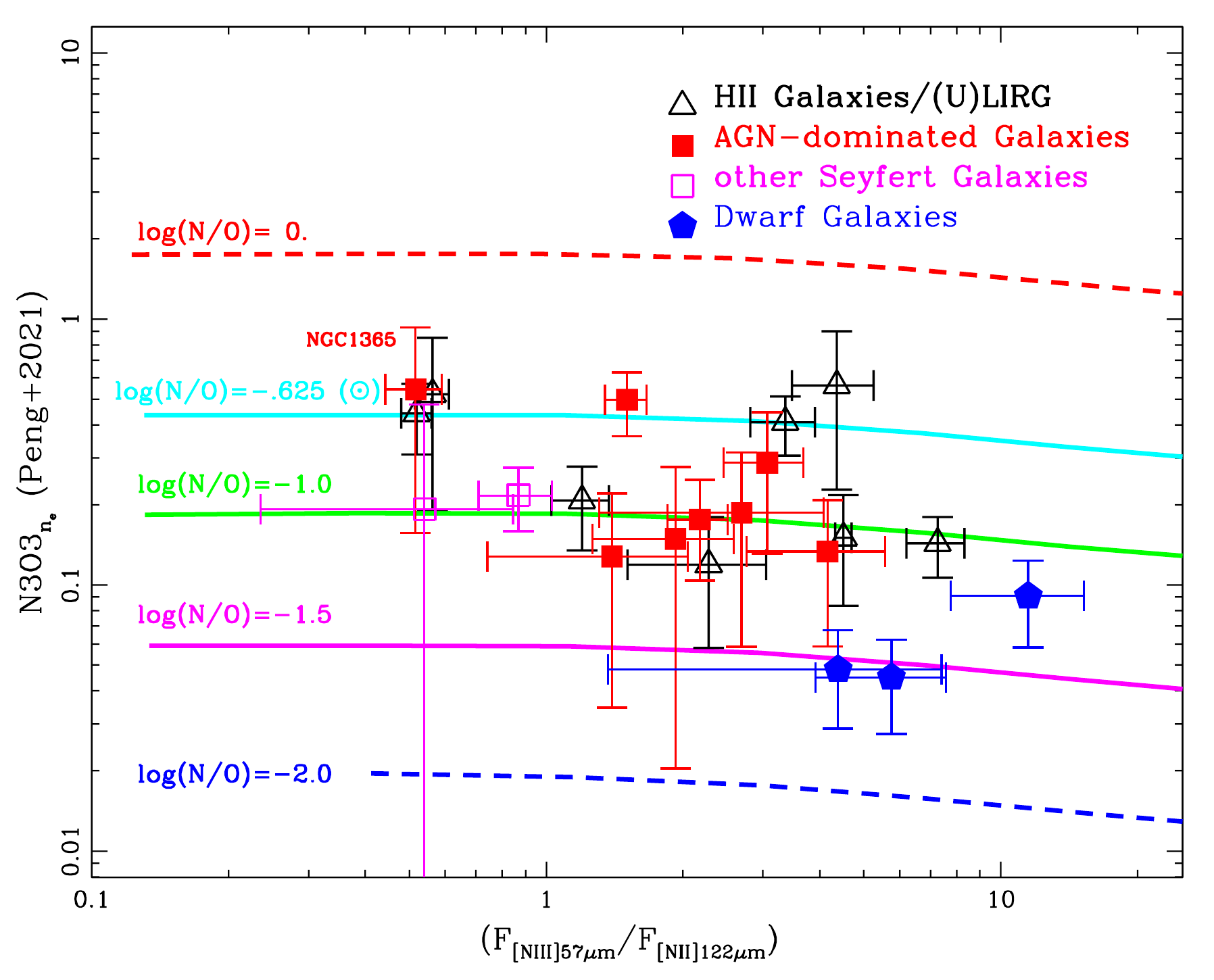}
\caption{Density-corrected N3O3$n_{e}$ \citep{peng2021} of the sample galaxies as a function of the [NIII]57$\mu$m/[NII]122$\mu$m line ratio. The galaxies are divided in the three classes, HII region galaxies/(U)LIRGs, Seyfert galaxies and Dwarf Galaxies. The horizontal lines give the positions of constant N/O ratio as computed from photoionization models.}\label{N3O3_2}
\end{figure}

Recently, \citet{peng2021} have presented the results of the analysis of the SOFIA observations of the far-IR lines of [NIII]57$\mu$m and [OIII]52$\mu$m of 8 galaxies, complemented by {\it Herschel}-PACS and {\it Spitzer}-IRS observations of [OIII]52$\mu$m  and [NeII]12.8$\mu$m and [NeIII]15.5$\mu$m,  respectively. They show that the [NIII]57$\mu$m/[OIII]52$\mu$m line ratio, denoted N3O3, is a physically robust probe of N/O, is insensitive to gas temperature and only weakly dependent on electron density. They also use the observations of the two lines of the [OIII]52$\mu$m and 88$\mu$m to correct this ratio for the effects of density. We have therefore used this work to derive the density corrected N3O3$_{n_e}$ ratio for our sample of galaxies, and confirm with an independent method our results. 
The value of N3O3$_{n_e}$ has been derived using the equations (2) and (4) of \citet{peng2021}.
We present in Figure \ref{N3O3} the N3O3$_{n_e}$ ratio as a function of the [NeIII]15.5$\mu$m/[NeII]12.8$\mu$m line ratio for each galaxy for which we have the observations of the fine-structure lines of [OIII]52 and 88$\mu$m, [NIII]57$\mu$m, [NeII]12.8$\mu$m and [NeIII]15.5$\mu$m.
In Fig. \ref{N3O3_2} we show the same N3O3$_{n_e}$ ratio as a function of the [NIII]57$\mu$m/[NII]122$\mu$m line ratio. Together with the data points of our sample of galaxies, we also show in these figures the lines of constant N/O ratio (log(N/O)=[-2, 0.]) in analogy to the Fig. 3 of \citet{peng2021}. The lines at fixed (N/O) ratio have been taken from CLOUDY photoionization models for stellar populations described in section \ref{abund}, adopting a solar O/H abundance, an ionization parameter in the range of log(U)=-4.0, -1.5 and electron density fixed at the value of log(n$_e$)=2.0 (cm$^{-3}$). 

\section{DISCUSSION}\label{discussion}


One of the main results in this study is the difference found between optical and IR determinations of the N/O abundance in galaxies. Although the scatter in Fig.\,\ref{NO_IR_opt} is relatively large, a systematic difference is observed for those galaxies with N/O abundances above $\gtrsim -0.6\, \rm{dex}$ determined from the optical lines. These galaxies, including both star-forming galaxies and AGN, show IR-based N/O abundances which are about a factor 2-3 lower when compared with the optical estimates, in agreement with the results obtained by \citet{peng2021} for a sample of 8 galaxies. This difference is in contrast with the overall agreement found between the optical and IR O/H abundances determinations in Fig.\,\ref{met_IR_opt}. Optical and IR methods are consistent within $\sim 0.2\, \rm{dex}$ scatter when O/H abundances are derived \citep{fernandez2021}, but differ in the N/O values obtained.

Among the different possibilities to explain this discrepancy we investigated three main scenarios that could affect the optical line tracers: differences in the ionization structure, contamination by Diffuse Ionized Gas (DIG; e.g. \citealt{vale2019}), and dust extinction. None of these hypothesis seem to explain the observed discrepancy. A difference in the ionization structure could result in lower N/O ratios for the IR tracers, because these lines probe higher ionization gas (O$^{2+}$ and N$^{2+}$) located closer to massive stars, where the primary production of nitrogen could be prevalent, resulting in lower N/O when compared to the optical lines that trace lower ionization gas (O$^+$ and N$^+$) with a higher nitrogen enrichment from a secondary origin. However Figs.\,\ref{deltaNO_vs_ne} and \ref{deltaNO_vs_N} show a flat distribution when the ratio between N/O optical and IR estimates is compared with the strength of the radiation field traced by the [NeIII]15.6$\mu$m/[NeII]12.8$\mu$m and [NIII]57$\mu$m/[NII]122$\mu$m ratios, respectively. That is, the star-forming galaxies with a strong radiation field (i.e. low-metallicity dwarf galaxies) are scattered around optical-to-IR N/O ratios of $\sim 1$ in Fig.\,\ref{deltaNO_vs_ne}, while solar-like starburst galaxies with the weakest radiation fields are tentatively shifted to ratios around $\sim2$.

An alternative explanation is the contamination by DIG affecting the optical lines in these galaxies. The presence of hot and low-density gas in the ISM was proposed by \citet{peng2021} as a possible cause of the optical-to-IR discrepancy. In this scenario, the DIG contamination would  be stronger for the lower excitation species of O$^+$ and N$^+$. As a matter of fact, the DIG emission can
account for $\sim 30\%$ of the optical line fluxes of the [OII]$\lambda \lambda 3727,3729$ and the [NII]$\lambda \lambda 6548,6584$ doublets.  As opposite, the DIG contamination on the higher ionization transitions in the IR would be negligible. Additionally, this difference would be enhanced for large galaxies with an old stellar population that could have enriched the DIG with secondary nitrogen. We test this hypothesis by comparing the N/O optical-to-IR ratio with the gas density derived from the [SII]$\lambda \lambda 6716,6731$ doublet in Fig.\,\ref{deltaNO_vs_SIIdens}. The sulfur doublet probes the gas density in the same ISM domain where [NII]$\lambda \lambda 6548,6584$ lines are produced, and therefore it should be sensitive to the presence of DIG due to the low densities associated with this component. However, no correlation is found in Fig.\,\ref{deltaNO_vs_SIIdens}, meaning that the discrepancies in the  $\Delta$(N/O) are not associated with a low-density gas component. Additionally, excluding the sulfur lines in the abundance computation with \textsc{HCm} does not result in significant differences in the N/O ratios obtained, as would be expected for DIG contamination affecting the sulfur line fluxes \citep[e.g.][]{pilyugin2018}. Thus, we conclude that DIG contamination does not significantly affect the optical line fluxes.

The third scenario is the possible effect of dust obscuration on the optical-based determinations. Although optical line fluxes were corrected by extinction estimated from the observed Balmer decrement values, uncorrected obscuration affecting the [OII]$\lambda \lambda 3727,3729$ doublet would particularly affect the optical N/O abundances, since the nitrogen lines are less affected by dust attenuation. This possibility is investigated in Fig.\,\ref{deltaNO_vs_ext}, where the {\rm $\Delta$(N/O)} is compared with the optical extinction $A_{\rm V}$ values derived from the Balmer decrement. No significant trend is observed, suggesting that uncorrected residual extinction cannot explain the differences seen in the N/O abundances.

\begin{figure}[ht!]
\centering
\includegraphics[width=\columnwidth]{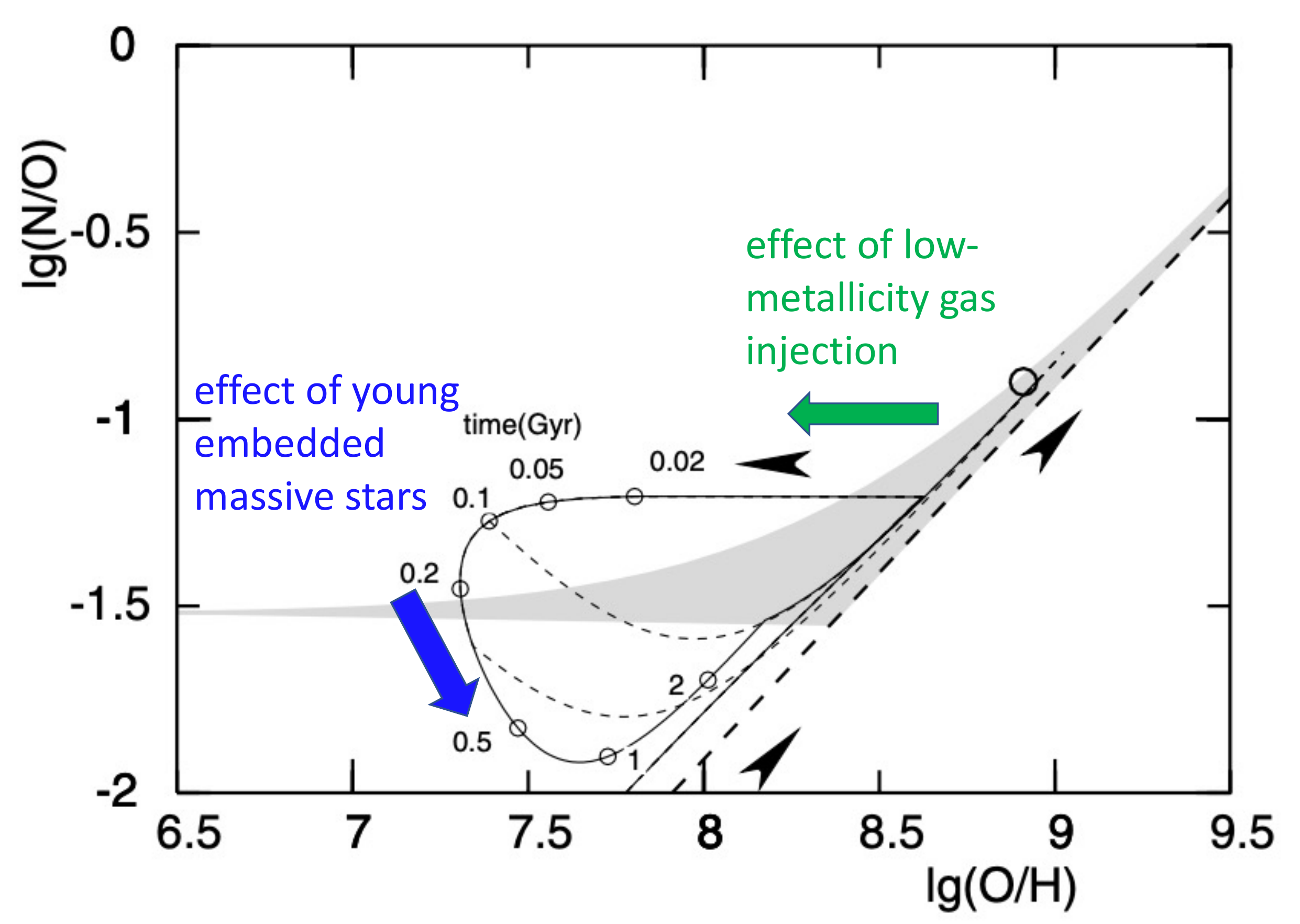}
\caption{The time evolution of the extreme model from \citet{koppen2005} with an infall rate of 100 galaxy masses per Gyr, starting at an age of 5 Gyr. The upper arrow indicated the effect of low-metallicity gas injection, while the lower arrow the effect of young embedded massive stars. The decrease in the N/O ratio (downward arrow) is due to the primary production of young massive stars, preferentially producing oxygen compared to nitrogen. Adapted from fig.\,3 of \citet{koppen2005}.}\label{kopper}
\end{figure}


An alternative explanation, more speculative, would imply that optical transitions are completely missing the embedded star forming regions for the galaxies in our sample. Consequently, the optical lines would trace preferentially the gas in more external regions, which would be more exposed to enrichment with secondary nitrogen and biased against dust embedded massive stars. As opposite, the gas traced by the IR transitions would be associated to lower N/O ratios due to the contribution from massive stars. In this scenario, the differences between optical and IR N/O abundances in galaxies would be a consequence of the chemical evolution process that may follow episodes of low-metallicity gas accretion from the circumgalactic medium or the external disk of the galaxy. This process has been modeled by \citet{koppen2005}, showing that the rapid decrease in oxygen abundance after the gas infall is followed by a sharp decrease in the N/O ratio due to the yields of the newly formed massive stars. The O/H and N/O increase at a later time due to the ongoing star formation and the secondary production of nitrogen from intermediate-mass stars, closing the loop in the N/O--O/H diagram (see Fig.\,\ref{kopper}). As a matter of fact, evidence of gas accretion affecting the N/O ratios in the disks of nearby star forming galaxies has been recently discovered \citep{luo2021}. Local variations in the star formation efficiency could also play a role to explain the different N/O ratios observed \citep{molla2006,florido2015,kumari2018}. If confirmed, this result would imply that the N/O ratios measured with IR lines are more genuine than those in the optical, since they reflect the chemical abundances in the ISM gas phase including also the obscured regions not detected by optical tracers.

Although the cause of the optical-to-IR discrepancy is still unknown, we favor the abundances derived from the IR lines because these are virtually independent of dust and temperature effects. The optical lines emissivities present a strong dependency with the gas temperature (e.g. fig.\,1 in \citealt{fernandez2021}) and could be affected by, e.g., inhomogeneities in the gas nebula \citep{peimbert1967}, while the IR lines can provide more robust abundance estimates \citep[e.g.][]{vermeij2002,dors2013}. On the other hand, our IR N/O determinations based on photoionization models are in agreement with the values obtained using the density-corrected N3O3$n_e$ parameter introduced by \citet{peng2021}, as shown in Figs.\,\ref{N3O3} and \ref{N3O3_2}. The optical-to-IR abundance discrepancy analyzed in this work warns on the use of optical tracers to study the N/O abundances in galaxies at high redshift, e.g. using future \textit{James Webb Space Telescope} observations of their restframe optical spectrum. These should be compared with measurements derived from the far-IR transitions redshifted in the submm range and observed with ALMA or future facilities in this spectral range, when available.

\section{SUMMARY}\label{sum}
We summarize here the main results of this paper.
\begin{enumerate}
    \item We have reduced and analyzed new and archival spectra of the [OIII]52$\mu$m and [NIII]57$\mu$m lines in 29 galaxies collected by the SOFIA FIFI-LS spectrometer. 
    \item Using literature data from {\it Herschel}-PACS and ISO-LWS, we have assembled a sample of 47 galaxies, including 21 HII region/(U)LIRG galaxies, 19 Seyfert galaxies, including one LINER and 7 dwarf galaxies,  for which we have full coverage of the [NIII]57$\mu$m line and at least one of the two [OIII] lines (at 52$\mu$m or 88$\mu$m).
    \item We have complemented the [NIII] and [OIII] far-IR lines observations of this sample with the mid- and far-IR lines useful for the characterization of the sample and the analysis carried out in this work aimed at the determination of the metallicity using IR spectra. In particular we used the {\it Spitzer}-IRS and the {\it Herschel}-PACS spectra where available for the galaxies of the sample. 
    \item We have exploited the use of various line ratios, namely the (2.2 $\times$ [OIII]88$\mu$m + [OIII]52$\mu$m)/[NIII]57$\mu$m ratio, the [OIII]88$\mu$m/[NIII]57$\mu$m ratio, the [OIII]88$\mu$m/[NII]122$\mu$m ratio, and the  ([NeII]12.8$\mu$m +[NeIII]15.6$\mu$m) to  ([SIII]18.7$\mu$m + [SIV]10.5$\mu$m) composite ratio, to map with the use of photoionization models their dependence with the metallicity, expressed as [12 + log(O/H)]. In other words, we assessed the use of these line ratios to measure the metallicity in galaxies.
    \item We find that the determination of the (O/H) ratio with optical emission lines is consistent, within the errors, with its determination through IR fine-structure lines.
    \item We find, as opposite, that the determination of the (N/O) abundance through optical lines is significantly different from its determination from the IR lines, especially for high values of (N/O) (N/O$\gtrsim$ -0.8).
    \item We explored if a difference in the ionization structure mapped by the IR lines, which trace higher ionization gas (O$^{2+}$ and N$^{2+}$) with respect to the one mapped in the optical (O$^{+}$ and N$^{+}$), could result in lower   N/O ratios for the IR lines. However we do not find a significant correlation between the $\rm  \Delta(N/O)= (N/O)_{OPT}-(N/O)_{IR}$ and the ionization indices. 
    \item We explored the possibility that the presence of diffuse ionized gas (DIG) in the ISM of galaxies could be responsible for this difference, by searching correlations between the $\rm {\Delta(N/O)}$ and the electron density as measured from different tracers (the [SII]{$\lambda$}{$\lambda$}6716,6731 lines, the [SIII]18$\mu$m and 33$\mu$m lines and the [OIII]52$\mu$m and 88$\mu$m lines), but we do not find any statistically significant correlation.
    \item We also searched the correlation between the $\rm  \Delta(N/O)$ and the optical extinction A$_V$, to search for a systematic underestimation of the correction for optical extinction of the optical lines used to determine the (N/O) ratio, but we did not find any correlation.
    \item We speculatively suggest that 
accretion of metal-poor gas from the circumgalactic medium could provide an explanation for the observed $\rm  \Delta(N/O)$, because the rapid decrease of the oxygen abundance during infall is followed by a decrease of the N/O ratio due to the primary production of young - possibly embedded - massive stars, which are preferentially traced by the IR diagnostics, while optical diagnostics would better trace the secondary production, when both N/O and O/H abundance ratios will increase.
\end{enumerate}
\facilities{SOFIA(FIFI-LS), Spitzer(IRS), Herschel(PACS)}

\begin{acknowledgments}
This work is based on observations made with the NASA/DLR Stratospheric Observatory for Infrared Astronomy (SOFIA). SOFIA is jointly operated by the Universities Space Research Association, Inc. (USRA), under NASA contract NNA17BF53C, and the Deutsches SOFIA Institut (DSI) under DLR contract 50 OK 0901 to the University of Stuttgart. We thank the anonymous referee, who helped to improve this paper. 
JAFO and LS acknowledge financial support by the Agenzia Spaziale Italiana (ASI) under the research contract 2018-31-HH0.
MPS acknowledges support from the Comunidad de Madrid through the Atracci\'on de Talento Investigador Grant 2018-T1/TIC-11035 and PID2019-105423GA-I00 (MCIU/AEI/FEDER,UE).  We acknowledge financial support from the NASA/SOFIA grant USRA 07-0239.
\end{acknowledgments}

\bibliography{biblio}{}
\bibliographystyle{aasjournal}

\newpage

\begin{table*}
\centering
\setlength{\tabcolsep}{3.pt}
\caption{Journal of the SOFIA FIFI-LS observations}
\label{tab:journal}
\scriptsize
\begin{tabular}{lccclccccc}
\hline\\[-0.2cm]
 Target     &  RA             &  dec             & $z$       & Type  & AOR-ID  & Mission-ID & PI             & line        &  $T_{exp}$ (s)  \\[0.05cm]
     (1)       & (2)                & (3)                 & (4)       & (5)   & (6)      & (7)    & (8)            & (9)         & (10)              \\[0.1cm]
\hline\\[-0.25cm]
MCG+12-02-001       & 00h\,54m\,03.6s & +73d\,05m\,12s & 0.015698 & ULIRG &  07\_0209\_11  & 2019-05-01\_FI\_F562   &  G. Stacey      & [OIII]$_{52{\mu}m}$  & 1320.96     \\
MCG+12-02-001       &  " "               & " "                 & "         &  "    & 07\_0209\_{12} & 2019-05-01\_FI\_F562   &  G. Stacey      & [NIII]$_{57{\mu}m}$  & 1351.68      \\
NGC1365       & 03h\,33m\,36.4s & -36d\,08m\,26s & 0.005457 & S1.8 &  05\_0111\_3  & 2017-07-28\_FI\_F424   &  G. Stacey      & [OIII]$_{52{\mu}m}$  & 1843.20     \\
IC342       & 03h\,46m\,48.5s & +68d\,05m\,47s & 0.000103 & HII &  03\_0135\_9  & 2015-03-26\_FI\_F205   &  K. Croxall      & [OIII]$_{52{\mu}m}$  & 967.68     \\
NGC1569       & 04h\,30m\,49.0s & +64d\,50m\,53s & -0.000347 & Dwarf &  87\_0005\_7  & 2014-04-22\_FI\_FO162   &  R. Klein      & [OIII]$_{52{\mu}m}$  & 14976.50     \\ 
NGC1569       &  " "               & " "                 & "         &  "    &  07\_0048\_2  & 2019-05-02\_FI\_F563  &  J. Spilker     & [NIII]$_{57{\mu}m}$  & 2805.76   \\
NGC1614       & 04h\,34m\,00.0s & -08d\,34m\,45s  &  0.015938 & HII   &  07\_0209\_14 & 2019-10-30\_FI\_F631  & G. Stacey       & [OIII]$_{52{\mu}m}$  &  522.24      \\
NGC1614       &  " "               & " "                 & "         &  "    &  07\_0209\_15 & 2019-10-30\_FI\_F631  & G. Stacey       & [NIII]$_{57{\mu}m}$  &   1136.64     \\
NGC1808       & 05h\,07m\,42.3s & -37d\,30m\,46s &  0.003319 & HII   &  05\_0111\_3  & 2017-07-28\_FI\_F424  & G. Stacey       & [OIII]$_{52{\mu}m}$  & 1751.04    \\
IIZw40        & 05h\,55m\,42.7s & +03d\,23m\,32s &  0.002632 & Dwarf &  06\_0225\_5  &  2018-11-06\_FI\_F524     & G. Stacey       & [OIII]$_{52{\mu}m}$  &    768.00   \\
IIZw40       &  " "               & " "                 & "         &  "    &  06\_0225\_10 & 2018-11-06\_FI\_F524  & G. Stacey       & [NIII]$_{57{\mu}m}$  &   1536.00    \\ 
NGC2146       & 06h\,18m\,37.8s & +78d\,21m\,25s &  0.002979 & HII &  03\_0135\_2  & 2015-03-12\_FI\_F199  & K. Croxall    & [OIII]$_{52{\mu}m}$  &  368.69      \\
NGC2146       &  " "               & " "                 & "         &  "    &  03\_0135\_17  & 2015-03-12\_FI\_F199  & K. Croxall    & [NIII]$_{57{\mu}m}$  &  307.228     \\
NGC2366       & 07h\,28m\,55.5s & +69d\,13m\,05s &  0.000267 & Dwarf &  07\_0239\_6  & 2019-05-14\_FI\_F570  & M.A. Malkan    & [OIII]$_{52{\mu}m}$  &  614.40      \\
NGC2366       &  " "               & " "                 & "         &  "    &  07\_0239\_7  & 2019-05-14\_FI\_F570  & M.A. Malkan    & [NIII]$_{57{\mu}m}$  &  614.40     \\
He2-10        & 08h\,36m\,15.2s & -26d\,24m\,34s &  0.002912 & Dwarf &  70\_0508\_6  & 2017-02-25\_FI\_F378  & A. Krabbe      & [OIII]$_{52{\mu}m}$  & 1658.88      \\
UGC5189        & 09h\,42m\,54.7s & +09d\,29m\,01s &  0.01072 & HII &  07\_0182\_1  & 2019-05-09\_FI\_F568  & T. Jones      & [OIII]$_{52{\mu}m}$  & 2211.84      \\
M82           & 09h\,55m\,52.2s & +69d\,40m\,48s &  0.000677 & HII   &  70\_0408\_1  & 2016-02-25\_FI\_F280 & A. Krabbe       & [OIII]$_{52{\mu}m}$  & 2918.40     \\
M82           &  " "               & " "                 & "         &  "    &  70\_0608\_10  & 2018-11-08\_FI\_F526 & A. Krabbe       & [NIII]$_{57{\mu}m}$  & 4177.92   \\
Haro3           & 10h\,45m\,22.4s & +55d\,57m\,38s &  0.003149 & Dwarf   &  06\_0225\_6  & 2018-11-07\_FI\_F525 & G. Stacey       & [OIII]$_{52{\mu}m}$  & 1413.12     \\
Mrk1271           & 10h\,56m\,09.1s & +06d\,10m\,22s &  0.00338 & HII   &  07\_0182\_3  & 2019-05-04\_FI\_F565 & T. Jones       & [OIII]$_{52{\mu}m}$  & 1720.32     \\
Arp299A           & 11h\,28m\,30.4s & +58d\,34m\,10s &  0.0103 & HII   &  05\_0111\_5  & 2017-02-28\_FI\_F379 & G. Stacey       & [OIII]$_{52{\mu}m}$  & 2949.12     \\
Arp299B\&C           &  " "               & " "                 & "         &  "    &  05\_0111\_5  & 2017-02-28\_FI\_F379 & G. Stacey       & [OIII]$_{52{\mu}m}$  & 2949.12   \\
Pox4           & 11h\,51m\,11.6s & -20d\,36m\,02s &  0.01197 & HII   &  07\_0182\_2  & 2019-05-03\_FI\_F564 & T. Jones       & [OIII]$_{52{\mu}m}$  & 1259.52     \\
Mrk193           & 11h\,55m\,28.3s & +57d\,39m\,52s &  0.017202 & HII   &  07\_0182\_4  & 2019-05-14\_FI\_F570 & T. Jones       & [OIII]$_{52{\mu}m}$  & 3164.16     \\
NGC4194           & 12h\,14m\,09.5s & +54d\,31m\,37s &  0.008342 & HII   &  07\_0209\_16  & 2019-05-09\_FI\_F568 & G. Stacey       & [OIII]$_{52{\mu}m}$  & 430.08     \\
NGC4194           &  " "               & " "                 & "         &  "    &  07\_0209\_18  & 2019-05-09\_FI\_F568 & G. Stacey       & [NIII]$_{57{\mu}m}$  & 860.16   \\
NGC4214           & 12h\,15m\,39.3s & +36d\,19m\,37s &  0.00097 & Dwarf   &  06\_0225\_8  & 2019-02-28\_FI\_F549 & G. Stacey       & [OIII]$_{52{\mu}m}$  & 1320.96     \\
NGC4214           &  " "               & " "                 & "         &  "    &  06\_0225\_9  & 2019-05-08\_FI\_F567 & G. Stacey       & [NIII]$_{57{\mu}m}$  & 1413.12   \\
NGC4536           & 12h\,34m\,27.1s & +02d\,11m\,17s &  0.006031 & HII   &  03\_0135\_6  & 2015-03-27\_FI\_F206 & K. Croxall       & [OIII]$_{52{\mu}m}$  & 399.36     \\
NGC4631       & 12h\,42m\,07.8s & +32d\,32m\,35s &  0.002021 & HII   &  07\_0239\_3  & 2019-05-10\_FI\_F569 &       M.A. Malkan & [OIII]$_{52{\mu}m}$  & 215.04   \\
NGC4631       &  " "               & " "                 & "         &  "    &  07\_0239\_3  & 2019-05-10\_FI\_F569 & M.A. Malkan     & [NIII]$_{57{\mu}m}$  & 1720.32   \\
NGC4670       & 12h\,45m\,17.1s & +27d\,07m\,31s &  0.003566 & Dwarf   &  06\_0222\_21  & 2019-03-02\_FI\_F551 &       T. Wiklind & [OIII]$_{52{\mu}m}$  & 491.52   \\
M83       & 13h\,37m\,00.9s & -29d\,51m\,56s &  0.001711 & HII   &  07\_0209\_8  & 2019-05-04\_FI\_F565 &       G. Stacey & [OIII]$_{52{\mu}m}$  & 2119.68   \\
NGC5253       & 13h\,39m\,56.0s & -31d\,38m\,24s &  0.001358 & Dwarf &  07\_0239\_4  & 2019-05-10\_FI\_F569 & M.A. Malkan     & [OIII]$_{52{\mu}m}$  & 307.20   \\
NGC5253       &  " "               & " "                 & "         &  "    &  07\_0239\_5  & 2019-05-10\_FI\_F569 & M.A. Malkan     & [NIII]$_{57{\mu}m}$  & 829.44   \\
\hline\\[-0.25cm]
\end{tabular}
\\
\begin{tablenotes}
\footnotesize

\end{tablenotes}
\end{table*}

\begin{table*}
\centering
\setlength{\tabcolsep}{3.pt}
\caption{Observed far-IR lines fluxes of the local galaxy sample. }
\label{tab:samplefir}
\scriptsize
\begin{tabular}{lccccccc}
\hline\\[-0.2cm]
\bf Name &  $\rm F_{[OIII]52\mu m\,}$ & $\rm F_{[NIII]57\mu m}$   & $ \rm F_{[OIII]88\mu m}$   & $ \rm F_{[NII]122\mu m}$     & $ \rm F_{[NII]205\mu m}$ & \bf Notes    & \bf Refs.\\[0.05cm]
         &   [$10^{-17}\, \rm{W/m^2}$] & [$10^{-17}\, \rm{W/m^2}$]  & [$10^{-17}\, \rm{W/m^2}$]   & [$10^{-17}\, \rm{W/m^2}$]     & [$10^{-17}\, \rm{W/m^2}$] &              &          \\[0.05cm]
  (1)    & (2)                        & (3)                       & (4)                        & (5)                          & (6)                      & (7)          & (8)      \\[0.1cm]
\hline\\[-0.25cm]
Haro11               &    ---            & 28.30$\pm$0.80       &    172.0$\pm$3.0   &   3.51$\pm$0.27  & 2.34$\pm$0.49    & $\star$ & C15 \\ 
NGC253               &    ---             &  599.5$\pm$178.3   &    625.2$\pm$53.5  &  802.7$\pm$28.1  & 175.3$\pm$11.2 & $\star$  & FO16 \\ 
MCG+12-02-001        & 305.0$\pm$30.9 &   52.9 $\pm$15.3    &    234.0$\pm$24.0  &   23.20$\pm$1.12   & ---             & $\ddag$  & P21,B08,AH10 \\ 
NGC1068              & 369.4$\pm$34.7  &  443.6$\pm$41.9    &    634.3$\pm$19.7  &  294.5$\pm$3.0   & 186.6$\pm$6.8 & $\star$ & FO16 \\ 
NGC1365              &  {72.43$\pm$23.57}   & 131.6$\pm$16.7     &  200.3$\pm$8.1     &  254.8$\pm$3.8   & 77.97$\pm$1.46  & $\dag$ $\#$ & FO16\\ 
NGC1569              & {1237.$\pm$245.}  &  {152.2$\pm$37.4}      &  2800.$\pm$10.   &  ---               &     ---         & $\ddag$ $\#$ & C15 \\ 
NGC1614              & 219.7$\pm$71.1   &   46.8$\pm$6.6     &  193.0$\pm$12.1    &  ---               & 10.62$\pm$0.39  & $\ddag$ $\#$ & FO16 \\ 
NGC1808              & {125.3$\pm$39.9}    & 152.6$\pm$10.8     &  204.4$\pm$11.0    &  271.2$\pm$4.3   &     ---       & $\dag$ $\#$ & FO16 \\ 
IIZw40               & {202.6$\pm$46.7}  &  {58.2$\pm$16.8}      &   359.0$\pm$4.0    &  $<$2.65  &  ---            & $\ddag$ $\#$ &  P21,C15 \\ 
Mrk3                 & 114.8$\pm$46.5  &   19.6$\pm$3.9     &     58.49$\pm$2.94   &   10.20$\pm$1.47   &     ---         & $\star$ & FO16 \\ 
NGC2146              & 1514.$\pm$201. &   551.$\pm$59.  &  1577.$\pm$65.   &   459.3$\pm$17.4   & 129.6$\pm$1.9 &  $\star$ &  B08,FO16\\ 
He2-10               &  {336.4$\pm$54.6}  &   98.5$\pm$10.9    &   338.0$\pm$5.0    &  ---               & 10.42$\pm$0.49  & $\dag$ $\#$ & C15\\ 
IRAS08572+3915       &     ---           &    2.4$\pm$0.37      &     5.1$\pm$0.26     &    0.74$\pm$0.15   &   ---           & $\star$ & DS17 \\ 
UGC5101              &        ---        &  10.13$\pm$6.28      &  14.33$\pm$3.36      &   13.16$\pm$1.81   & 6.04$\pm$0.44   & $\star$ & FO16 \\ 
M82                  & {2915.$\pm$171.6} &  {1620.$\pm$244.3}    &  1991.$\pm$16.7    &  482.3$\pm$5.4  & 437.6$\pm$8.3 & $\ddag$ $\#$ & FO16 \\ 
NGC3256              &        ---        & 169.8$\pm$13.8     &   461.1$\pm$6.9    &  140.6$\pm$2.0   &  45.27$\pm$0.97 & $\star$ & FO16 \\ 
Haro3                & 124.4$\pm$17.8   &   12.3$\pm$1.7      &   185.0$\pm$4.0    &   2.14$\pm$0.39   &    ---          & $\dag$ $\#$ & C15 \\ 
IRAS10565+2448       &   ---             & 9.10$\pm$0.40        &   15.6$\pm$0.40      &     8.1$\pm$0.3    &    2.44$\pm$0.13 & $\star$ &  PS17,P16 \\ 
IRAS11095-0238       &   4.80$\pm$1.21   & 1.13$\pm$0.77        &      ---             &    1.05$\pm$0.43   &  0.49$\pm$0.16  & $\star$ &  FO16 \\ 
Arp299\,A=IC694      & {328.1$\pm$37.0}  &   73.0$\pm$5.0     &    280.0$\pm$3.2   &   10.05$\pm$0.78   &     ---         & $\dag$ $\#$ & C19 \\ 
Arp299\,B+C=NGC3690  & {246.8$\pm$63.7}  &   72.0$\pm$1.3     &    300.0$\pm$2.6   &   15.96$\pm$0.38   &     ---        & $\dag$ $\#$ & C19 \\ 
NGC4151              &  37.56$\pm$9.56   &   21.73$\pm$2.00     &   48.40$\pm$2.87     &    7.09$\pm$0.76   &  6.63$\pm$0.68  & $\star$ & FO16 \\ 
IRAS12112+0305       &   ---             &   5.6$\pm$0.9        &   7.4$\pm$0.5        &     1.4$\pm$0.4    &       ---       & $\star$ & PS17, DS17    \\ 
NGC4194             &  282.5$\pm$14.6  &   65.0$\pm$22.0     &   206.0$\pm$14.0   &  $<$19.0            &      ---        & $\dag$ $\#$ &  B08  \\ 
NGC4214-reg.1       &  130.9$\pm$11.7  &  {19.30$\pm$4.52}       &   319.0$\pm$6.2    &    4.4$\pm$2.0  &  17.6$\pm$0.53   & $\ddag$ $\#$ &  C19   \\ 
NGC4631             &   {114.6$\pm$30.7}  & {160.1$\pm$21.5}    &  204.7$\pm$4.3 &  36.74$\pm$2.56  & 58.97$\pm$1.47   & $\ddag$ $\#$ & FO16 \\ 
NGC4945             &        $<$714     &  177.$\pm$16.        & 253.$\pm$60.         &   347.6$\pm$9.7   &     ---          & $\star$ & FO16 \\ 
NGC5033             &        ---         &  12.46$\pm$2.10      &  25.79$\pm$2.58      &   57.44$\pm$1.71  &   ---            & $\star$ & FO16 \\ 
IRAS13120-5453      &        ---         &   16.00$\pm$3.84     &   30.08$\pm$5.92     &  33.66$\pm$3.00   &  10.62$\pm$0.63 & $\star$ & FO16, PS17 \\ 
CenA=NGC5128        &       $<$462        &   53.34$\pm$10.28    & 173.47$\pm$6.68      &  85.15$\pm$2.81  &  51.66$\pm$1.47 & $\star$ & FO16 \\ 
M83                 &  205.4$\pm$18.7  &  166.5$\pm$10.3    & 216.9$\pm$6.97      &  321.0$\pm$4.59   &  92.59$\pm$3.90 & $\dag$ $\#$ & FO16 \\ 
NGC5253             &  412.0$\pm$40.0  &   {106.36$\pm$20.6}    & 901.0$\pm$4.0      &   9.25$\pm$1.21  &    ---          & $\ddag$ $\#$ & C15 \\ 
Mrk273              &  ---               &   13.78$\pm$5.75     & 33.00$\pm$5.53       &   8.58$\pm$1.31   &  3.84$\pm$0.28  & $\star$ & FO16 \\ 
IC4329A             &  26.26$\pm$2.59     &  9.31$\pm$1.38     &  30.16$\pm$1.31       &  3.31$\pm$0.45    &    ---           & $\star$ & FO16 \\ 
Mrk463E             & 45.95$\pm$7.92     &    9.61$\pm$2.11     & 39.68$\pm$4.13       &   2.31$\pm$0.27   &  0.51$\pm$0.20  & $\star$ & FO16 \\ 
Circinus            &       ---          &  268.3$\pm$15.3    & 565.5$\pm$9.8      &   299.4$\pm$4.0 &    ---         & $\star$ & FO16 \\ 
NGC5506             & 101.71$\pm$20.12   &   30.79$\pm$2.14     & 102.3$\pm$3.3      &   14.14$\pm$1.15  &      ---        & $\star$ & FO16 \\ 
NGC6240             & 46.57$\pm$26.39    & 12.50$\pm$5.86       & 33.10$\pm$5.57       &   23.15$\pm$2.22  &  18.47$\pm$0.39 & $\star$ & FO16 \\ 
IRAS17208–0014      &    ---             &  10.72$\pm$2.85      & 25.94$\pm$4.61       &    9.84$\pm$1.28  &   3.25$\pm$0.24 & $\star$ & FO16, PS17 \\ 
3C405=Cyg A         & 78.58$\pm$15.69    &    9.49$\pm$3.8      & 28.04$\pm$1.3        &    6.80$\pm$0.46  &   1.68$\pm$0.15 & $\star$ & FO16 \\ 
IRAS20551-4250      &       ---          &   4.0$\pm$0.7        & 13.1$\pm$0.4         &     2.3$\pm$0.33  & 0.76$\pm$0.11 & $\star$ & PS17,DS17,P16 \\ 
NGC7130=IC5135      &        ---         & 22.00$\pm$5.49       & 19.58$\pm$3.01       &   34.02$\pm$1.86  &  13.94$\pm$0.29 & $\star$ & FO16 \\ 
NGC7172             &        ---         &  6.96$\pm$2.13       & 14.27$\pm$2.49       &   11.74$\pm$1.04  &  21.83$\pm$0.58 & $\star$ & FO16 \\ 
NGC7314             &  ---               & 3.59$\pm$0.44        & 15.19$\pm$0.62       &    1.90$\pm$0.19  & --- & $\star$ & FO16 \\ 
NGC7469             &        ---         &  38.80$\pm$9.88      & 37.03$\pm$5.12       &   41.43$\pm$1.48  &  11.45$\pm$0.29 & $\star$ & FO16 \\
IRAS23128-5919      &        ---         &  17.1$\pm$0.9        & 44.4$\pm$0.7         &    4.4$\pm$0.5    & 1.83$\pm$0.10 & $\star$ & P16,DS17 \\
NGC7582             & 139.8$\pm$8.1    &  65.58$\pm$10.23     & 201.9$\pm$5.3      &  75.45$\pm$2.04   &  19.59$\pm$0.73 & $\star$ & FO16 \\ 
\hline\\[-0.25cm]
\end{tabular}
\\

\begin{tablenotes}

\footnotesize
Notes: From left to right, the table columns show: (1) object name; (2), (3), (4), (5), (6) line fluxes, in units of $10^{-17}\, \rm{W/m^2}$, of the lines: $\rm F_{[OIII]52\mu m\,}$; $\rm F_{[NIII]57\mu m\,}$;  $\rm F_{[OIII]88\mu m}$ ; $\rm F_{[NII]122\mu m}$; $\rm F_{[NII]205\mu m}$; (7): origin of far-IR spectroscopy: All measurements of  $\rm F_{[NII]205\mu m}$ are from Herschel-SPIRE; $\star$: all data from Herschel-PACS; $\dag$: $\rm F_{[OIII]52\mu m\,}$ from  SOFIA FIFI-LS; $\ddag$: $\rm F_{[OIII]52\mu m\,}$ and $\rm F_{[NIII]57\mu m\,}$ from  SOFIA FIFI-LS; $\#$:  SOFIA data reduced in this work; (11) reference for the line fluxes: P21: \citet{peng2021}, C15: \citet{cormier2015}, C19: \citet{cormier2019}, B08: ISO-LWS data from \citet{brauher2008};  DS17: \citet{diaz2017}; FO16: Herschel-PACS data from \citet{fernandez2016} and references therein; P16: \citet{pearson2016}, Pearson priv. comm. ; PS17: \citet{pereira2017}. 
\end{tablenotes}
\end{table*}

\newpage

\begin{table*}
\centering
\setlength{\tabcolsep}{3.pt}
\caption{Observed mid-IR fluxes of the local galaxy sample.}
\label{tab:sample2}
\scriptsize
\begin{tabular}{lccccccccc}
\hline\\[-0.2cm]
 Name & Type  &  $\rm F_{[SIV]10.5\mu m\,}$ & $\rm F_{[NeII]12.8\mu m}$   & $ \rm F_{[NeIII]15.5\mu m}$    & $ \rm F_{[SIII]18.7\mu m}$    & $ \rm F_{[OIV]25.9\mu m}$    & $ \rm F_{[SIII]33.5\mu m}$    & $ \rm F_{[SiII]34.8\mu m}$     &  Refs.\\[0.02cm]
      & &  [$10^{-17}\, \rm{W/m^2}$]  & [$10^{-17}\, \rm{W/m^2}$]  &  [$10^{-17}\, \rm{W/m^2}$]       & [$10^{-17}\, \rm{W/m^2}$] & [$10^{-17}\, \rm{W/m^2}$] & [$10^{-17}\, \rm{W/m^2}$]   & [$10^{-17}\, \rm{W/m^2}$]  &      \\[0.02cm]
  (1) & (2) & (3)  & (4) & (5) & (6)  & (7)   & (8) & (9)         & (10)           \\[0.1cm]
\hline\\[-0.25cm]
Haro11        & Dwarf & 49.4$\pm$1.1    & 32.7$\pm$0.9      & 112.$\pm$5.      & 53.1$\pm$2.9     & 4.390$\pm$0.784   & 81.7$\pm$7.0      & 55.8$\pm$4.5 & C15 \\
NGC253        & HII   &   $<$10.50      & 2832.30$\pm$64.20 &  204.60$\pm$9.60 & 666.40$\pm$14.90 & 154.70$\pm$26.90  & 1538.00$\pm$30.10 & 2412.00$\pm$48.00 & FO16 \\
MCG+12-02-001 & LIRG  &   3.14$\pm$0.81  &  201.10$\pm$2.12 &   36.99$\pm$0.67 & 73.46$\pm$1.55   & $<$6.36           & 167.60$\pm$5.15   & 181.70$\pm$9.64   & I13 \\
NGC1068       & S1h  & 536.10$\pm$12.80 & 458.10$\pm$13.80 & 1371.00$\pm$10.20 & 240.60$\pm$14.00 & 2030.00$\pm$27.30 & 374.10$\pm$23.20  &  604.40$\pm$17.10 & FO16 \\
NGC1365       & S1    & 18.60$\pm$0.78   & 143.00$\pm$3.79  &   61.30$\pm$0.51 &  51.20$\pm$0.57  &  365.00$\pm$26.90 & 720.00$\pm$102.00 & 1303.00$\pm$81.00 & FO16 \\
IC342         & HII   & 4.76$\pm$0.66    & 615.46$\pm$10.52 & 37.20$\pm$0.90   &  320.03$\pm$5.96 &      $<$7.70      &  672.46$\pm$7.28  &  985.73$\pm$10.29 &  FO16 \\  
NGC1569       & Dwarf &  247.$\pm$3.     &  30.5$\pm$2.7    &   324.$\pm$13.   &   131.$\pm$5.    &  31.9$\pm$1.5     &  185.$\pm$6.      &  22.95$\pm$1.19   & C15 \\
NGC1614       & HII   &  6.89$\pm$0.54   & 249.00$\pm$7.00  & 63.32$\pm$1.59   &   83.03$\pm$2.64 &    8.68$\pm$0.85  &  101.06$\pm$2.11  &  148.60$\pm$4.11  & FO16 \\  
NGC1808       & HII   &  1.47$\pm$0.00   & 177.36$\pm$16.36 & 17.26$\pm$0.68   &   46.20$\pm$2.29 &      $<$9.54      &  205.83$\pm$20.27 &  354.26$\pm$15.15 & FO16 \\    
IIZw40        & Dwarf &  200.$\pm$10.    &  7.35$\pm$0.79   & 141.$\pm$9.      &   52.1$\pm$2.2   &   7.94$\pm$1.15   &  78.2$\pm$2.1     &  36.8$\pm$3.5    & C15 \\
Mrk3          & S1h   & 59.30$\pm$0.62   &  98.00$\pm$1.02  & 175.00$\pm$1.10  &   53.60$\pm$4.49 &  196.00$\pm$2.40  &   52.40$\pm$6.89  &   84.60$\pm$3.28  &  FO16 \\  
NGC2146       & LIRG  &   6.30$\pm$0.44  & 625.00$\pm$15.35 &  91.16$\pm$0.95  &  190.12$\pm$4.15 &   19.33$\pm$4.48  &  848.02$\pm$36.93 & 1209.35$\pm$21.70 & FO16 \\   
He2-10        & Dwarf &   32.7$\pm$1.3   & 380.0$\pm$13.0   & 156.0$\pm$5.0    &  267.0$\pm$19.0  & 7.915$\pm$2.514   &   317.0$\pm$10.0  &  193.0$\pm$4.0    & C15 \\
IRAS08572+3915 & ULIRG &   $<$0.5        &  8.36$\pm$0.69   & 2.46$\pm$0.50    &   1.84$\pm$0.51  &      $<$2.1       &   $<$7.7          &  ---              & A07,V09 \\
UGC5101       & ULIRG &  1.32$\pm$0.32   & 37.43$\pm$0.45   &  14.05$\pm$0.18  &   8.46$\pm$0.20  & 7.34$\pm$0.83     &   13.91$\pm$1.39  &  27.47$\pm$3.31   & I13  \\ 
NGC2976       & HII   &	    ---             & 6.3$\pm$0.6      &  2.2$\pm$0.2     & 5.1$\pm$0.3      & 0.3$\pm$0.1       & 7.5$\pm$0.2       & 6.9$\pm$0.4      & D06 \\
M82           & HII   &  5.65$\pm$0.74   & 506.22$\pm$45.04 &  80.98$\pm$1.99  & 172.06$\pm$5.80  &   45.67$\pm$6.61  & 1812.90$\pm$36.79 & 2166.00$\pm$25.52 & FO16 \\ 
NGC3256        & HII  &  5.25$\pm$0.57   & 514.19$\pm$9.34  &  64.42$\pm$0.86  & 171.83$\pm$2.07  &   12.23$\pm$3.02  &  484.64$\pm$13.79 &  623.37$\pm$9.81  & ? \\
Haro3         & Dwarf &  40.8$\pm$1.7    & 35.2$\pm$1.3     &   98.4$\pm$7.4   & 50.3$\pm$3.9     &   1.79$\pm$0.5    &  85.1$\pm$2.4     &  39.9$\pm$2.4     & C15 \\
IRAS10565+2448 & ULIRG & $<$1.25         & 61.75$\pm$0.81   & 7.67$\pm$0.42    & 12.35$\pm$1.01  & $<$2.40          & 19.31$\pm$1.53    &  39.88$\pm$5.59 & I13 \\
IRAS11095-0238 & ULIRG & $<$1.20         & 6.08$\pm$0.61    &  1.89$\pm$0.19   & 1.22$\pm$0.24    &  $<$0.90          &  ---              &    ---          & FO16  \\
Arp299\,B=NGC3690\_W\footnote{11h\,28m\,31.0s ~58d\,33m\,43} &  LIRG & 5.52$\pm$1.11 & 103.80$\pm$2.71 & 54.42$\pm$0.98 & 47.71$\pm$1.79  & 28.90$\pm$3.78  & 204.00$\pm$11.63 & 228.80$\pm$6.27 & I10 \\
Arp299\,A=NGC3690\_E\footnote{11h\,28m\,33.7s ~58d\,33m\,49} & LIRG  & 5.06$\pm$0.79 & 237.40$\pm$2.65 &  57.05$\pm$0.98 & 61.61$\pm$2.04 & 16.07$\pm$14.29 & 271.90$\pm$22.72 & 267.00$\pm$18.23 & I10 \\
NGC4151       & S1    &  84.80$\pm$5.51  & 134.00$\pm$5.84  & 214.50$\pm$11.20 &   71.00$\pm$5.05 &  244.00$\pm$9.07  &  68.50$\pm$5.90   &  135.00$\pm$3.77  &  FO16 \\ 
IRAS12112+0305 & ULIRG &  0.47$\pm$0.11  &  14.0$\pm$0.14   &  3.70$\pm$0.04   &   5.40$\pm$0.16  &  $<$1.1           &  10.0$\pm$0.5     &  ---              & I10, V09 \\
NGC4194       & LIRG  &  9.15$\pm$0.53   & 175.70$\pm$1.40  &  56.20$\pm$0.60  &  71.26$\pm$0.91  &  25.10$\pm$1.85   &  157.40$\pm$8.53  &  185.50$\pm$3.99  &  I13  \\
NGC4214-reg.1 & Dwarf &  56.80$\pm$2.10  &  89.80$\pm$2.20  & 187.00$\pm$1.40  & 117.80$\pm$2.00  &  $<$9.78          &  187.10$\pm$2.70  &   ---             & C15 \\
NGC4536       & HII   &  0.95$\pm$0.20   &  35.46$\pm$0.39  &   6.11$\pm$0.06  &   16.34$\pm$0.91 &    1.72$\pm$0.25  &   100.48$\pm$4.53 &  114.02$\pm$1.53  &  FO16 \\
NGC4631       & HII   &  1.39$\pm$0.42   &  45.92$\pm$0.71  &  10.15$\pm$0.13  &   39.69$\pm$2.12 &      $<$1.47      &    85.16$\pm$0.76 &  114.74$\pm$1.89  &  FO16 \\
NGC4945       & S     &       ---        &  698.40$\pm$60.40 &    68.10$\pm$2.27  &       ---         &     28.35$\pm$1.39  &   359.60$\pm$20.40  &  732.80$\pm$7.84  & FO16 \\
NGC 5033      & S2    &    2.83$\pm$0.20 &   13.26$\pm$0.18  &     5.08$\pm$0.15  &   14.88$\pm$0.50  &      5.08$\pm$0.51  &    17.38$\pm$0.85   &   45.35$\pm$1.52  &  FO16 \\  
IRAS13120-5453& ULIRG &    0.50$\pm$0.10 &  150.00$\pm$15.00 &    18.46$\pm$1.85  &   19.18$\pm$1.92  &      6.42$\pm$1.28  &    60.64$\pm$6.06   &  107.10$\pm$10.70 &  FO16 \\
CenA=NGC5128  & S2    &   14.01$\pm$3.55 &  189.10$\pm$17.50 &   148.00$\pm$6.08  &   48.50$\pm$2.75  &    129.00$\pm$7.05  &   148.80$\pm$10.20  &  285.40$\pm$8.86  & FO16 \\
M83-nucleus   & HII   &  3.31$\pm$0.23   &  503.33$\pm$19.88 &    29.30$\pm$0.77  &  227.66$\pm$16.51 &      5.75$\pm$1.08  &   263.50$\pm$9.21   &  391.40$\pm$8.55  & FO16 \\
NGC5253       & Dwarf &  541.$\pm$211.   &  121.1$\pm$59.6   &     656.$\pm$39.   &   261.$\pm$124.   &      9.80$\pm$2.41  &    626.$\pm$398.    &  329.$\pm$231.    & FO16\\
Mrk273        & S2    &  9.58$\pm$0.96   &    41.90$\pm$4.19 &    33.57$\pm$3.36  &   13.35$\pm$1.33  &     56.36$\pm$5.64  &    42.56$\pm$4.26   &   14.66$\pm$2.90  & FO16 \\  
IC4329A       & S1    & 29.10$\pm$1.32   &    27.60$\pm$0.73 &    57.00$\pm$0.97  &   15.00$\pm$1.44  &    117.00$\pm$1.42  &    16.00$\pm$2.19   &   32.50$\pm$3.06  & FO16 \\     
Mrk463E       & S1h   & 29.86$\pm$2.99   &     9.25$\pm$0.92 &    40.78$\pm$4.08  &   15.85$\pm$1.59  &     69.17$\pm$6.92  &    15.50$\pm$1.55   &   29.79$\pm$2.98  & FO16 \\        
Circinus      & S1h   & 123.20$\pm$53.20 &  393.00$\pm$67.50 &   385.40$\pm$29.90 &  194.50$\pm$26.20 &    897.50$\pm$48.90 &   594.40$\pm$89.20  &  729.90$\pm$51.70 & FO16 \\    
NGC5506       & S1h   & 73.50$\pm$1.56   &    85.10$\pm$1.43 &   153.70$\pm$1.10  &   58.60$\pm$7.79  &    226.20$\pm$4.00  &    91.70$\pm$24.40  &  137.00$\pm$6.92  & FO16 \\  
NGC6240       & LIN   &  2.68$\pm$0.27   &  171.00$\pm$17.10 &    60.60$\pm$6.10  &   17.10$\pm$1.71  &     26.75$\pm$2.67  &    38.11$\pm$3.81   &  265.90$\pm$26.60 & FO16 \\   
IRAS17208–0014 & ULIRG & $<$0.4          &   38.0$\pm$0.38   &    7.90$\pm$0.16    &    7.30$\pm$0.22  &     $<$3.2          &      $<$13.0        &  60.0$\pm$1.2     & V09 \\
3C405         & S2    & 16.20$\pm$0.70   &    21.70$\pm$0.70 &    47.90$\pm$0.50  &   24.20$\pm$0.60  &     78.50$\pm$0.70  &    29.00$\pm$1.00   &        ---        & FO16 \\
IRAS20551-4250 & ULIRG &   $<$1.28       &    13.49$\pm$0.95 &     2.73$\pm$0.61  &    6.74$\pm$0.43  &     $<$4.18         &     9.98$\pm$4.49   &   23.30$\pm$8.81  & I13 \\
NGC7130=IC5135& S2    &  5.27$\pm$0.84   &    79.30$\pm$0.93 &    29.40$\pm$0.77  &   19.60$\pm$0.33  &     19.70$\pm$0.84  &    48.20$\pm$2.59   &   93.90$\pm$4.90  & FO16 \\  
NGC7172       & S2    &  5.87$\pm$0.61   &    33.00$\pm$1.01 &    17.10$\pm$0.68  &   11.90$\pm$1.00  &     45.40$\pm$0.48  &    26.90$\pm$1.51   &   59.30$\pm$2.42  & FO16 \\  
NGC7314       & S1h   & 15.90$\pm$0.53   &     8.08$\pm$0.39 &    23.20$\pm$0.53  &    9.97$\pm$0.71  &     67.00$\pm$0.41  &    15.00$\pm$1.71   &   14.20$\pm$1.76  & FO16 \\ 
NGC7469       & S1    &  9.00$\pm$0.79   &   191.00$\pm$2.70 &    35.80$\pm$0.75  &   75.40$\pm$4.52  &     34.00$\pm$3.80  &    63.60$\pm$9.21   &  194.00$\pm$19.10 & FO16 \\
NGC7582       & S1h   & 21.30$\pm$1.43   &   322.00$\pm$6.41 &   105.00$\pm$2.05  &   87.30$\pm$1.99  &    262.00$\pm$5.54  &   244.00$\pm$7.85   &        ---        & FO16 \\   
IRAS23128-5919 & ULIRG & 4.80$\pm$0.28   &    32.12$\pm$0.45 &    21.72$\pm$0.46  &   24.64$\pm$1.70  &     14.97$\pm$2.47  &   19.35$\pm$2.28    &  34.03$\pm$7.47   & I13 \\

\\[0.1cm]
\hline\\[-0.25cm]
\end{tabular}
\\
\begin{tablenotes}
\footnotesize
Notes: From left to right, the table columns show: (1) object name; (2) Galaxy type (dwarf Galaxy, Seyfert, Starburst (HII), and ULIRG); (3), (4), (5), (6), (7), (8), (9) fluxes of the given mid-IR fine structure lines, in units of $\rm {10^{-17} W/m\,^2}$, from the following references: (10):  FO16: \citet{fernandez2016} and references therein; I13: \citet{inami2013}; C15: \citet{cormier2015}; D06: \citet{dale2006}: fluxes averaged over $\sim$23"$\times$15" and listed in units of $10^{-9} {\rm W m^{-2} sr^{-1}}$; A07: \citet{armus2007}; V09: \citet{veilleux2009}; I10: \citet{imanishi2010}. 
\end{tablenotes}
\end{table*}

\begin{table*}
\centering
\setlength{\tabcolsep}{3.pt}
\caption{Observed properties of other galaxies observed by SOFIA in the [OIII]52$\mu$m line. }
\label{tab:sample1bis}
\scriptsize
\begin{tabular}{lccclcccccc}
\hline\\[-0.2cm]
 \bf Name & \bf RA  & \bf dec                                & $z$       & \bf Type  & \bf 12+log(O/H) &  $\rm F_{[OIII]52\mu m\,}$ &  $ \rm F_{[OIII]88\mu m}$     & \bf Notes    & \bf refs.\\[0.05cm]
      &     &                                    &      &     &             & [$10^{-17}\, \rm{W/m^2}$] & [$10^{-17}\, \rm{W/m^2}$] &      &      \\[0.15cm]
  (1) & (2) & (3)                                & (4)       & (5) & (6)               & (7)             & (8)                   & (9)   
  & (10)                       \\[0.1cm]
\hline\\[-0.25cm]
IC342         & 03h\,46m\,48.5s & +68d\,05m\,47s &  0.000103 & HII  &   8.71$\pm$0.01    &              {98.21$\pm$18.99}        &   171.14$\pm$7.92 & $\dag$ $\#$ & \\ 
NGC2366       & 07h\,28m\,55.6s & +69d\,13m\,05s &  0.000267 & Dwarf & 7.64$\pm$0.03  &  {155.84$\pm$39.85}  &      226.00$\pm$1.00 & $\dag$ $\#$  & M13, C15  \\ 
UGC5189       & 09h\,42m\,54.7s & +09d\,29m\,01s &  0.010720 & HII  &   8.29$\pm$0.07      &   
{30.75$\pm$4.45}  & ---   &  $\dag$ $\#$ &  PT07 \\ 
NGC3077       & 10h\,03m\,19.1s & +68d\,44m\,02s &  0.000047 & HII  &   8.60$\pm$0.01              &    745.65$\pm$112.98            &      ---            & $\dag$ $\#$ & C04 \\
Mrk1271       & 10h\,56m\,09.1s & +06d\,10m\,22s &  0.003380 & HII  &  7.99$\pm$0.04  &       $<$ 12\footnote{3$\sigma$ upper limit, very noisy line profile} & ---                  & $\dag$ $\#$ & I98 \\ 
Pox4          & 11h\,51m\,11.6s & -20d\,36m\,02s &  0.011970 & HII &  7.97$\pm$0.06   &    46.85$\pm$6.44      & ---                & $\dag$ $\#$ & K96 \\ 
Mrk193        & 11h\,55m\,28.3s & +57d\,39m\,52s &  0.017202 & HII  & 7.79$\pm$0.27  &    $<$ 12\footnote{3$\sigma$ upper limit, line profile not consistent with the expected line width of $\sim$300 $\rm km s^{-1}$}  &    ---                       & $\dag$ $\#$ & I94\\ 
NGC4536       & 12h\,34m\,27.1s & +02d\,11m\,17s &  0.006031 & HII  &  8.83$\pm$0.01    &    {136.90$\pm$29.26}   & 166.61$\pm$7.63 & $\dag$ $\#$   \\ 
NGC4670       & 12h\,45m\,17.1s & +27d\,07m\,31s &  0.003566 & Dwarf &  8.47$\pm$0.02                  &  {38.17$\pm$4.2}    & ---     & $\dag$ $\#$ & \\ 
\\[0.1cm]
\hline\\[-0.25cm]
\end{tabular}
\\
\begin{tablenotes}
\footnotesize
Notes: From left to right, the table columns show: (1) object name; (2,3) 2MASS coordinates; (4) NED redshift ($z$); (5) Galaxy type: dwarf Galaxy, Seyfert, Starburst (HII), Luminous IR Galaxy (LIRG) or Ultra Luminous OR Galaxy (ULIRG); (6) optical metallicity; (7), (8) line fluxes, in units of $\rm {10^{-17} W/m\,^2}$, of the lines:  $\rm F_{[NIII]57\mu m\,}$  $\rm F_{[OIII]88\mu m}$ ; (9): origin of far-IR spectroscopy: 
$\dag$: $\rm F_{[OIII]52\mu m\,}$ from  SOFIA FIFI-LS; $\#$: SOFIA data reduced in this work;
the $\rm F_{[OIII]88\mu m}$ line fluxes are from Herschel-PACS \citep[][and references therein]{fernandez2016};
(10) reference for the metallicity data: M13: \citet{madden2013}; C15: \citet{cormier2015}; PT07: \citet{pilyugin2007}; C04: \citet{calzetti2004}; I98: \citet{izotov1998}; K96: \citet{kobulnicky1996}; I94: \citet{izotov1994}.
\end{tablenotes}
\end{table*}

\begin{table*}
\centering
\setlength{\tabcolsep}{3.pt}
\caption{Observed properties of the local galaxy sample: coordinates, types \& metallicities }\label{tab:sample1}
\scriptsize
\begin{tabular}{lccccccccc}
    \hline\\[-0.2cm]
\bf Name & \bf RA  & \bf dec & $z$ & \bf Type & \bf 12+log(O/H) & \bf log(N/O)             &  \bf Refs.    & \bf \bf 12+log(O/H)$_{\rm IR}$ & \bf log(N/O)$_{\rm IR}$  \\[0.05cm]
      &     &                     &               &           &      &  &    &     &      \\[0.05cm]
(1) & (2) & (3) & (4) & (5) & (6) & (7) & (8) & (9) & (10)\\[0.1cm]
\hline\\[-0.25cm]
Haro11        & 00h\,36m\,52.5s & -33d\,33m\,17s &  0.020598 & Dwarf & 8.23$\pm$0.03 & -0.92  & M13,C19 & 8.51$\pm$0.07 & -1.2$\pm$0.2 \\ 
NGC253        & 00h\,47m\,33.1s & -25d\,17m\,19s &  0.000811 & HII   & 8.61$\pm$0.01 & -0.43$\pm$0.04   & TW & 8.7$\pm$0.2 & -0.4$\pm$0.2  \\ 
MCG+12-02-001 & 00h\,54m\,03.6s & +73d\,05m\,12s &  0.015698 & LIRG  & 8.7 & ... & AH10 & 8.7$\pm$0.1 & -1.0$\pm$0.1  \\ 
NGC1068       & 02h\,42m\,40.7s & -00d\,00m\,48s &  0.003793 & S1h   & 8.60$\pm$0.06 & -0.02$\pm$0.02   & TW & ... & -0.35$\pm$0.05 \\ 
NGC1365       & 03h\,33m\,36.4s & -36d\,08m\,26s &  0.005457 & S1    & 8.88$\pm$0.02 & ... & TW & ... & -0.3$\pm$0.05 \\ 
NGC1569       & 04h\,30m\,49.0s & +64d\,50m\,53s & -0.000347 & Dwarf & 8.19$\pm$0.03 & -1.39 & M13,C19 & 7.7$\pm$0.2 & -1.6$\pm$0.1 \\ 
NGC1614       & 04h\,34m\,00.0s & -08d\,34m\,45s &  0.015938 & HII   & 8.60$\pm$0.01 & -0.56$\pm$0.03 & TW & 8.69$\pm$0.08 & -1.0$\pm$0.1 \\ 
NGC1808       & 05h\,07m\,42.3s & -37d\,30m\,46s &  0.003319 & HII   & 8.71$\pm$0.01 & -0.4$\pm$0.1 & TW & 8.5$\pm$0.3 & -0.8$\pm$0.3 \\ 
IIZw40        & 05h\,55m\,42.6s & +03d\,23m\,32s &  0.002632 & Dwarf & 8.09$\pm$0.02 & -1.44 & M13,C19 & 7.7$\pm$0.2 & ... \\ 
Mrk3          & 06h\,15m\,36.4s & +71d\,02m\,15s &  0.013509 & S1h   & 8.69$\pm$0.07 & -0.55$\pm$0.03 & TW & ... & -0.9$\pm$0.1 \\ 
NGC2146       & 06h\,18m\,37.8s & +78d\,21m\,25s &  0.002979 & LIRG  & 8.71$\pm$0.02 & -0.77$\pm$0.03 & TW & 8.7$\pm$0.1 & -0.9$\pm$0.1 \\ 
NGC2366       & 07h\,28m\,55.6s & +69d\,13m\,05s &  0.000267 & Dwarf & 7.64$\pm$0.03 & ... & M13 & & ... \\ 
He2-10        & 08h\,36m\,15.2s & -26d\,24m\,34s &  0.002912 & Dwarf & 8.65$\pm$0.01 & -0.75$\pm$0.01 & TW & 8.64$\pm$0.08 & -1.0$\pm$0.1 \\ 
IRAS08572+3915 & 09h\,00m\,25.4s & +39d\,03m\,54s & 0.05835  & ULIRG & 8.62$\pm$0.02 & -0.96$\pm$0.05 & TW & 8.5$\pm$0.2 & -0.7$\pm$0.2 \\ 
UGC5101       & 09h\,35m\,51.6s & +61d\,21m\,11s &  0.01615  & ULIRG & 8.78$\pm$0.02 & -0.7$\pm$0.1 & TW & 8.6$\pm$0.1 & -0.6$\pm$0.4 \\ 
M82           & 09h\,55m\,52.7s & +69d\,40m\,46s &   0.00068 & HII   &  8.70$\pm$0.01 & -0.46$\pm$0.03 & TW & 8.6$\pm$0.1 & -0.8$\pm$0.2 \\ 
NGC3256       & 10h\,27m\,51.3s & -43d\,54m\,13s &   0.00935 & HII   & 8.39$\pm$0.07 & -0.39$\pm$0.02 & TW & 8.70$\pm$0.09 & -0.80$\pm$0.09 \\ 
Haro3         & 10h\,45m\,22.4s & +55d\,57m\,38s &  0.003149 & Dwarf & 8.37$\pm$0.02 & -1.35 & M13,C19 & 8.30$\pm$0.07 & -1.4$\pm$0.1 \\ 
IRAS10565+2448 & 10h\,59m\,18.2s & +24d\,32m\,34s &  0.04310 & ULIRG &  8.77$\pm$0.02 & -0.50$\pm$0.01 & TW & 8.6$\pm$0.1 & -0.6$\pm$0.1 \\ 
IRAS11095-0238 & 11h\,12m\,03.4s & -02d\,54m\,23s & 0.106634 & ULIRG & 8.53$\pm$0.02 & -0.77$\pm$0.09 & TW & 8.7$\pm$0.1 & -0.9$\pm$0.3 \\ 
Arp299\,A=IC694 & 11h\,28m\,33.7s & 58d\,33m\,49s &  0.01041 & LIRG   & 8.56$\pm$0.02 & -0.74$\pm$0.02 & TW & 8.5$\pm$0.1 & -1.0$\pm$0.1  \\ 
Arp299\,B+C=NGC3690 & 11h\,28m\,31.0s & 58d\,33m\,43s & 0.01022 & LIRG & 8.79$\pm$0.02 & -0.50$\pm$0.01 & TW & 8.6$\pm$0.1 & -1.0$\pm$0.1 \\ 
NGC4151       & 12h\,10m\,32.6s & +39d\,24m\,21s &  0.003319 & S1    & 8.5$\pm$0.1 & -0.71$\pm$0.07 & TW & ... & -0.59$\pm$0.06 \\ 
IRAS12112+0305 & 12h\,13m\,46.0s & +02d\,48m\,38s & 0.073317 & ULIRG & 8.66$\pm$0.02 & -0.75$\pm$0.01 & TW & 8.4$\pm$0.2 & -0.5$\pm$0.2 \\ 
NGC4194       & 12h\,14m\,09.5s & +54d\,31m\,37s &  0.008342 & LIRG  & 8.52$\pm$0.01 & -0.63$\pm$0.03 & TW & 8.5$\pm$0.2 & -0.9$\pm$0.1 \\ 
NGC4214-reg.1 & 12h\,15m\,39.3s & +36d\,19m\,37s &   0.00097 & Dwarf & 8.20$\pm$0.03  & ... & M13 & 8.6$\pm$0.1 & -1.4$\pm$0.1 \\ 
NGC4631       & 12h\,42m\,07.8s & +32d\,32m\,35s &  0.002021 & HII   & 8.52$\pm$0.01 & -1.11$\pm$0.03 & TW & 8.69$\pm$0.07 & ... \\ 
NGC4945       & 13h\,05m\,27.5s & -49d\,28m\,06s &  0.00188  & S2    & ... & ... &  & 8.5$\pm$0.4 & -0.6$\pm$0.2 \\ 
NGC5033       & 13h\,13m\,27.5s & +36d\,35m\,38s &  0.00292  & S2    & 8.85$\pm$0.07 & -0.25$\pm$0.01 & TW & ... & -0.73$\pm$0.07 \\ 
IRAS13120-5453 & 13h\,15m\,06.4s & -55d\,09m\,23s & 0.03076  & ULIRG & ... &  ... & & 8.7$\pm$0.1 & -0.7$\pm$0.2 \\
CenA=NGC5128  & 13h\,25m\,27.6s & -43d\,01m\,09s &  0.00183  & S2    & 8.85$\pm$0.02 & -0.91$\pm$0.04 & TW & ... & -0.92$\pm$0.08 \\ 
M83           & 13h\,37m\,00.9s & -29d\,51m\,56s &  0.001711 & HII   & 8.6$\pm$0.2 & -0.34$\pm$0.02 & TW & 8.7$\pm$0.1 & -0.7$\pm$0.2 \\ 
NGC5253       & 13h\,39m\,56.0s & -31d\,38m\,24s &  0.001358 & Dwarf & 8.16$\pm$0.03 & -1.45 & M13,C19 & 8.2$\pm$0.1 & -1.2$\pm$0.1 \\ 
Mrk273        & 13h\,44m\,42.1s & +55d\,53m\,13s &  0.03778  & S2    & 8.85$\pm$0.06 & -1.06$\pm$0.07 & TW & ... & -0.8$\pm$0.2 \\
IC4329A       & 13h\,49m\,19.3s & -30d\,18m\,34s &  0.016054 & S1    & 8.4$\pm$0.2   & -1.0$\pm$0.4 & TW & ... & -0.77$\pm$0.07 \\
Mrk463E       & 13h\,56m\,02.9s & +18d\,22m\,18s &  0.05035 & S1h    & 8.5$\pm$0.2   & -1.0$\pm$0.4 & TW & ... & -0.9$\pm$0.1 \\ 
Circinus      & 14h\,13m\,09.9s & -65d\,20m\,21s &  0.00145  & S1h   & 8.8$\pm$0.1   & -0.4$\pm$0.1 & TW & ... & -0.74$\pm$0.02 \\
NGC5506       & 14h\,13m\,14.9s & -03d\,12m\,28s &  0.006181 & S1h   & 8.76$\pm$0.07 & -1.1$\pm$0.3 & TW & ... & -0.81$\pm$0.05 \\ 
NGC6240       & 16h\,52m\,58.9s & +02d\,24m\,04s &   0.02448 & LIN   & 8.85$\pm$0.01 & -1.1$\pm$0.1 & TW & ... & -0.9$\pm$0.2 \\ 
IRAS17208-0014 & 17h\,23m\,21.9s & -00\,d17m\,01s & 0.042810 & ULIRG & 8.69$\pm$0.01 & -0.61$\pm$0.03 & TW & 8.7$\pm$0.1 & -0.8$\pm$0.2 \\ 
3C405=Cyg A   & 19h\,59m\,28.3s & +40d\,44m\,02s &  0.056075 & S2    & 8.78$\pm$0.04 & -0.50$\pm$0.06 & TW & ... & -1.0$\pm$0.2 \\ 
IRAS20551-4250 & 20h\,58m\,26.8s & -42d\,39m\,00s & 0.042996 & ULIRG & 8.50$\pm$0.01 & -0.83$\pm$0.05 & TW & 8.70$\pm$0.09 & -0.9$\pm$0.2 \\ 
NGC7130=IC5135 & 21h\,48m\,19.5s & -34d\,57m\,05s & 0.01615  & S2    & 8.60$\pm$0.09 & -0.6$\pm$0.3 & TW & ... & -0.4$\pm$0.1 \\ 
NGC7172       & 22h\,02m\,01.9s & -31d\,52m\,11s & 0.00868   & S2    & 8.8$\pm$0.2   & ... & TW & ... & -0.7$\pm$0.1 \\ 
NGC7314       & 22h\,35m\,46.2s & -26d\,03m\,02s &  0.004763 & S1h   & 8.88$\pm$0.02 & -0.2$\pm$0.2 & TW & ... & -1.03$\pm$0.05  \\ 
NGC7469       & 23h\,03m\,15.6s & +08d\,52m\,26s  & 0.01632   & S1    & 8.4$\pm$0.1   & -0.38$\pm$0.03 & TW & ... & -0.4$\pm$0.1 \\
IRAS23128-5919 & 23h\,15m\,46.8s & -59d\,03m\,16s &  0.04460 & ULIRG & 8.40$\pm$0.07 & -0.75$\pm$0.01 & TW & 8.48$\pm$0.07 &  -0.8$\pm$0.1 \\ 
NGC7582       & 23h\,18m\,23.7s & -42d\,22m\,14s &  0.005254 & S1h   & 8.5$\pm$0.1   & -0.47$\pm$0.02 & TW & ... & -0.71$\pm$0.06 \\ 
    \hline\\[-0.25cm]
\end{tabular}
\begin{tablenotes}
\footnotesize
Notes: From left to right, the table columns show: (1) object name; (2,3) 2MASS coordinates; (4) NED redshift ($z$); (5) Galaxy type: dwarf, Seyfert, starburst (HII), Luminous IR Galaxy (LIRG) or Ultra Luminous IR Galaxy (ULIRG); (6,7) Oxygen abundance and nitrogen-to-oxygen relative abundance determined from optical nebular lines; (8) References for the optical abundances: AH10: \citealt{alonso2010}, C19: \citealt{cormier2019}, M13: \citealt{madden2013}, TW: this work; (9) Oxygen abundance and nitrogen-to-oxygen relative abundance determined from IR nebular lines using \textsc{HCm-IR} \citep{fernandez2021}. 
\end{tablenotes}
\end{table*}

\clearpage
\appendix

\section{Appendix information}

We present below our reductions for various galaxies included in Table \ref{tab:samplefir}.


\begin{figure*}[ht!]
    \centering
    \subfigure[]{\includegraphics[width=0.498\textwidth]{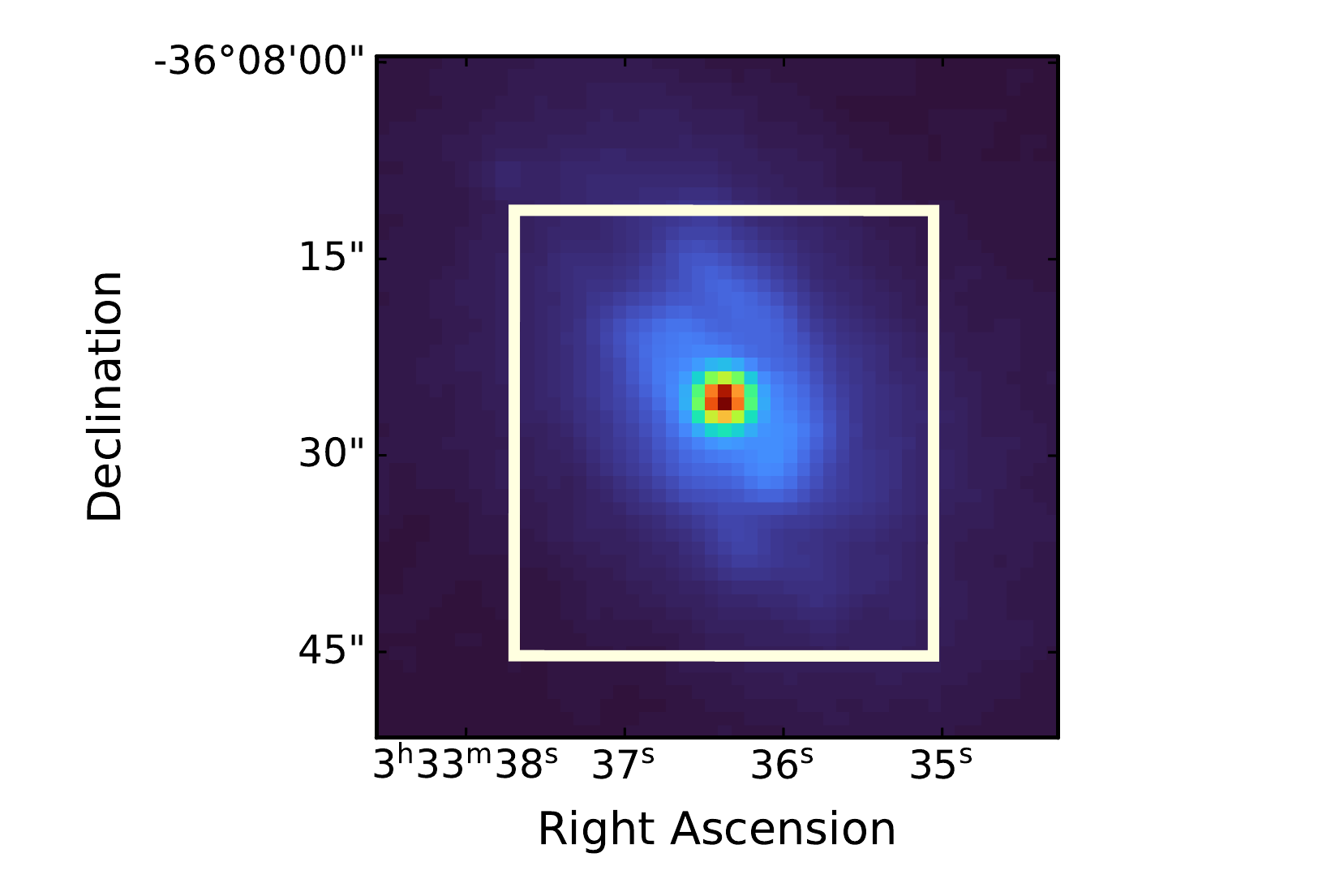}\label{fig:NGC1365-A}}\\
    \subfigure[]{\includegraphics[width=0.498\textwidth]{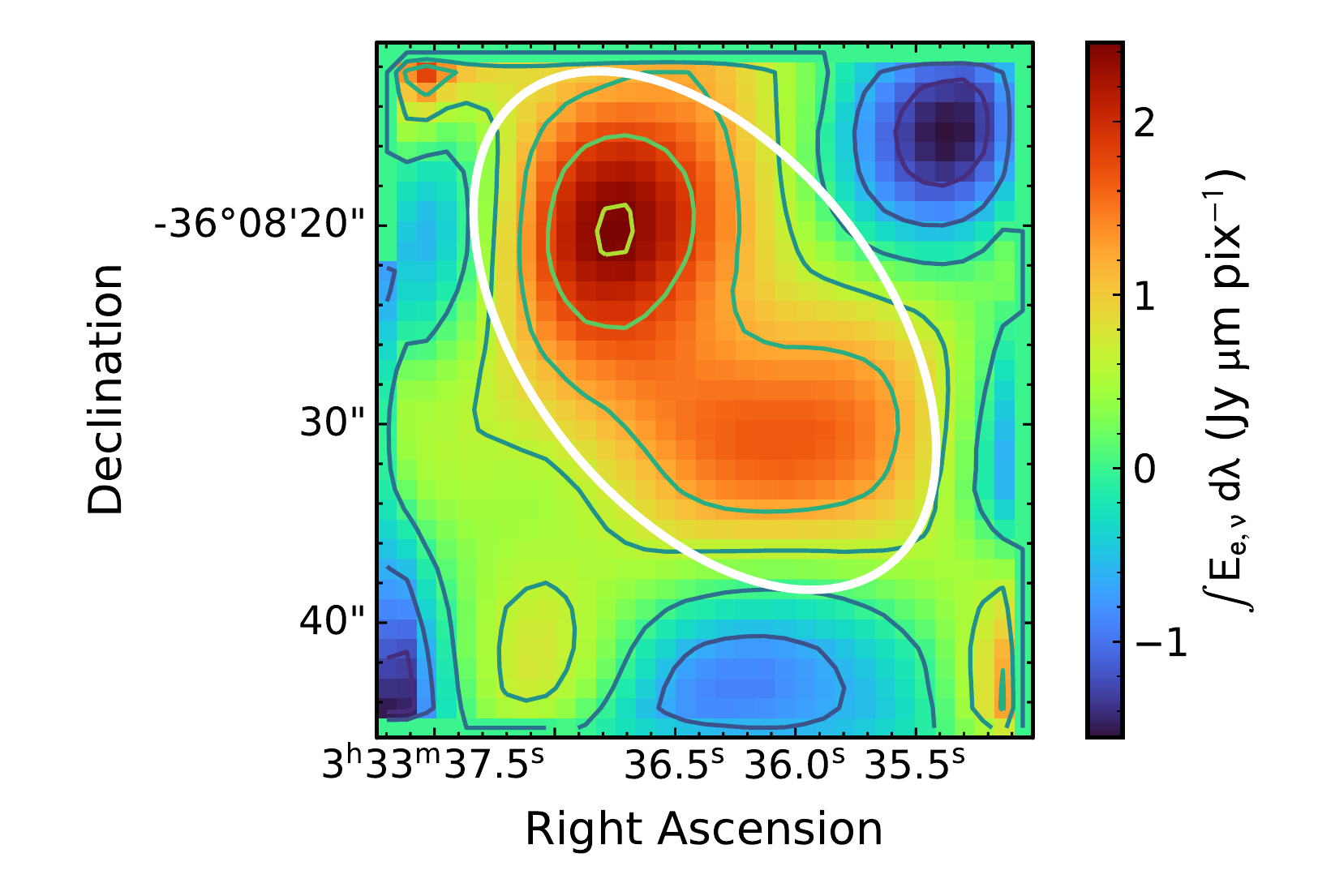}\label{fig:NGC1365-B}}~
    \subfigure[]{\includegraphics[width=0.49\textwidth]{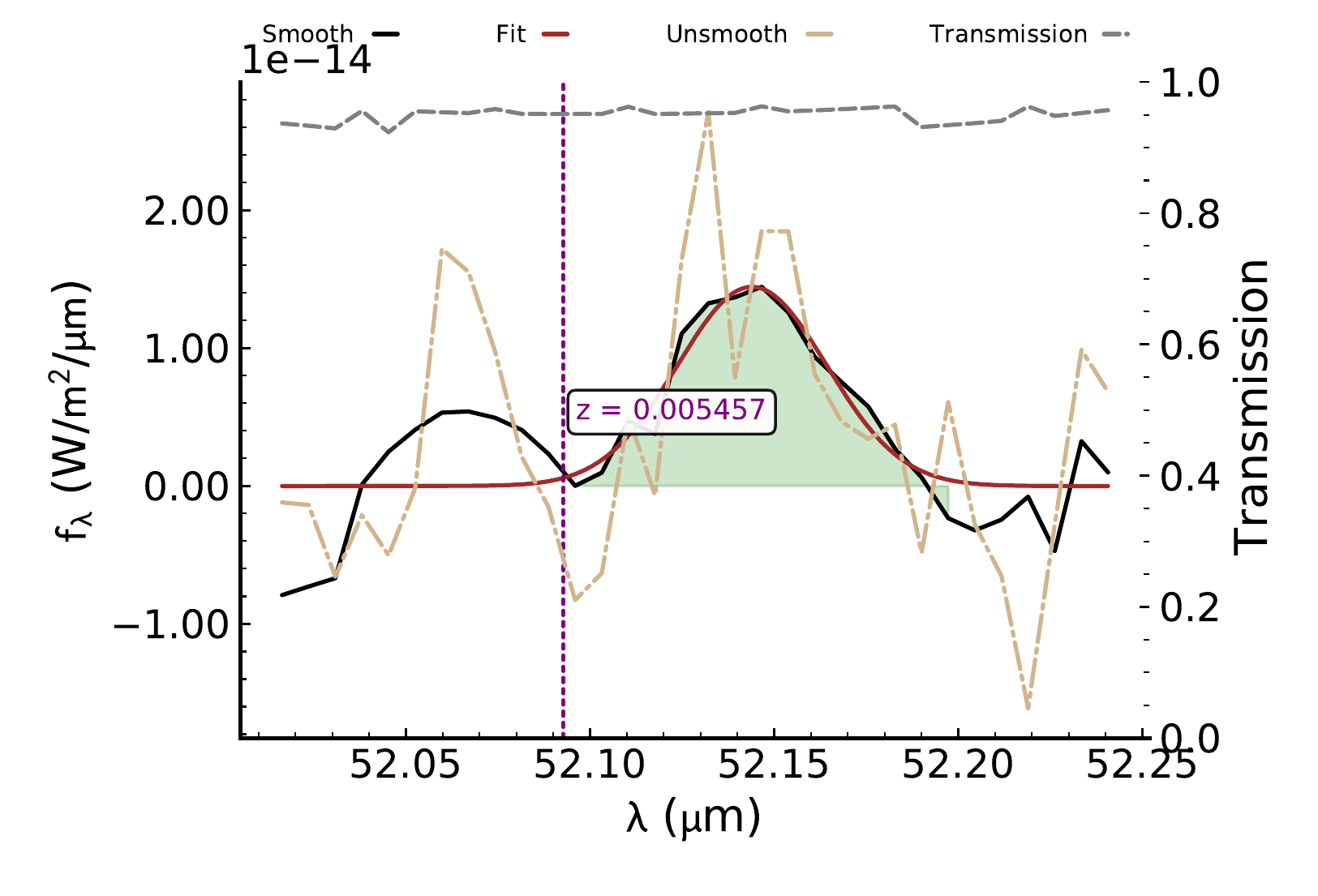}\label{fig:NGC1365-C}}
    \caption{The 2MASS image (Figure \ref{fig:NGC1365-A}) as well as 2-D linemap, and 1-D spectrum for [OIII]52$\mu$m  (Figures  \ref{fig:NGC1365-B} and \ref{fig:NGC1365-C}, respectively) in NGC1365. In this case the optical center, located at the active galactic nucleus lies in between the two peaks of the far-IR emisison which are on either side of the galactic disk.}
    \label{fig:NGC1365}
\end{figure*}

\begin{figure*}[ht!]
    \centering
    \subfigure[]{\includegraphics[width=0.498\textwidth]{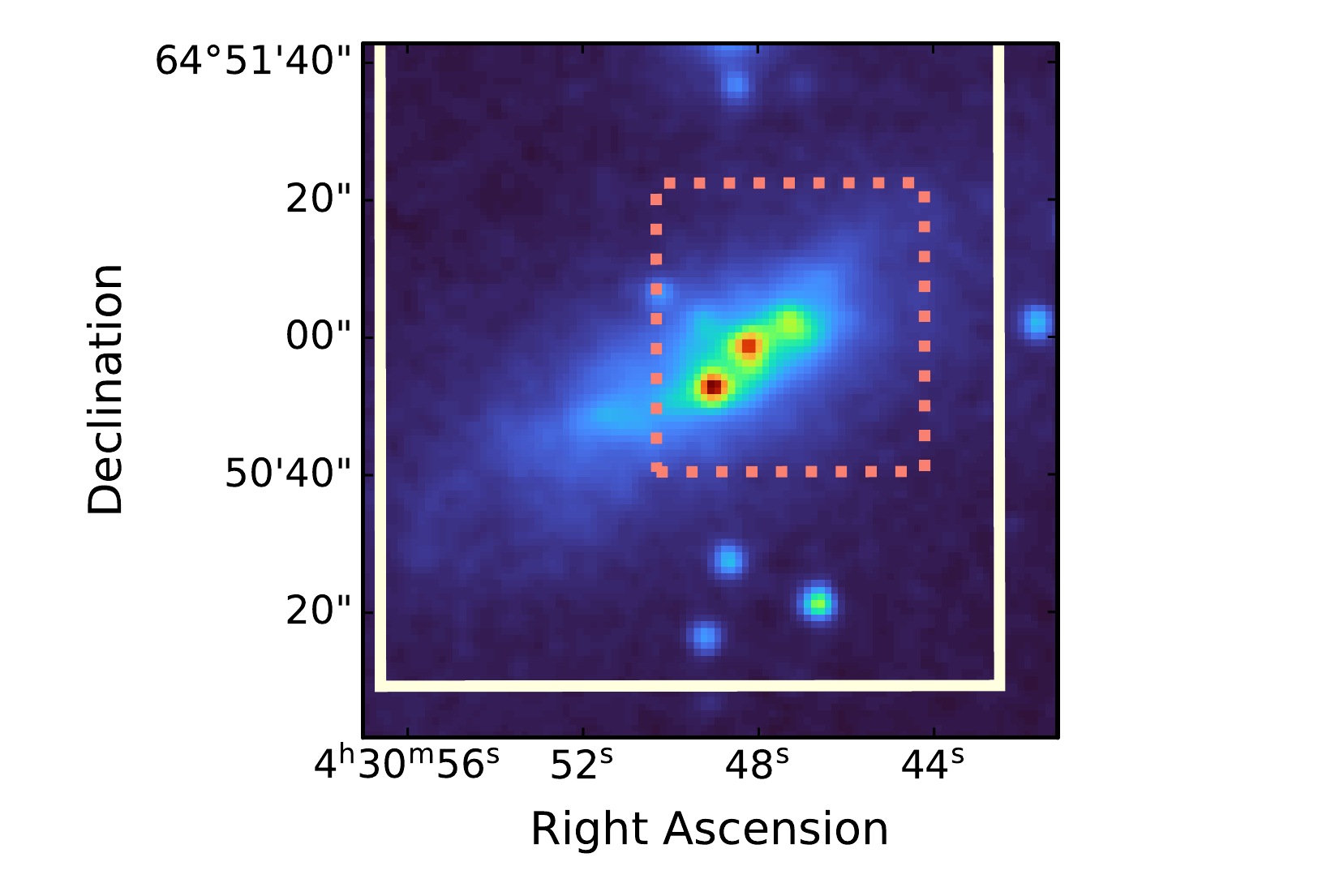}\label{fig:NGC1569-A}}~\\
    \subfigure[]{\includegraphics[width=0.498\textwidth]{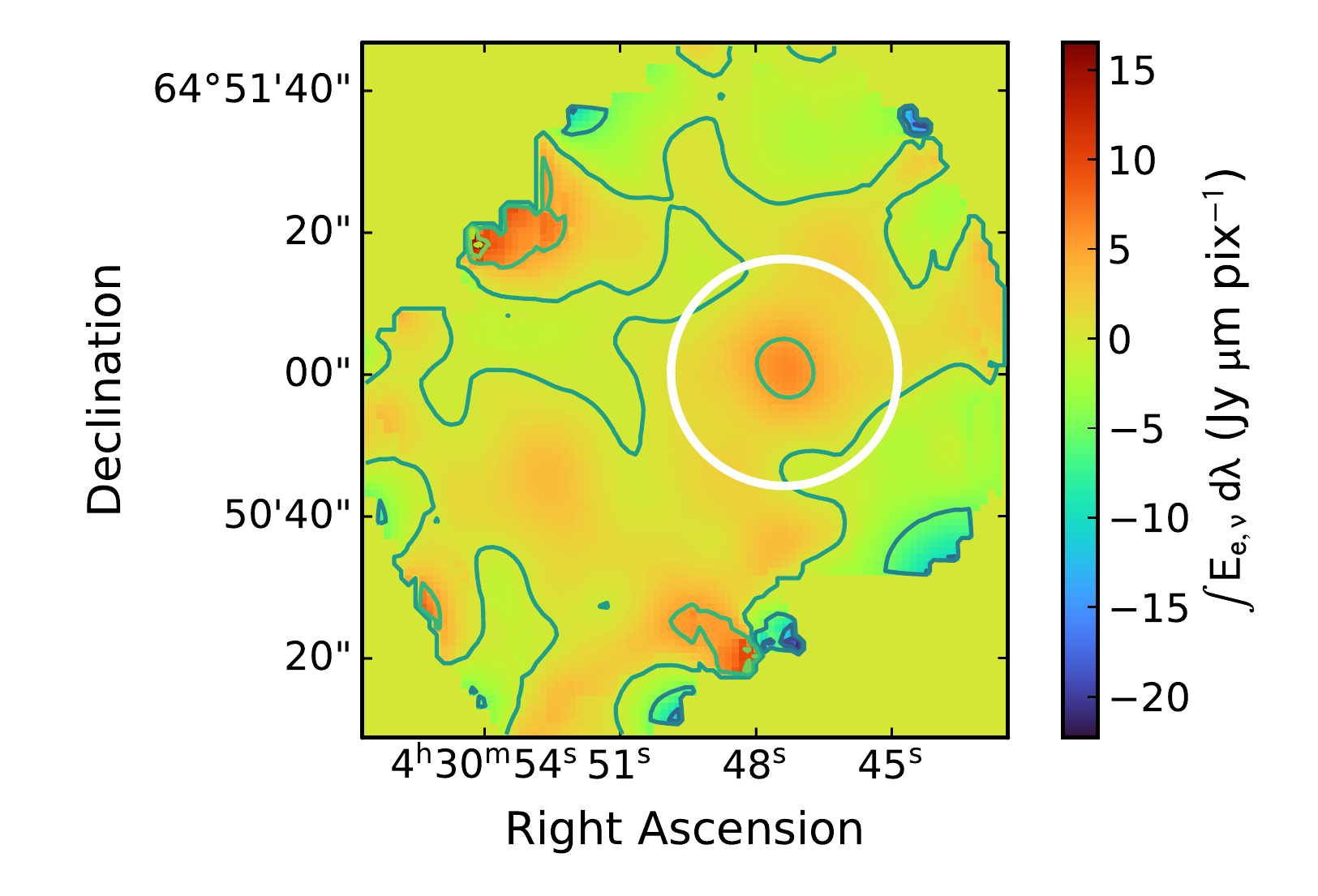}\label{fig:NGC1569-B}}~
    \subfigure[]{\includegraphics[width=0.49\textwidth]{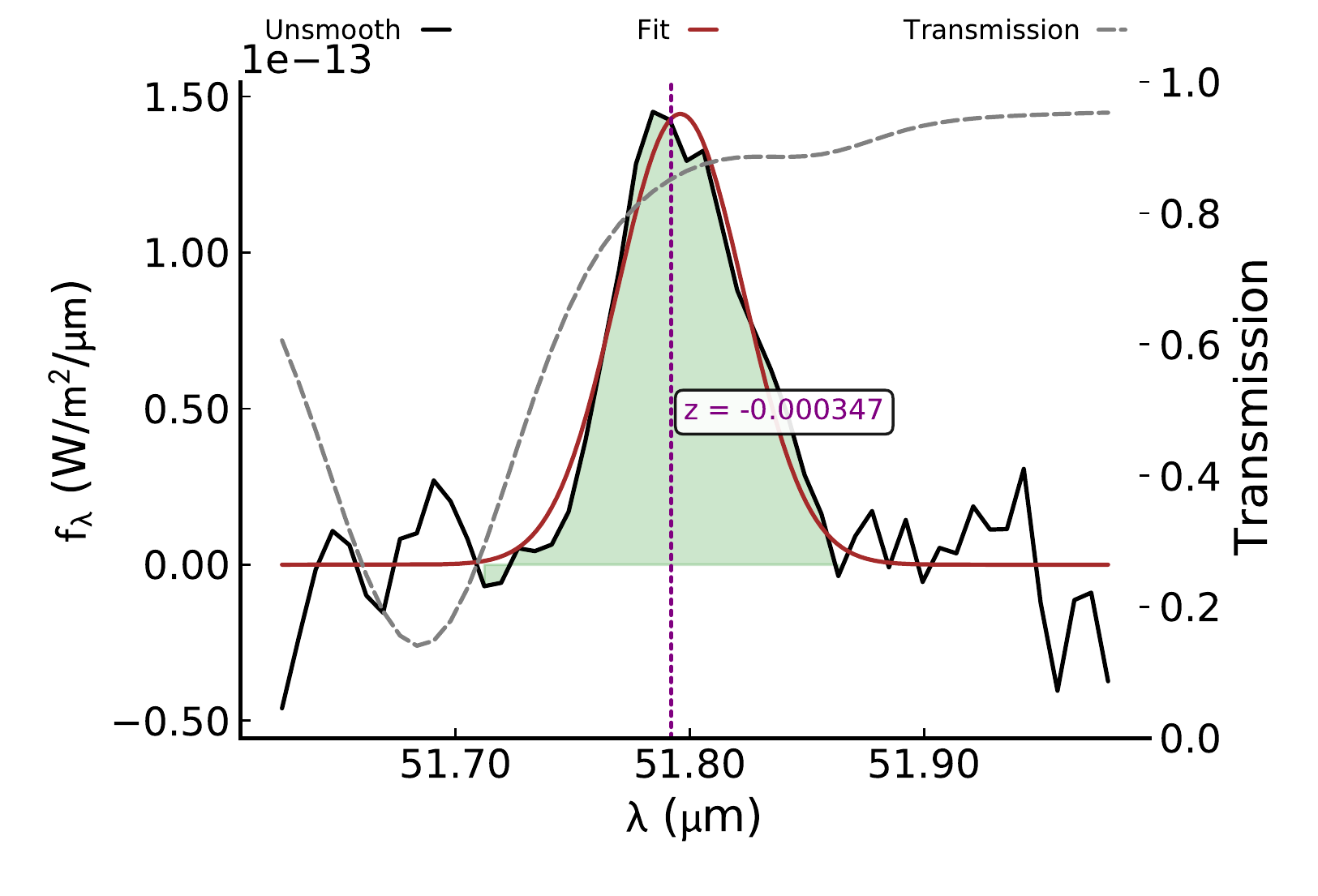}\label{fig:NGC1569-C}}\\
    \subfigure[]{\includegraphics[width=0.498\textwidth]{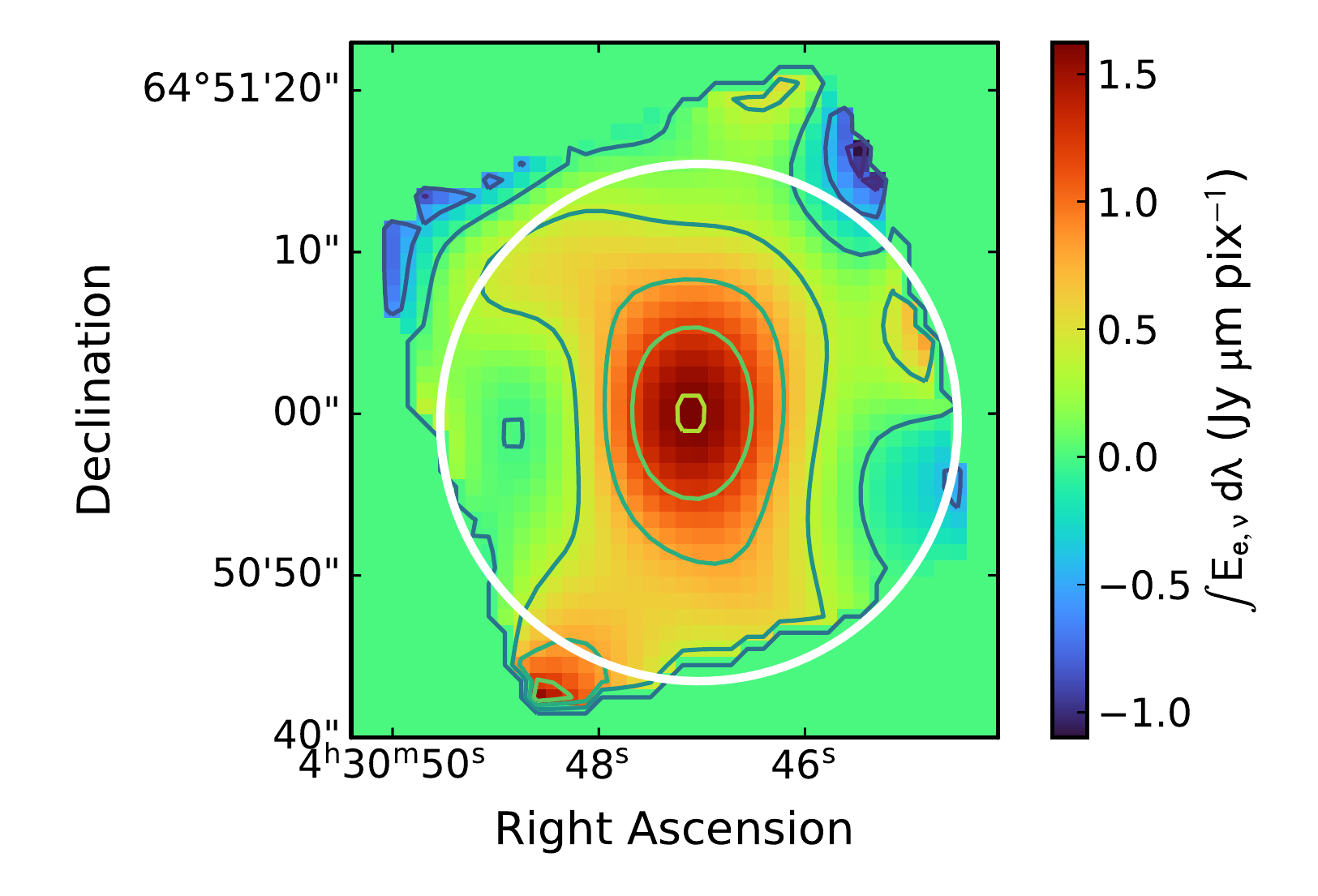}\label{fig:NGC1569-D}}~
    \subfigure[]{\includegraphics[width=0.49\textwidth]{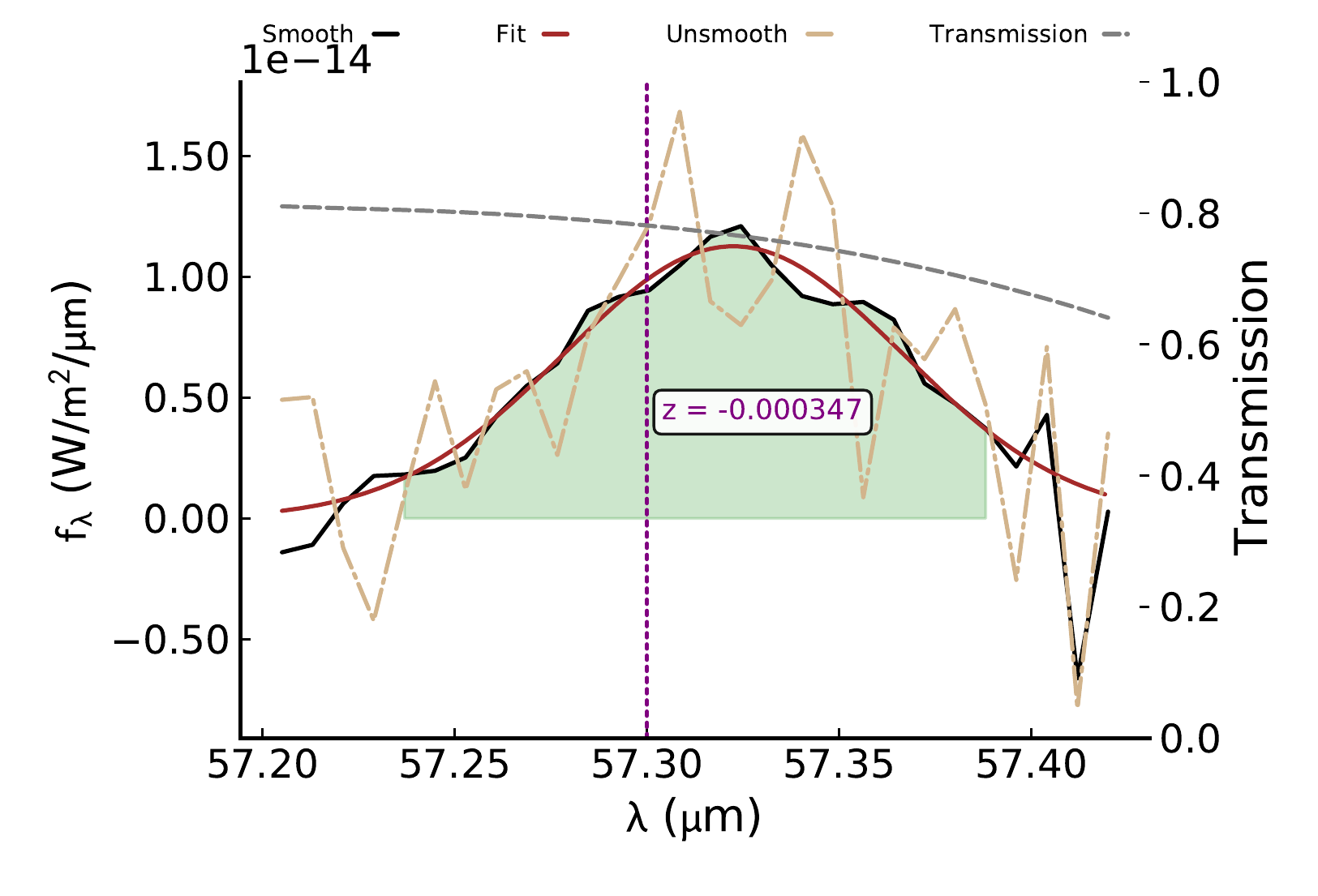}\label{fig:NGC1569-E}}
    \caption{The 2MASS image (Figure \ref{fig:NGC1569-A}), 2-D linemaps and 1-D spectra for [OIII]52$\mu$m  (Figures \ref{fig:NGC1569-B} and \ref{fig:NGC1569-C}, respectively) and [NIII]57$\mu$m  (Figures \ref{fig:NGC1569-D} and \ref{fig:NGC1569-E}, respectively) in NGC1569. The profiles for both lines have not been corrected for atmospheric transmission, and the large apertures are meant to be comparable to the Herschel apertures used for calculating [OIII]88$\mu$m. The [OIII]52$\mu$m linemap is a scan of a larger area, but apertures for both [OIII]52$\mu$m and [NIII]57$\mu$m have the same size, and are centered around the same coordinate.}
    \label{fig:NGC1569}
\end{figure*}

\begin{figure*}[ht!]
    \centering
    \subfigure[]{\includegraphics[width=0.498\textwidth]{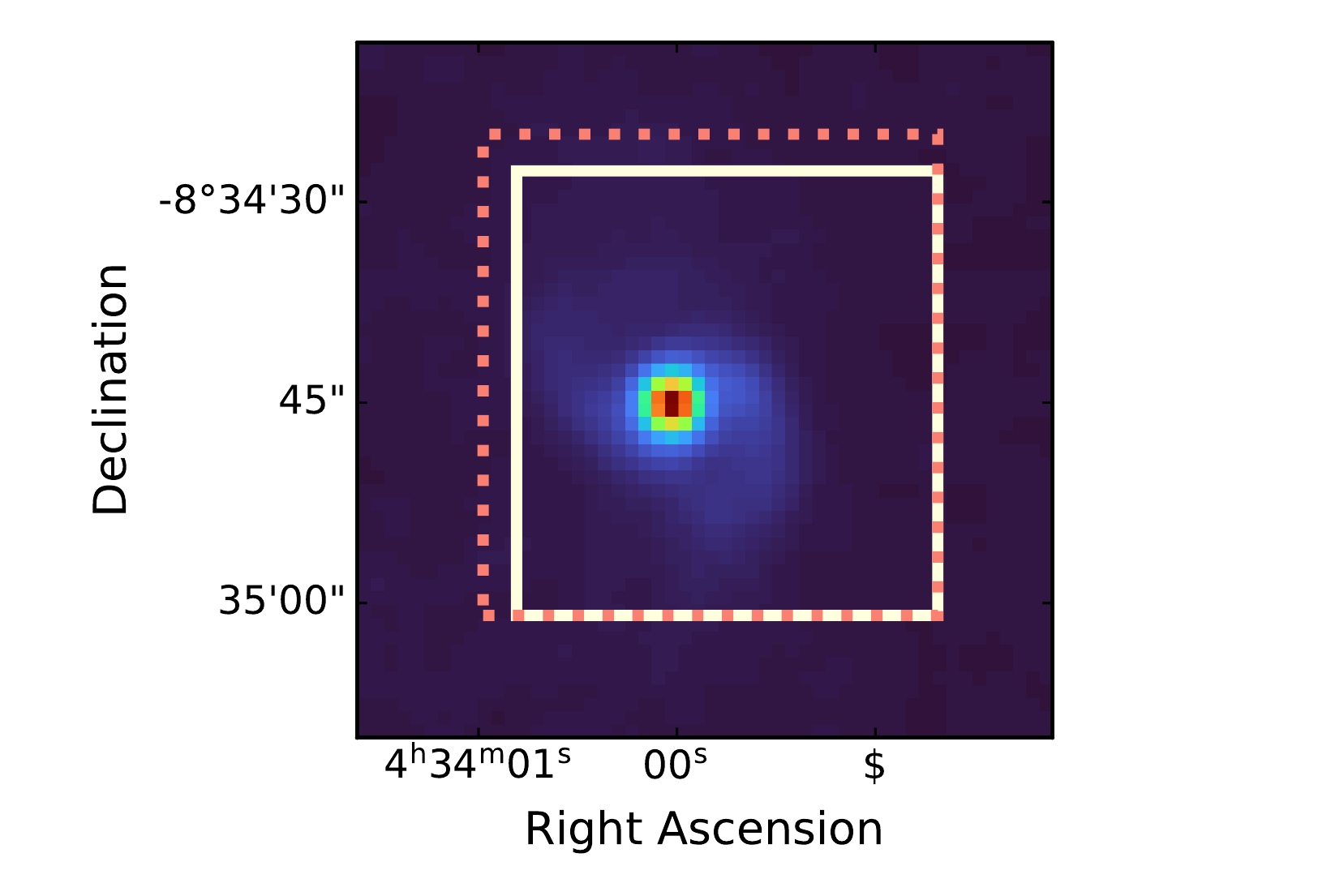}\label{fig:NGC1614-A}}\\
    \subfigure[]{\includegraphics[width=0.498\textwidth]{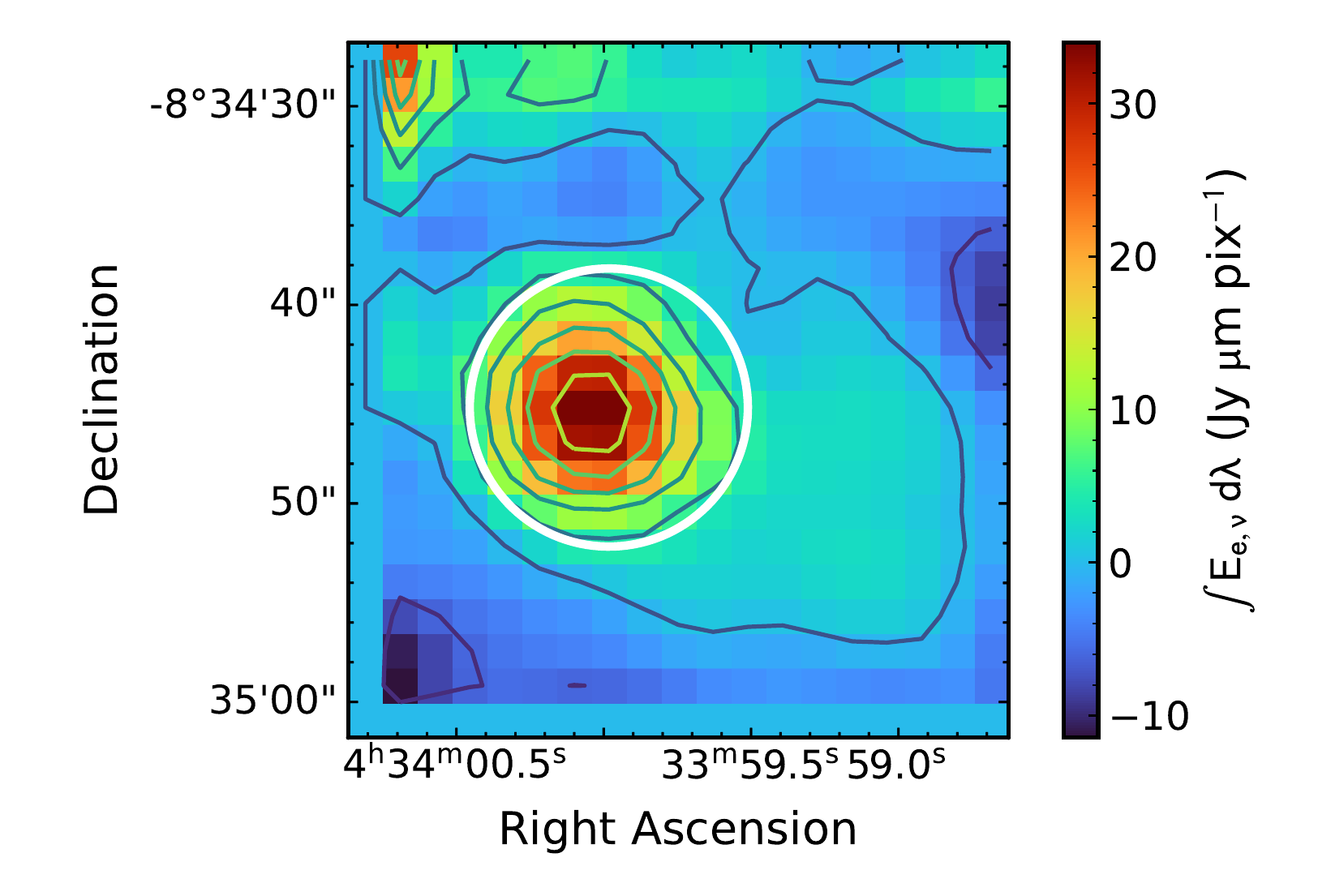}\label{fig:NGC1614-B}}~
    \subfigure[]{\includegraphics[width=0.49\textwidth]{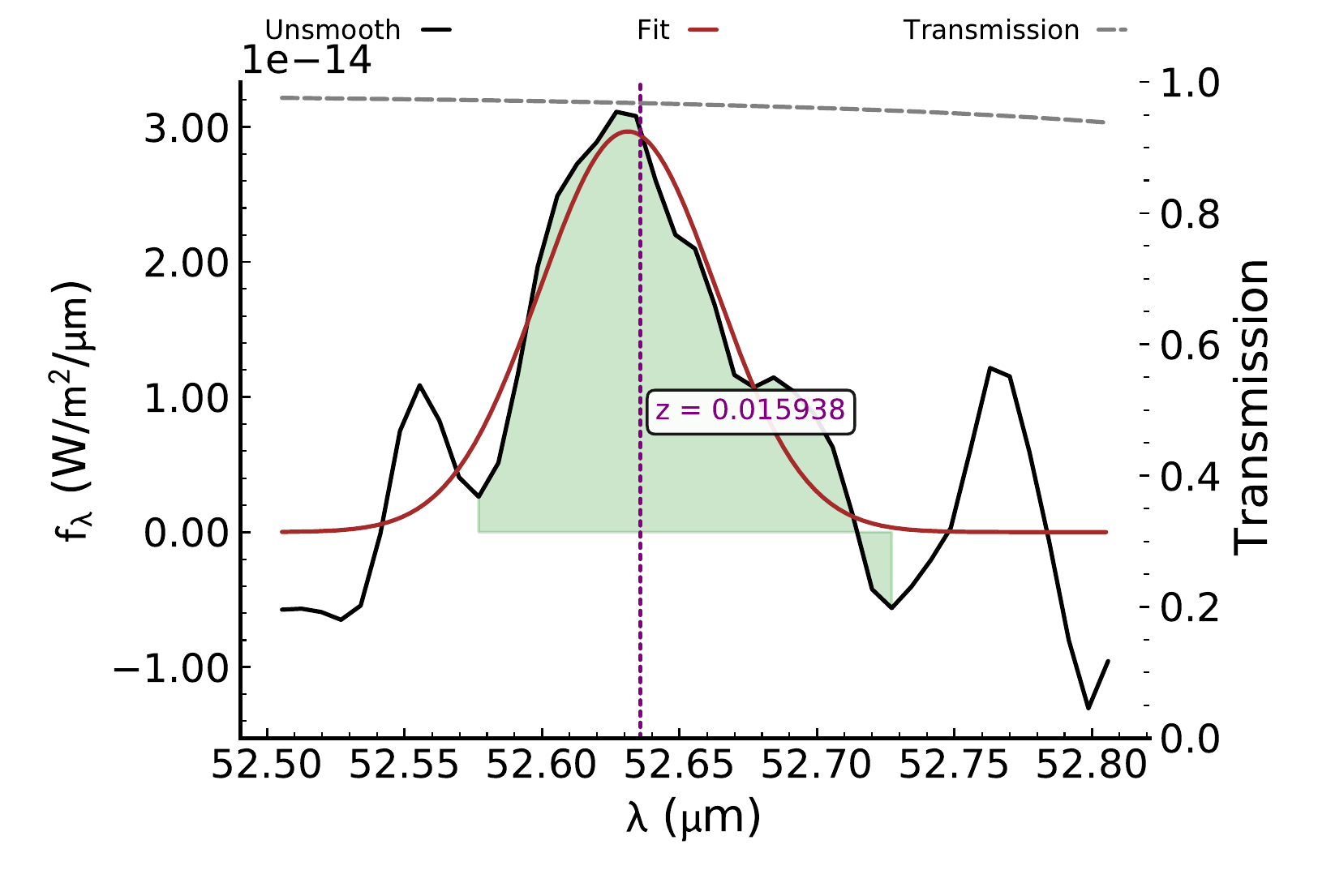}\label{fig:NGC1614-C}}\\
    \subfigure[]{\includegraphics[width=0.498\textwidth]{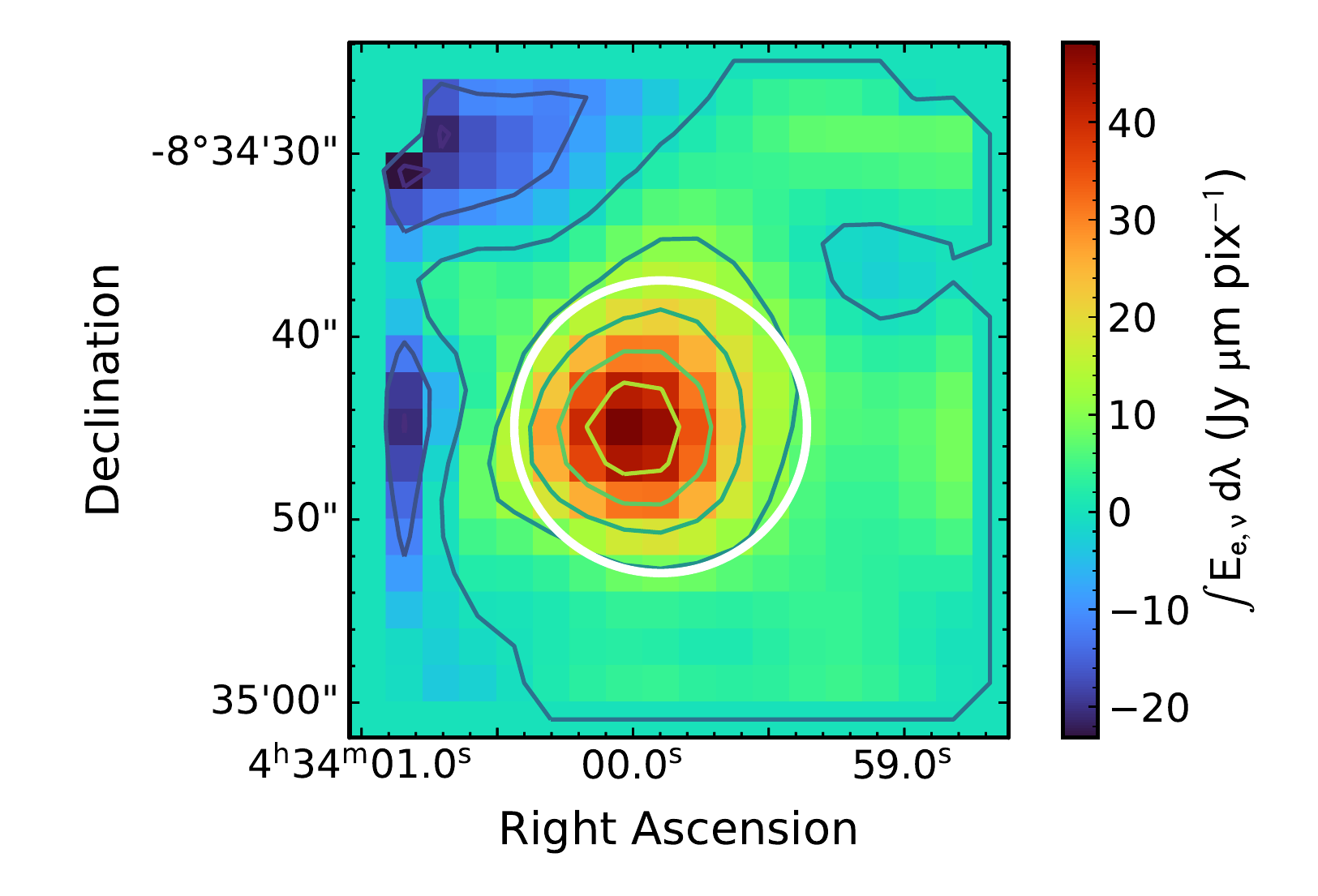}\label{fig:NGC1614-D}}~
    \subfigure[]{\includegraphics[width=0.49\textwidth]{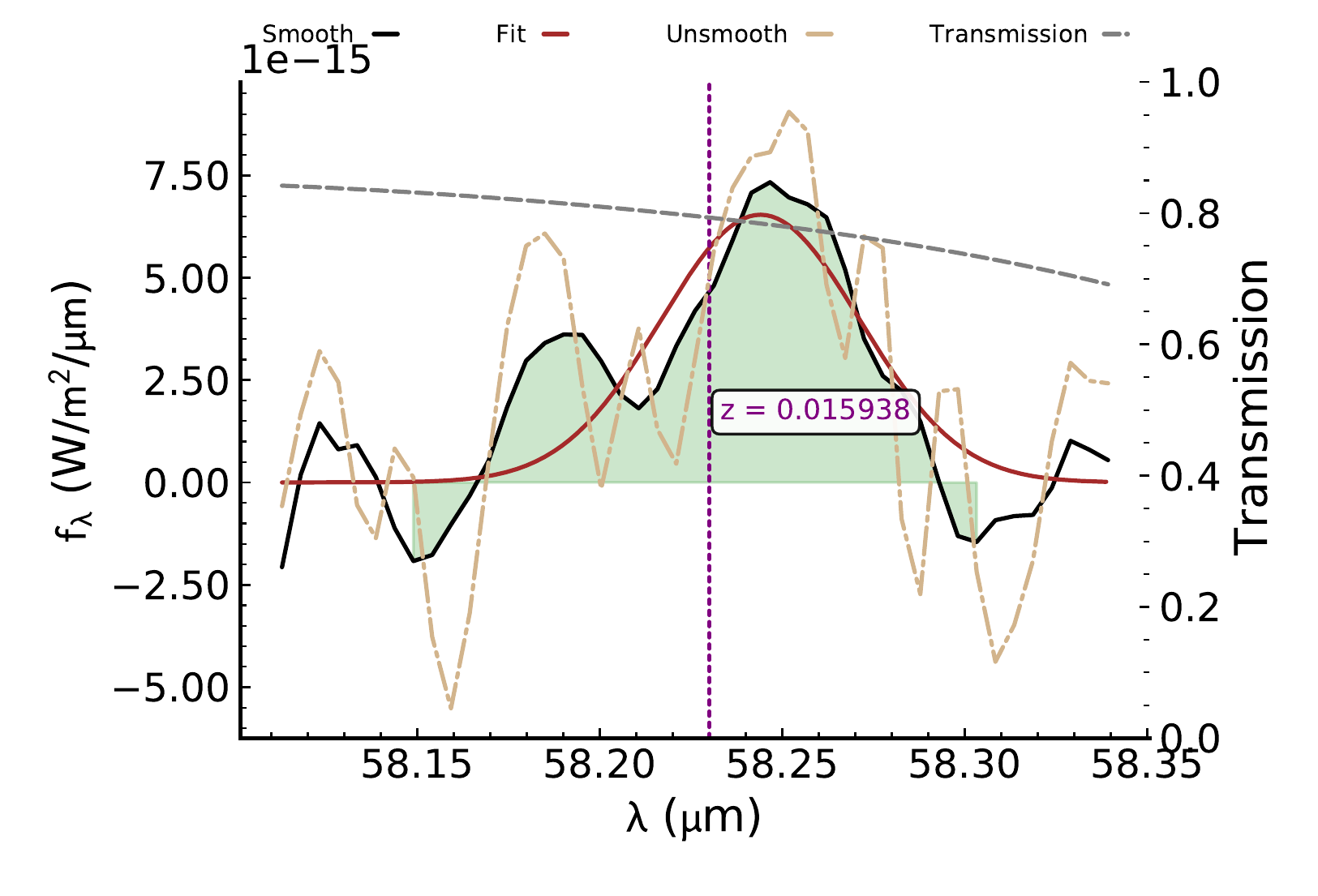}\label{fig:NGC1614-E}}
    \caption{The 2MASS image (Figure \ref{fig:NGC1614-A}), 2-D linemaps and 1-D spectra for [OIII]52$\mu$m  (Figures \ref{fig:NGC1614-B} and \ref{fig:NGC1614-C}, respectively) and [NIII]57$\mu$m  (Figures \ref{fig:NGC1614-D} and \ref{fig:NGC1614-E}, respectively) in NGC1614.}
    \label{fig:NGC1614}
\end{figure*}

\begin{figure*}[ht!]
    \centering
    \subfigure[]{\includegraphics[width=0.498\textwidth]{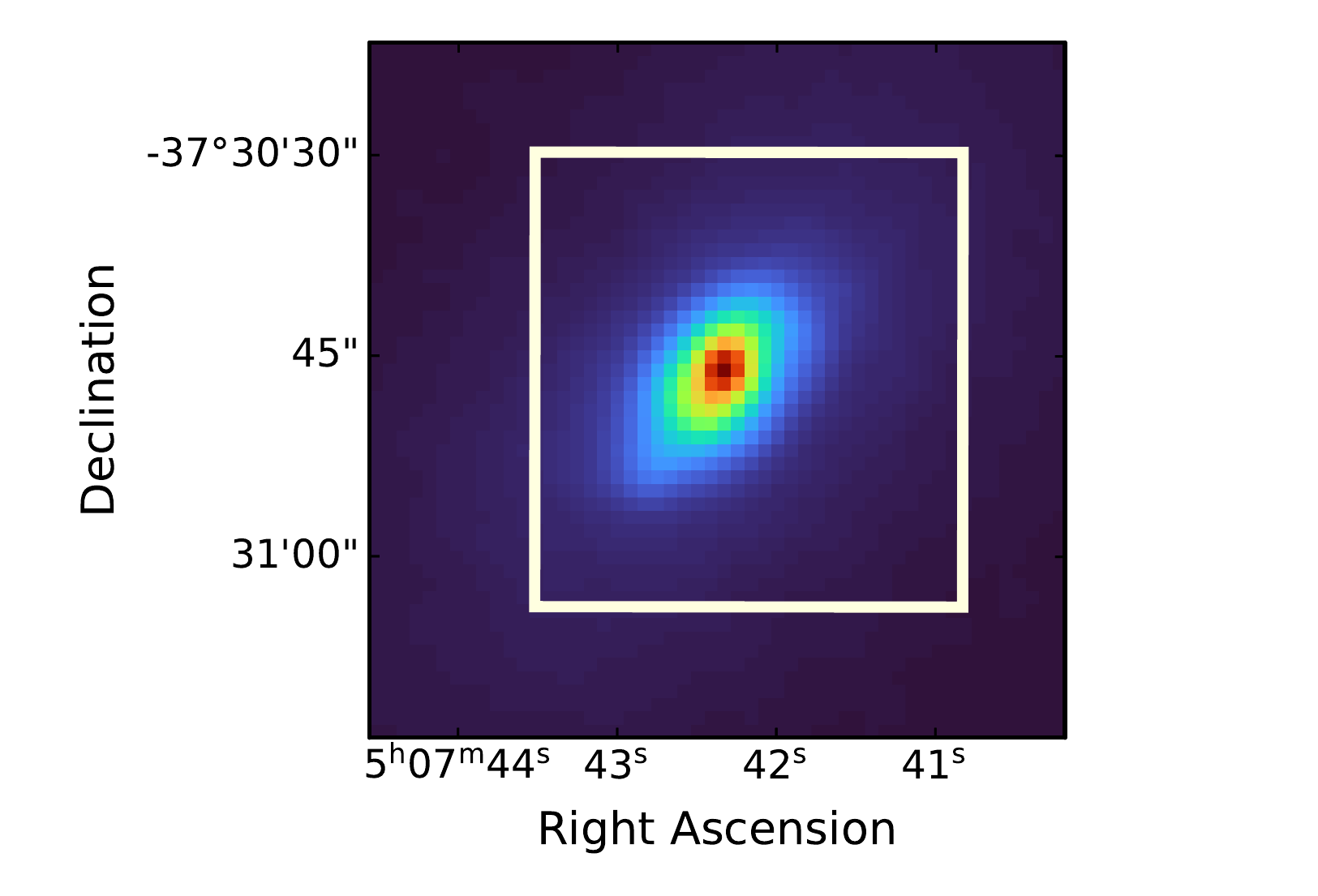}\label{fig:NGC1808-A}}\\
    \subfigure[]{\includegraphics[width=0.498\textwidth]{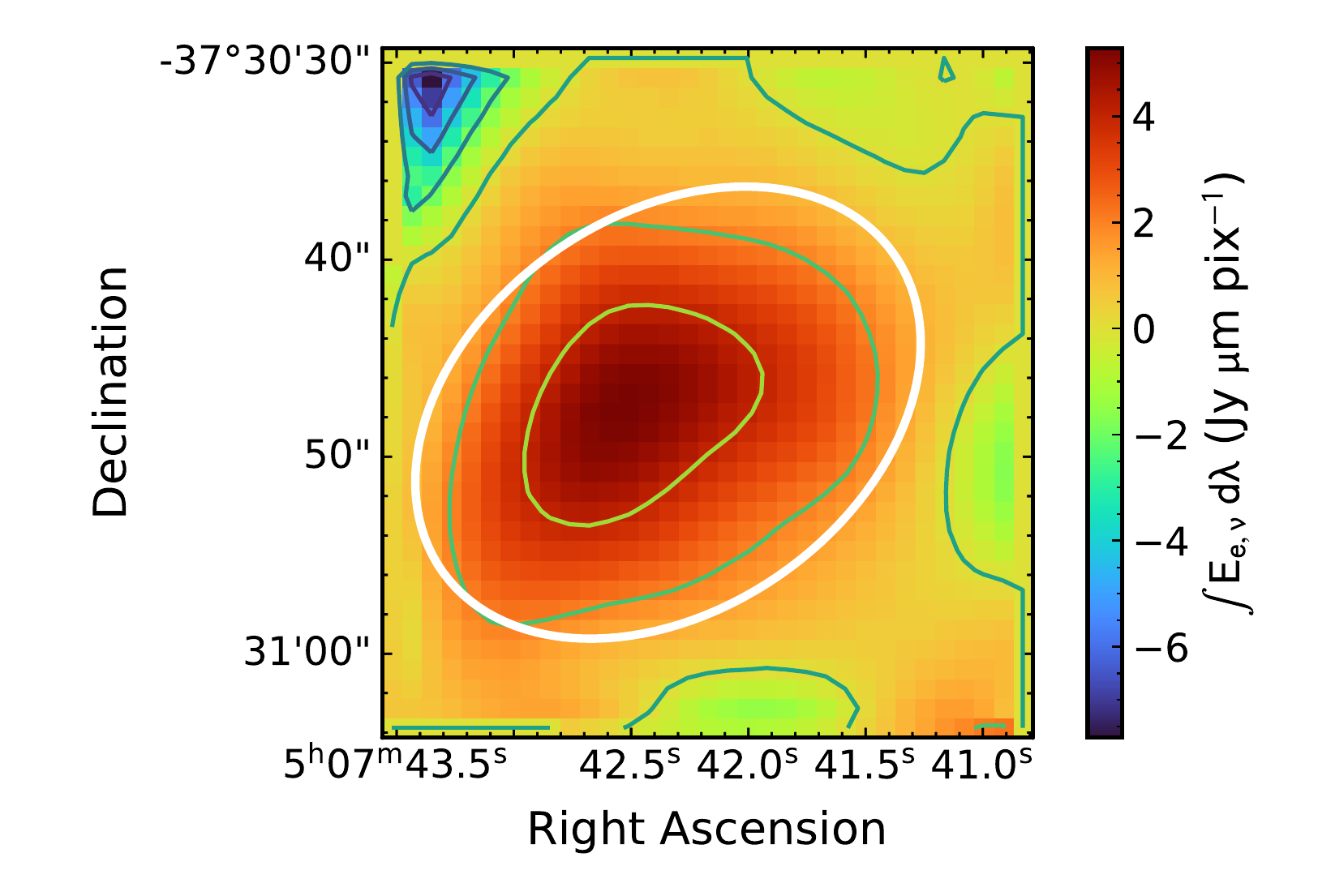}\label{fig:NGC1808-B}}~
    \subfigure[]{\includegraphics[width=0.49\textwidth]{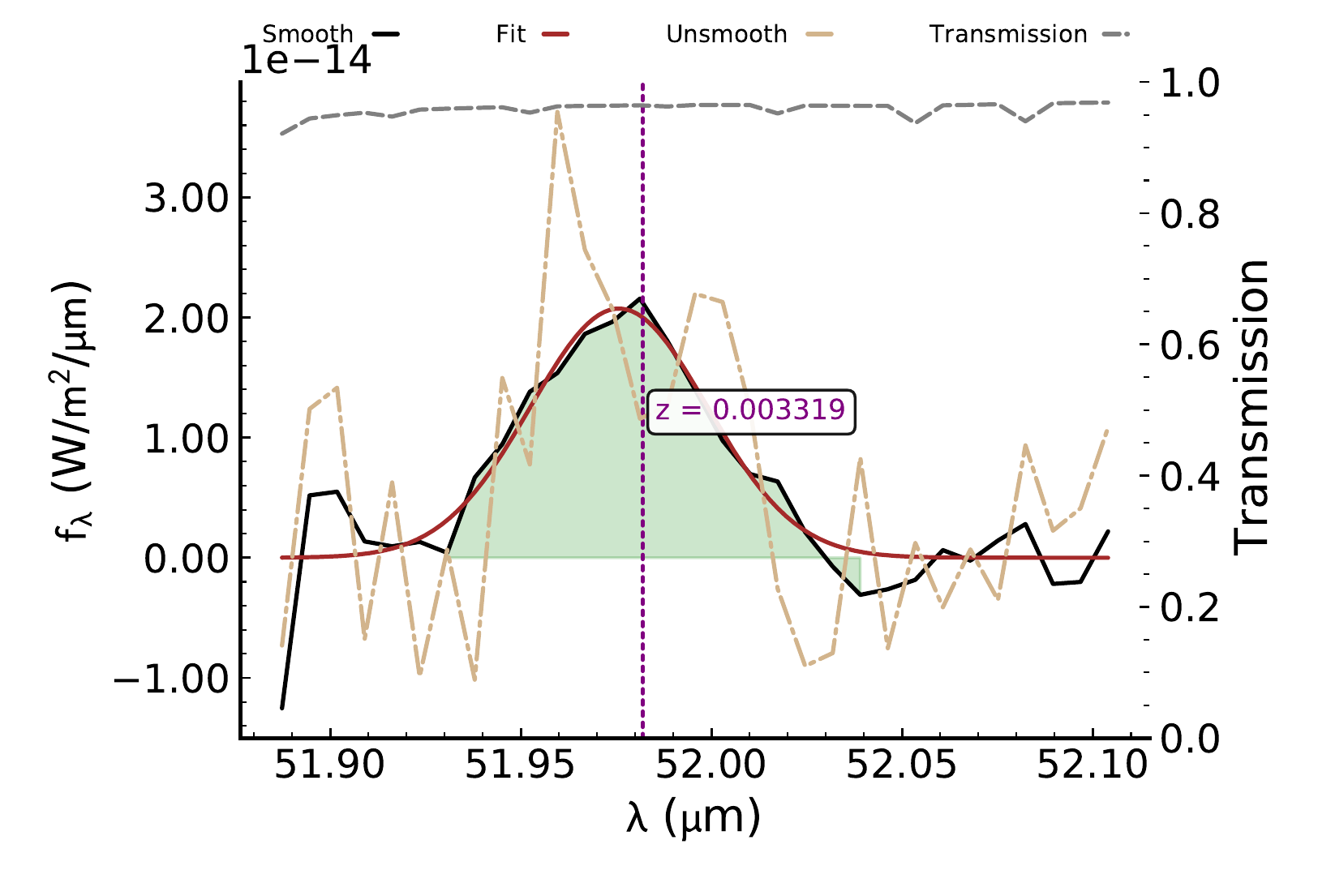}\label{fig:NGC1808-C}}
    \caption{The 2MASS image (Figure \ref{fig:NGC1808-A}) as well as 2-D linemap and 1-D spectrum for [OIII]52$\mu$m  (Figures \ref{fig:NGC1808-B} and \ref{fig:NGC1808-C}, respectively) in NGC1808.}
    \label{fig:NGC1808}
\end{figure*}

\begin{figure*}[ht!]
    \centering
    \subfigure[]{\includegraphics[width=0.498\textwidth]{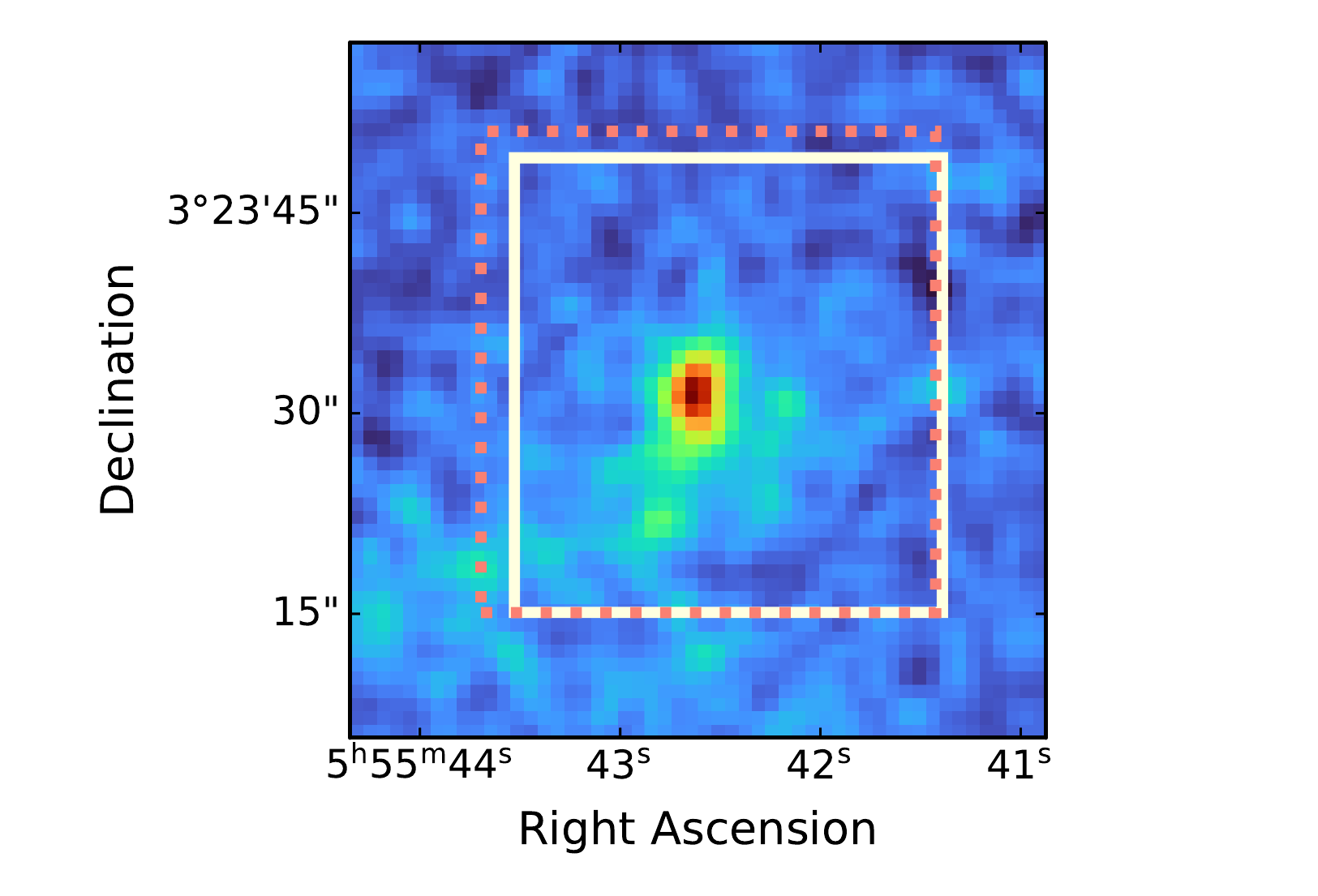}\label{fig:IIZw40-A}}~\\
    \subfigure[]{\includegraphics[width=0.498\textwidth]{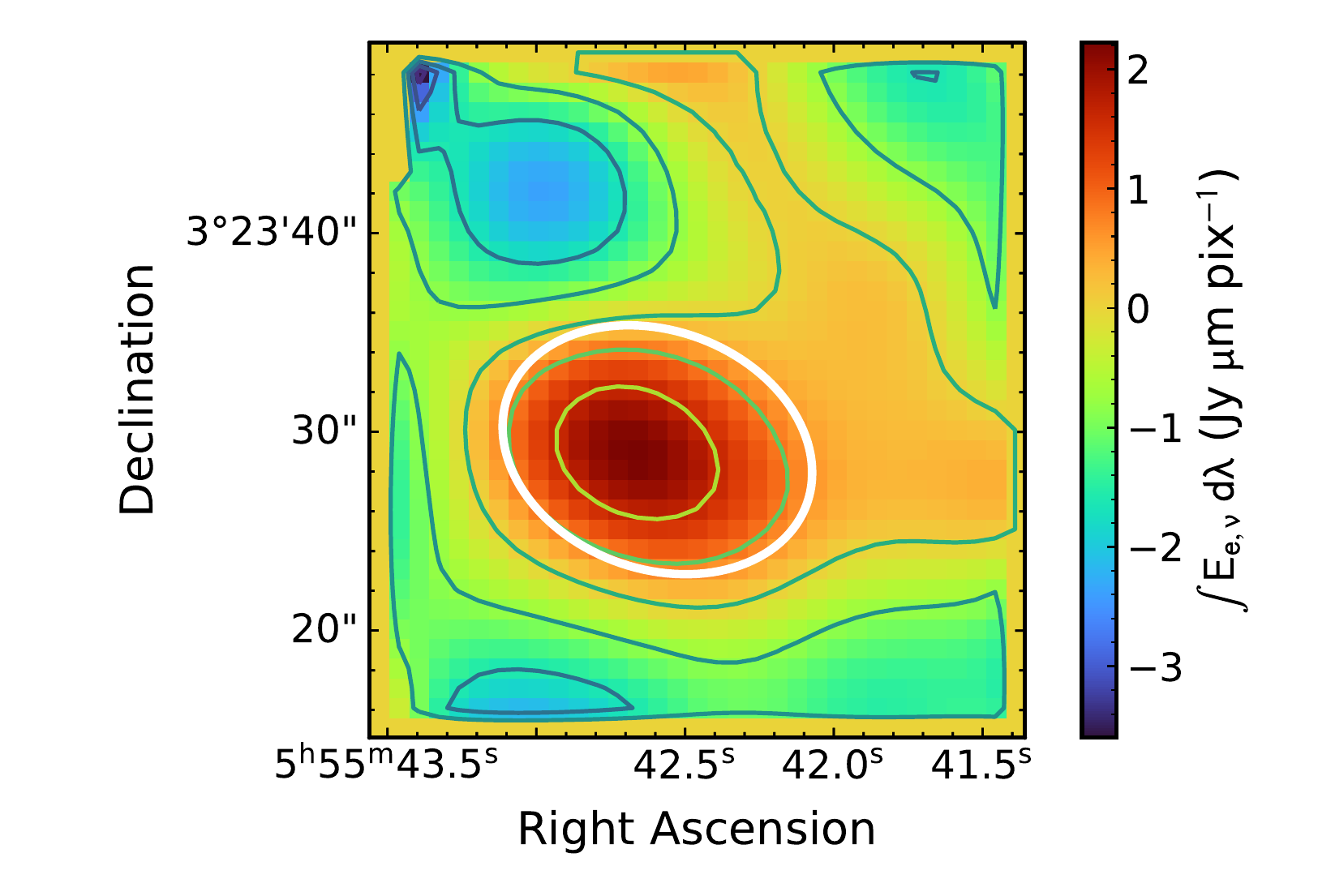}\label{fig:IIZw40-B}}~
    \subfigure[]{\includegraphics[width=0.49\textwidth]{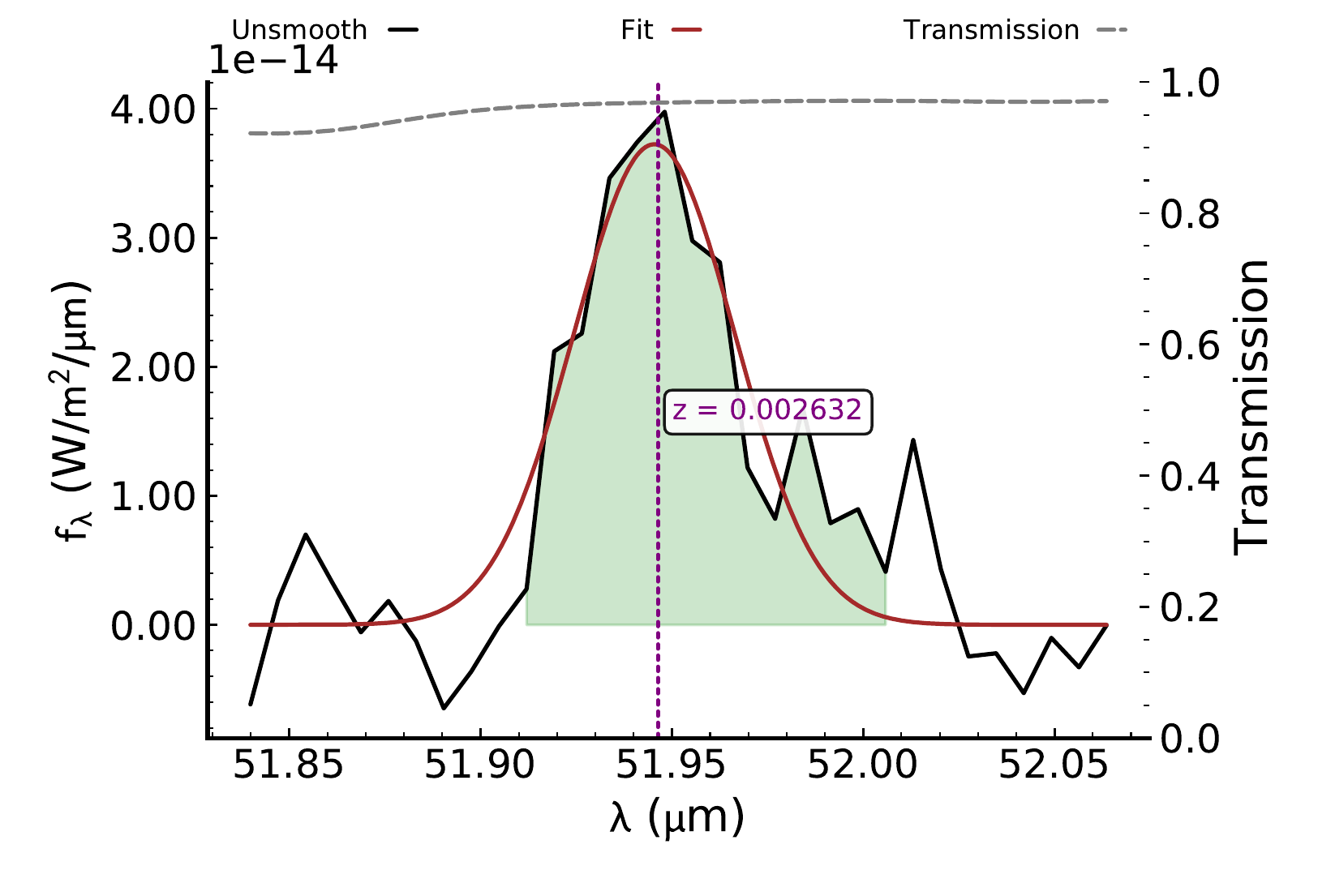}\label{fig:IIZw40-C}}\\
    \subfigure[]{\includegraphics[width=0.498\textwidth]{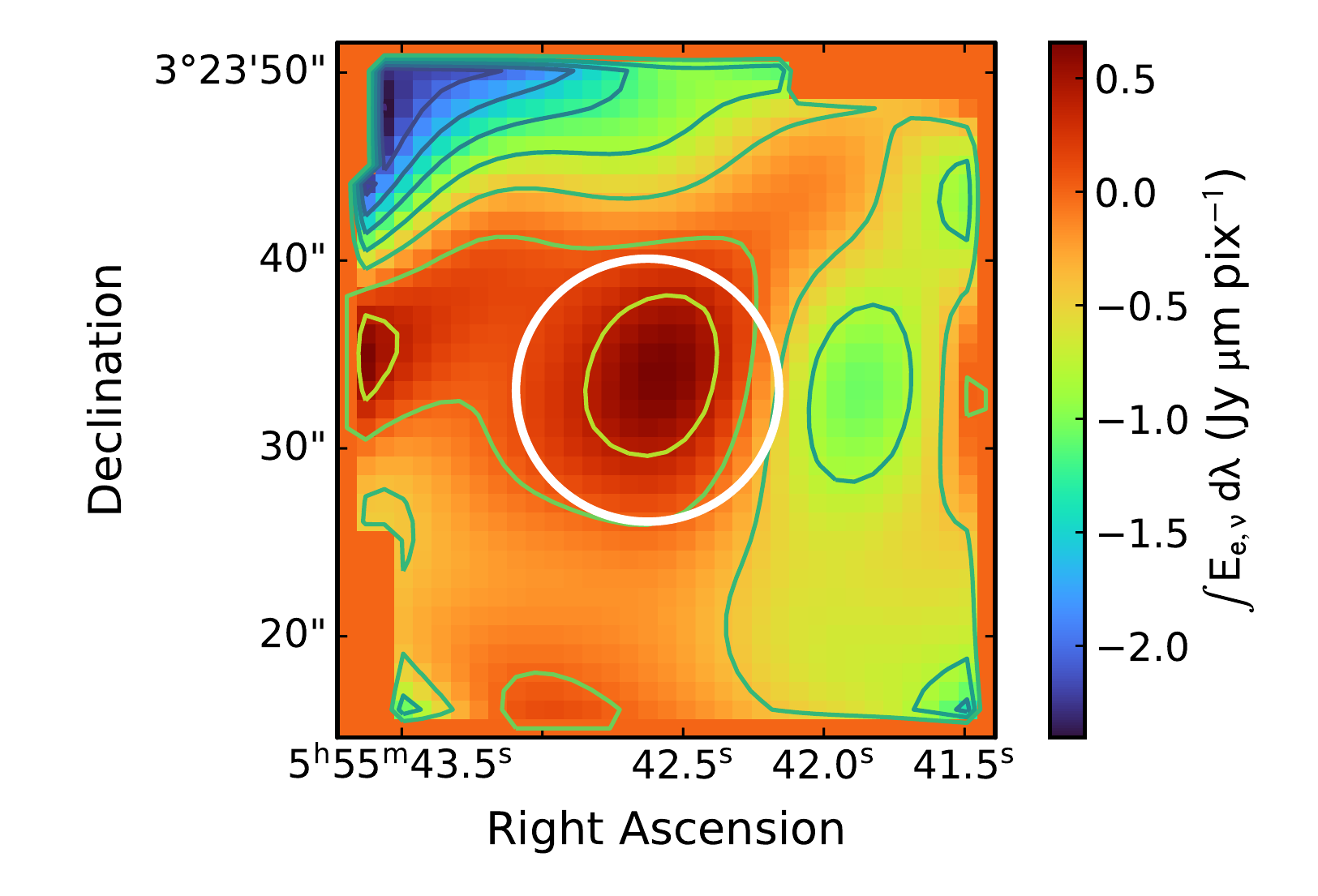}\label{fig:IIZw40-D}}~
    \subfigure[]{\includegraphics[width=0.49\textwidth]{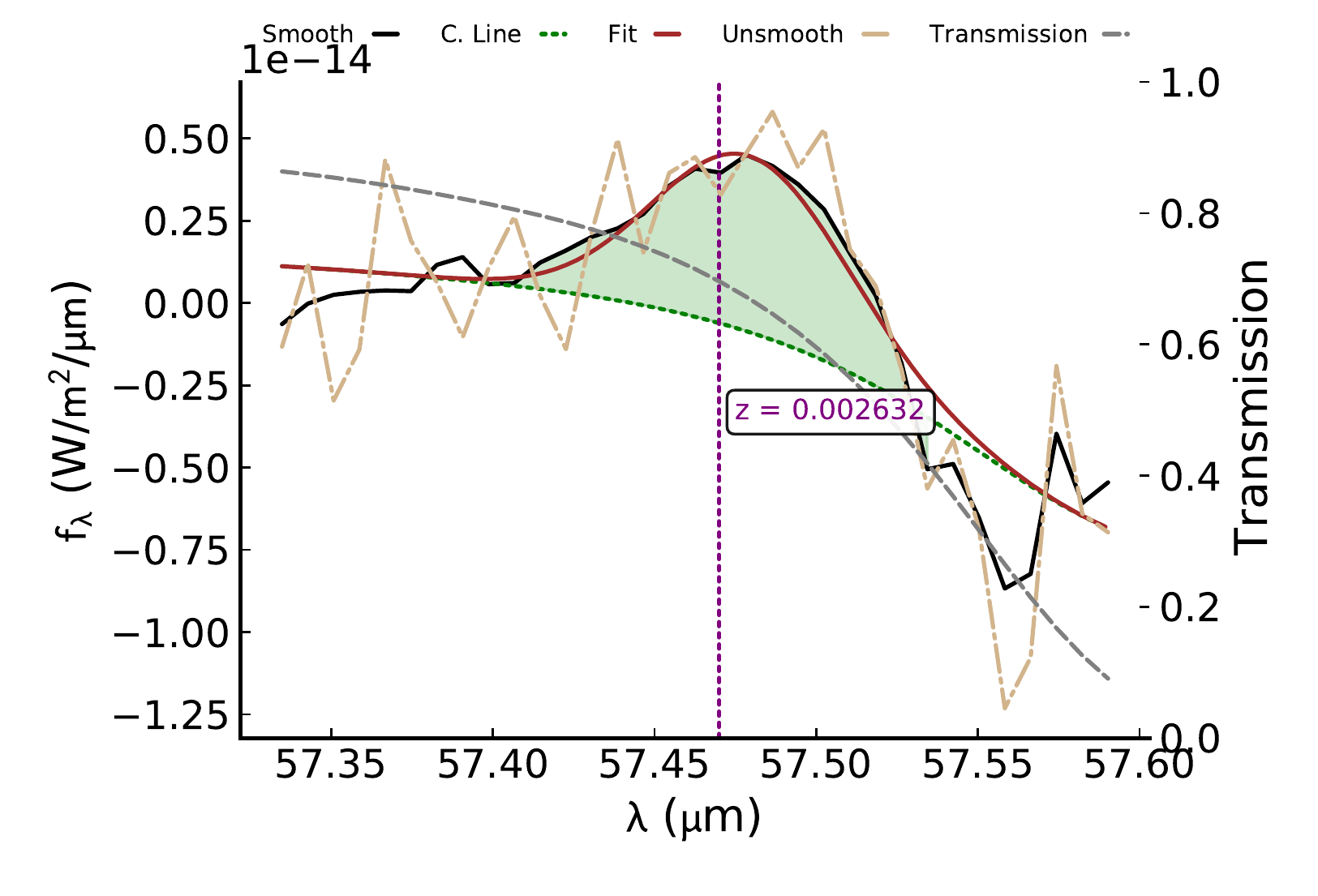}\label{fig:IIZw40-E}}
    \caption{The 2MASS image (Figure \ref{fig:IIZw40-A}) 2-D linemaps and 1-D spectra for [OIII]52$\mu$m  (Figures \ref{fig:IIZw40-B} and \ref{fig:IIZw40-C}, respectively) and [NIII]57$\mu$m  (Figures \ref{fig:IIZw40-D} and \ref{fig:IIZw40-E}, respectively) in IIZw40. The [NIII]57$\mu$m profile has not been corrected for atmospheric transmission. Further, to account for the poor transmission towards the red wavelengths, the continuum has been chosen as the best fit of a straight line times the transmission line shape (green dotted line in Figure \ref{fig:IIZw40-D}).}
    \label{fig:IIZw40}
\end{figure*}


\begin{figure*}[ht!]
    \centering
    \subfigure[]{\includegraphics[width=0.498\textwidth]{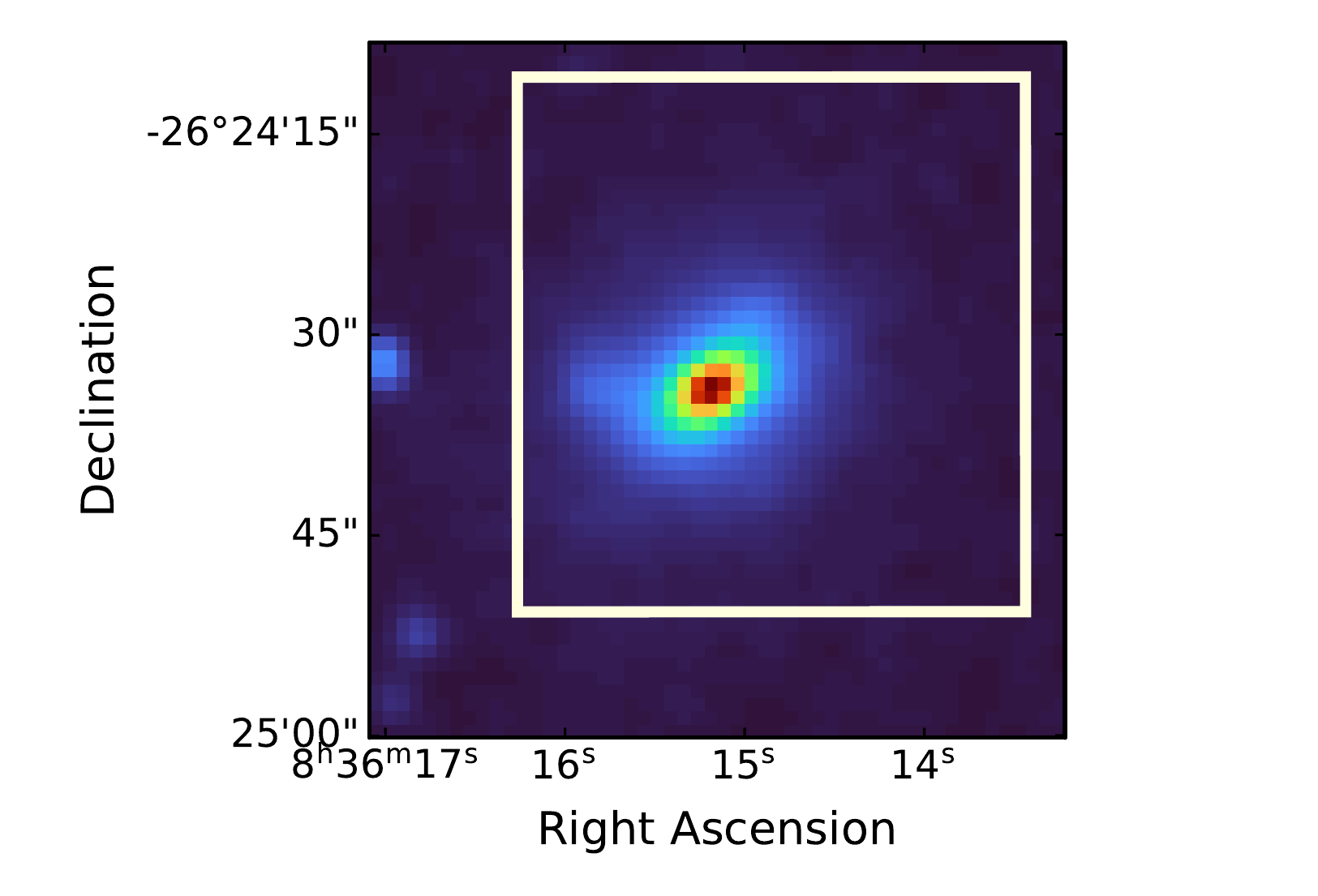}\label{fig:He2-10-A}}\\
    \subfigure[]{\includegraphics[width=0.498\textwidth]{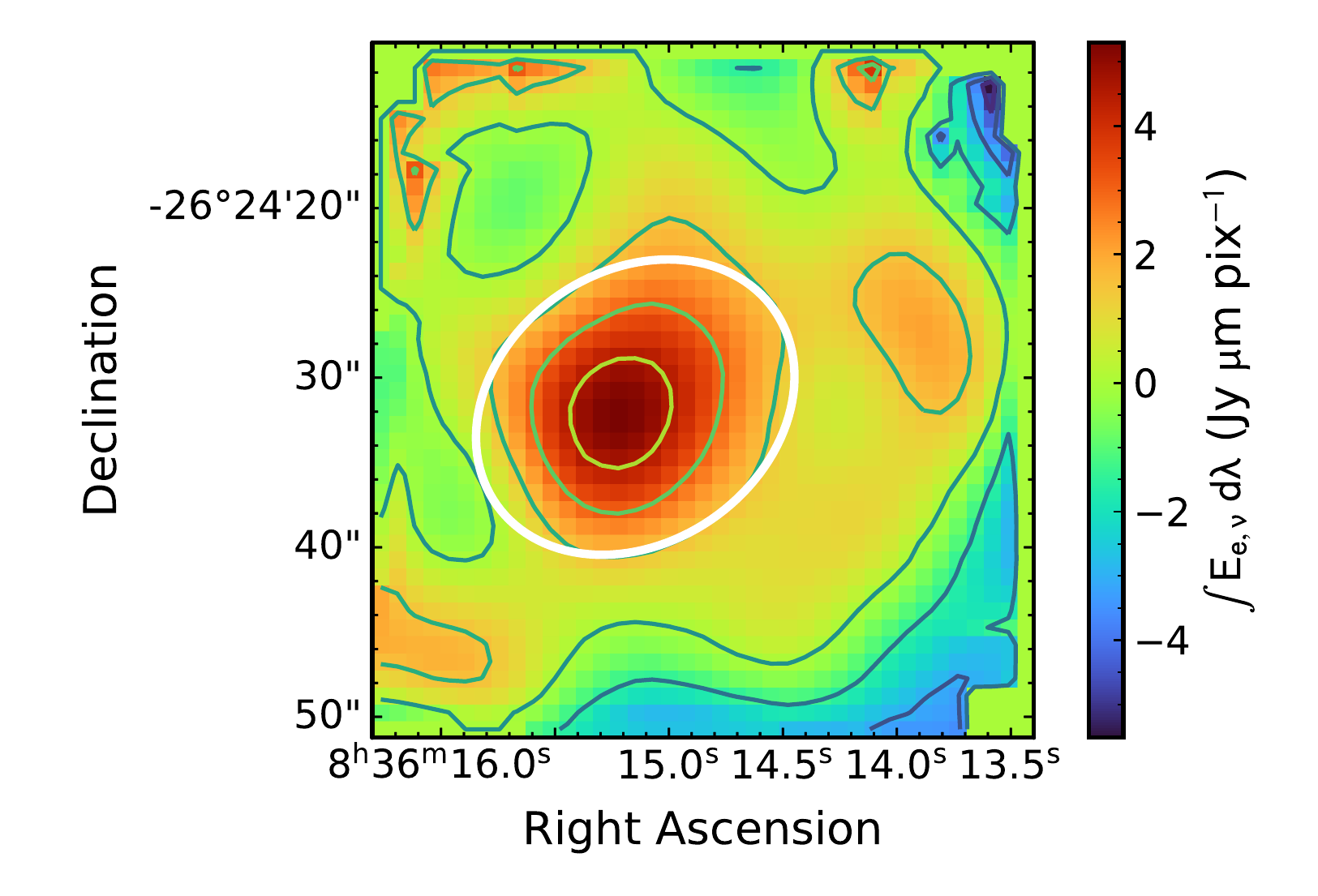}\label{fig:He2-10-B}}~
    \subfigure[]{\includegraphics[width=0.49\textwidth]{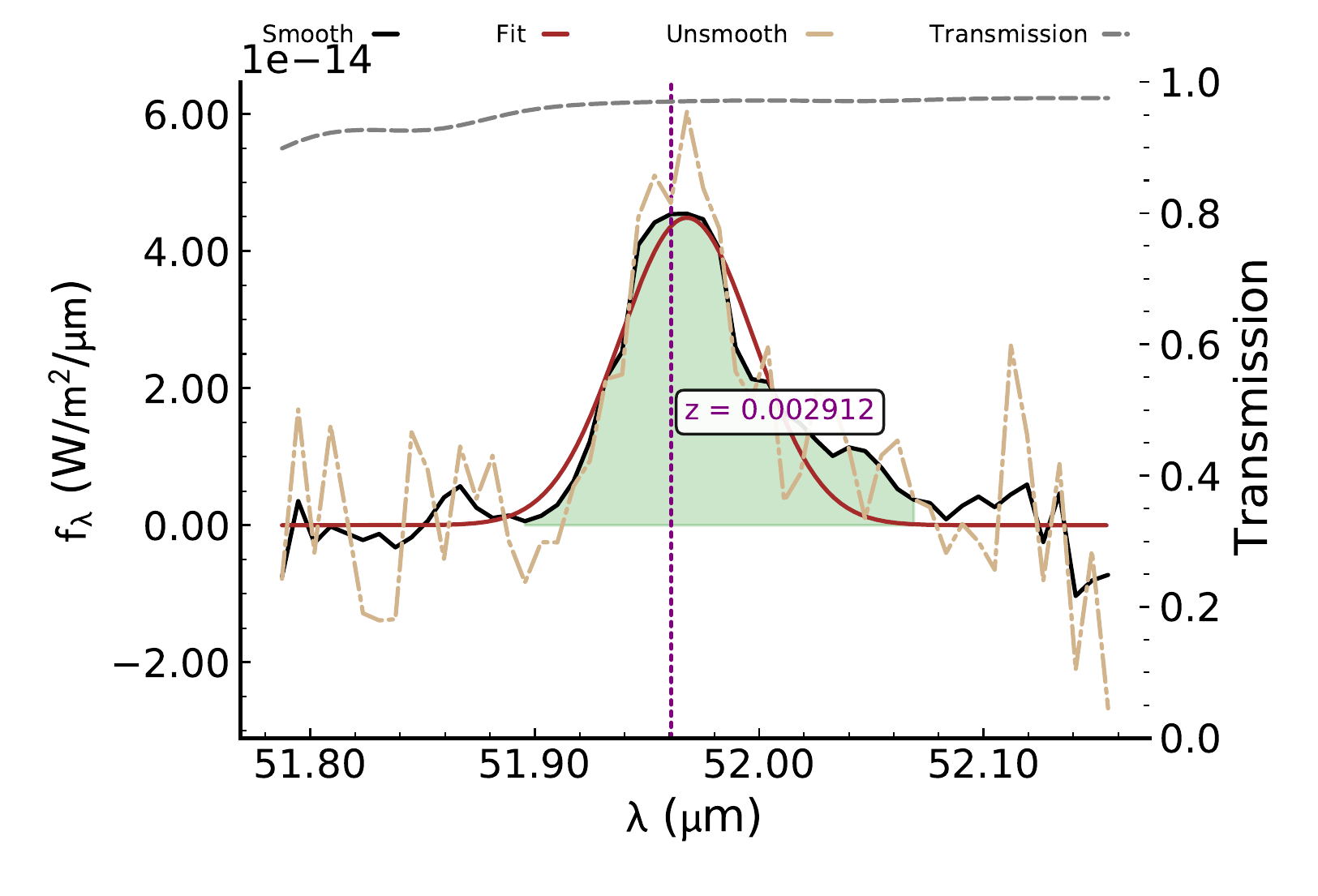}\label{fig:He2-10-C}}
    \caption{The 2MASS image (Figure \ref{fig:He2-10-A}) as well as 2-D linemap and 1-D spectrum for [OIII]52$\mu$m  (Figures \ref{fig:He2-10-B} and \ref{fig:He2-10-C}, respectively) in He2-10.}
    \label{fig:He2-10}
\end{figure*}

\begin{figure*}[ht!]
    \centering
    \subfigure[]{\includegraphics[width=0.498\textwidth]{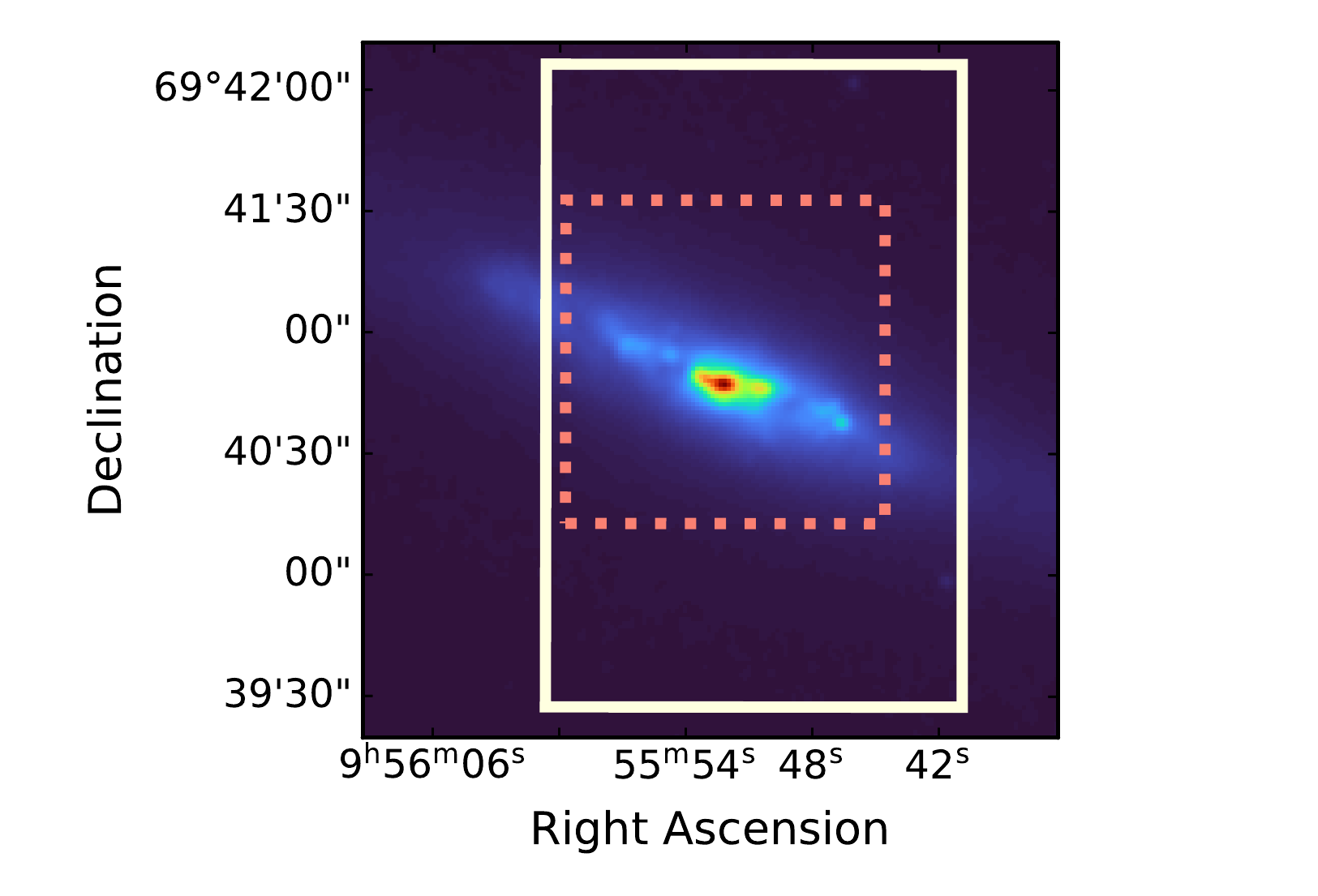}\label{fig:M82-A}}~\\
    \subfigure[]{\includegraphics[width=0.498\textwidth]{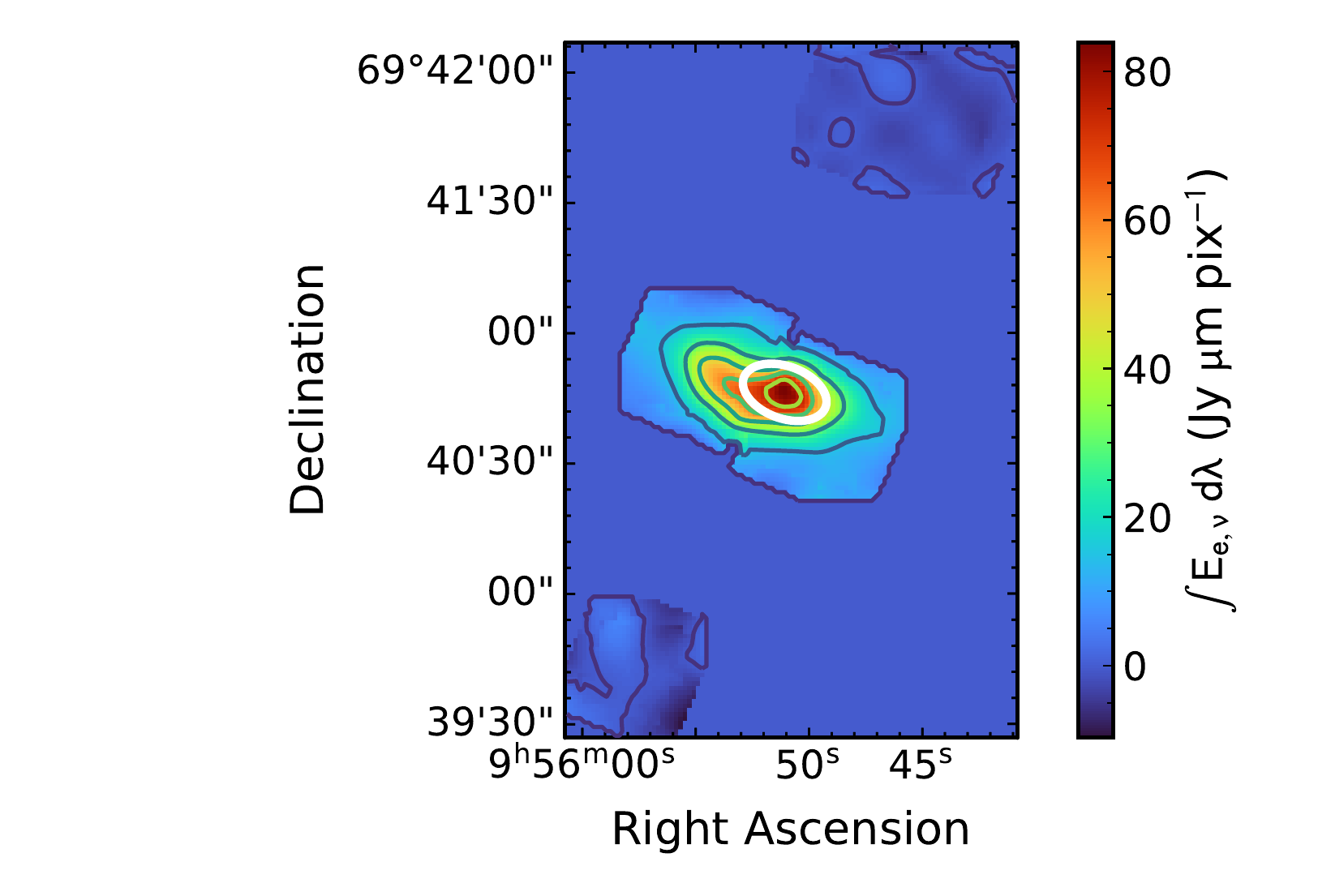}\label{fig:M82-B}}~
    \subfigure[]{\includegraphics[width=0.49\textwidth]{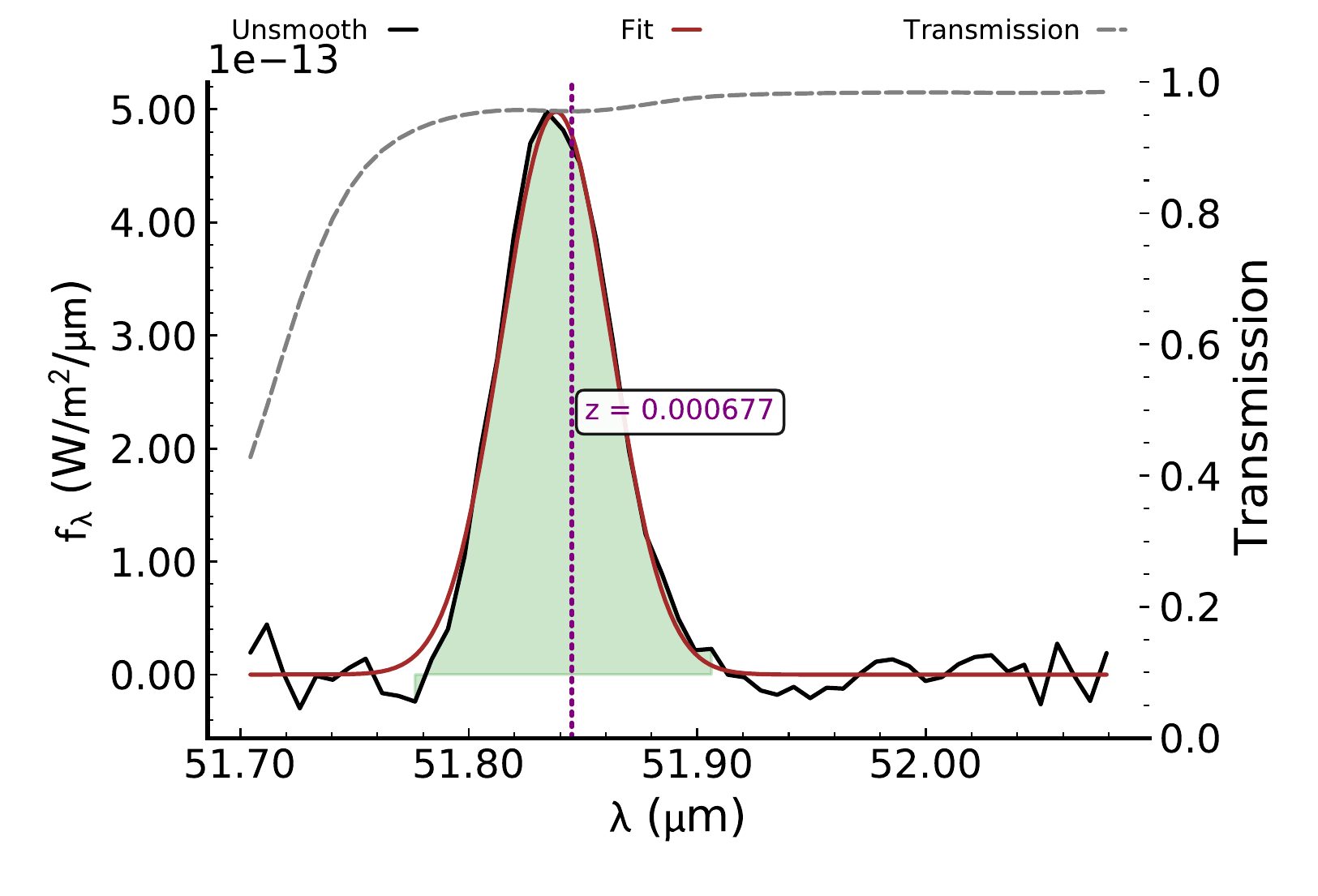}\label{fig:M82-C}}\\
    \subfigure[]{\includegraphics[width=0.498\textwidth]{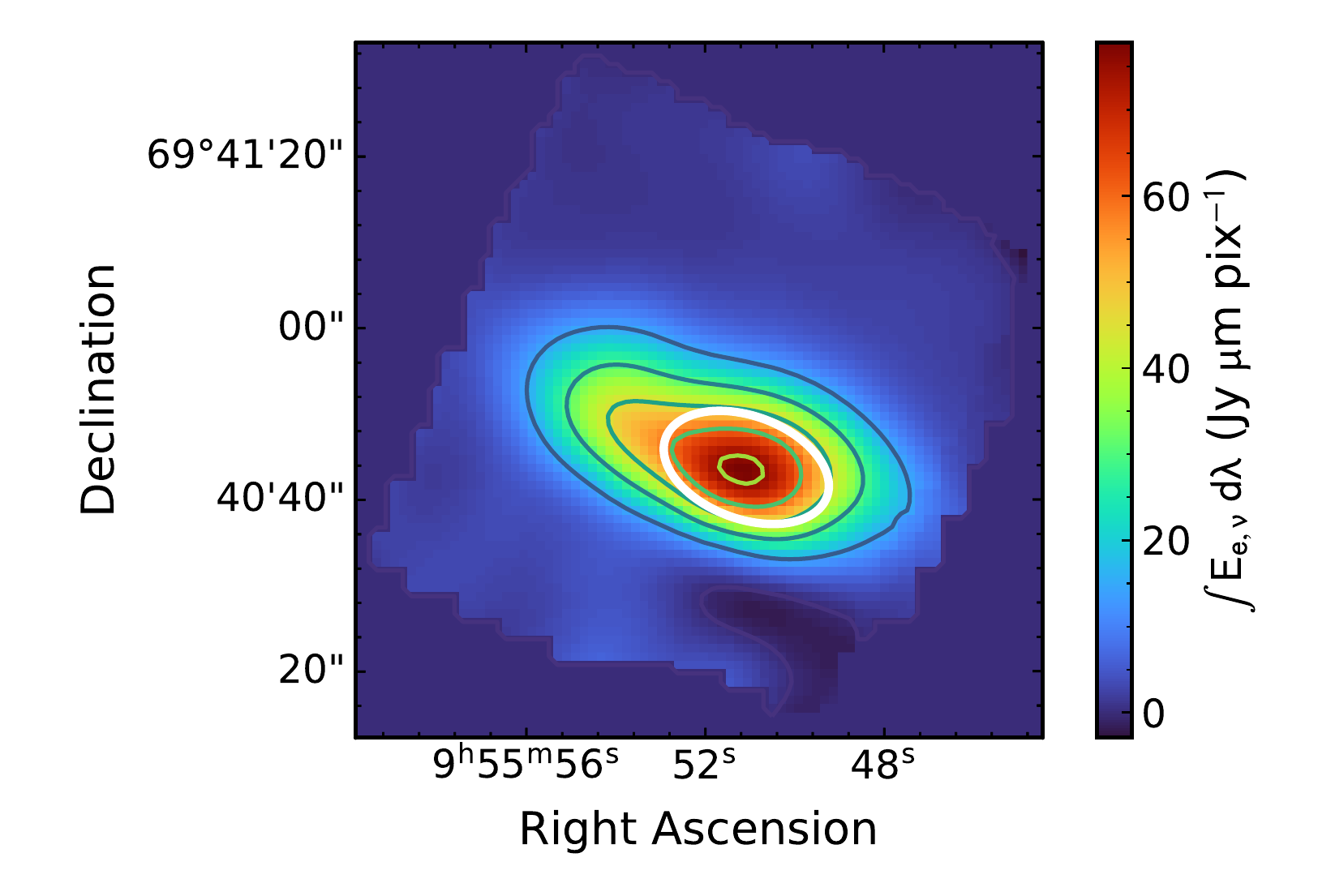}\label{fig:M82-D}}~
    \subfigure[]{\includegraphics[width=0.49\textwidth]{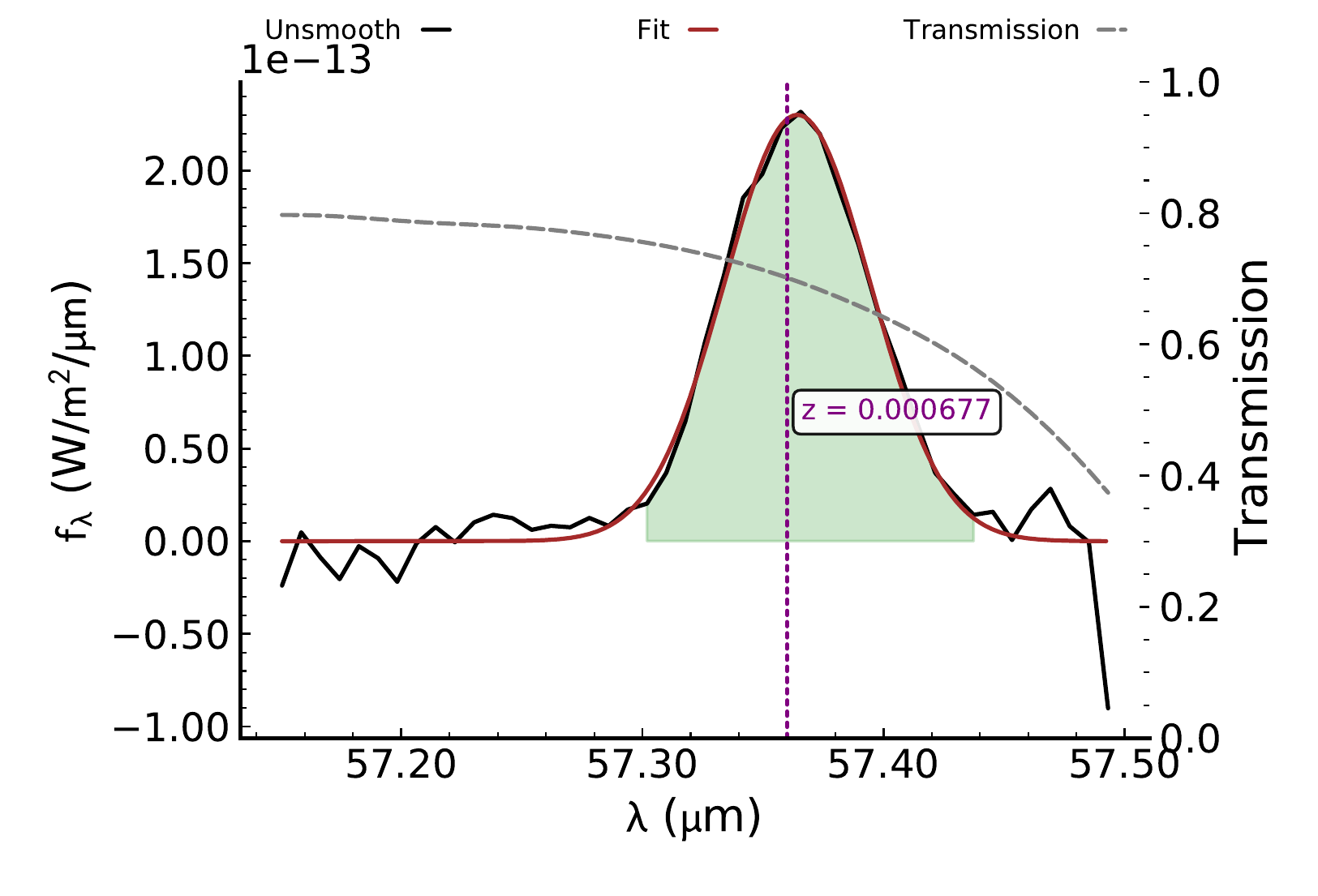}\label{fig:M82-E}}
    \caption{The 2MASS image (Figure \ref{fig:M82-A}), 2-D linemaps and 1-D spectra for [OIII]52$\mu$m  (Figures \ref{fig:M82-B} and \ref{fig:M82-C}, respectively) and [NIII]57$\mu$m (Figures \ref{fig:M82-D} and \ref{fig:M82-E}, respectively) in M82.}
    \label{fig:M82}
\end{figure*}

\begin{figure*}[ht!]
    \centering
    \subfigure[]{\includegraphics[width=0.498\textwidth]{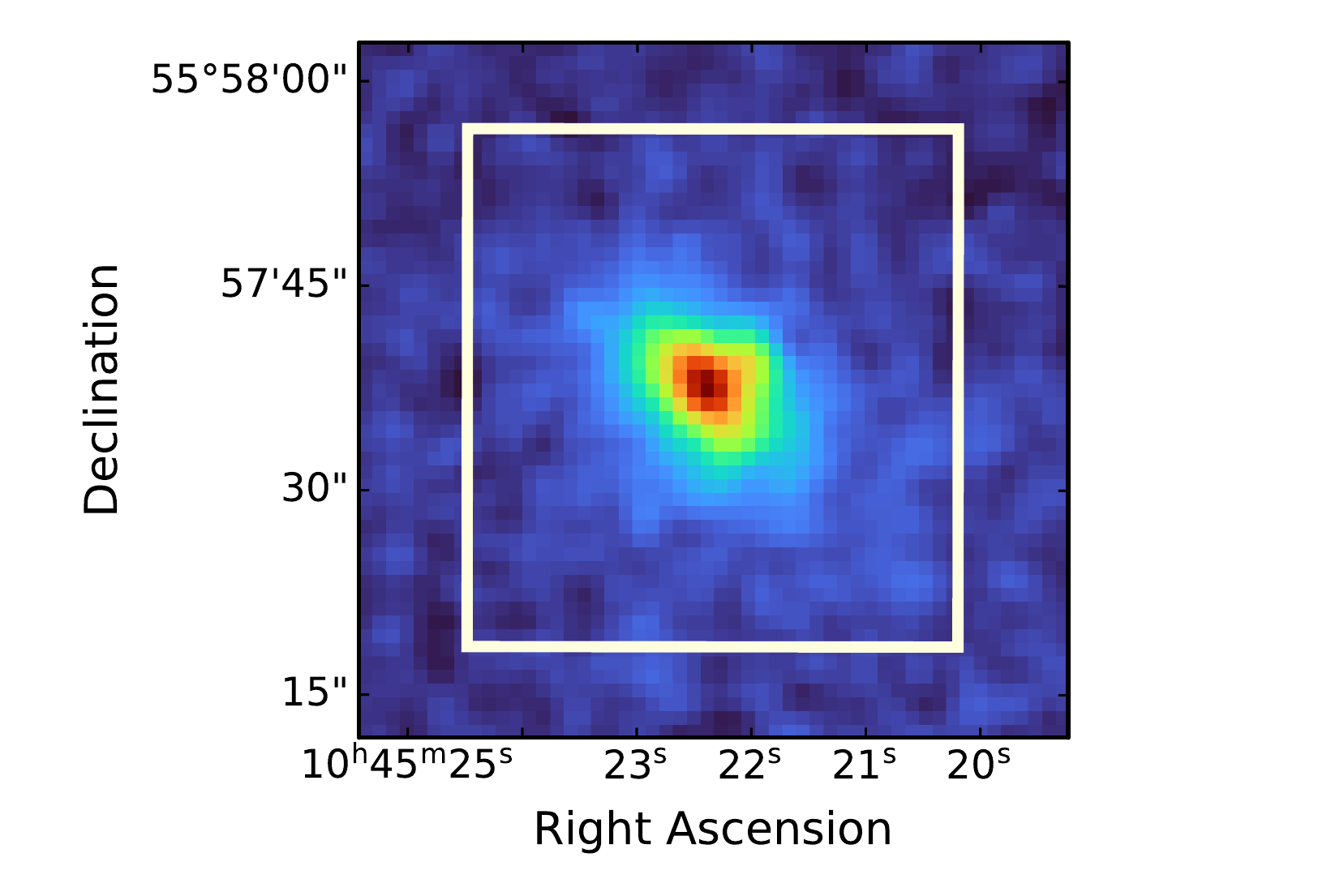}\label{fig:Haro3-A}}\\
    \subfigure[]{\includegraphics[width=0.498\textwidth]{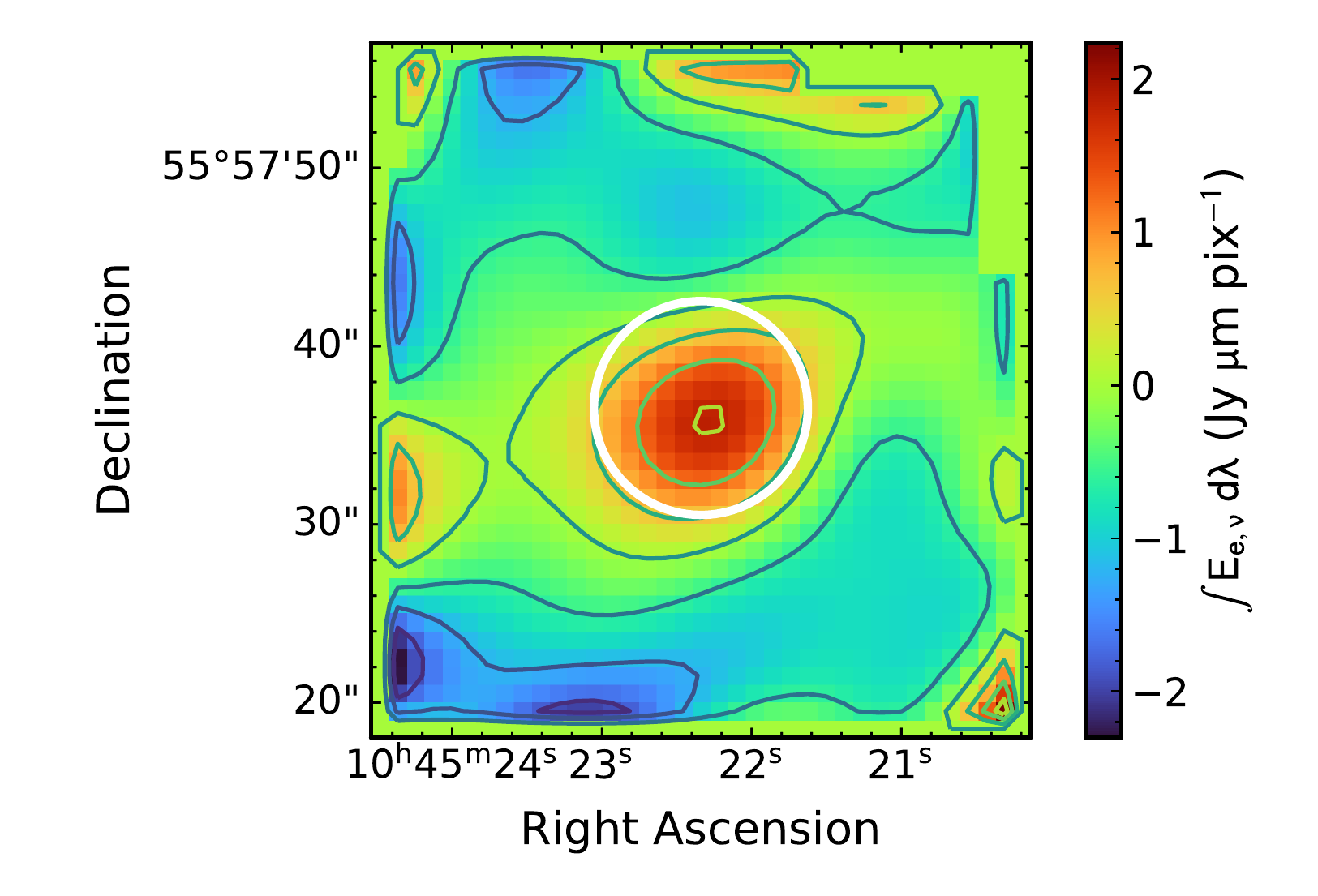}\label{fig:Haro3-B}}~
    \subfigure[]{\includegraphics[width=0.49\textwidth]{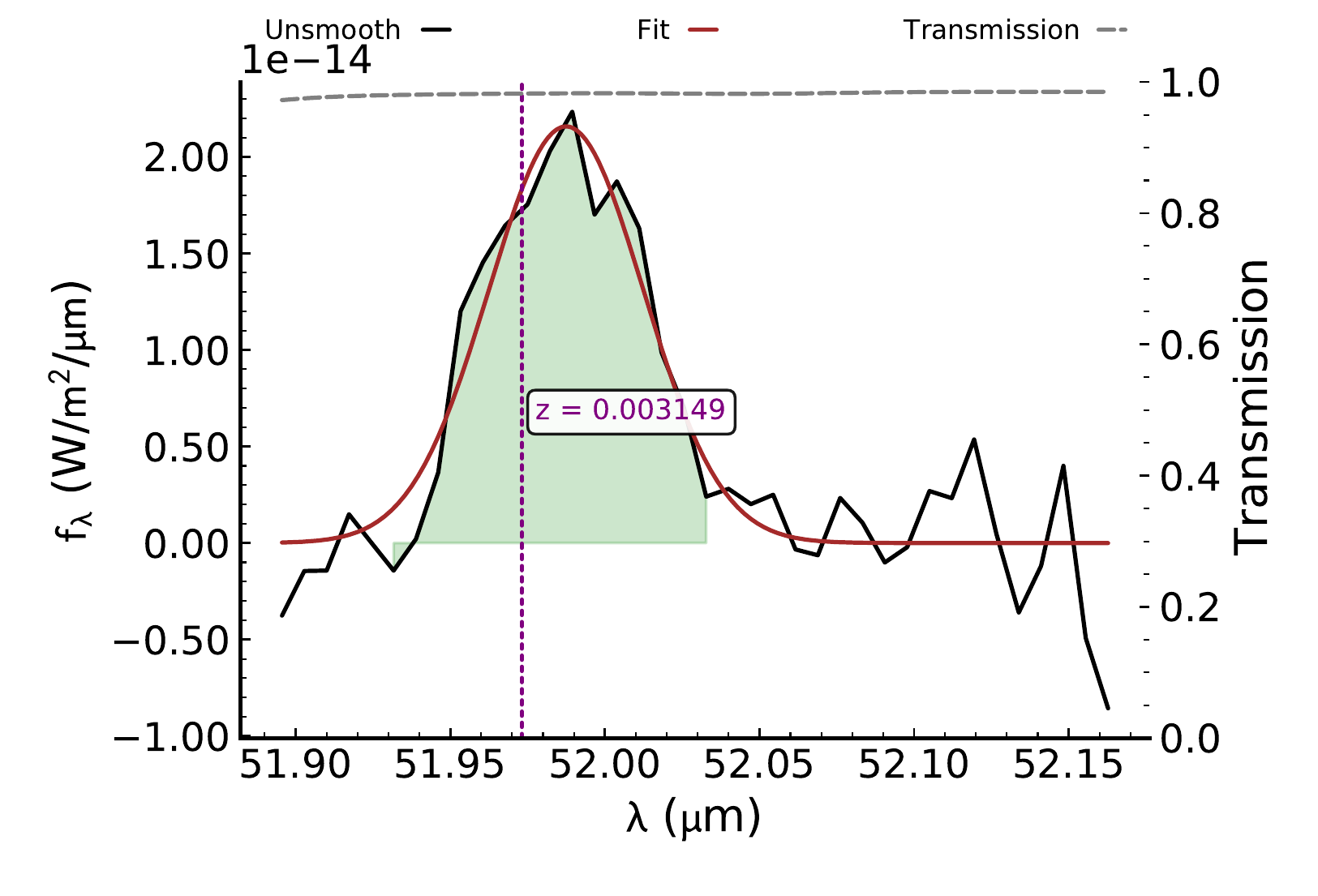}\label{fig:Haro3-C}}
    \caption{The 2MASS image (Figure \ref{fig:Haro3-A}) 2-D linemap and 1-D spectrum for [OIII]52$\mu$m  (Figures \ref{fig:Haro3-B} and \ref{fig:Haro3-C}, respectively) in Haro3.}
    \label{fig:Haro3}
\end{figure*}

\begin{figure*}[ht!]
    \centering
    \subfigure[]{\includegraphics[width=0.498\textwidth]{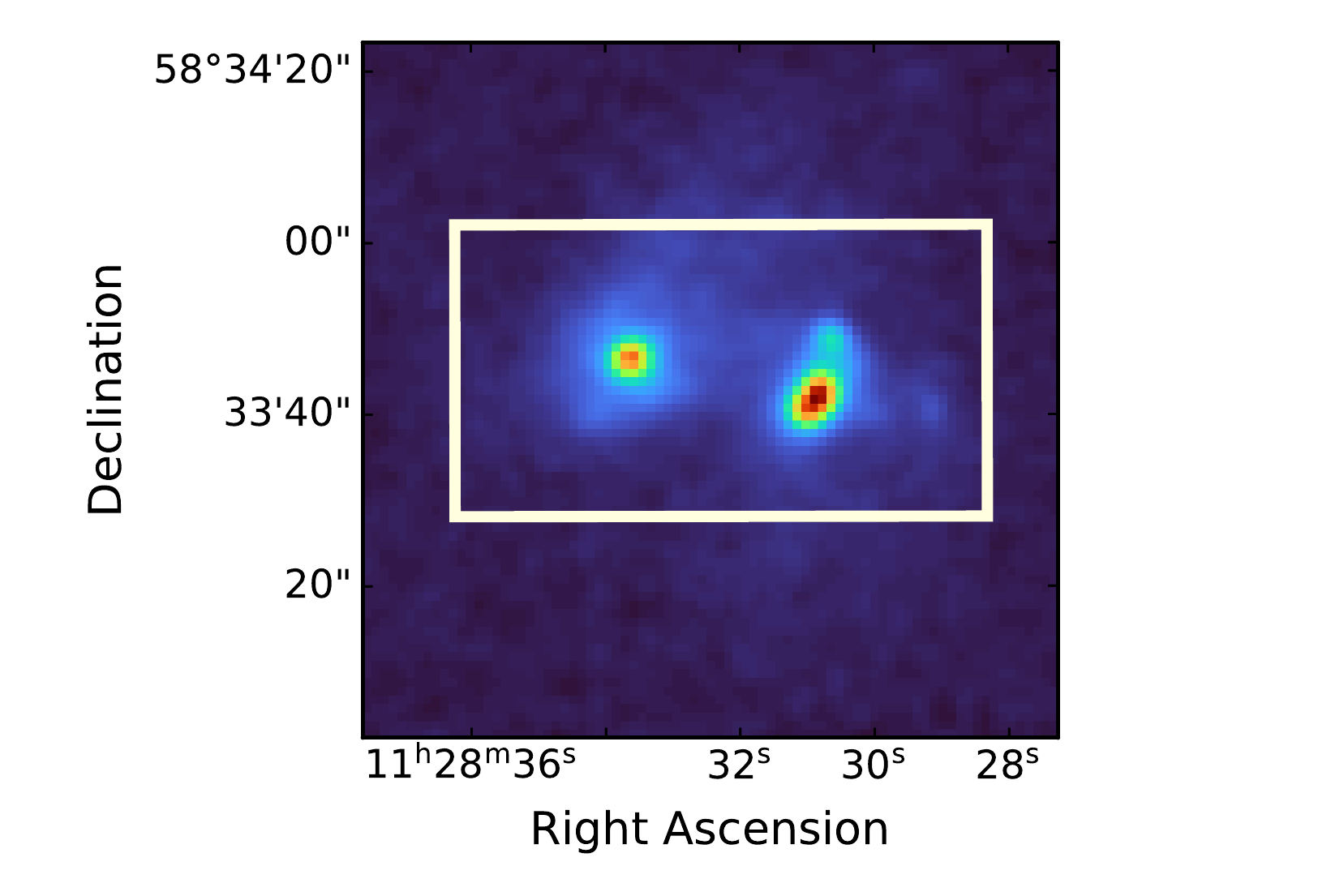}\label{fig:Arp299-2MASS}}\\
    \subfigure[]{\includegraphics[width=0.498\textwidth]{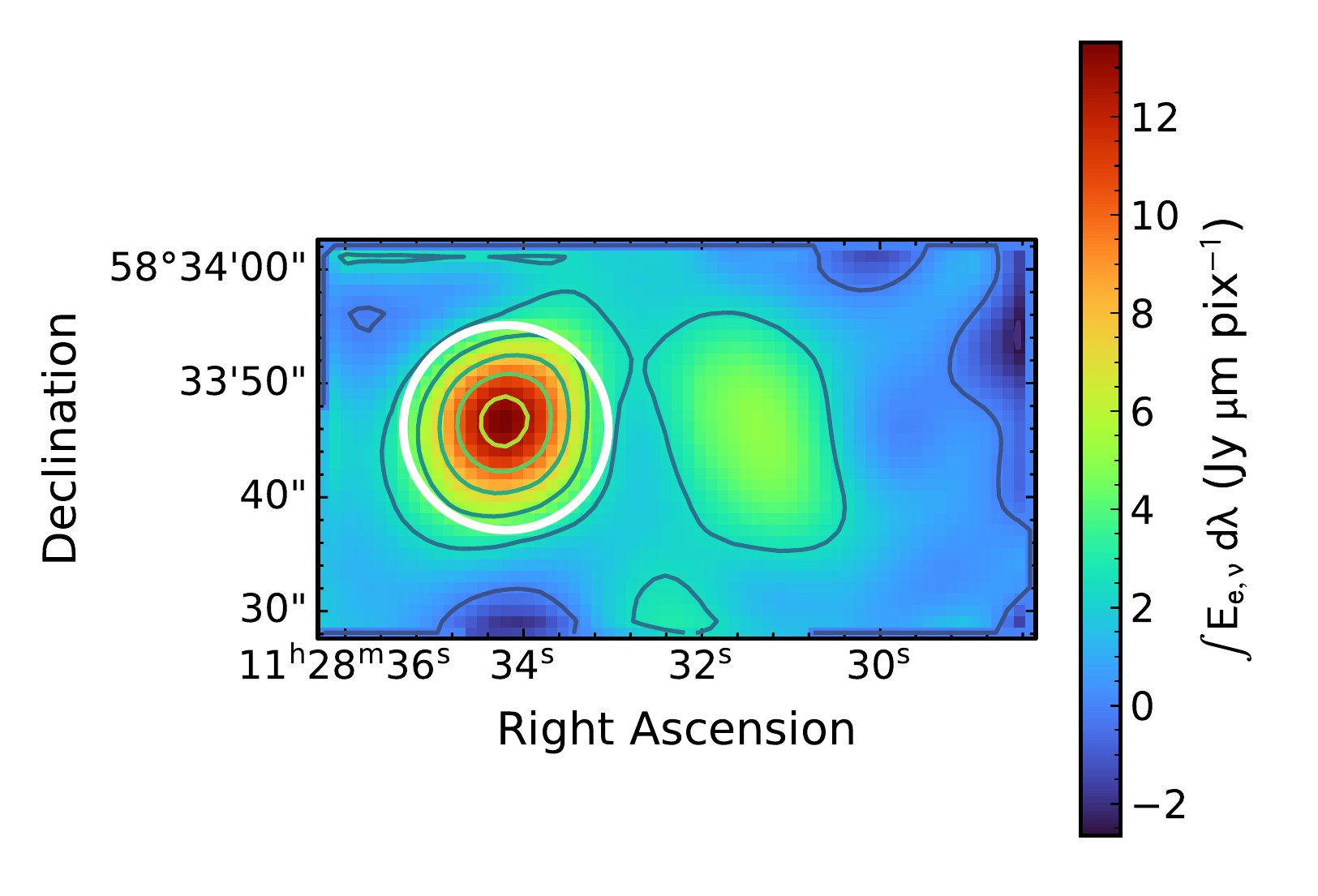}\label{fig:Arp299A-A}}~
    \subfigure[]{\includegraphics[width=0.49\textwidth]{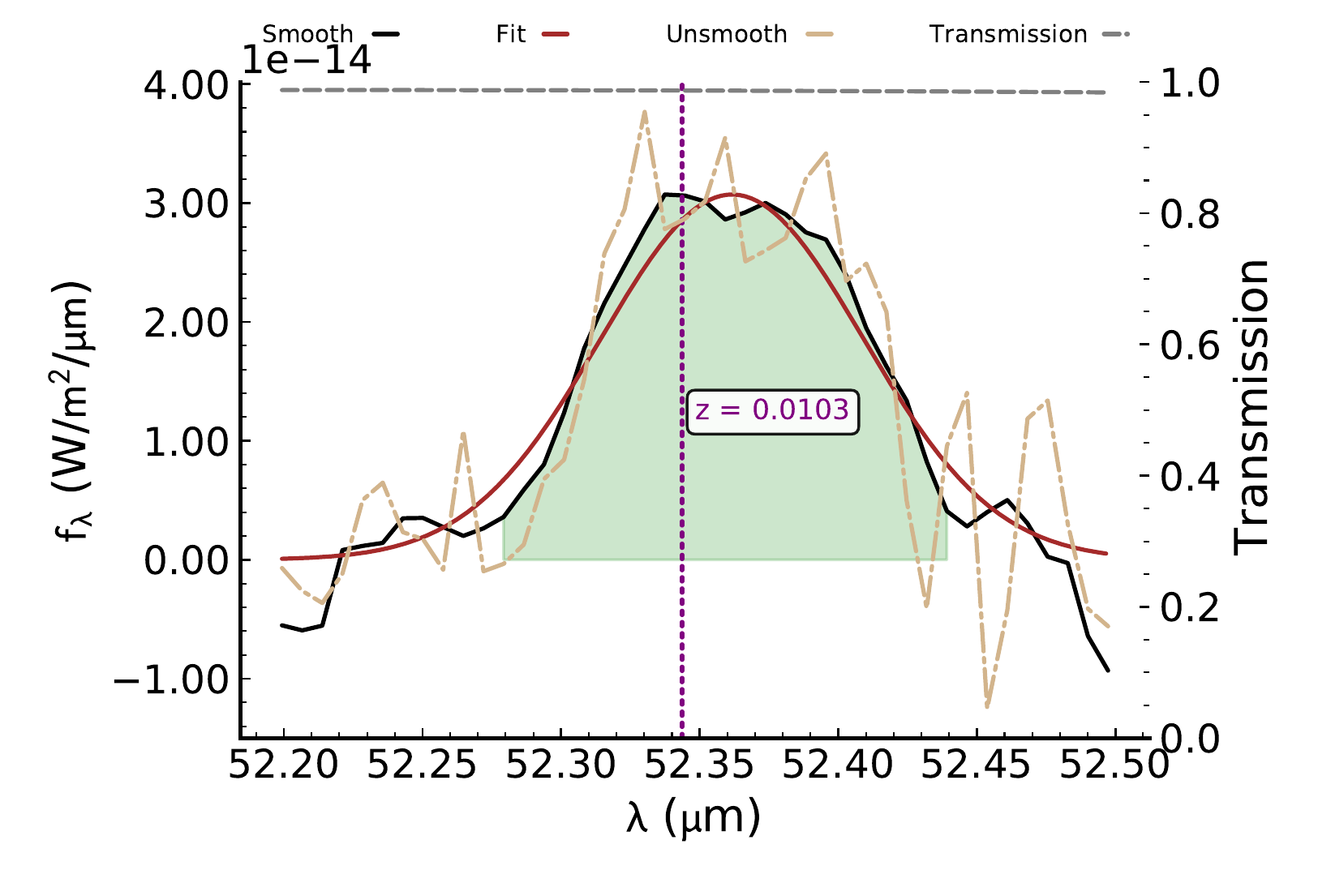}\label{fig:Arp299A-B}}\\
    \subfigure[]{\includegraphics[width=0.498\textwidth]{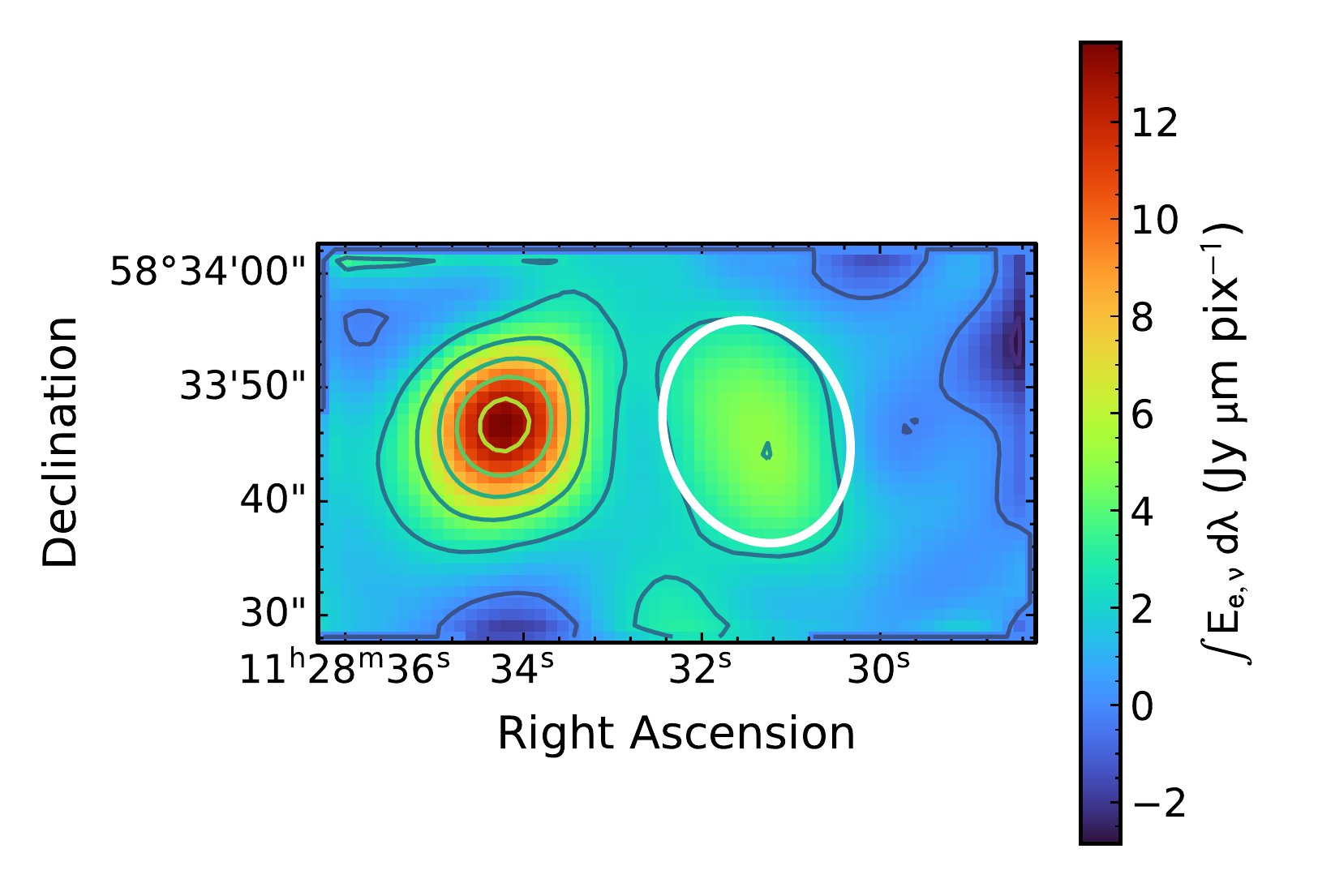}\label{fig:Arp299B&C-A}}~
    \subfigure[]{\includegraphics[width=0.49\textwidth]{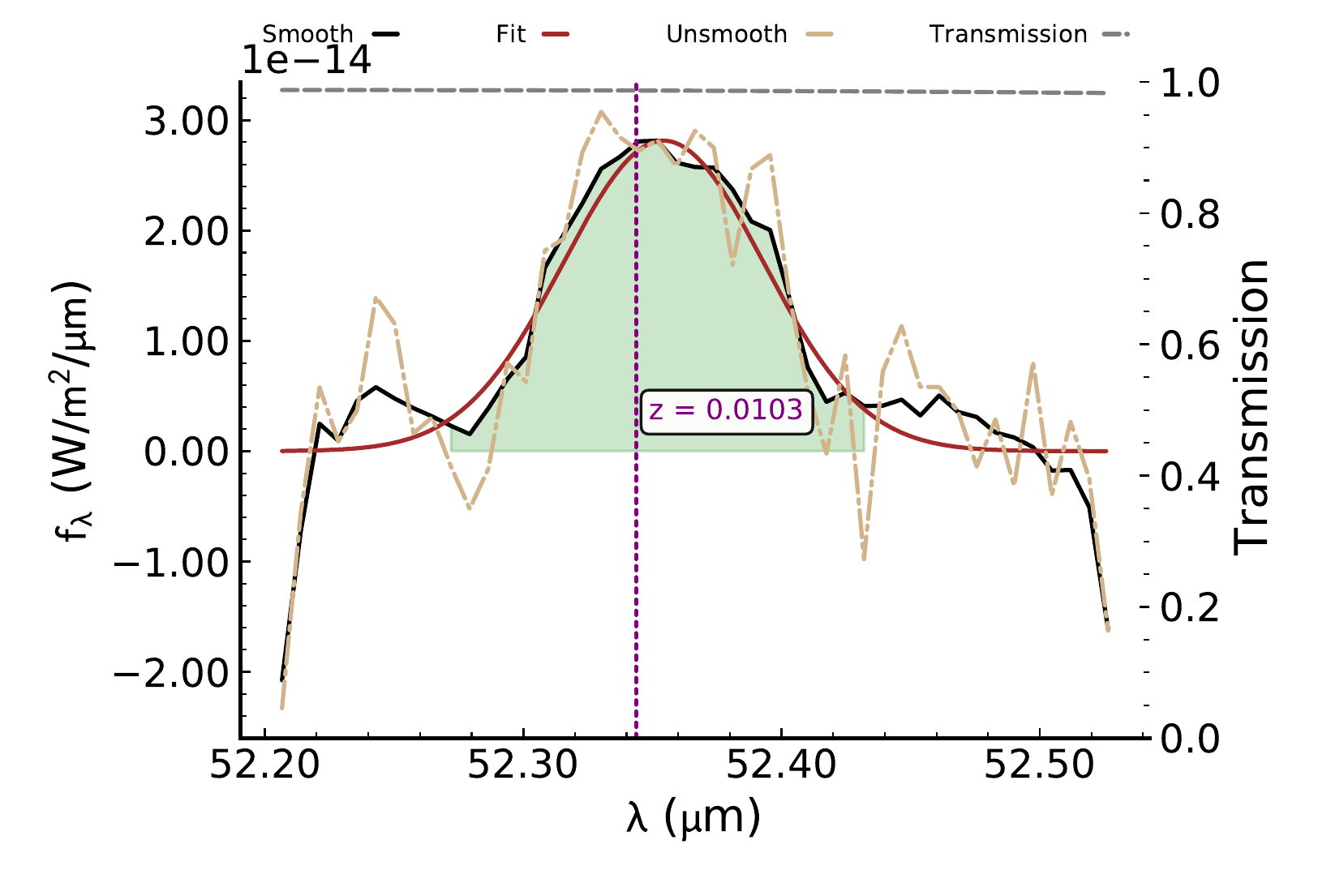}\label{fig:Arp299B&C-B}}
    \caption{The 2MASS image of Arp299 (Figure \ref{fig:Arp299-2MASS}), the [OIII]52$\mu$m 2-D linemap and 1-D spectrum of Arp299A (Figures \ref{fig:Arp299A-A} and \ref{fig:Arp299A-B}, respectively), [OIII]52$\mu$m 2-D linemap and 1-D spectrum of Arp299B\&C (Figures \ref{fig:Arp299B&C-A} and \ref{fig:Arp299B&C-B}), respectively}.
    \label{fig:Arp299}
\end{figure*}

\begin{figure*}[ht!]
    \centering
    \subfigure[]{\includegraphics[width=0.498\textwidth]{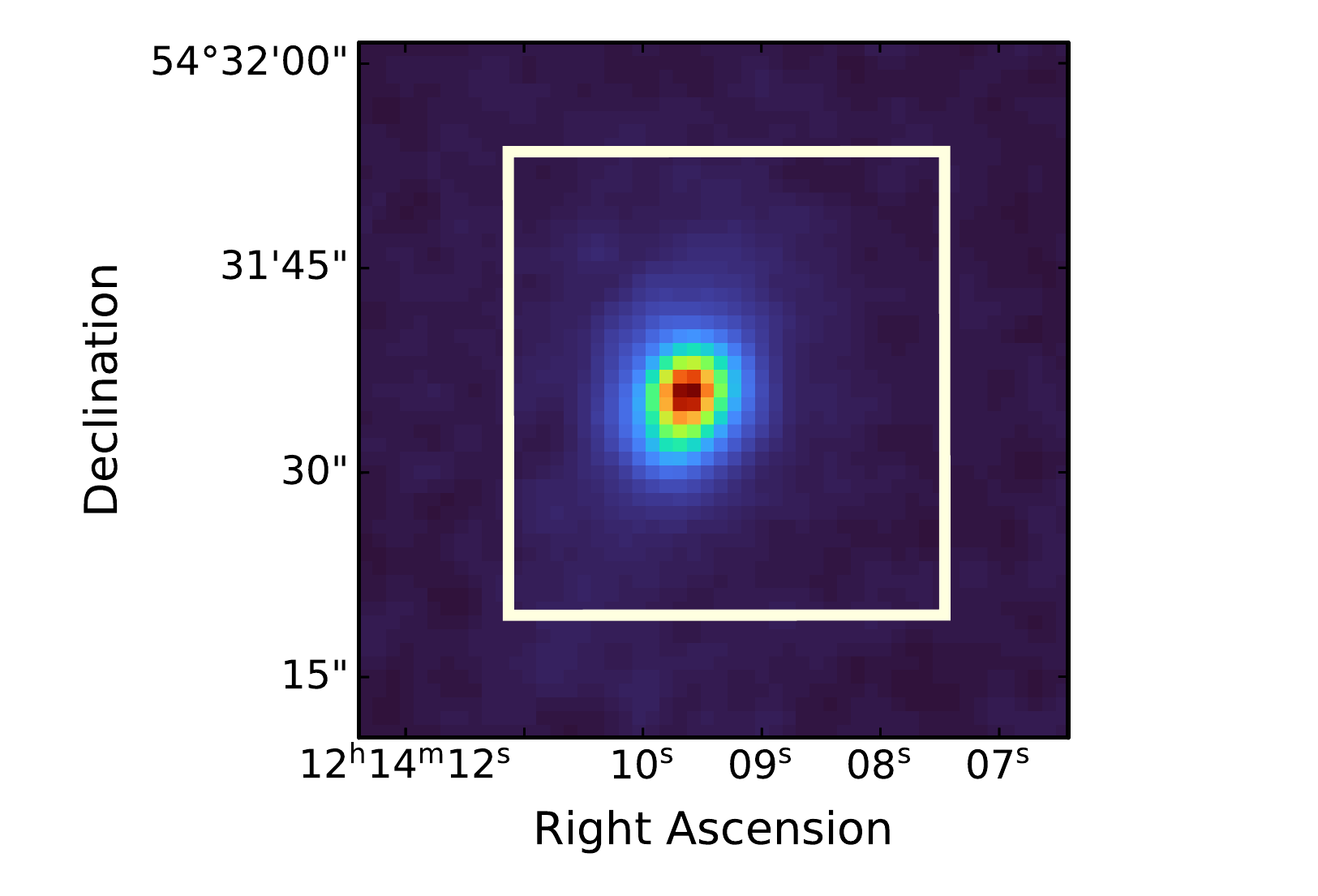}\label{fig:NGC4194-A}}\\
    \subfigure[]{\includegraphics[width=0.498\textwidth]{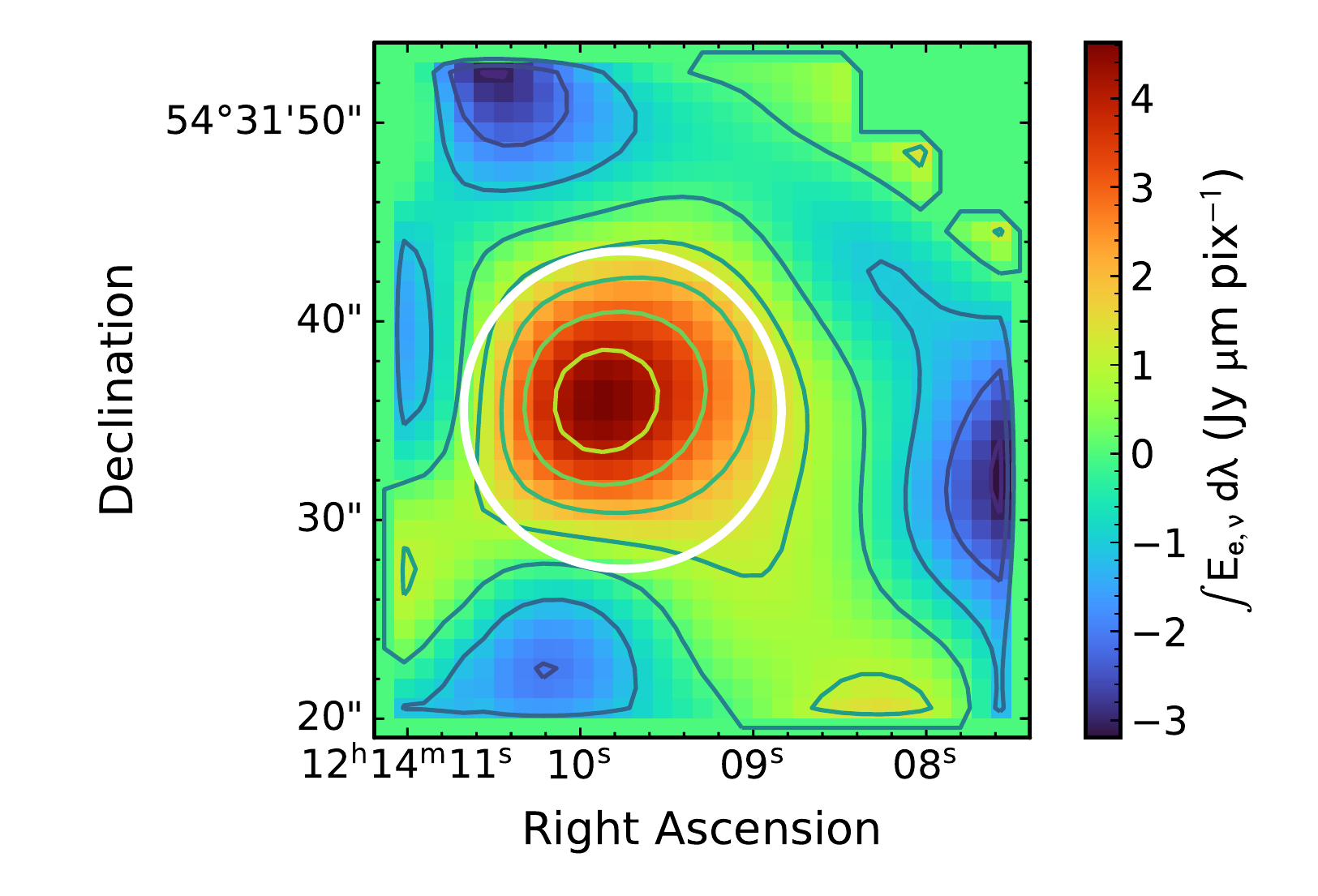}\label{fig:NGC4194-B}}~
    \subfigure[]{\includegraphics[width=0.49\textwidth]{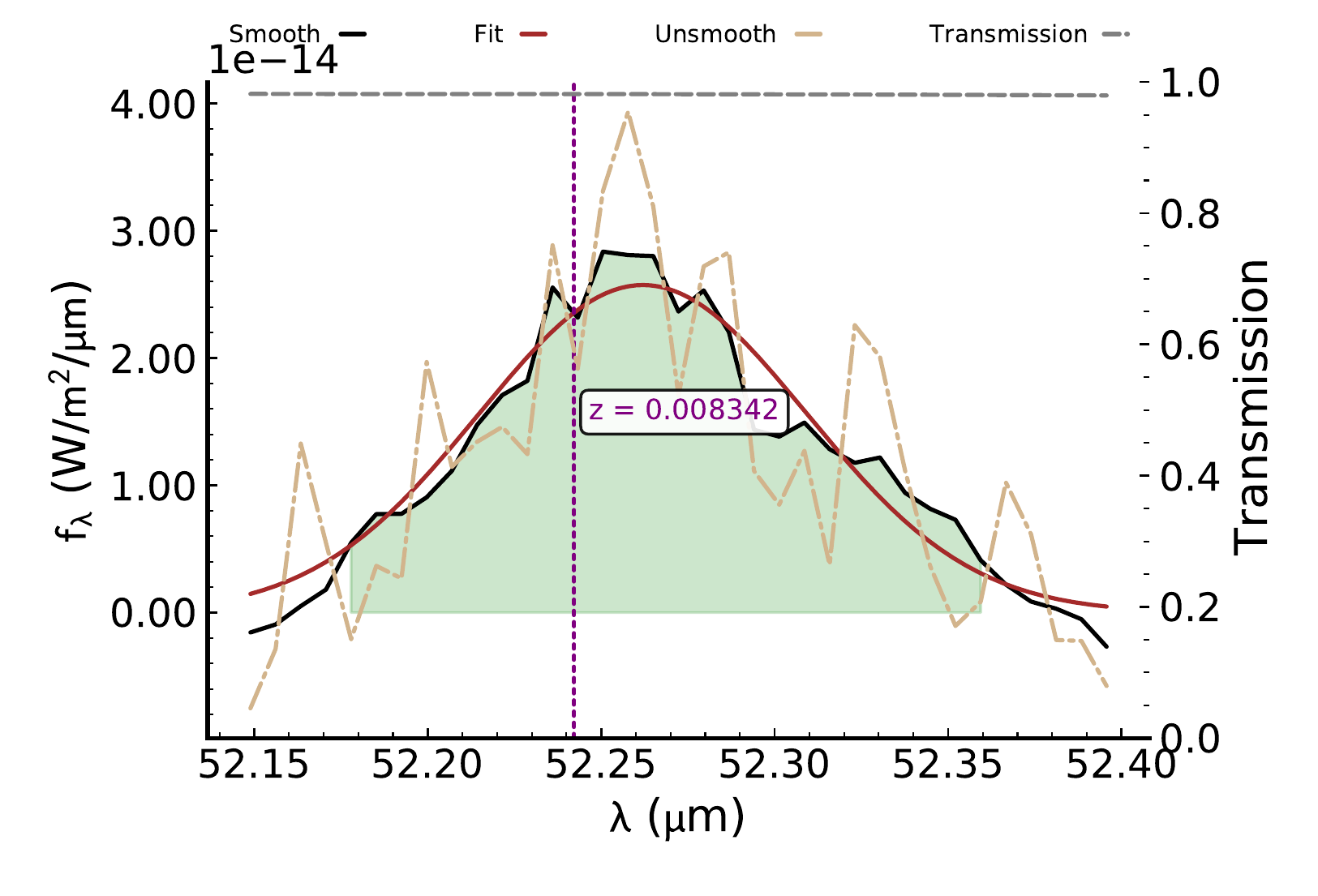}\label{fig:NGC4194-C}}
    \caption{The 2MASS image (Figure \ref{fig:NGC4194-A}), 2-D linemap and 1-D spectrum for [OIII]52$\mu$m  (Figures \ref{fig:NGC4194-B} and \ref{fig:NGC4194-C}, respectively) in NGC4194.}
    \label{fig:NGC4194}
\end{figure*}

\begin{figure*}[ht!]
    \centering
    \subfigure[]{\includegraphics[width=0.498\textwidth]{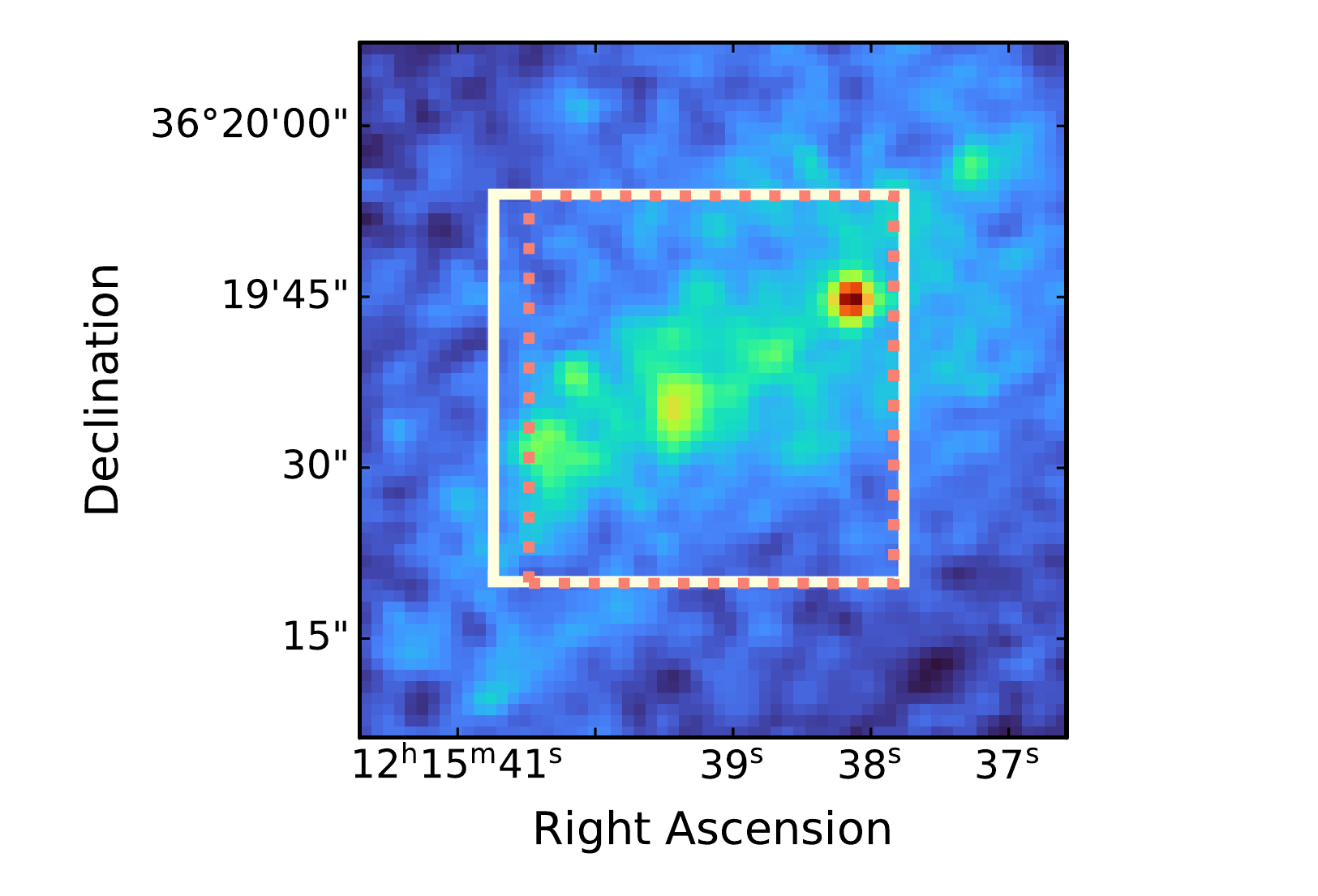}\label{fig:NGC4214-A}}~\\
    \subfigure[]{\includegraphics[width=0.498\textwidth]{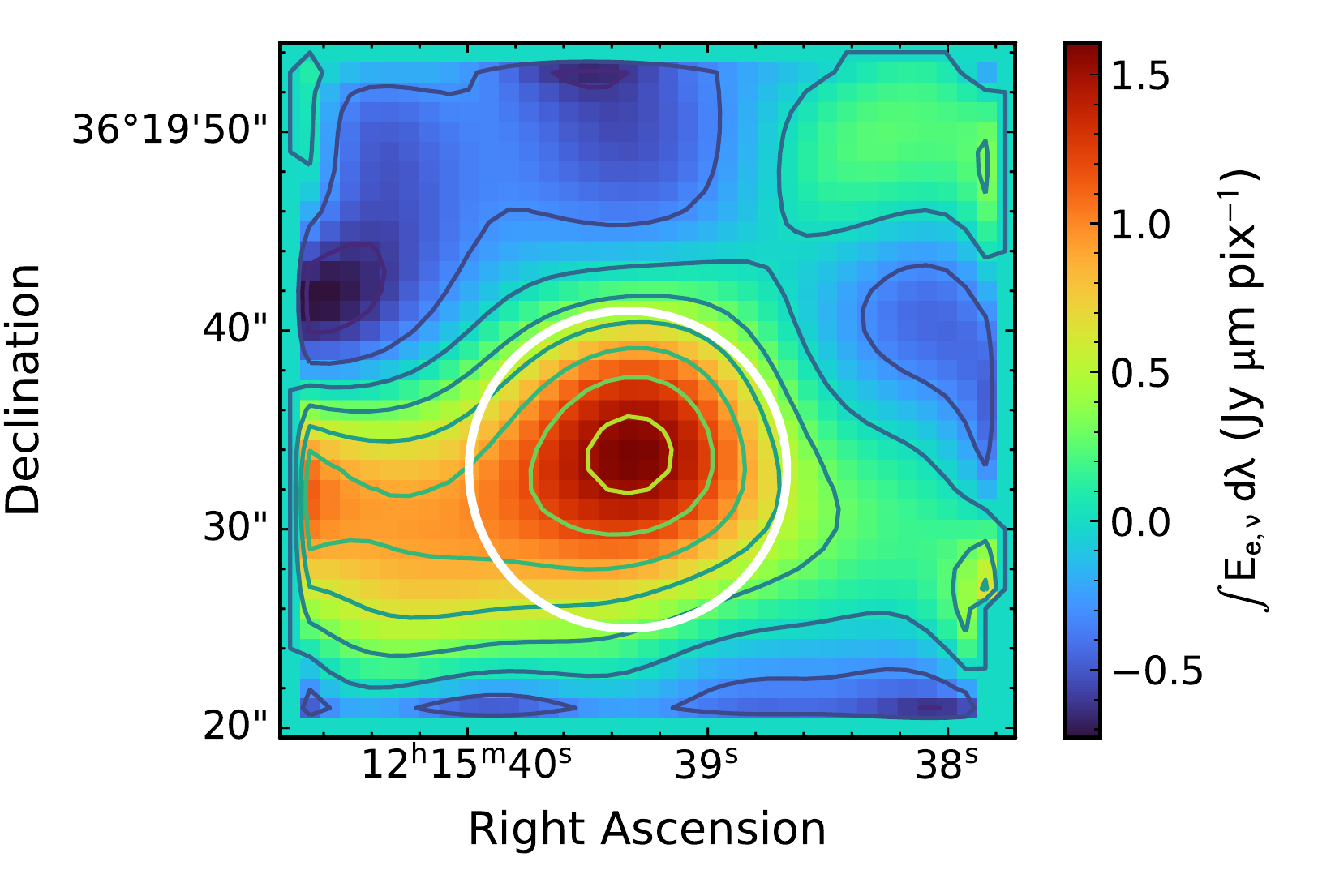}\label{fig:NGC4214-B}}~
    \subfigure[]{\includegraphics[width=0.49\textwidth]{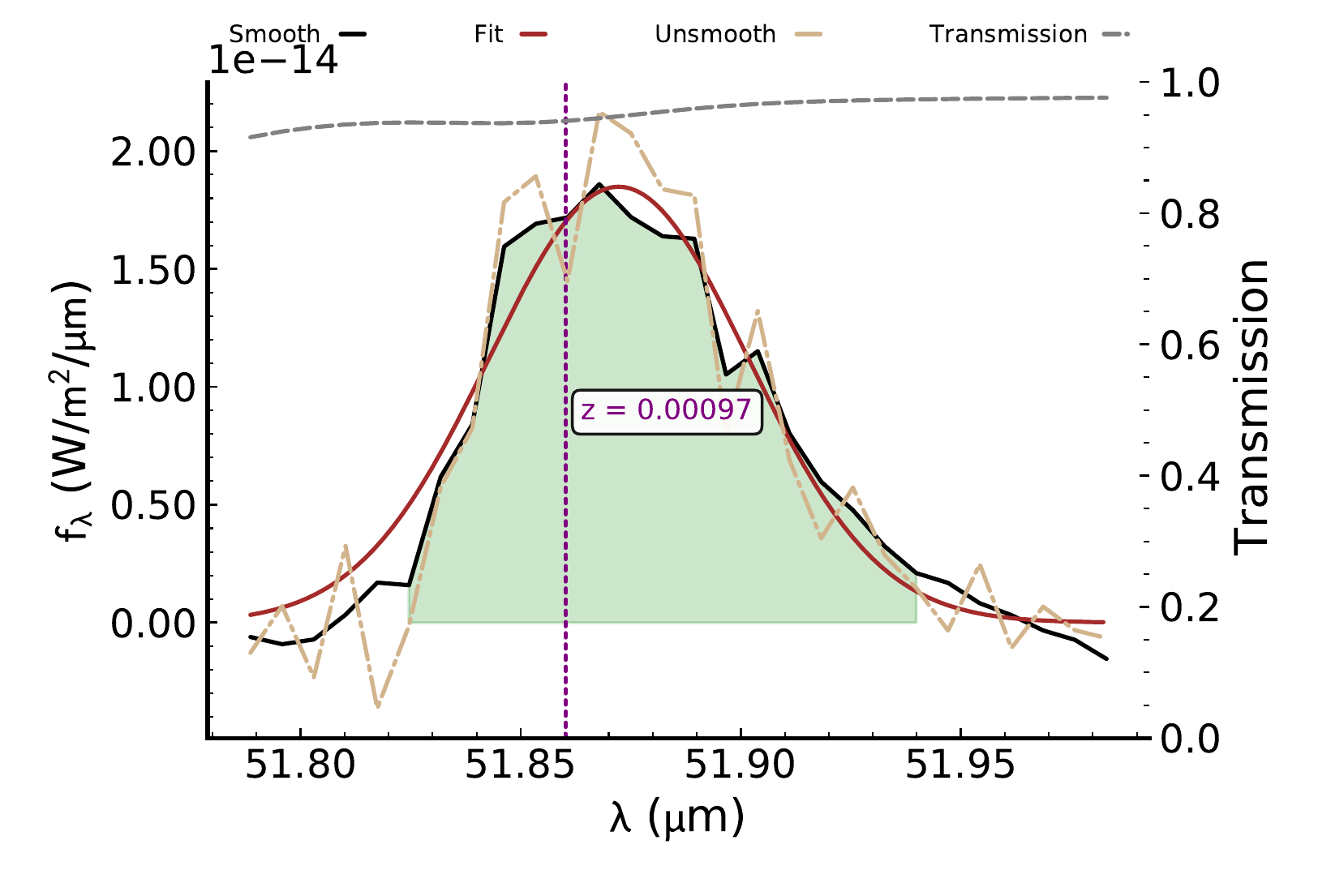}\label{fig:NGC4214-C}}\\
    \subfigure[]{\includegraphics[width=0.498\textwidth]{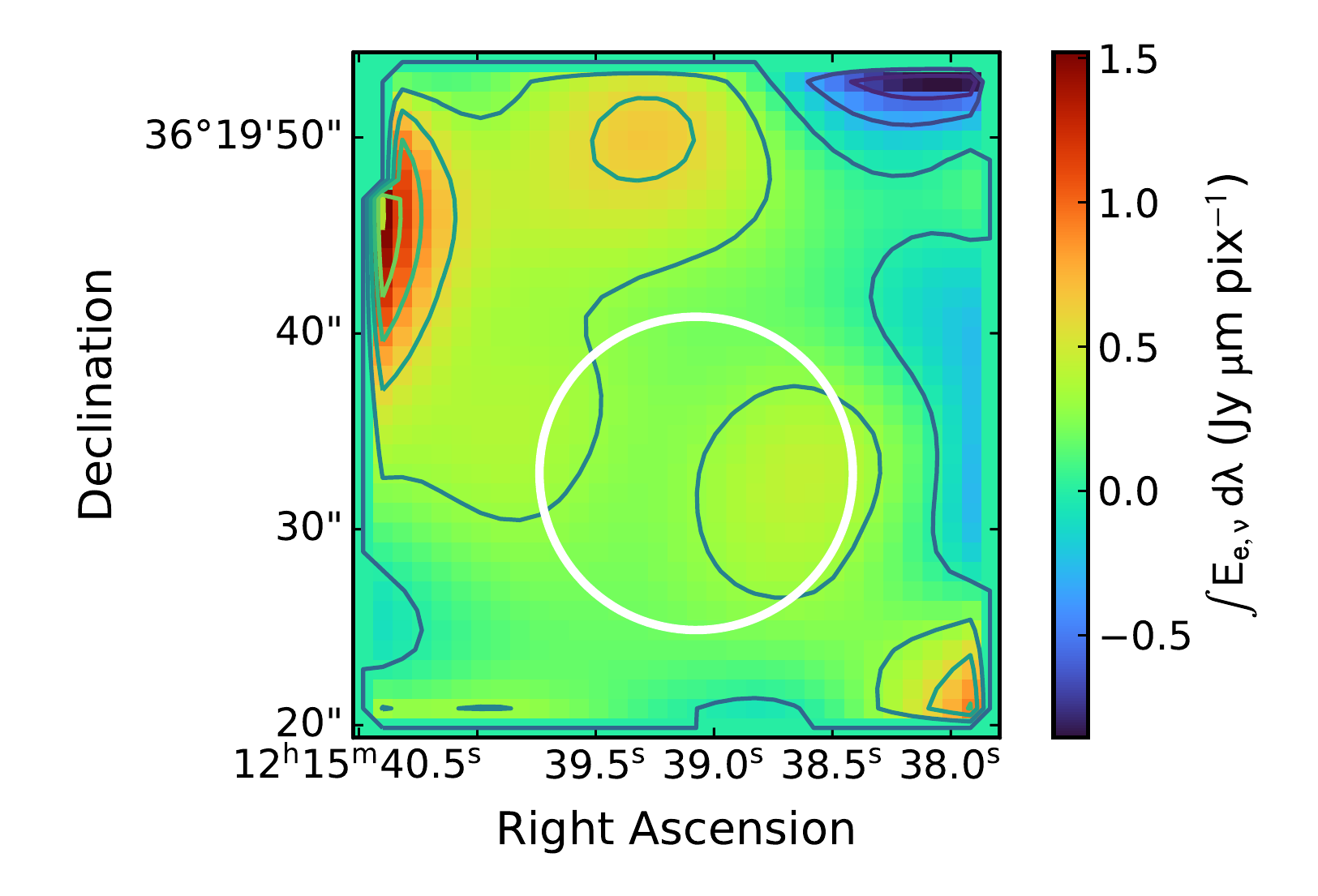}\label{fig:NGC4214-D}}~
    \subfigure[]{\includegraphics[width=0.49\textwidth]{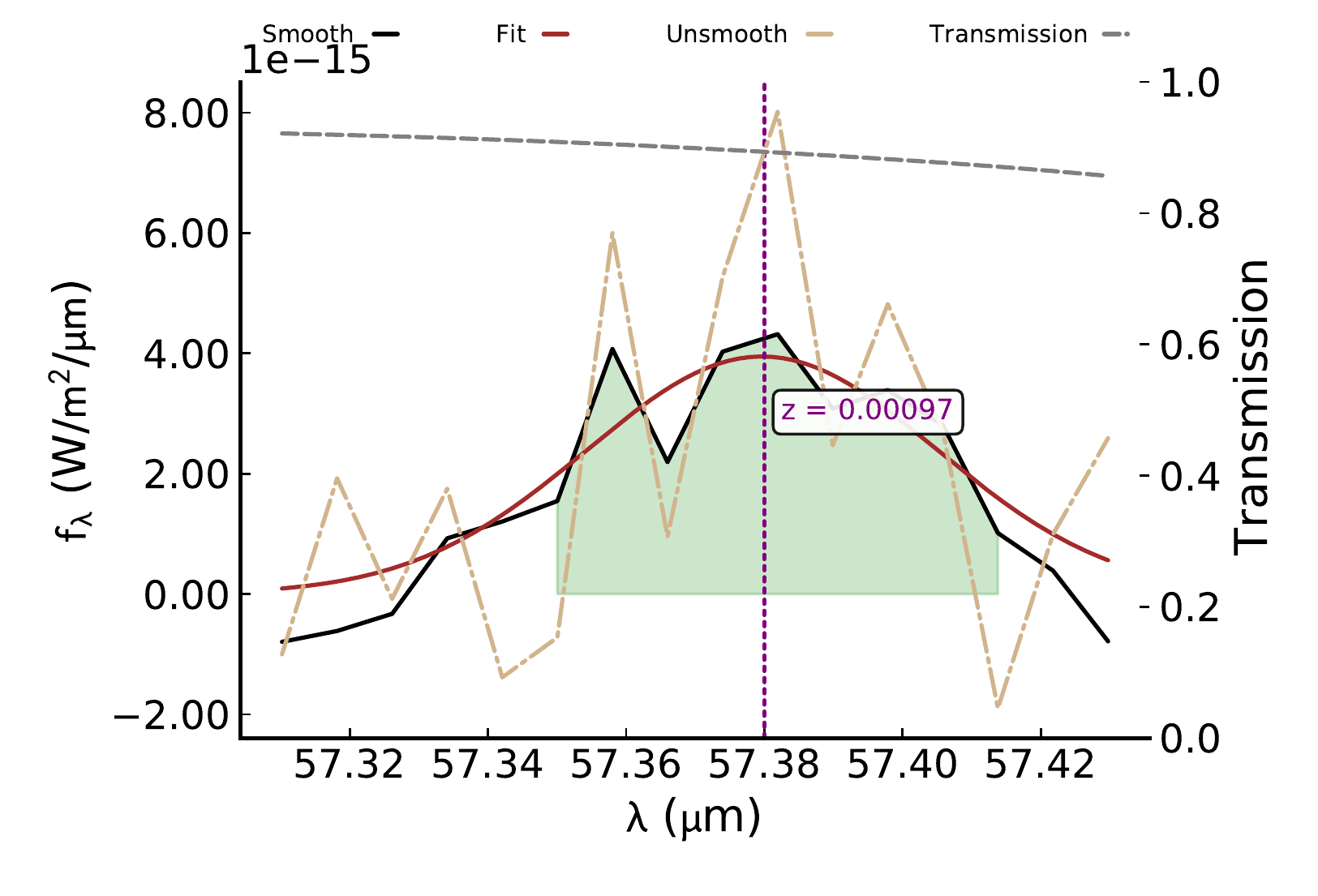}\label{fig:NGC4214-E}}
    \caption{The 2MASS image (Figure \ref{fig:NGC4214-A}), 2-D linemaps and 1-D spectra for [OIII]52$\mu$m  (Figures \ref{fig:NGC4214-B} and \ref{fig:NGC4214-C}, respectively) and [NIII]57$\mu$m  (Figures \ref{fig:NGC4214-D} and \ref{fig:NGC4214-E}, respectively) in NGC4214. While the [NIII]57$\mu$m map does not show a clear peak, we centered the circular extraction aperture around the same location as for the [OIII]52$\mu$m map.}
    \label{fig:NGC4214}
\end{figure*}

\begin{figure*}[ht!]
    \centering
    \subfigure[]{\includegraphics[width=0.498\textwidth]{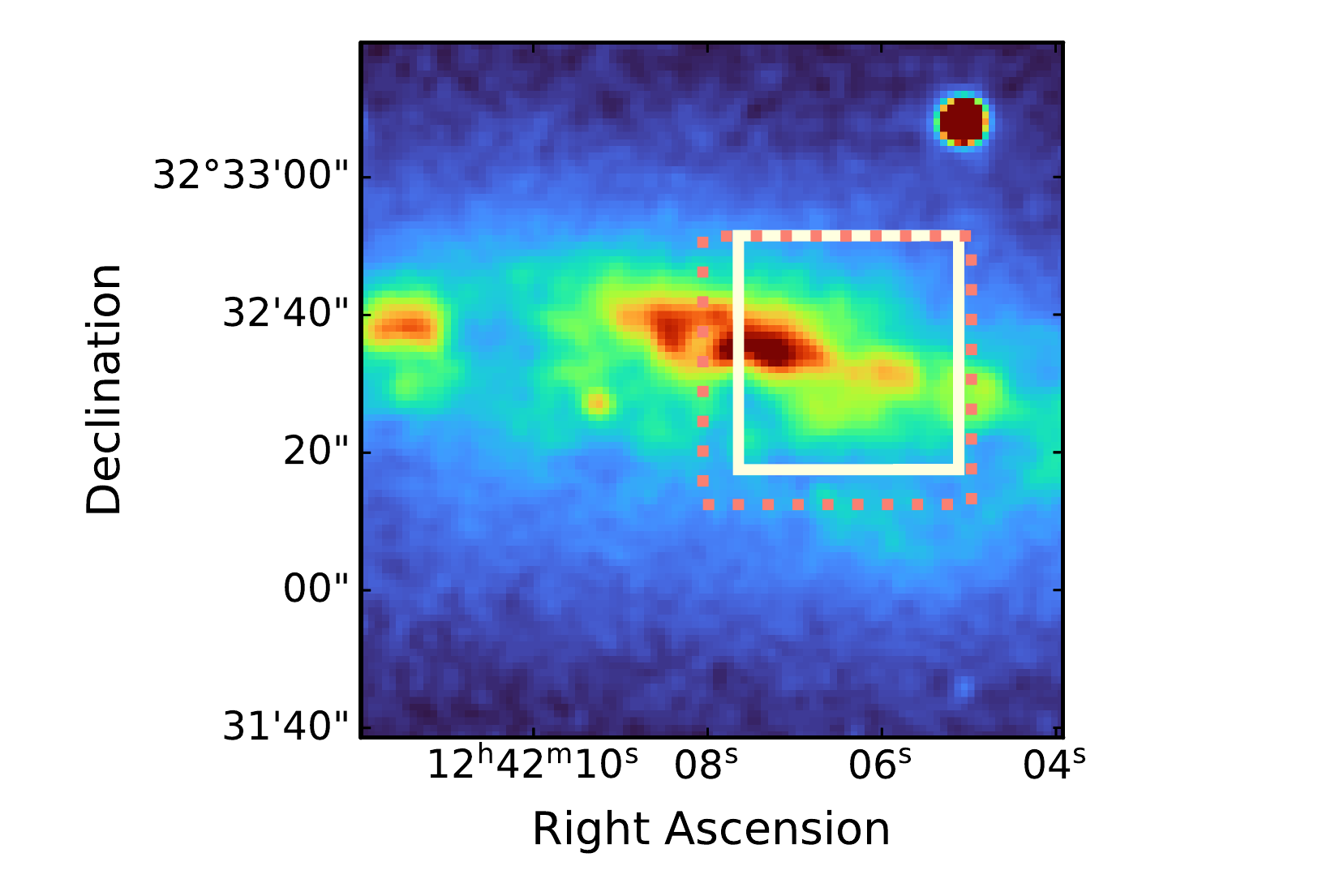}\label{fig:NGC4631-A}}~\\
    \subfigure[]{\includegraphics[width=0.498\textwidth]{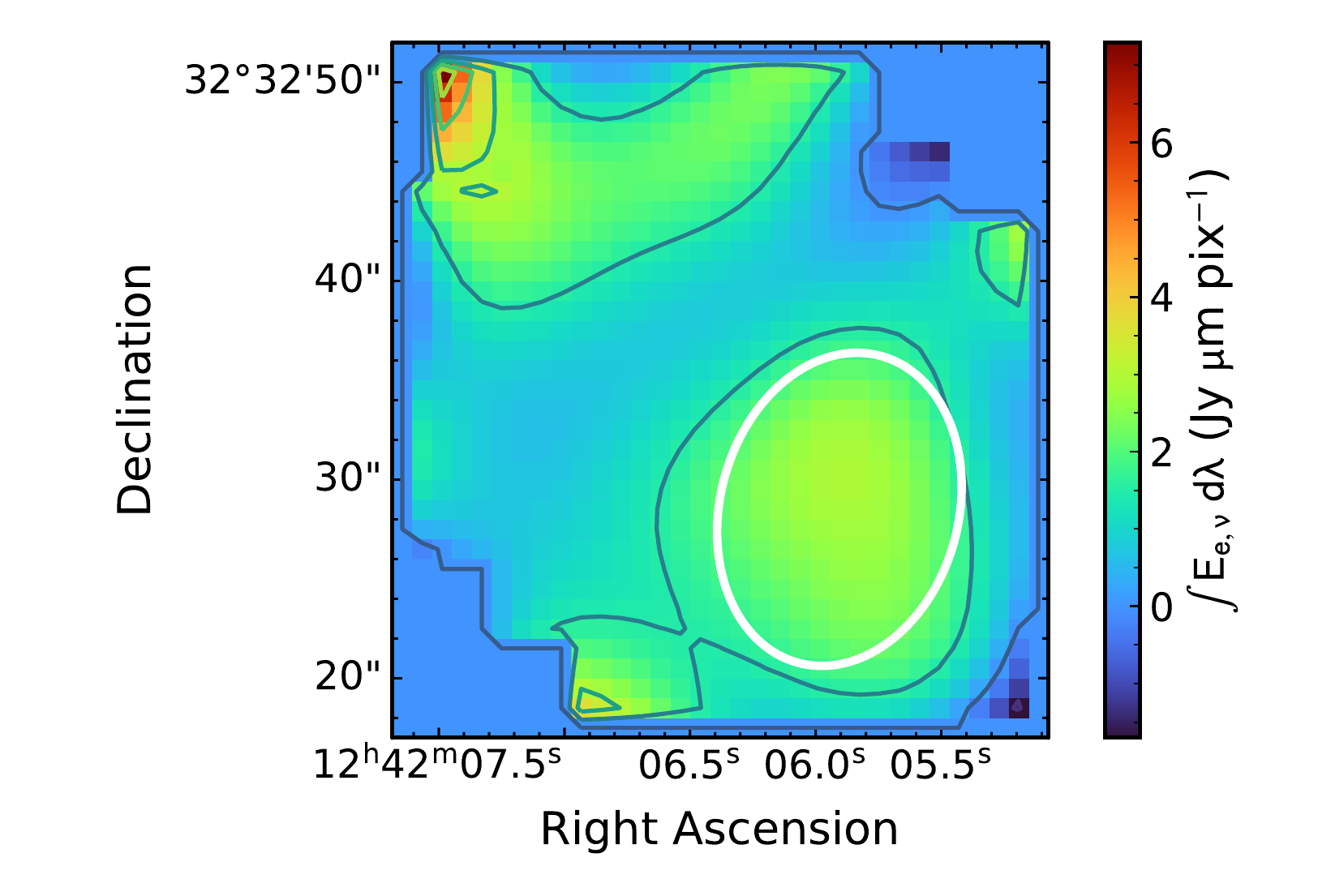}\label{fig:NGC4631-B}}~
    \subfigure[]{\includegraphics[width=0.49\textwidth]{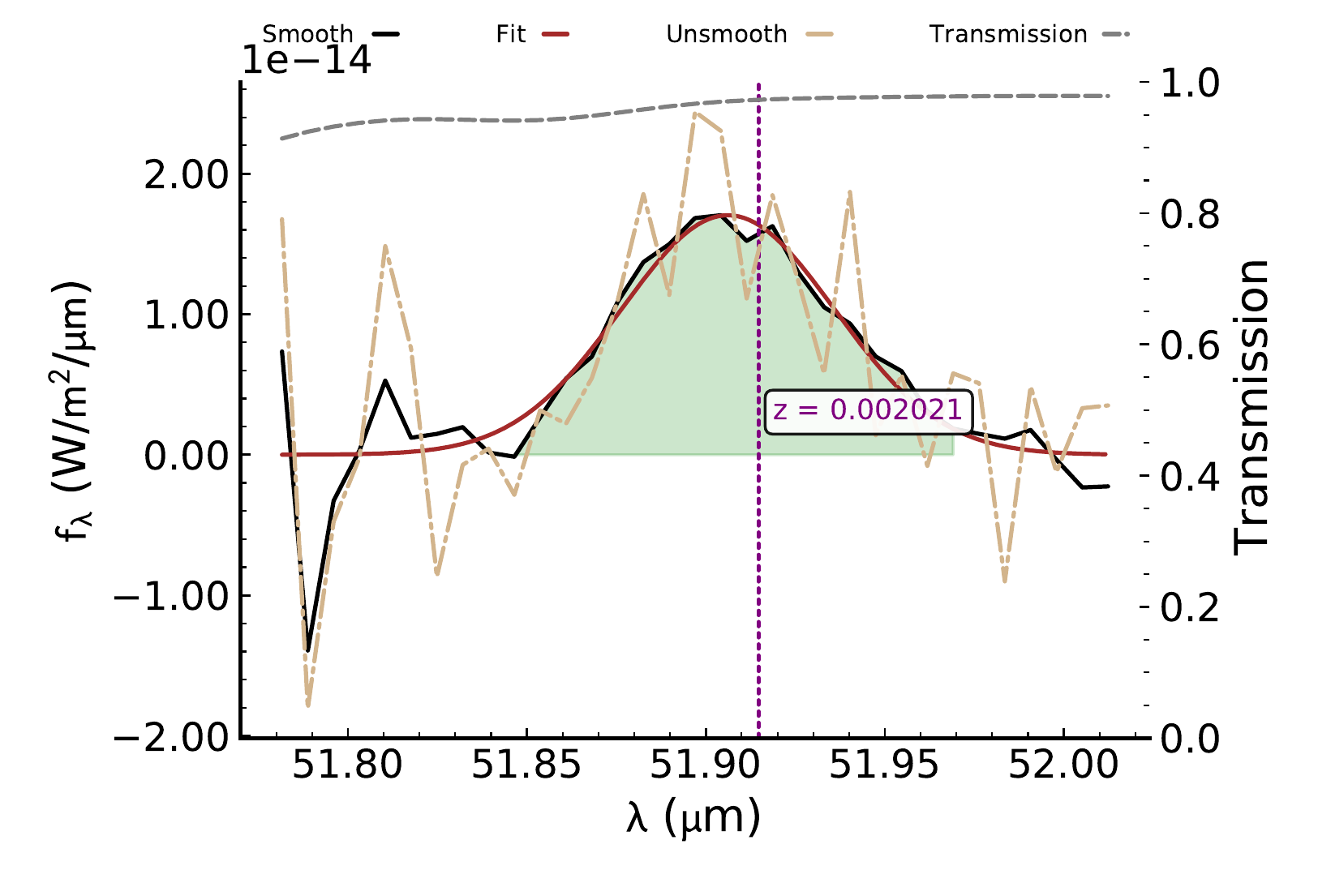}\label{fig:NGC4631-C}}\\
    \subfigure[]{\includegraphics[width=0.498\textwidth]{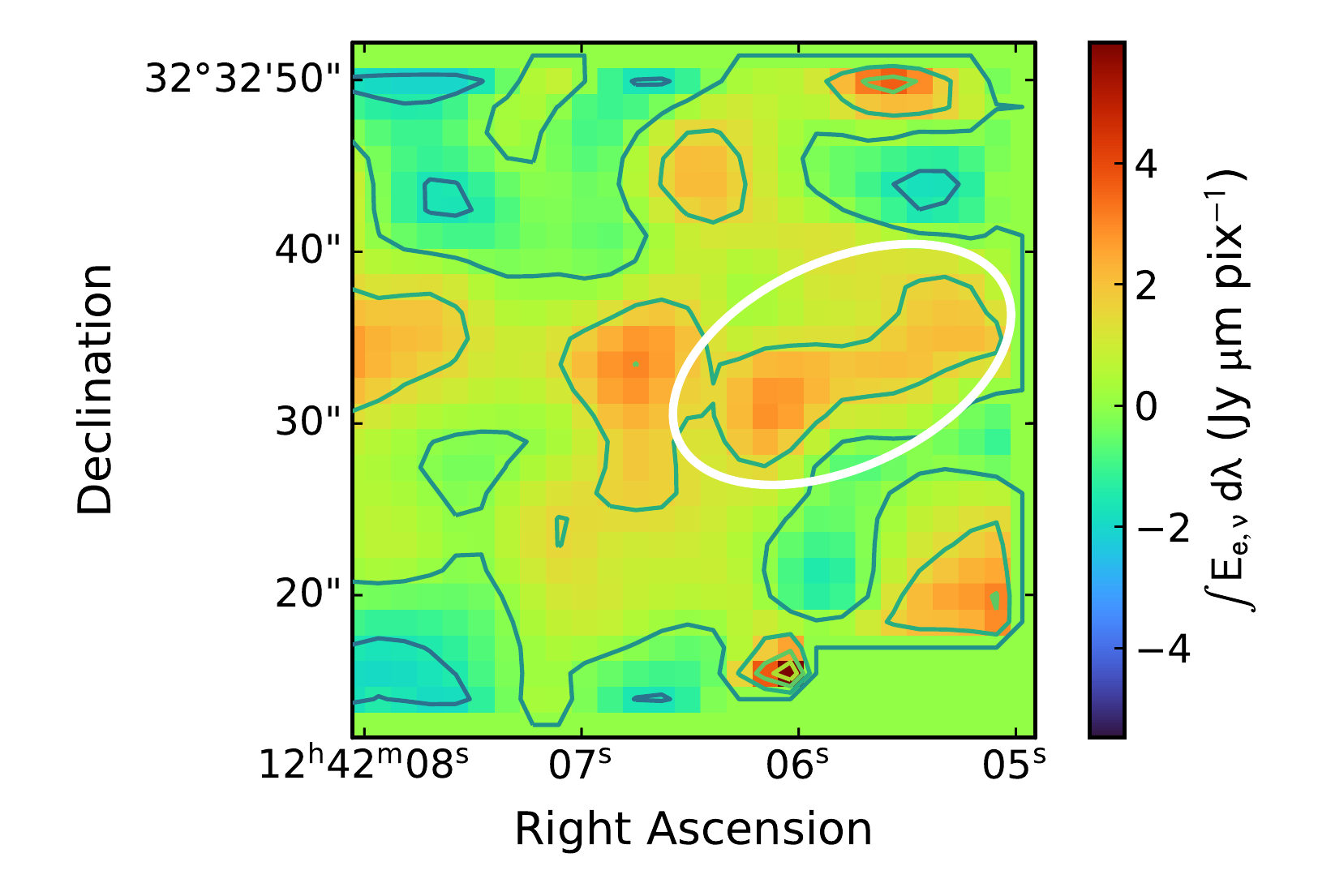}\label{fig:NGC4631-D}}~
    \subfigure[]{\includegraphics[width=0.49\textwidth]{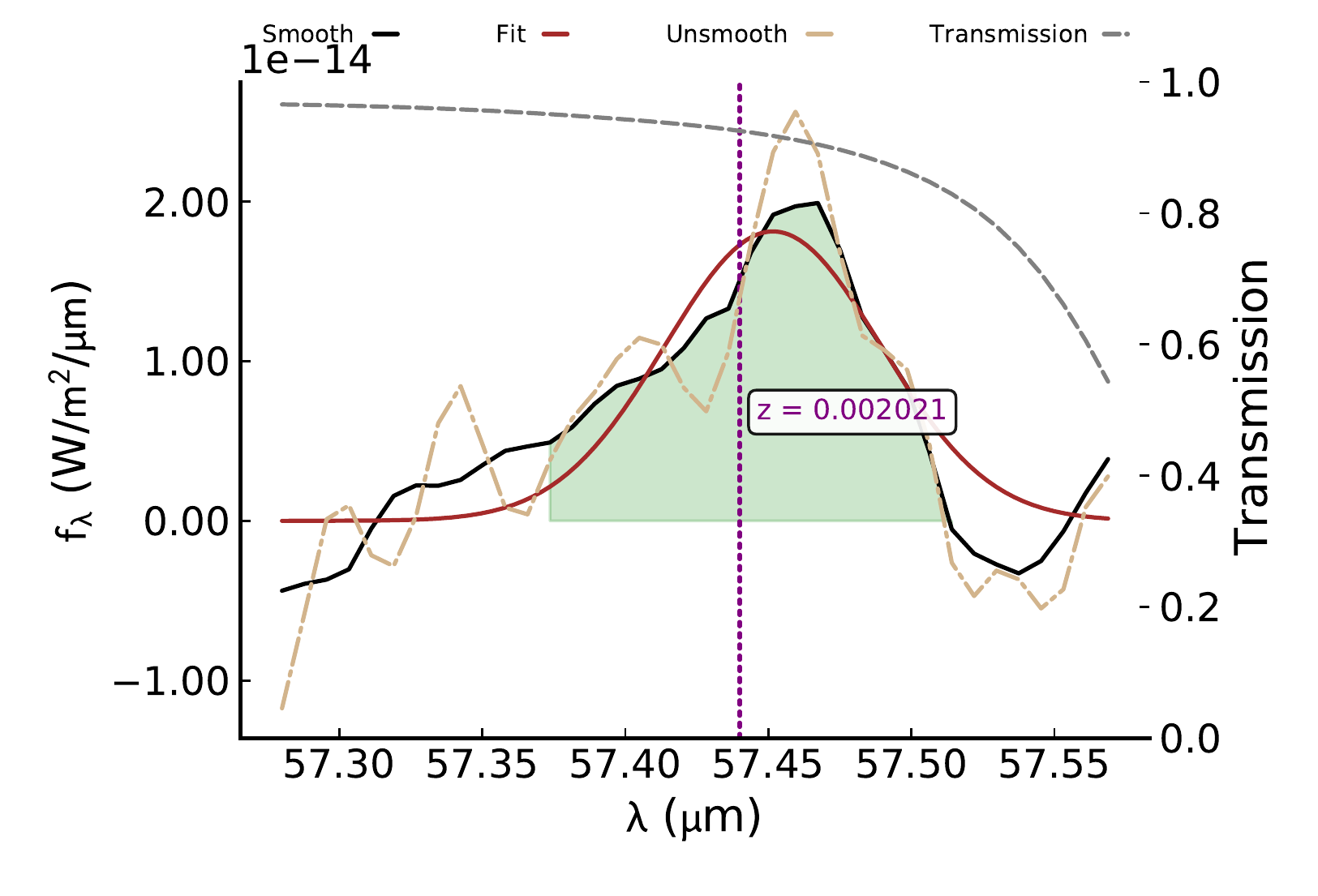}\label{fig:NGC4631-E}}
    \caption{The 2MASS image (Figure \ref{fig:NGC4631-A}), 2-D linemaps and 1-D spectra for [OIII]52$\mu$m  (Figures \ref{fig:NGC4631-B} and \ref{fig:NGC4631-C}, respectively) and [NIII]57$\mu$m  (Figures \ref{fig:NGC4631-D} and \ref{fig:NGC4631-E}, respectively) in NGC4631. The datacube for [NIII]57 was processed independently by us to deal with the low transmission at the edge of the spectral range.}
    \label{fig:NGC4631}
\end{figure*}

\begin{figure*}[ht!]
    \centering
    \subfigure[]{\includegraphics[width=0.498\textwidth]{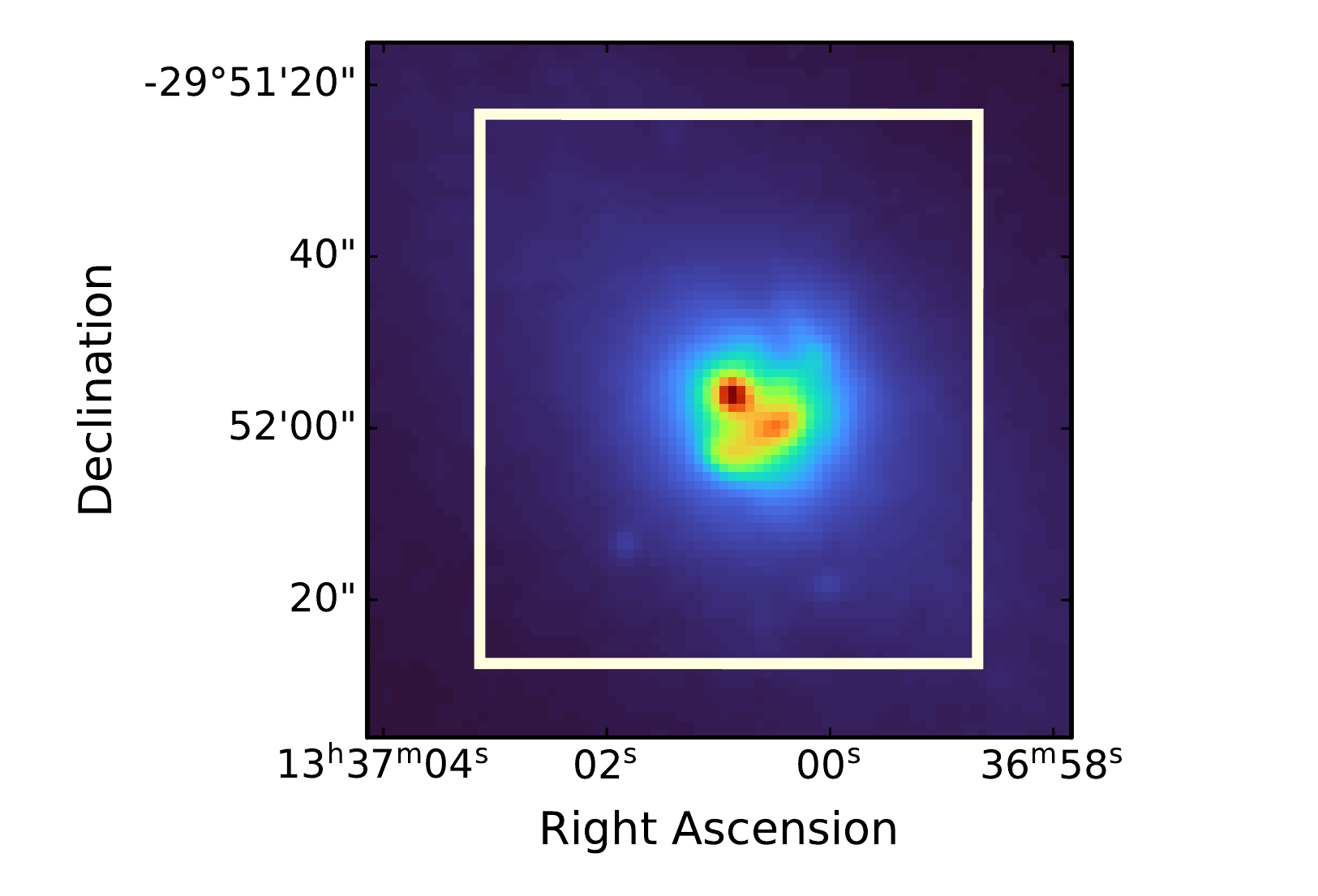}\label{fig:M83-A}}\\
    \subfigure[]{\includegraphics[width=0.498\textwidth]{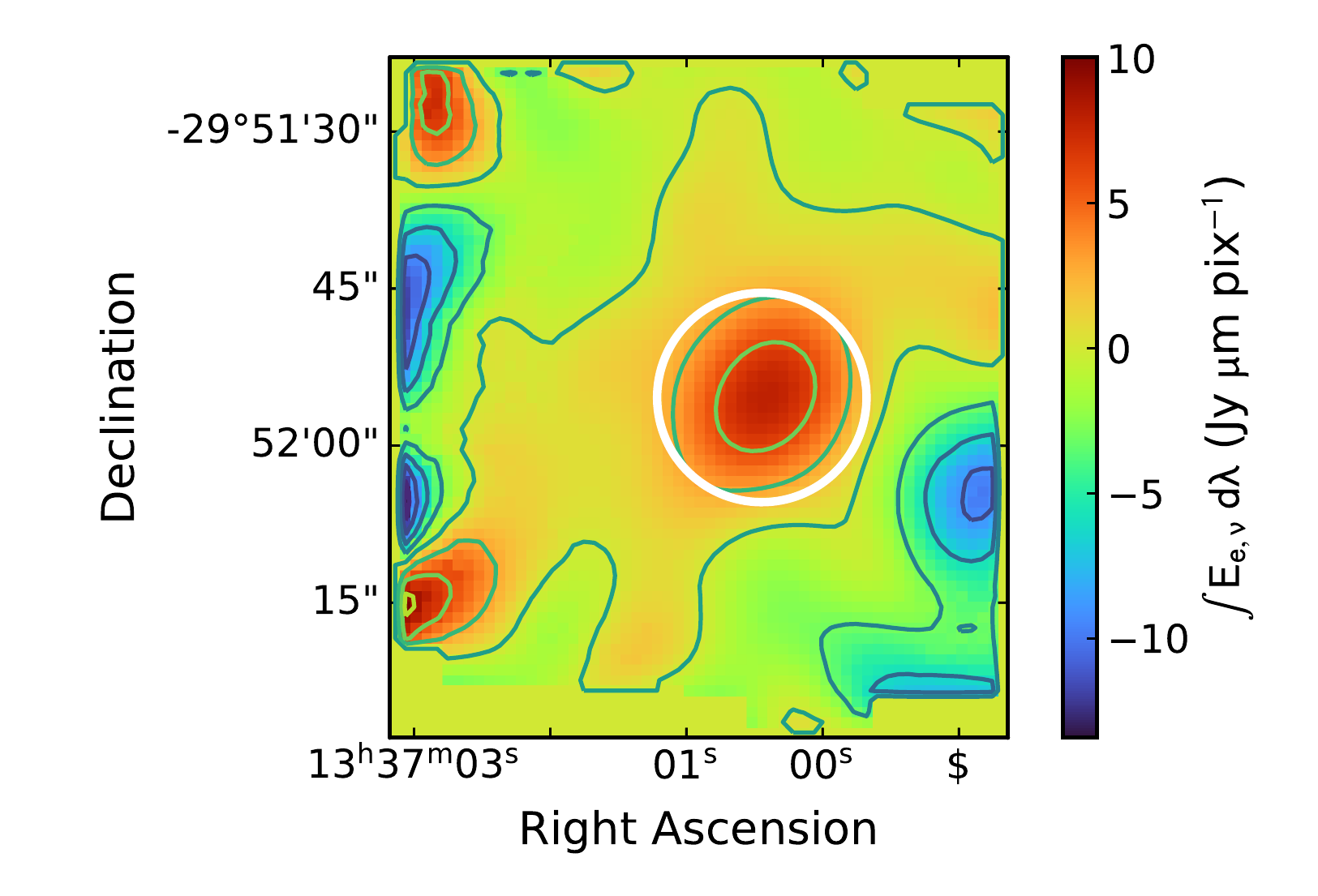}\label{fig:M83-B}}~
    \subfigure[]{\includegraphics[width=0.49\textwidth]{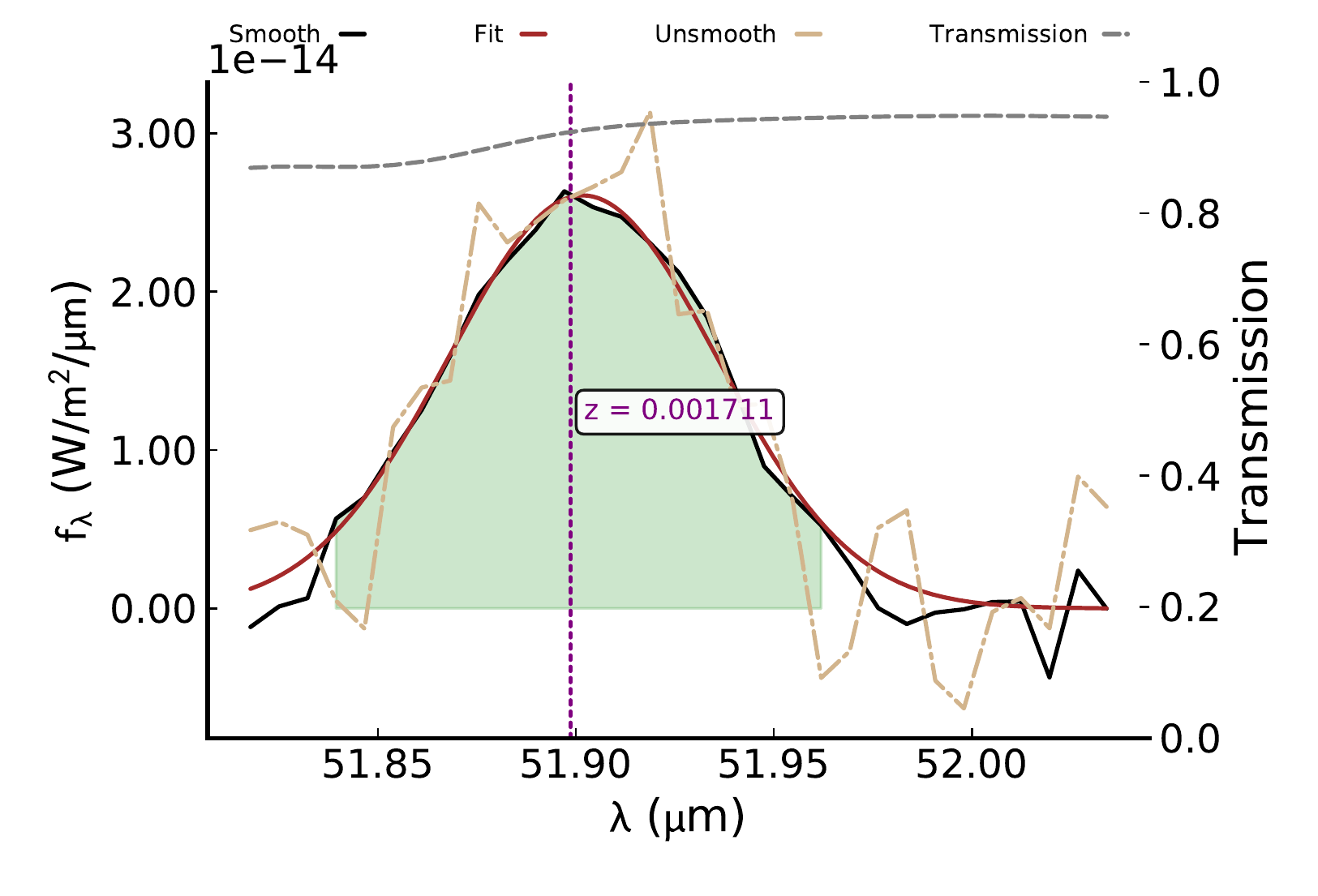}\label{fig:M83-C}}
    \caption{The 2MASS image (Figure \ref{fig:M83-A}), 2-D linemap and 1-D spectrum for [OIII]52$\mu$m  (Figures \ref{fig:M83-B} and \ref{fig:M83-C}, respectively) in M83.}
    \label{fig:M83}
\end{figure*}

\clearpage

We now show the linemaps and spectra for the galaxies mentioned in Table \ref{tab:sample1bis}, for which only the [OIII]52$\mu$m line was measured.

\begin{figure*}[ht!]
    \centering
    \subfigure[]{\includegraphics[width=0.498\textwidth]{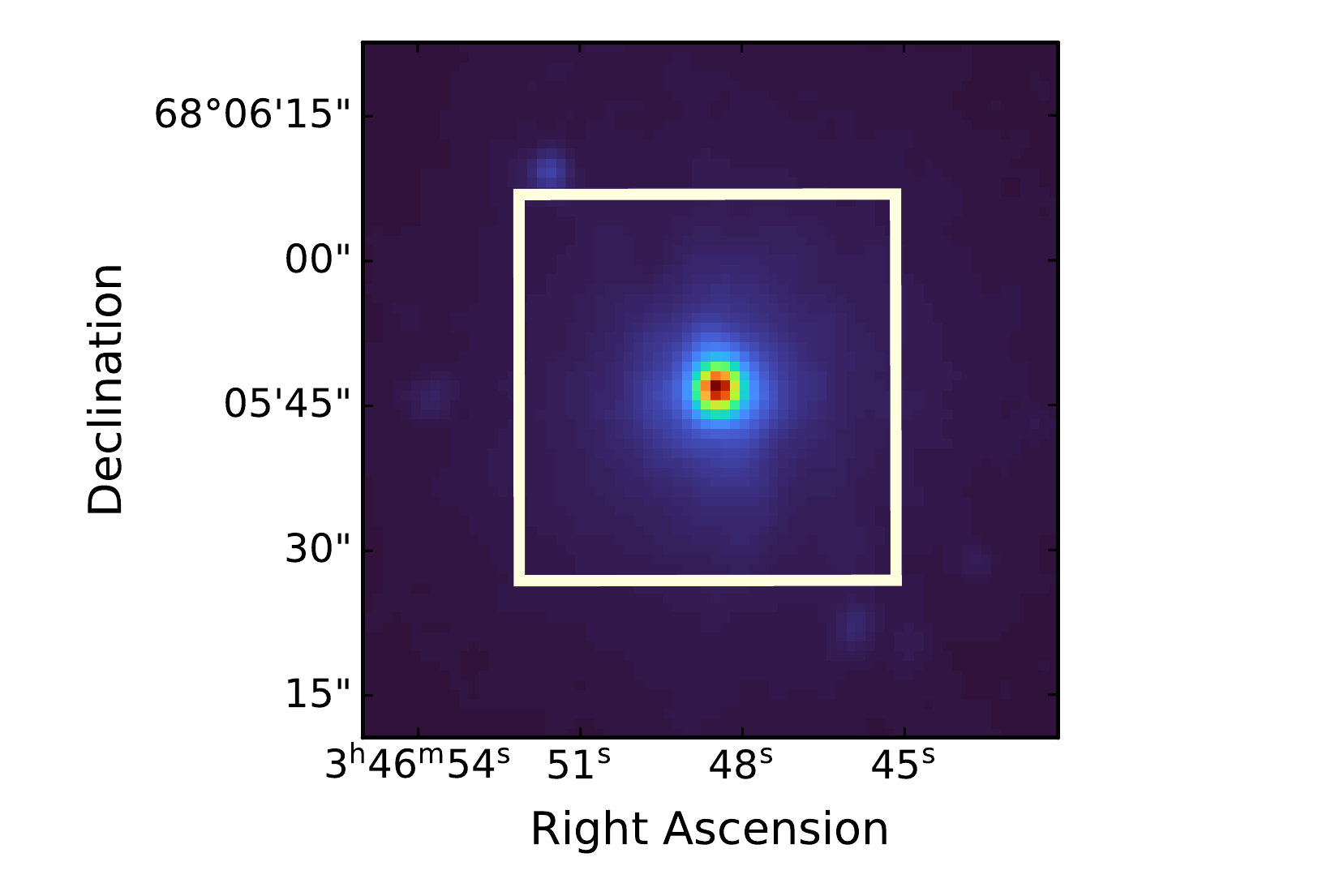}\label{fig:IC342-A}}\\
    \subfigure[]{\includegraphics[width=0.498\textwidth]{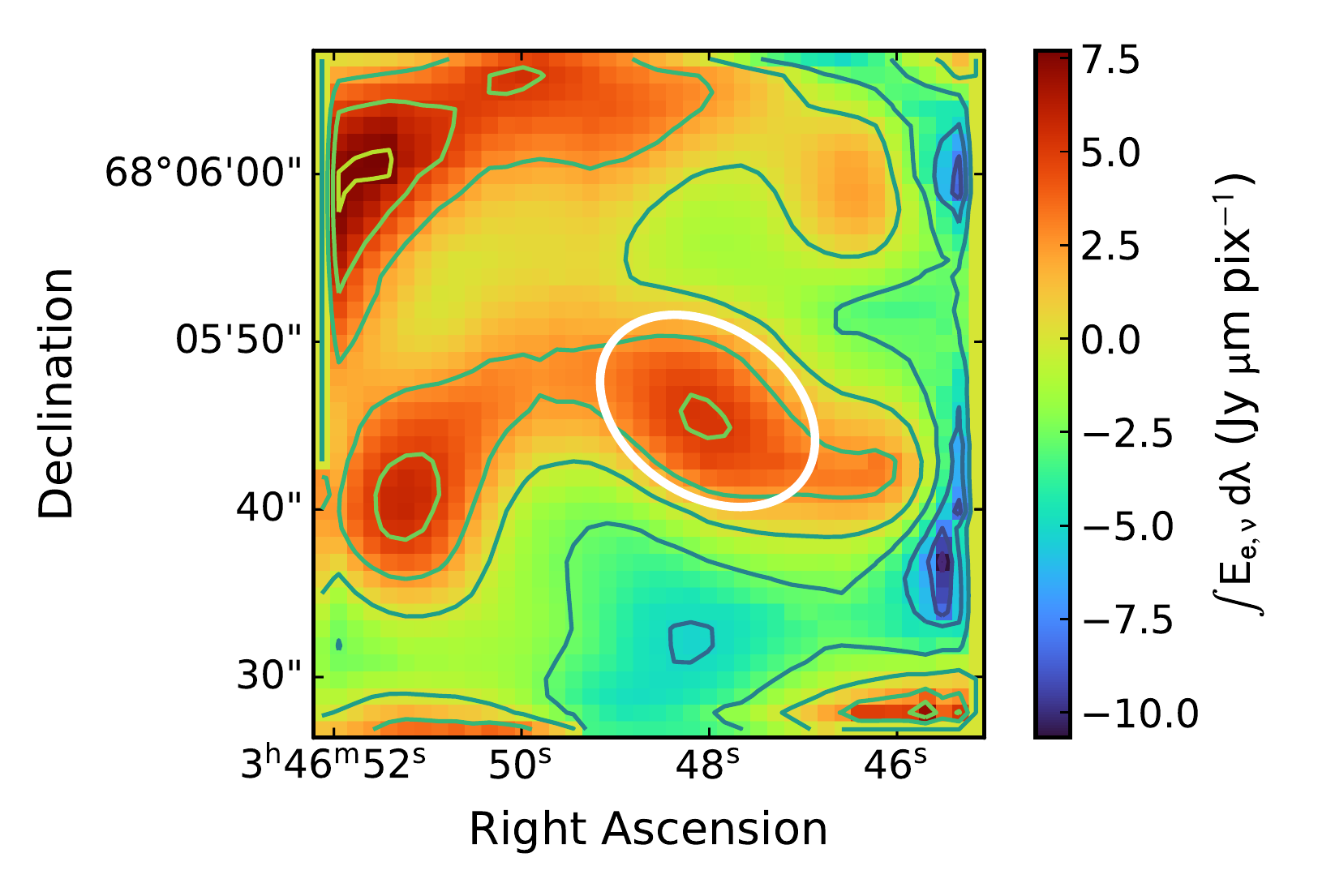}\label{fig:IC342-B}}~
    \subfigure[]{\includegraphics[width=0.49\textwidth]{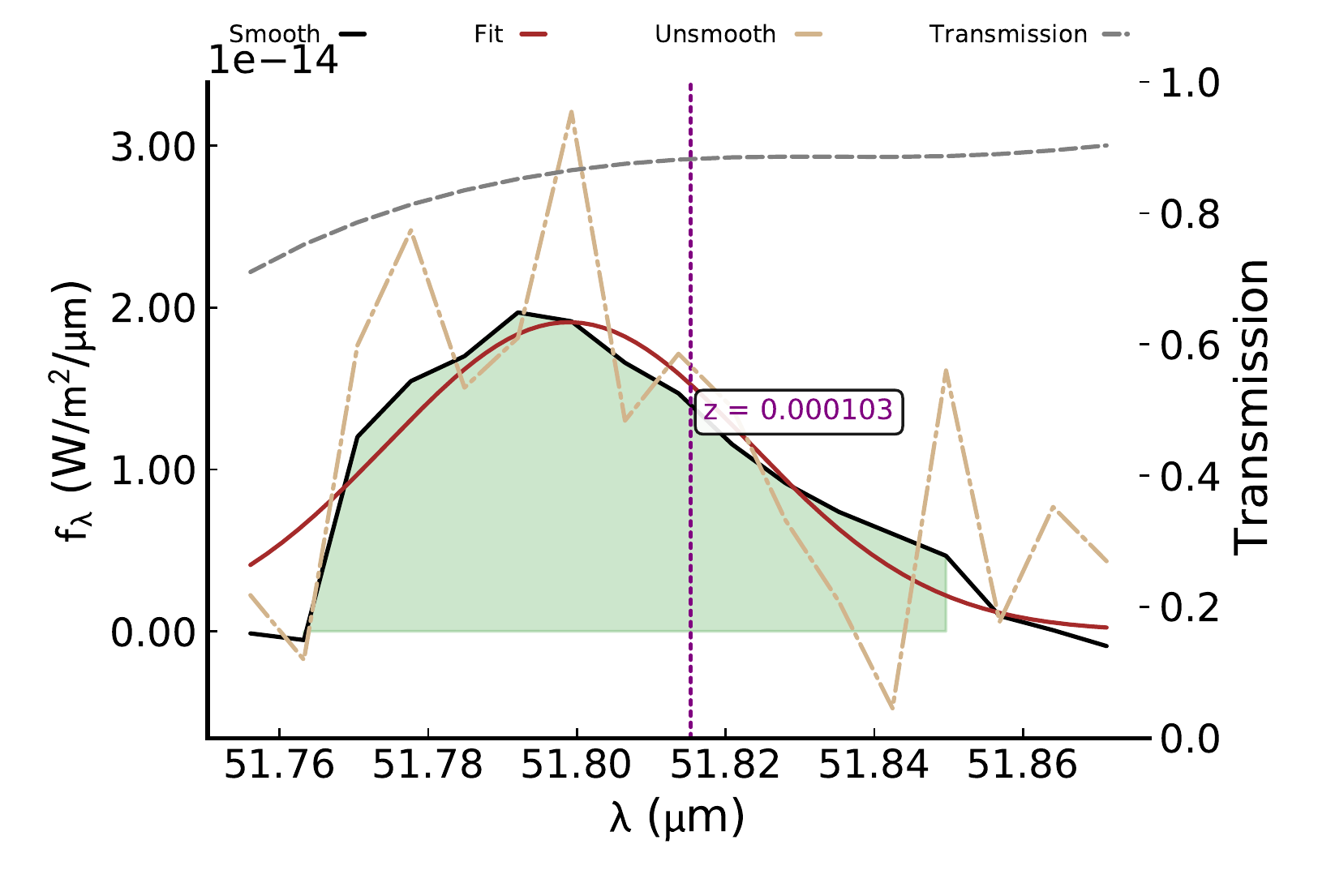}\label{fig:IC342-C}}
    \caption{The 2MASS image (Figure \ref{fig:IC342-A}), 2-D linemap and 1-D spectrum for [OIII]52$\mu$m  (Figures \ref{fig:IC342-B} and \ref{fig:IC342-C}) in IC342.}
    \label{fig:IC342}
\end{figure*}


\begin{figure*}[ht!]
    \centering
    \subfigure[]{\includegraphics[width=0.498\textwidth]{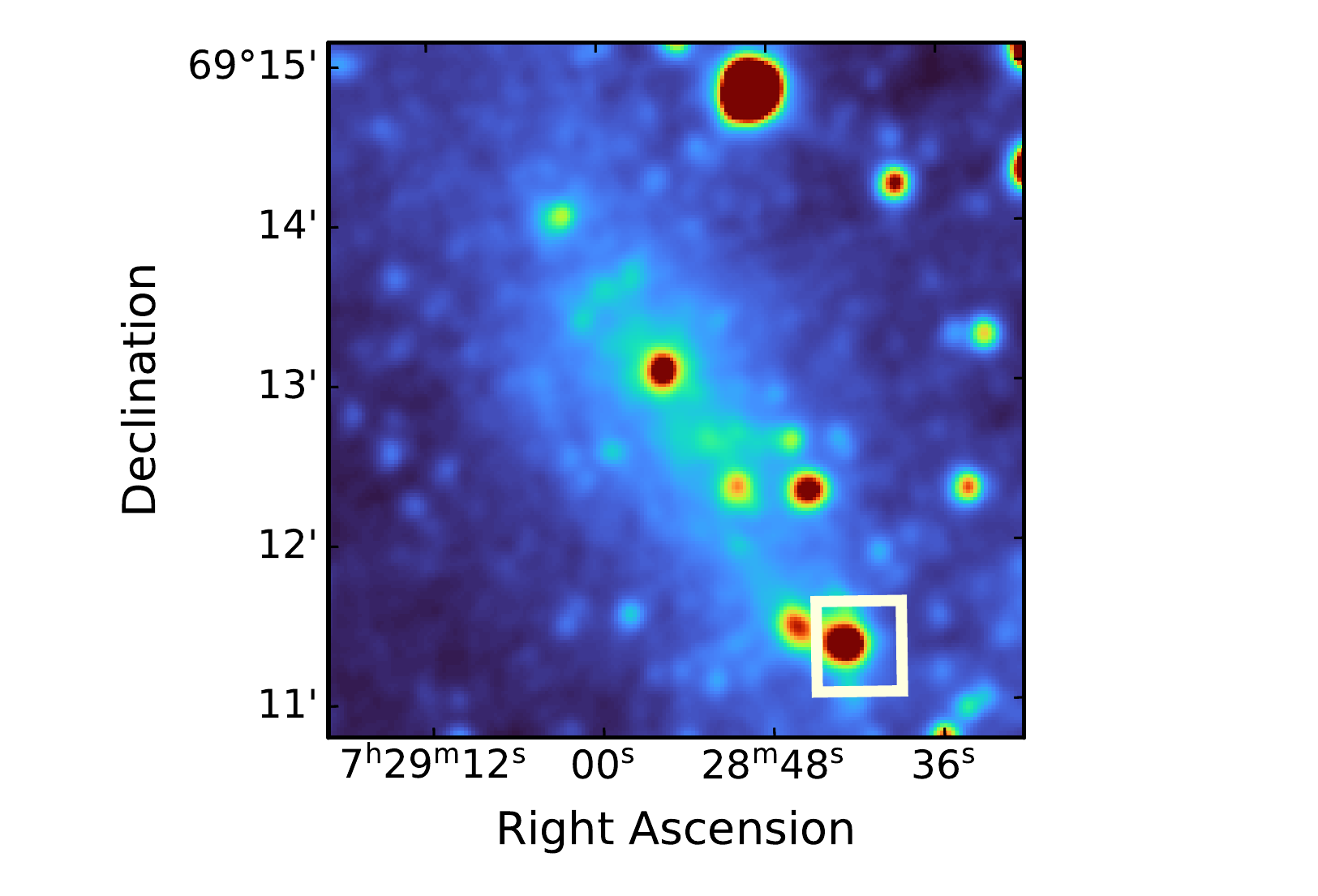}\label{fig:NGC2366-A}}\\
    \subfigure[]{\includegraphics[width=0.498\textwidth]{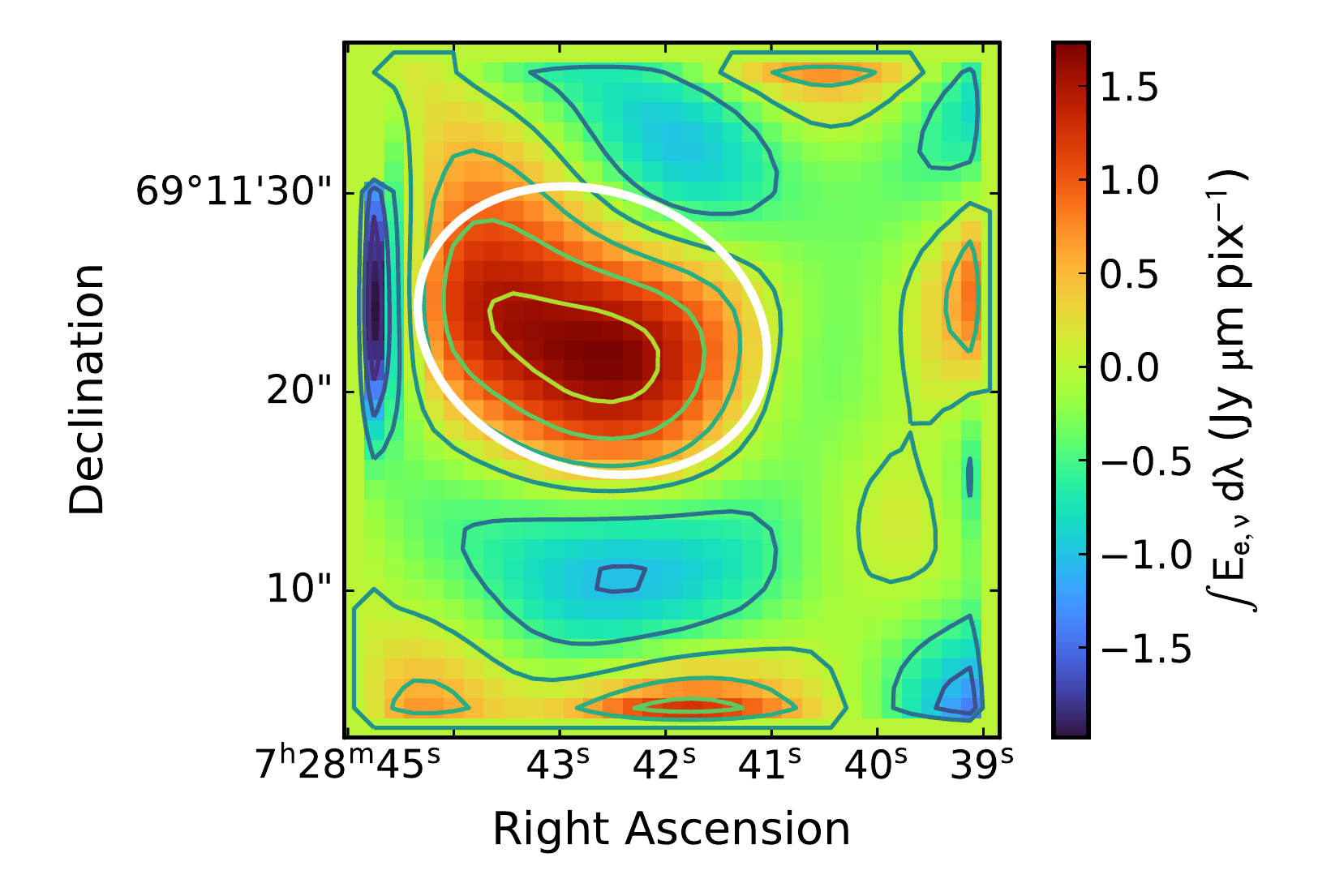}\label{fig:NGC2366-B}}~
    \subfigure[]{\includegraphics[width=0.49\textwidth]{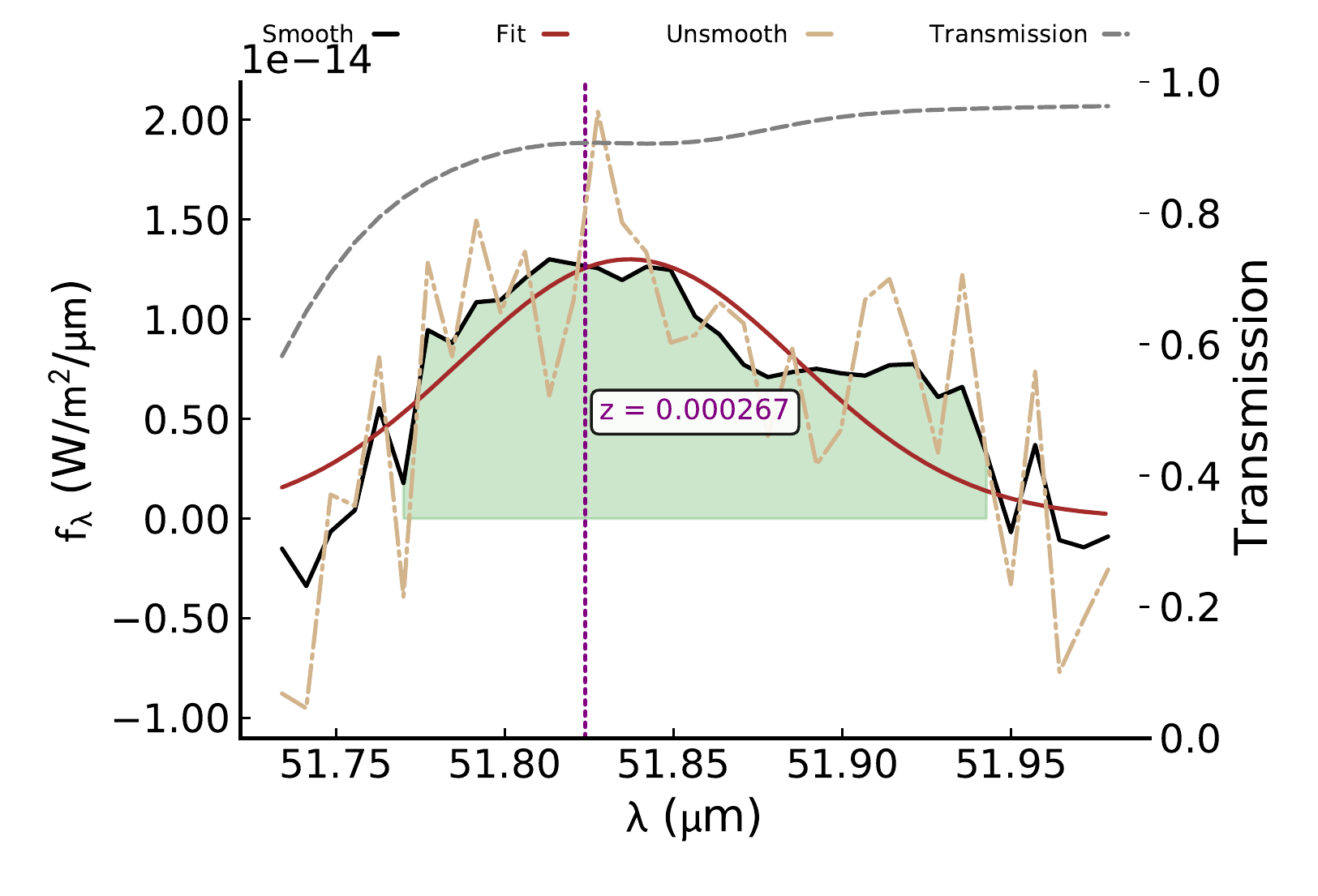}\label{fig:NGC2366-C}}
    \caption{The WISE W1 band image (Figure \ref{fig:NGC2366-A}; the 2MASS map in this case did not show appreciable detection), 2-D linemap and 1-D spectrum for [OIII]52$\mu$m  (Figures \ref{fig:NGC2366-B} and \ref{fig:NGC2366-C}) in NGC2366. The profiles have not been corrected for atmospheric transmission. Once again, the SOFIA survey region shows significant departure from the optical center, just like NGC2366. This is because the SOFIA in this case was not pointed to the optical center, as displayed in the WISE map.}
    \label{fig:NGC2366}
\end{figure*}

\begin{figure*}[ht!]
    \centering
    \subfigure[]{\includegraphics[width=0.498\textwidth]{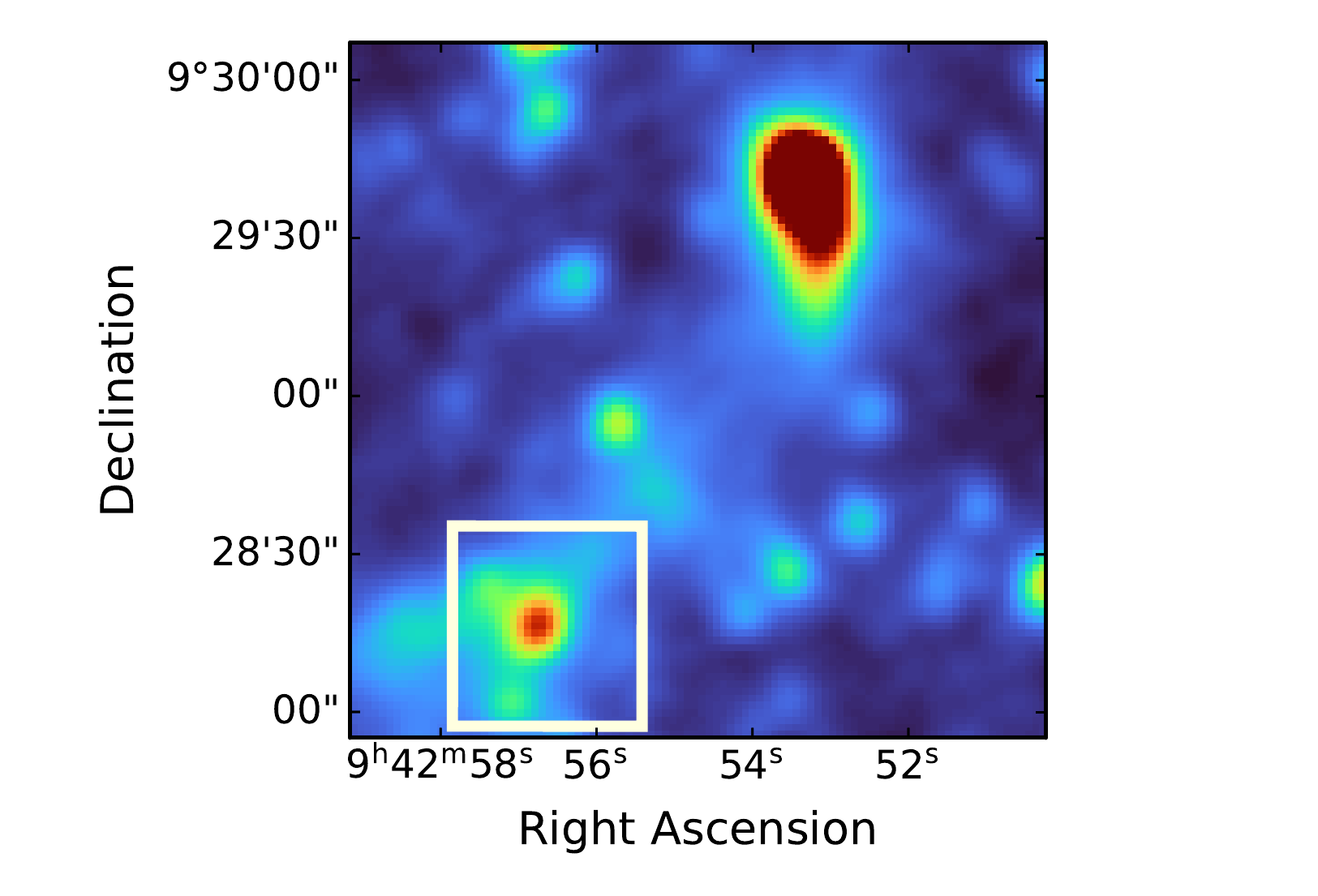}\label{fig:UGCS5189-A}}\\
    \subfigure[]{\includegraphics[width=0.498\textwidth]{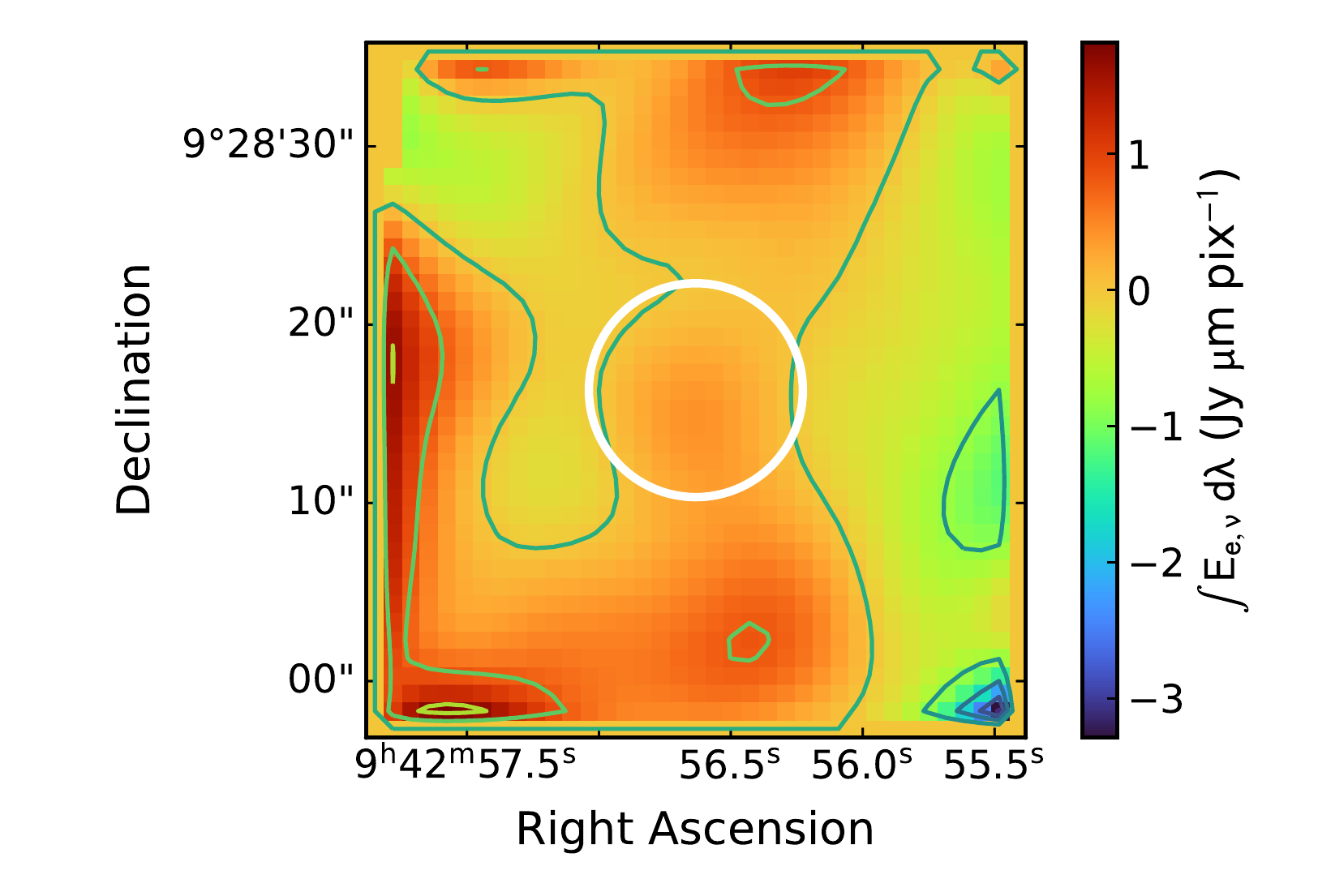}\label{fig:UGCS5189-B}}~
    \subfigure[]{\includegraphics[width=0.49\textwidth]{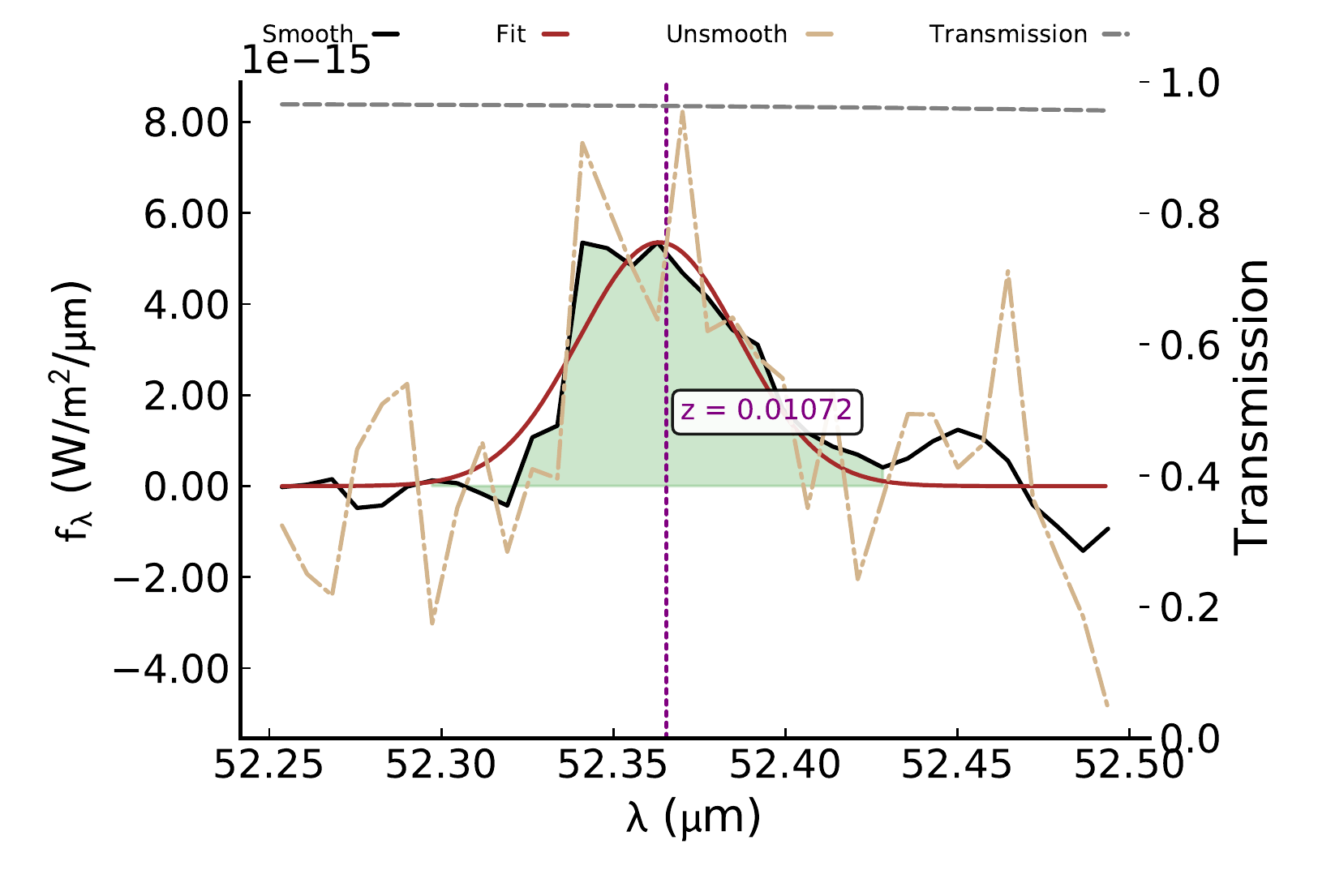}\label{fig:UGCS5189-C}}
    \caption{The WISE W1 band image (Figure \ref{fig:UGCS5189-A}; the 2MASS map did not show appreciable detection), 2-D linemaps and 1-D spectrum for [OIII]52$\mu$m  (Figures \ref{fig:UGCS5189-A} and \ref{fig:UGCS5189-B}, respectively) in UGC5189. Once again, like NGC2366, one may notice the significant deviation of the SOFIA survey region from the optical center, just like NGC2366.}
    \label{fig:UGCS5189}
\end{figure*}


\begin{figure*}[ht!]
    \centering
    \subfigure[]{\includegraphics[width=0.498\textwidth]{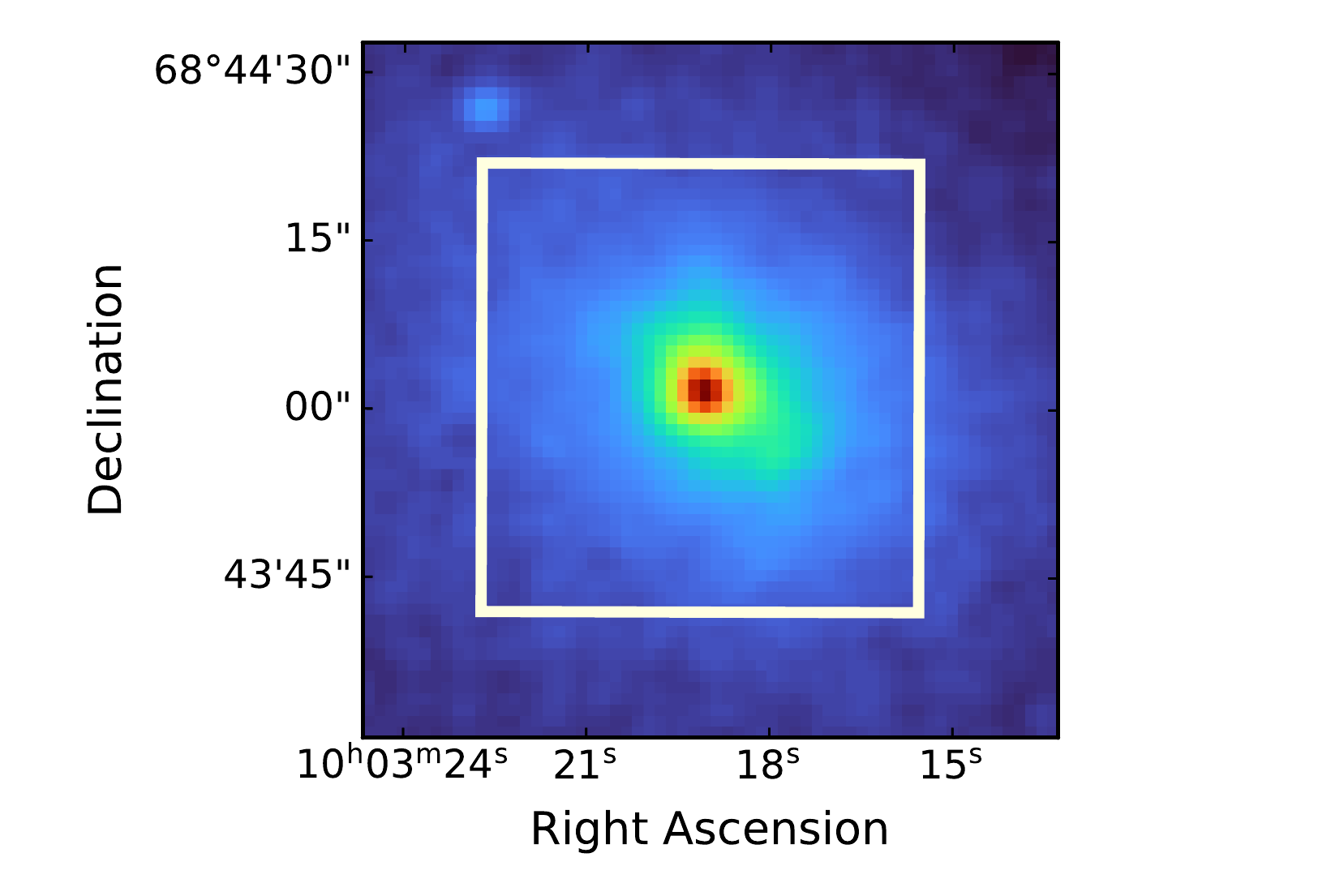}\label{fig:NGC3077-A}}\\
    \subfigure[]{\includegraphics[width=0.498\textwidth]{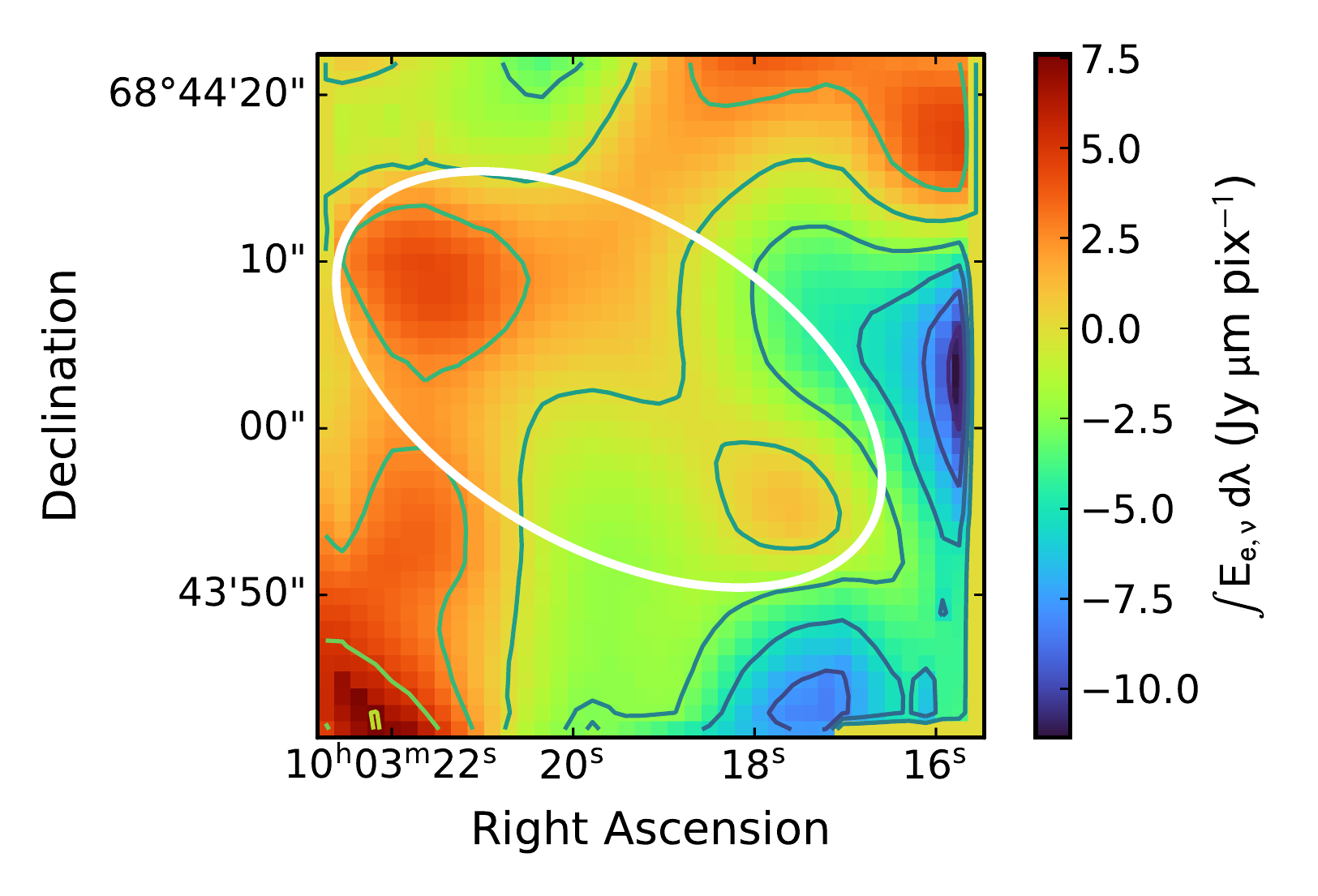}\label{fig:NGC3077-B}}~
    \subfigure[]{\includegraphics[width=0.49\textwidth]{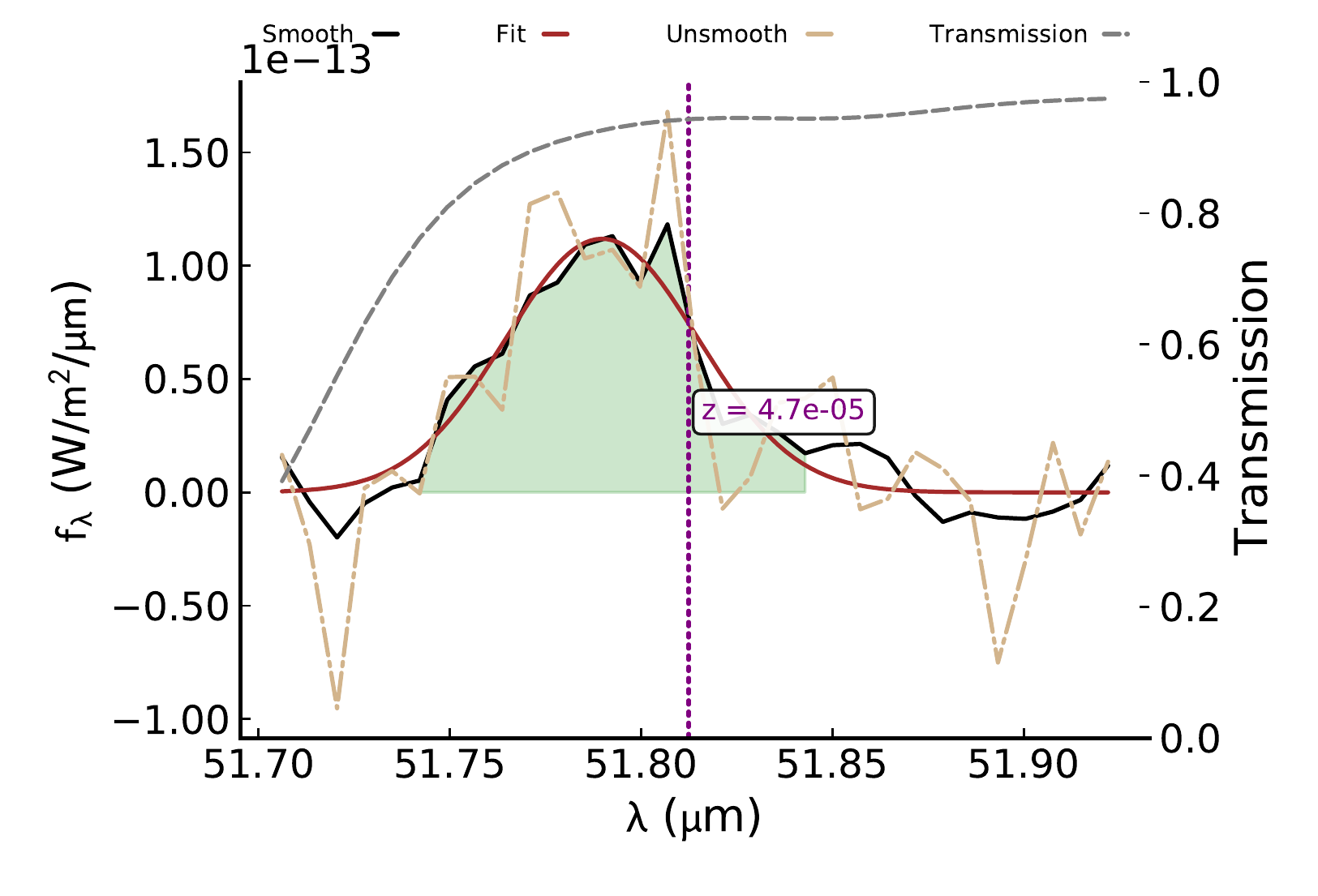}\label{fig:NGC3077-C}}
    \caption{The 2MASS image (Figure \ref{fig:NGC3077-A}), 2-D linemap and 1-D spectrum for [OIII]52$\mu$m  (Figures \ref{fig:NGC3077-B} and \ref{fig:NGC3077-C}) in NGC3077. The profiles have not been corrected for atmospheric transmission.}
    \label{fig:NGC3077}
\end{figure*}

\begin{figure*}[ht!]
    \centering
    \subfigure[]{\includegraphics[width=0.498\textwidth]{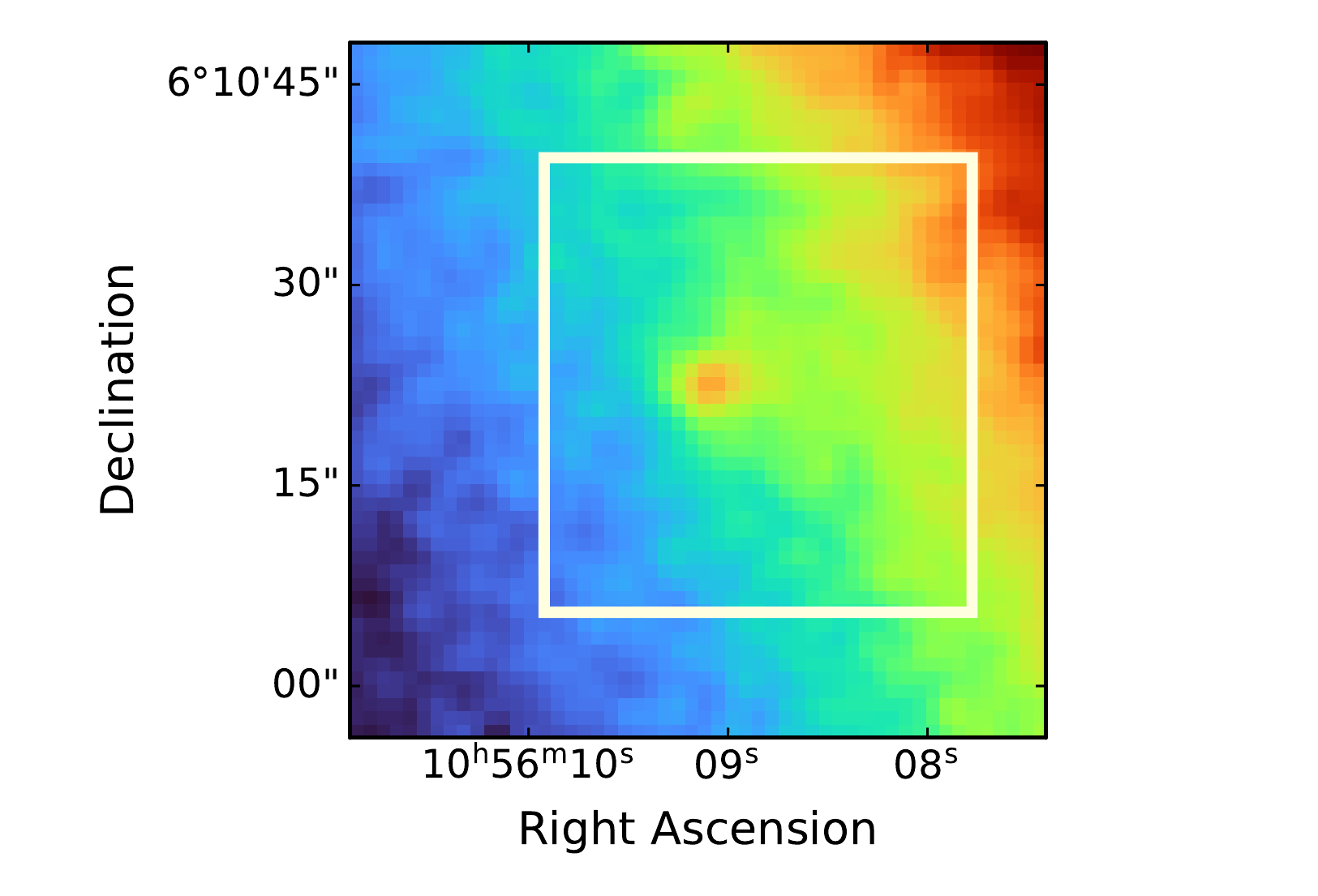}\label{fig:Mrk1271-A}}\\
    \subfigure[]{\includegraphics[width=0.498\textwidth]{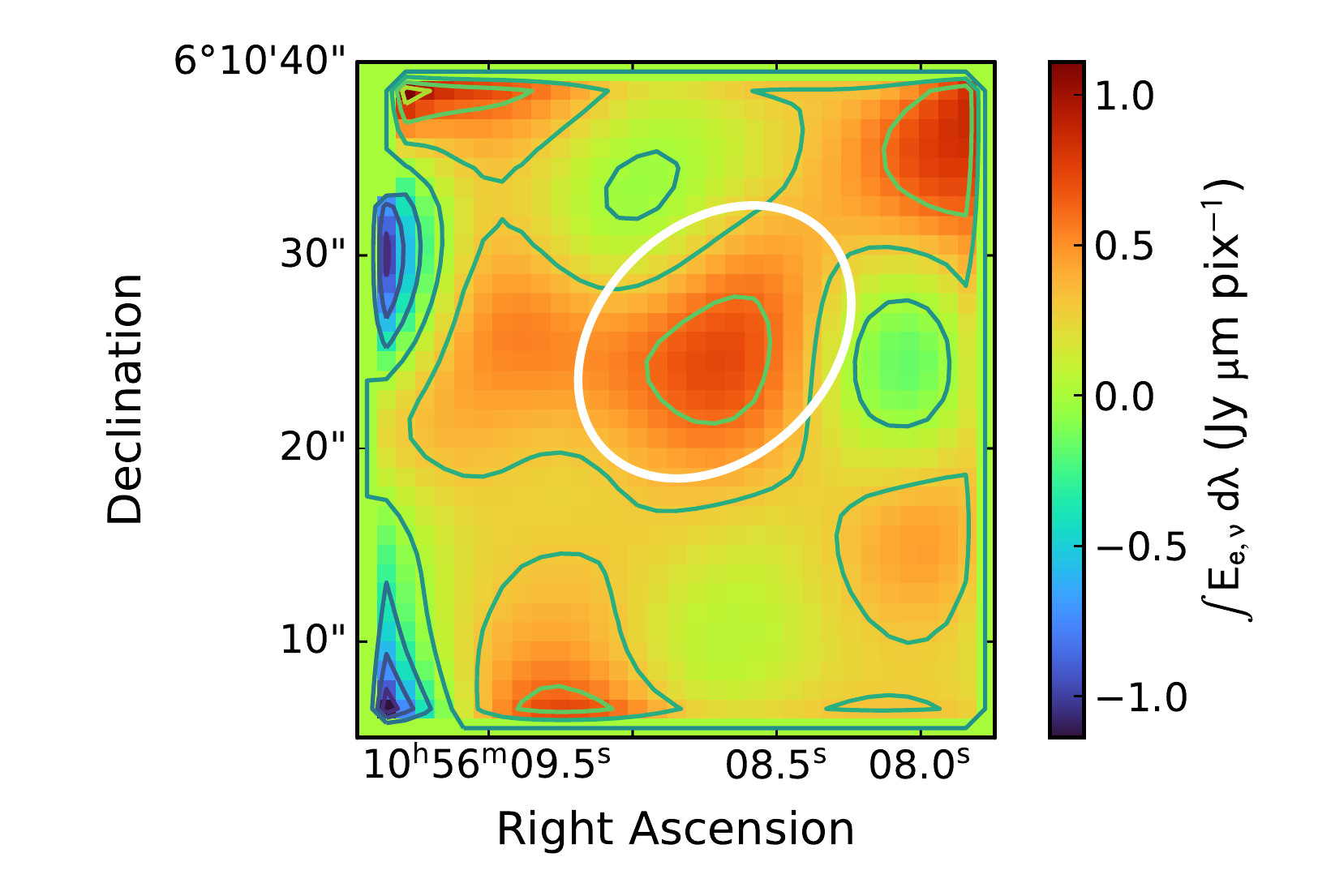}\label{fig:Mrk1271-B}}~
    \subfigure[]{\includegraphics[width=0.49\textwidth]{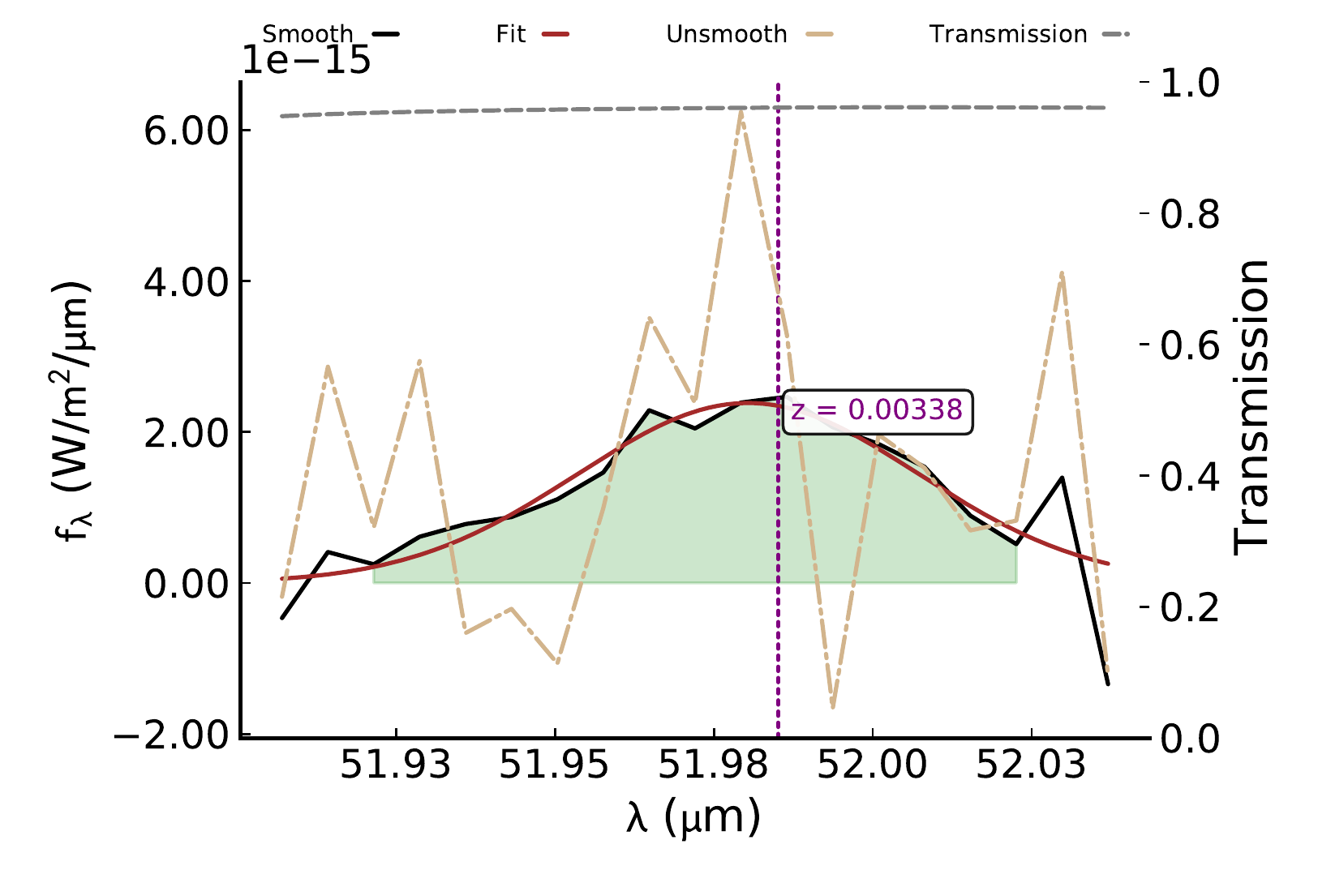}\label{fig:Mrk1271-C}}   
    \caption{The 2MASS image (Figure \ref{fig:Mrk1271-A}),2-D linemap and 1-D spectrum for [OIII]52$\mu$m  (Figures \ref{fig:Mrk1271-B} and \ref{fig:Mrk1271-C}) in Mrk1271. The profiles have not been corrected for atmospheric transmission.}
    \label{fig:Mrk1271}
\end{figure*}

\begin{figure*}[ht!]
    \centering
    \subfigure[]{\includegraphics[width=0.498\textwidth]{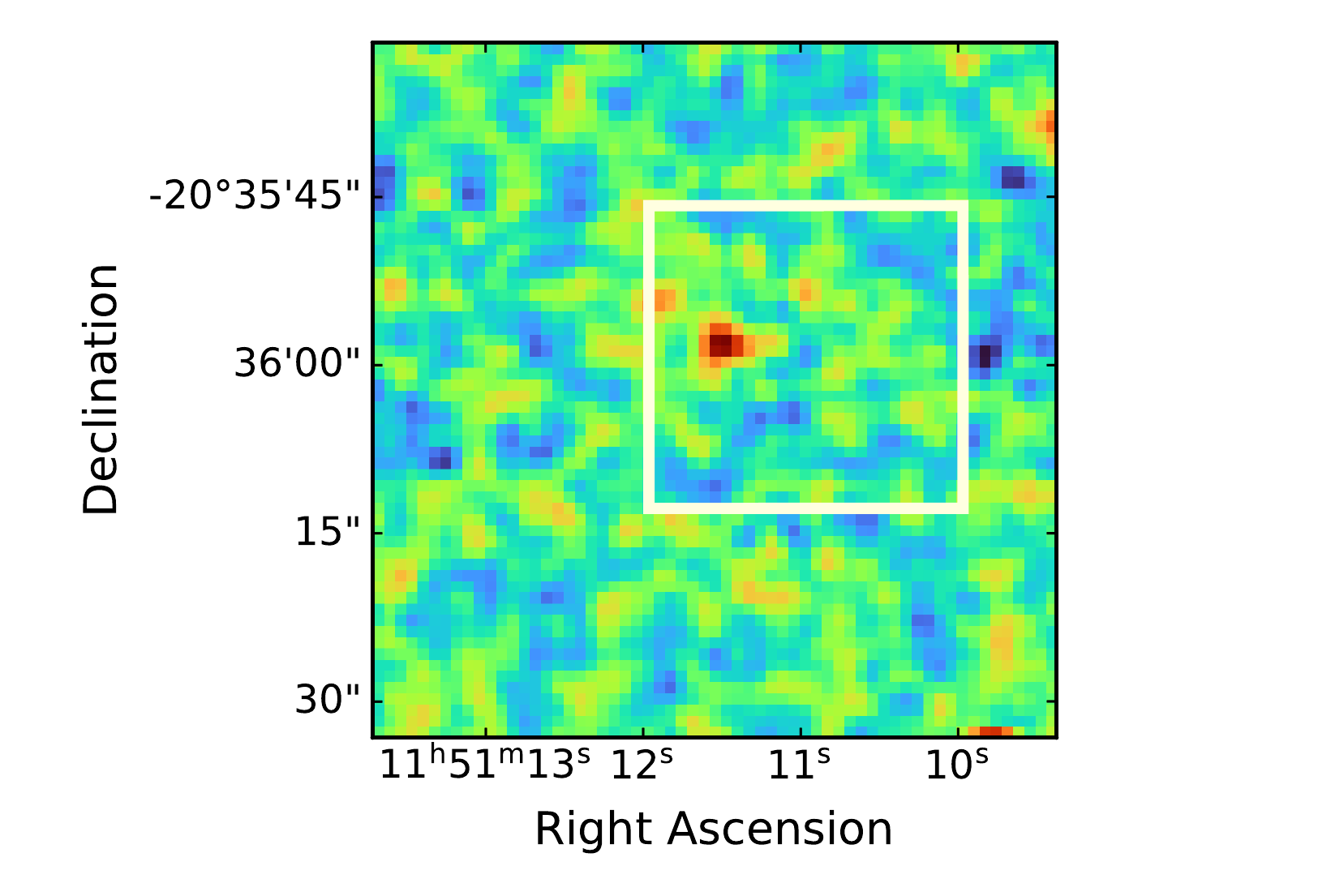}\label{fig:Pox4-A}}\\
    \subfigure[]{\includegraphics[width=0.498\textwidth]{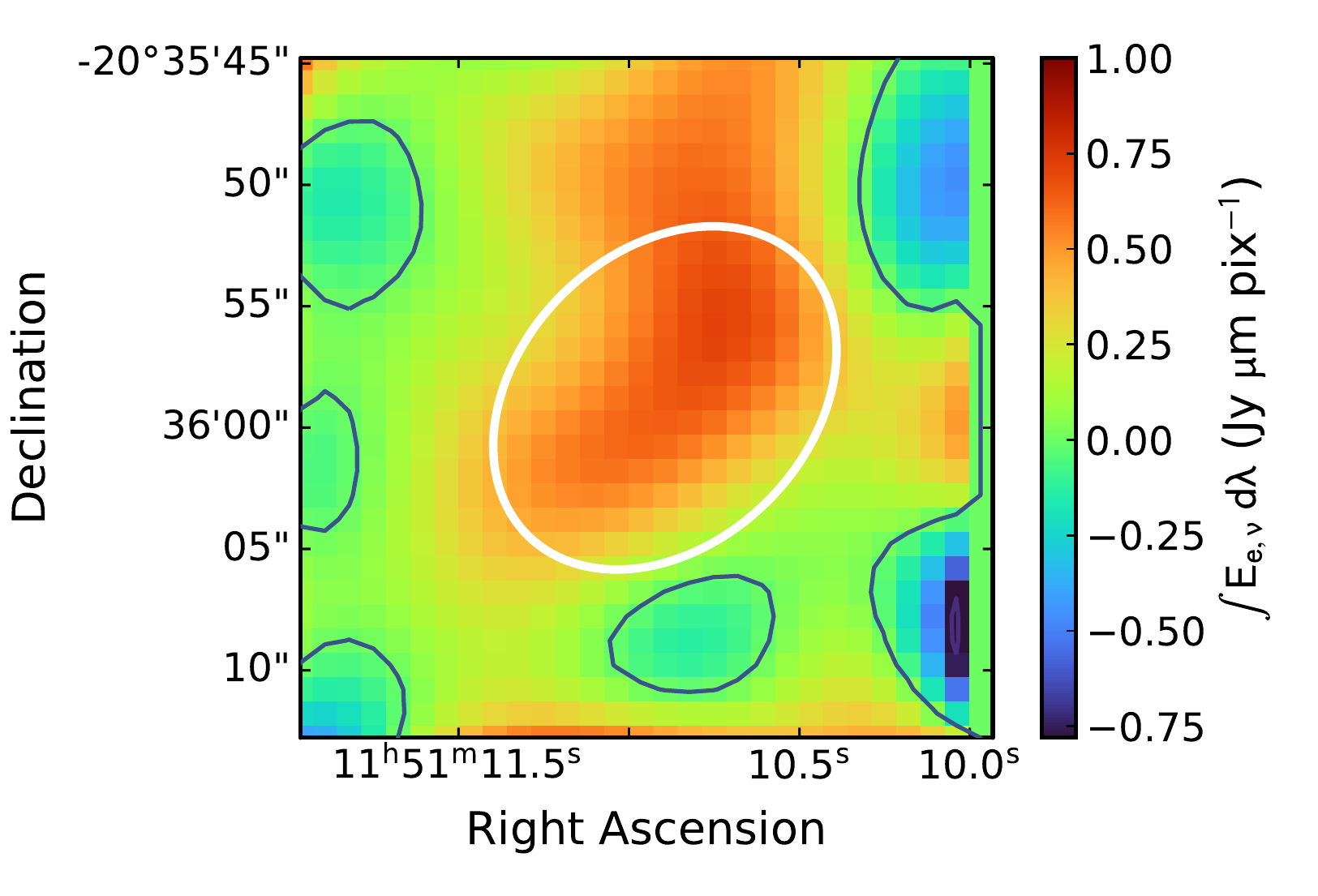}\label{fig:Pox4-B}}~
    \subfigure[]{\includegraphics[width=0.49\textwidth]{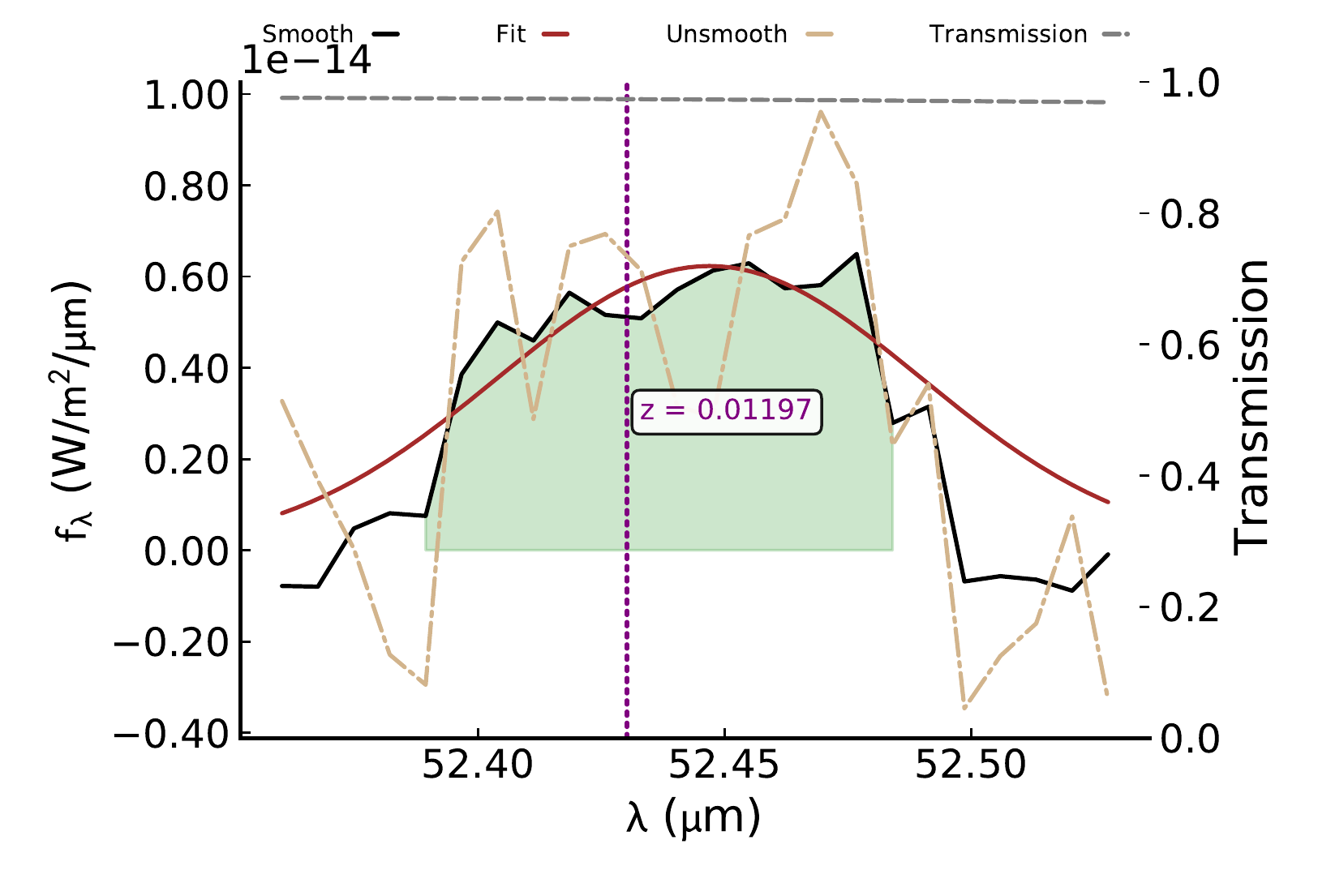}\label{fig:Pox4-C}}
    \caption{The 2MASS image (Figure \ref{fig:Pox4-A}), 2-D linemap and 1-D spectrum for [OIII]52$\mu$m  (Figures \ref{fig:Pox4-B} and \ref{fig:Pox4-C}, respectively) in Pox4.}
    \label{fig:Pox4}
\end{figure*}

\begin{figure*}[ht!]
    \centering
    \subfigure[]{\includegraphics[width=0.498\textwidth]{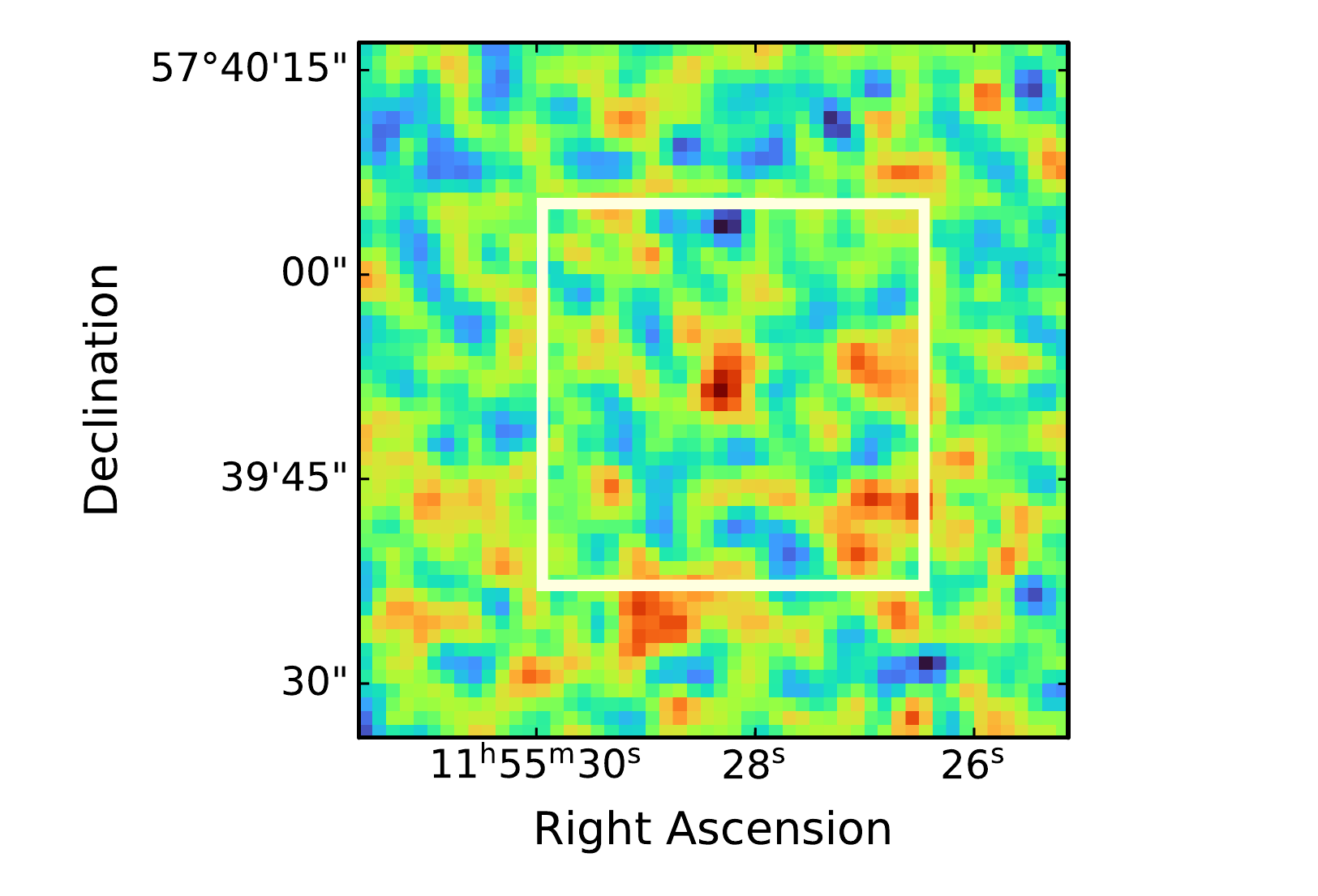}\label{fig:Mrk193-A}}\\
    \subfigure[]{\includegraphics[width=0.498\textwidth]{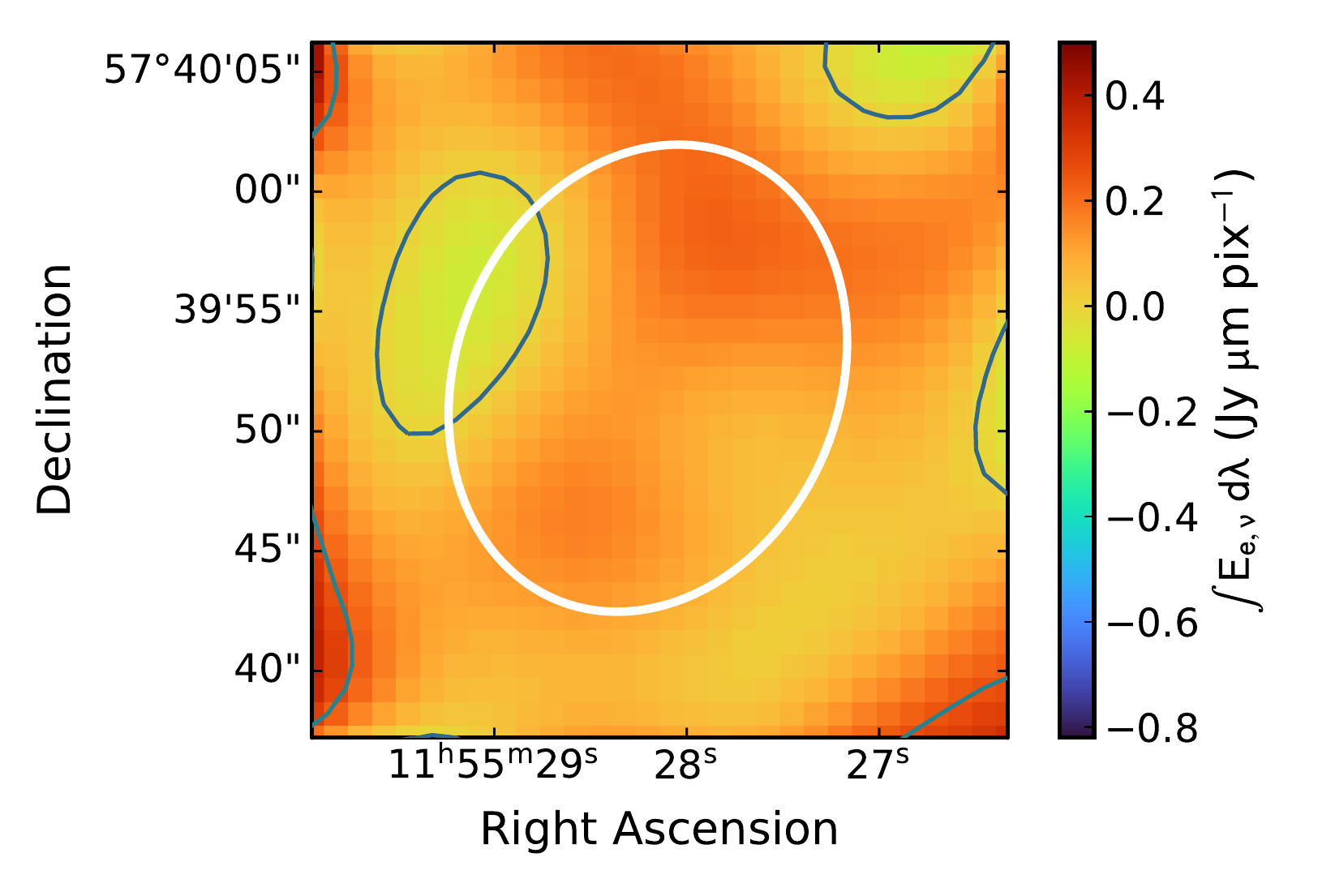}\label{fig:Mrk193-B}}~
    \subfigure[]{\includegraphics[width=0.49\textwidth]{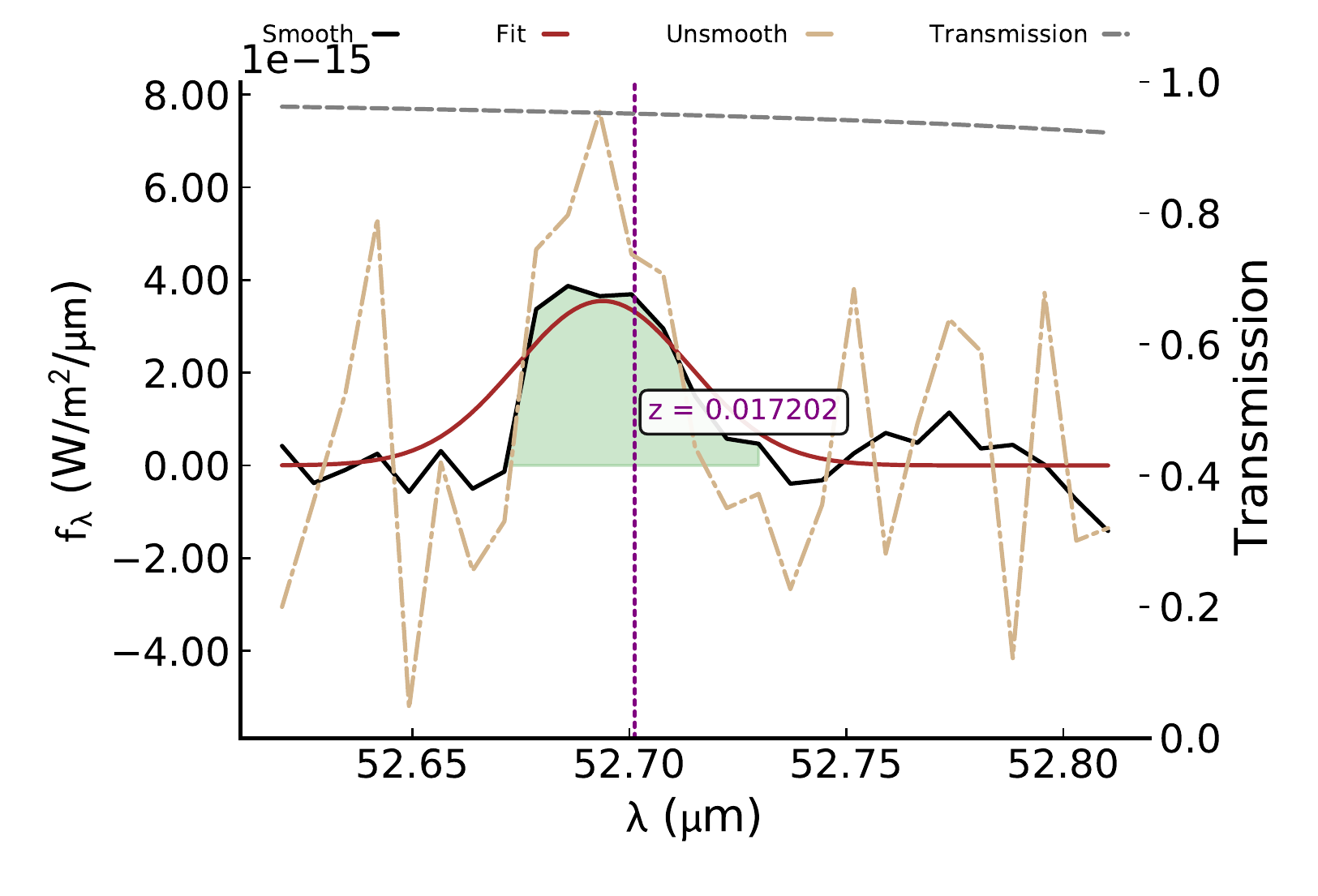}\label{fig:Mrk193-C}}
    \caption{The 2MASS image (Figure \ref{fig:Mrk193-A}), 2-D linemap and 1-D spectrum for [OIII]52$\mu$m  (Figures \ref{fig:Mrk193-B} and \ref{fig:Mrk193-C}) in Mrk193.}
    \label{fig:Mrk193}
\end{figure*}

\begin{figure*}[ht!]
    \centering
    \subfigure[]{\includegraphics[width=0.498\textwidth]{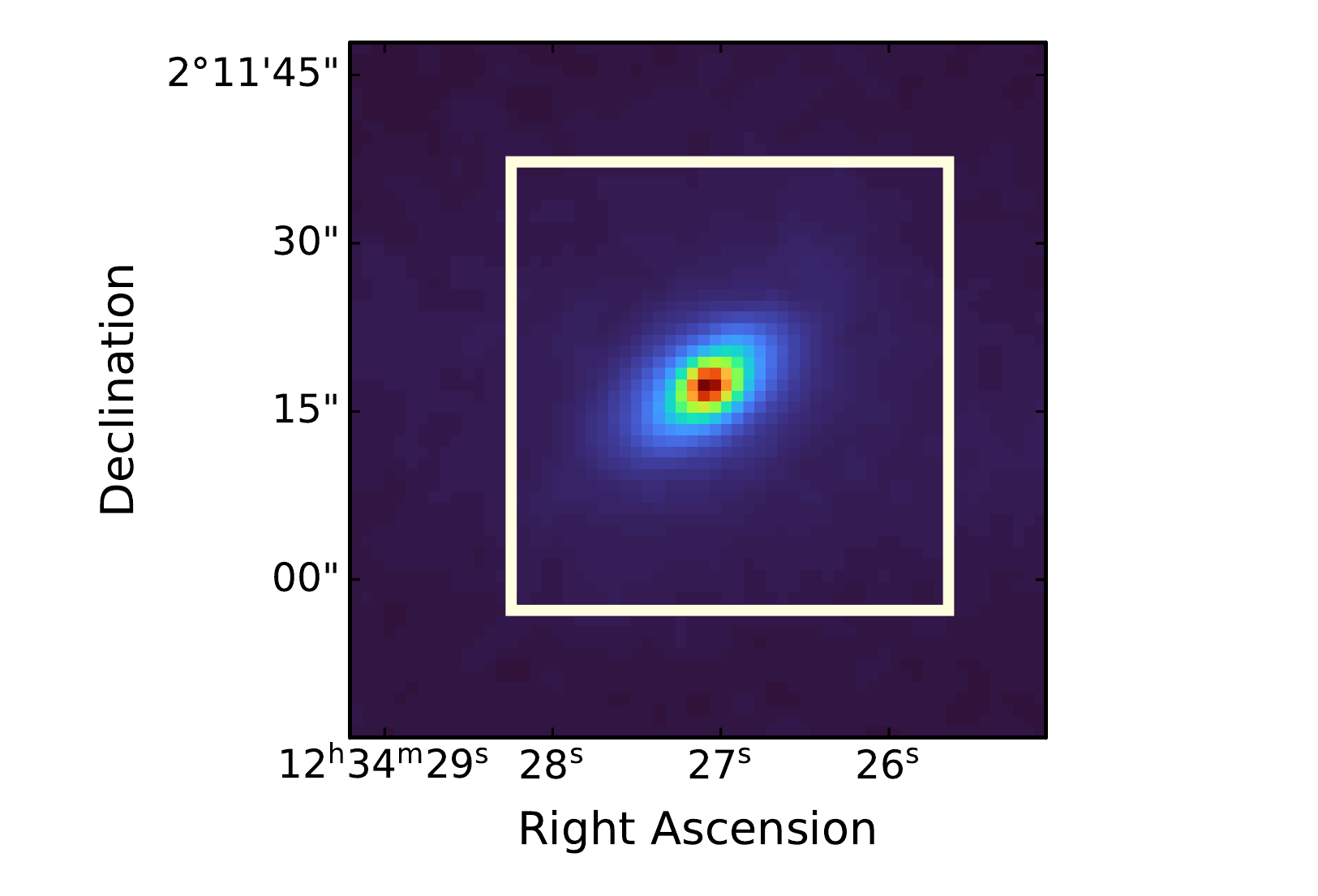}\label{fig:NGC4536-A}}\\
    \subfigure[]{\includegraphics[width=0.498\textwidth]{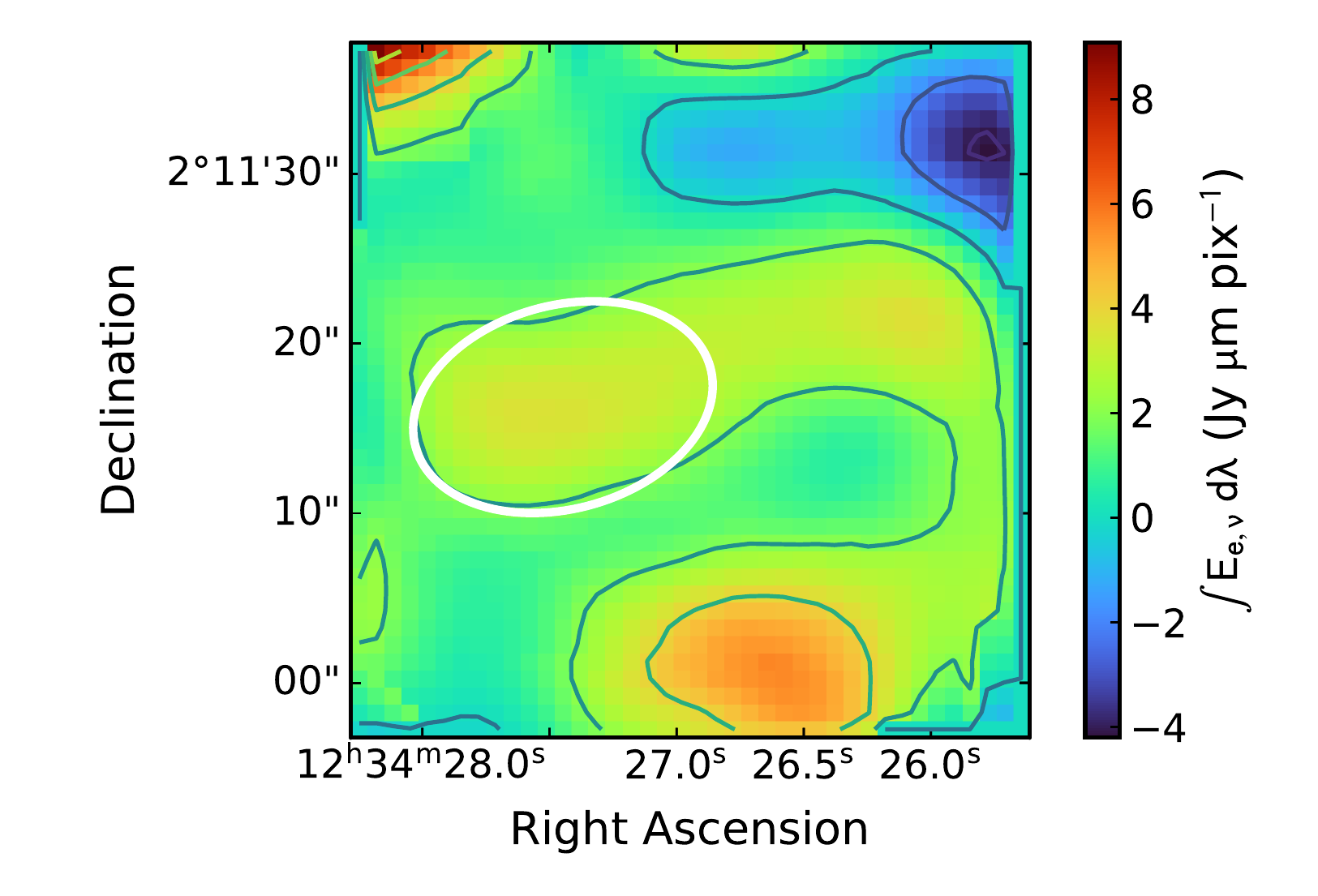}\label{fig:NGC4536-B}}~
    \subfigure[]{\includegraphics[width=0.49\textwidth]{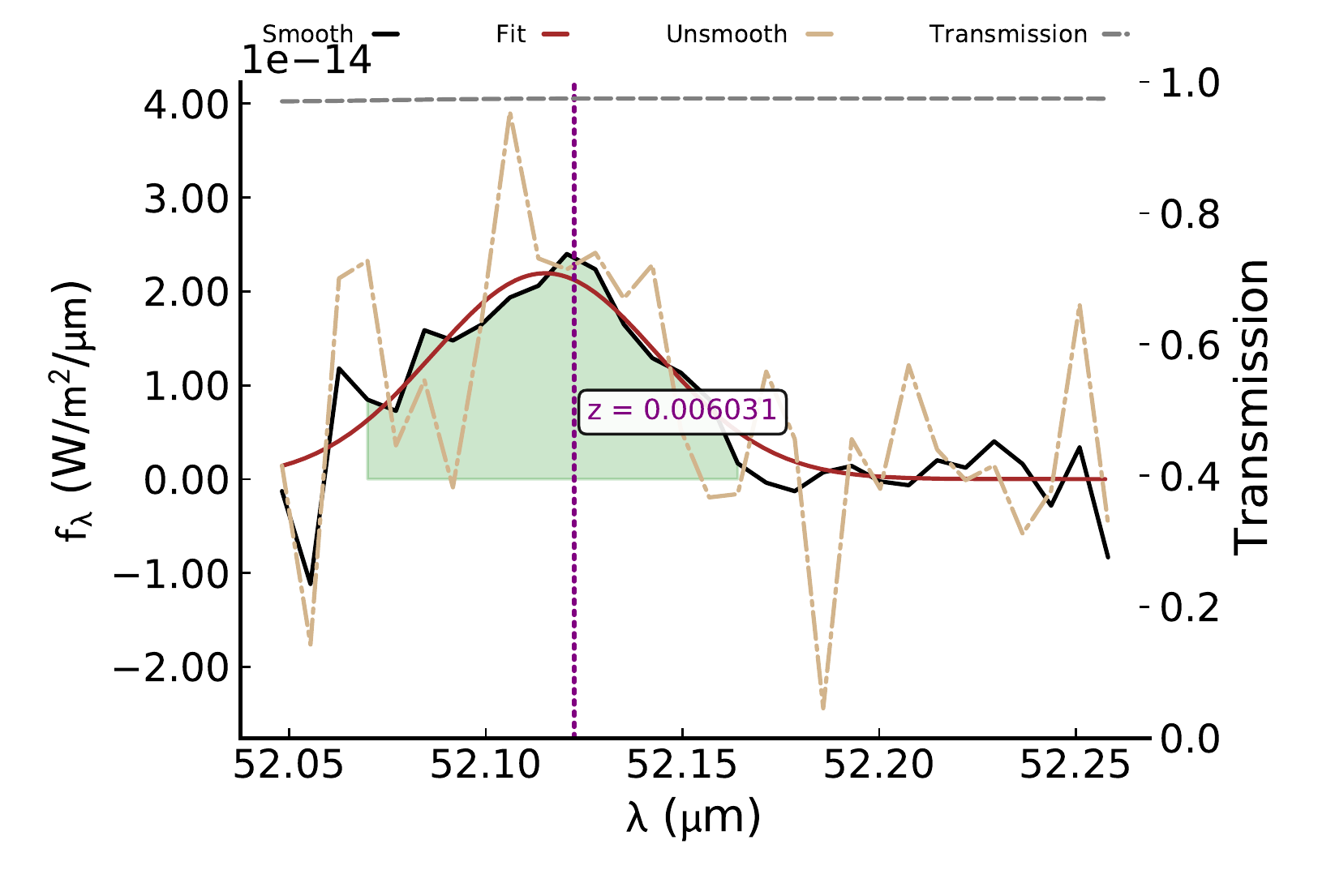}\label{fig:NGC4536-C}}
    \caption{The 2MASS image (Figure \ref{fig:NGC4536-A}), 2-D linemap and 1-D spectrum for [OIII]52$\mu$m  (Figures \ref{fig:NGC4536-B} and \ref{fig:NGC4536-C}) in NGC4536.}
    \label{fig:NGC4536}
\end{figure*}

\begin{figure*}[ht!]
    \centering
    \subfigure[]{\includegraphics[width=0.498\textwidth]{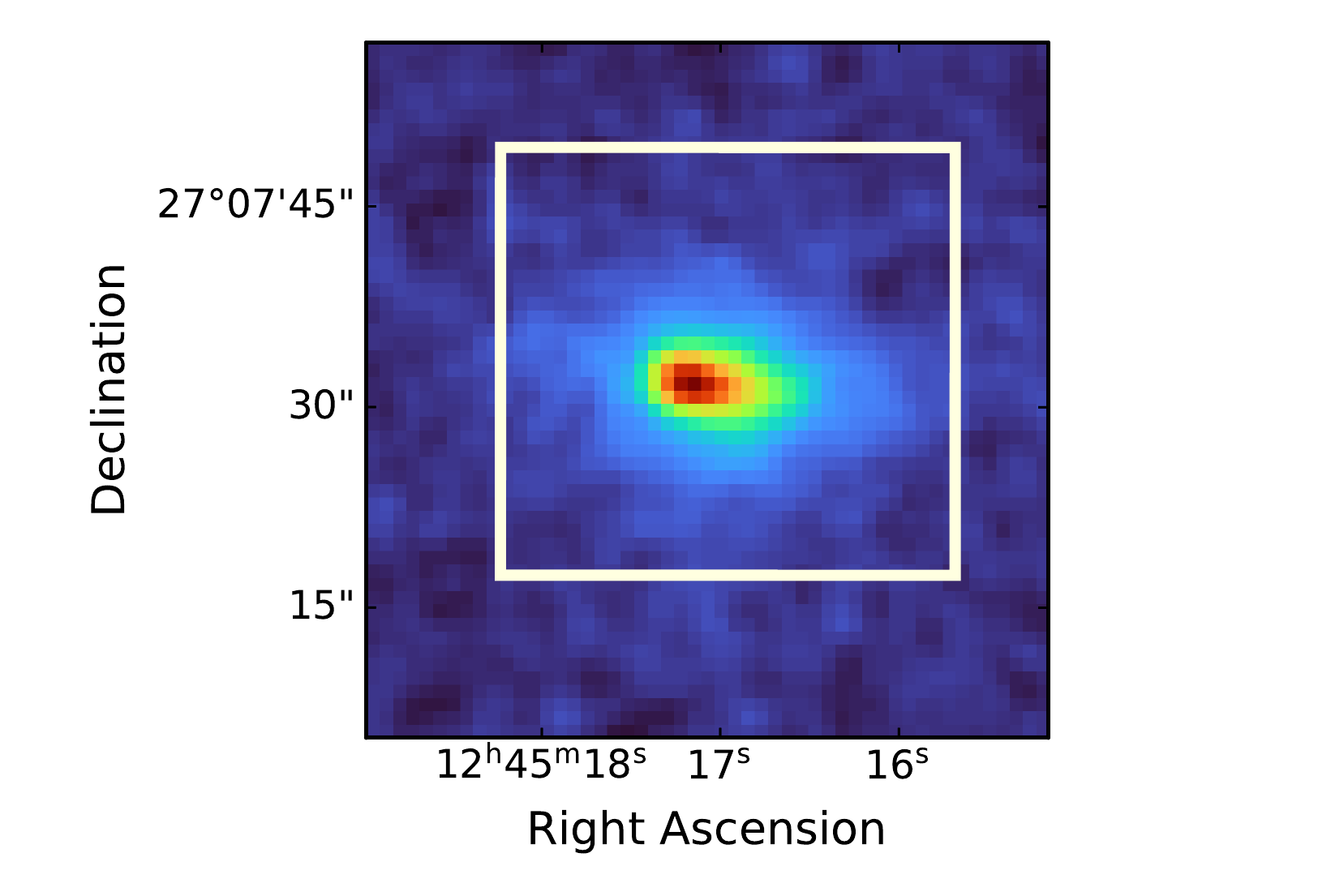}\label{fig:NGC4670-A}}\\
    \subfigure[]{\includegraphics[width=0.498\textwidth]{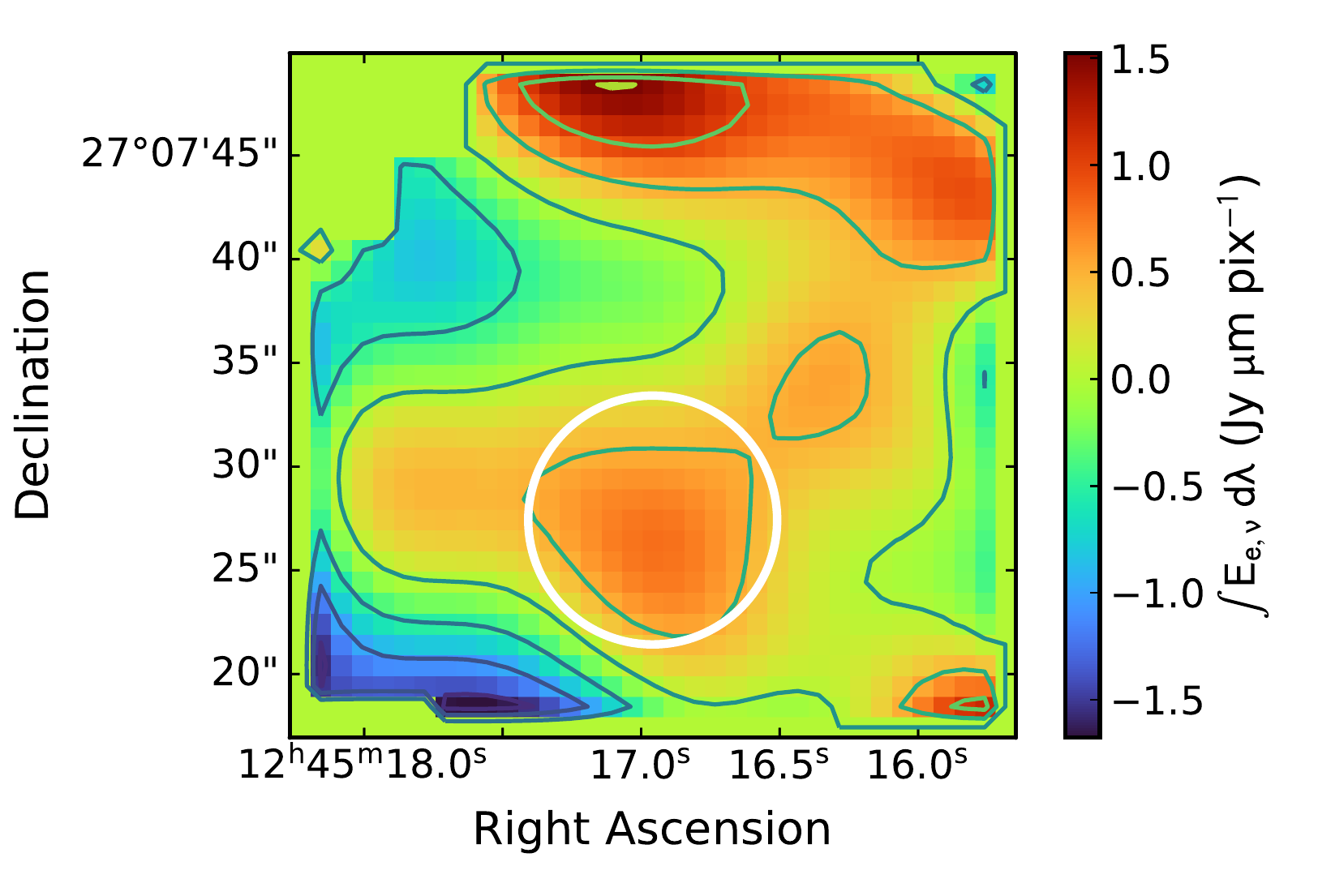}\label{fig:NGC4670-B}}~
    \subfigure[]{\includegraphics[width=0.49\textwidth]{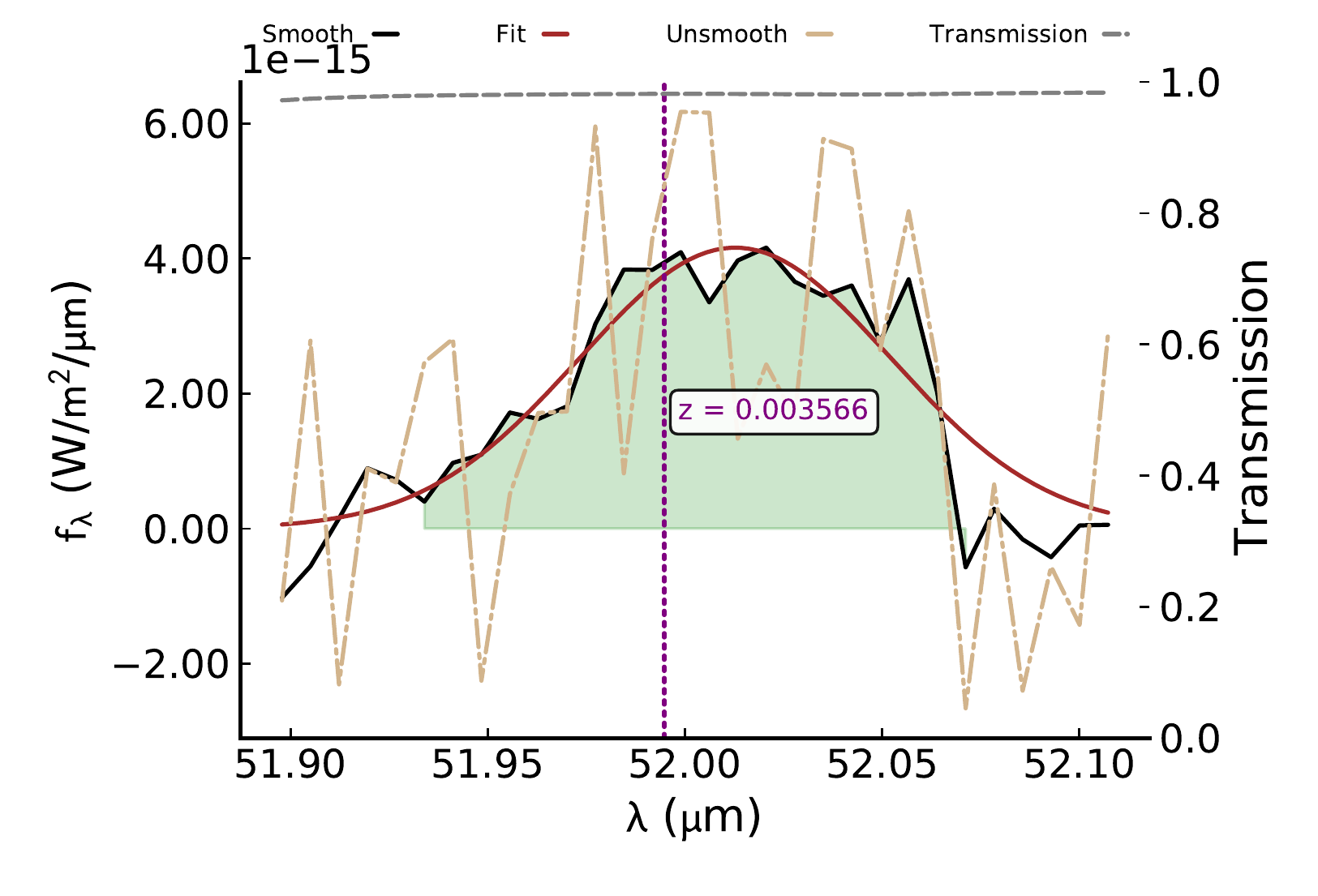}\label{fig:NGC4670-C}}
    \caption{The 2MASS image (Figure \ref{fig:NGC4670-A}), 2-D linemap and 1-D spectrum for [OIII]52$\mu$m  (Figures \ref{fig:NGC4670-B} and \ref{fig:NGC4670-C}) in NGC4670.}
    \label{fig:NGC4670}
\end{figure*}


\clearpage

\newpage

\section{N/O abundances for AGN from IR photoionization models}\label{app_NO}

\begin{figure}[h!!!!]
\centering
\includegraphics[width=0.7\columnwidth]{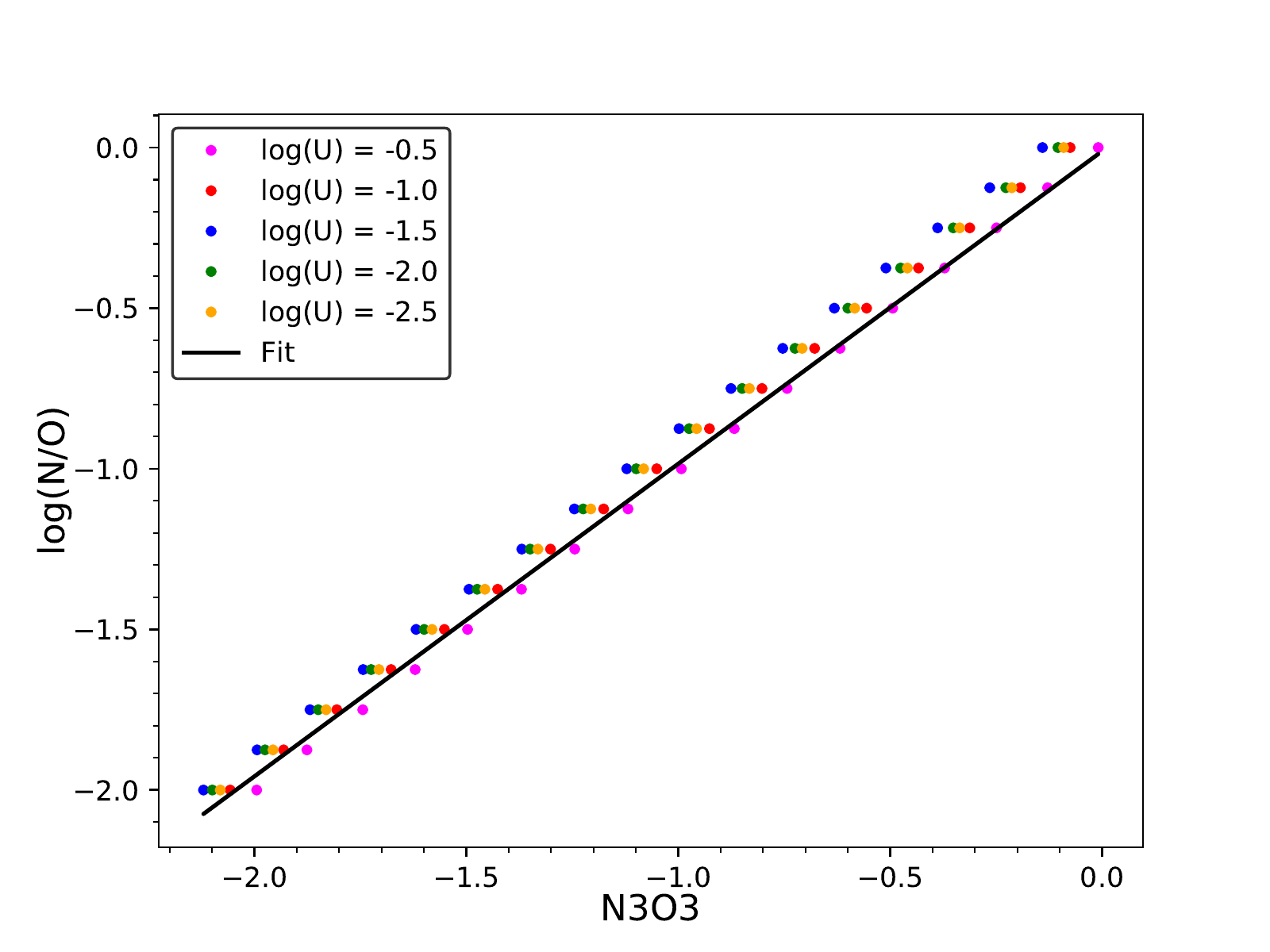}
\caption{Predicted dependency of the inferred N/O abundances on the N3O3 parameter for AGN models with $n_{\rm e} = 500\, \rm{cm^{-3}}$ and $\log U$ in the $-2.5$ to $-0.5$ range. The solid black line shows a linear regression derived for the whole grid of models (see Section\,\ref{abund}).}\label{fig_NO_n3o3}
\end{figure}

Fig.\,\ref{fig_NO_n3o3} shows the tight dependence of the N3O3 parameter, based on the ratio of [NIII]$57\mu$m and [OIII]$52,88\mu$m lines, with the global N/O abundance ratio for the grid of AGN photoionization models using a power law continuum index of $\alpha_{\rm OX} = -0.8$ ($F_\nu \propto \nu^{\alpha_{\rm OX}}$). Due to the similar ionization structure of nitrogen and oxygen elements, variations in $\log \rm{U}$ have a small effect in the N3O3 parameter, allowing us use a linear regression fit (black solid line; see Eq.\,\ref{eq_NO_n3o3}) to determine the relative N/O abundances.

\newpage
\section{$\Delta$(N/O) versus density: IR lines diagnostics}\label{extra_dens}

In Fig. \ref{deltaNO_vs_SIIdens} we have explored the dependence of the $\rm  \Delta(N/O)$ with the electron density of the gas as measured from the optical [SII]{$\lambda$}{$\lambda$}6716,6731 doublet. We have also explored if the 
$\rm  \Delta(N/O)$ has some dependence on the electron density as derived from the IR fine structure lines of [SIII] and [OIII]. We presents the results here.

\begin{figure}[hb!]
\centering
\includegraphics[width=0.45\columnwidth]{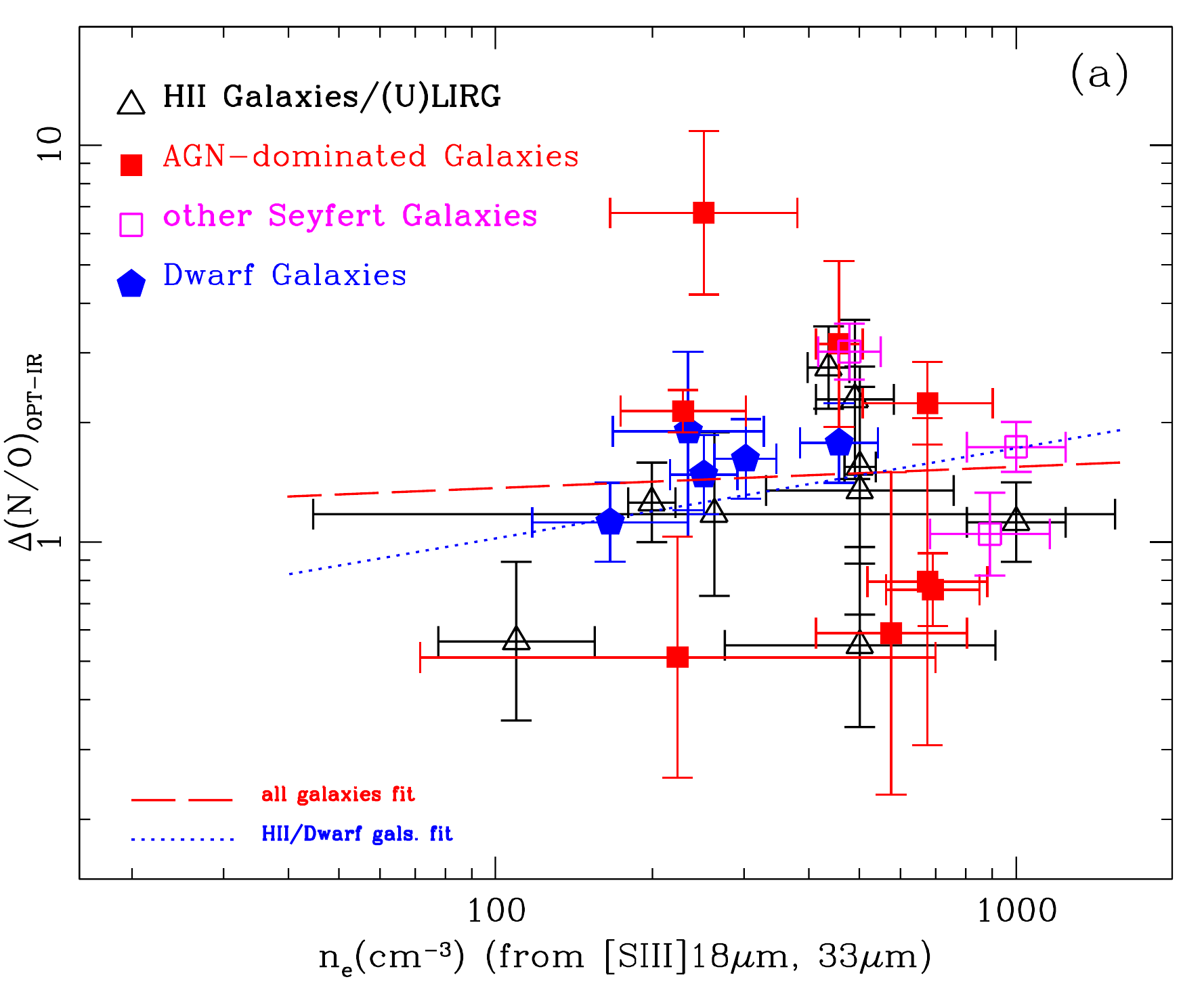}
\includegraphics[width=0.45\columnwidth]{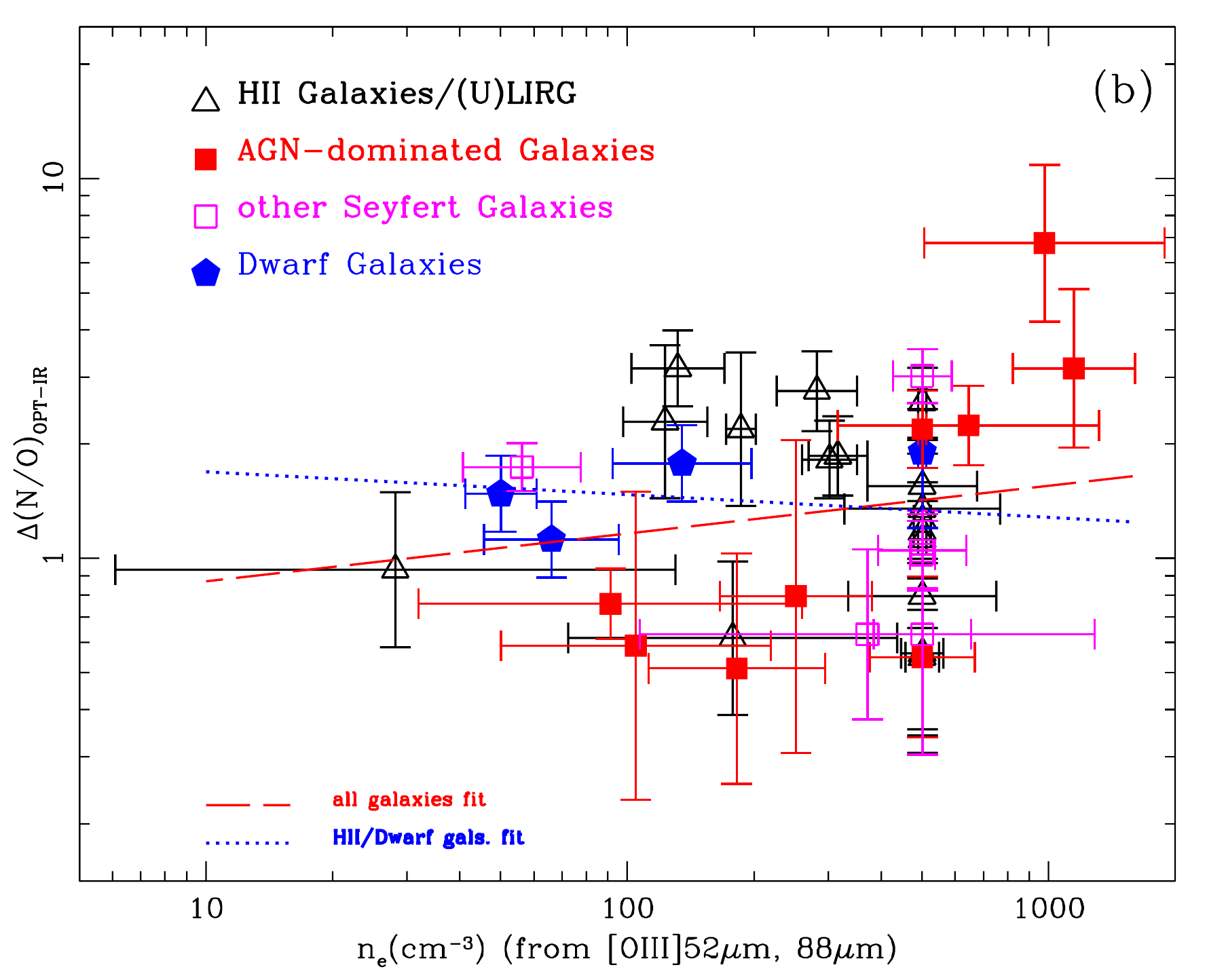}

\caption{{\bf(a: left)} Logarithmic difference between the N/O ratio computed from IR emission lines and the N/O from optical lines versus the electron density of the gas, as measured from the IR [SIII]18.7$\mu$m and 33.5$\mu$m lines.
The dotted line gives the fit for HII/ULIRG and dwarf galaxies: $y = (0.23 \pm 0.22)\cdot x -(0.45 \pm 0.56) (\chi^2=0.47, R=0.29)$. A fit to all the data gives (shown as a dashed line): $y = (0.05 \pm 0.23)\cdot x +(0.03 \pm 0.60) (\chi^2=1.97, R=0.05)$.
{\bf(b: right)} Logarithmic difference between the N/O ratio computed from IR emission lines and the N/O from optical lines versus the electron density of the gas, as measured from the IR [OIII]52$\mu$m and 88$\mu$m lines.
The dotted line gives the fit for HII/ULIRG and dwarf galaxies: $y = -(0.06 \pm 0.13)\cdot x +(0.29 \pm 0.32) (\chi^2=0.98, R=-0.10)$. A fit to all the data gives (shown as a dashed line): $y = (0.12 \pm 0.12)\cdot x -(0.19 \pm 0.30) (\chi^2=2.58, R=0.17)$.
}\label{deltaNO_vs_IIIdens}
\end{figure}

A similar trend to what is seen in Fig. \ref{deltaNO_vs_SIIdens} is also present in Fig. \ref{deltaNO_vs_IIIdens}(a), where the $\rm  \Delta(N/O)$ value is plotted as a function of the electron density as derived from the IR fine structure lines of [SIII]18.7$\mu$m and 33.5$\mu$m. Also in this plot we see a higher $\rm  \Delta(N/O)$ value for higher gas densities, however no correlation is present either for HII region/ULIRG galaxies and dwarf galaxies or for the total sample that includes Seyfert galaxies.

Fig. \ref{deltaNO_vs_IIIdens}(b) shows the $\rm  \Delta(N/O)$ value plotted as a function of the electron density as derived from the IR fine structure lines of [OIII]52$\mu$m and 88$\mu$m. In this figure, if one considers only the HII region/ULIRG and dwarf galaxies and excludes the Seyfert galaxies, a decreasing trend of the  $\rm  \Delta(N/O)$ is found as a function of density. However, because of the large spread of the data and the poor statistics, no statistically significant correlation is found.
We notice, however, that a decreasing trend of the $\rm  \Delta(N/O)$ value with the electron density, related to a higher value of (N/O)$_{OPT}$ at low densities, would indeed be expected by the presence of diffuse ionized gas (DIG). This is hot and low density gas in the disk of galaxies and can be excited by relatively old stars, which have enough time to increase the N abundance through secondary production,  and thus would be also associated to higher (N/O)$_{OPT}$ values, while the IR lines would be less contaminated due to the higher excitation. This scenario would be in agreement with the conclusion by \citet{peng2021}.

\end{document}